\documentclass[prb,aps,nofootinbib,amssymb,twocolumn,superscriptaddress,10pt]{revtex4-2}
\usepackage{project_macros-public}
\usepackage{project_macros-private}

\crefname{appendix}{Appendix}{Appendices}
\crefname{equation}{Eq.}{Eqs.}
\crefname{figure}{Fig.}{Figs.}
\crefname{table}{Table}{Tables}
\crefname{section}{Section}{Sections}
\creflabelformat{appendix}{[#2#1#3]}

\makeatletter
\AtBeginDocument{\let\LS@rot\@undefined}
\makeatother

\makeatletter
\renewcommand\onecolumngrid{
\do@columngrid{one}{\@ne}%
\def\set@footnotewidth{\onecolumngrid}
\def\footnoterule{\kern-6pt\hrule width 1.5in\kern6pt}%
}

\global\long\def\bsl#1{\boldsymbol{{\bf #1}}}%

\global\long\def\ket#1{\left| #1\right\rangle }%

\global\long\def\bra#1{\left\langle #1 \right|}%

\global\long\def\kket#1{\left\Vert #1\right\rangle }%

\global\long\def\bbra#1{\left\langle #1\right\Vert }%

\global\long\def\braket#1#2{\left\langle #1\right. \left| #2 \right\rangle }%

\global\long\def\bbrakket#1#2{\left\langle #1\right. \left\Vert #2\right\rangle }%

\global\long\def\av#1{\left\langle #1 \right\rangle }%

\global\long\def\tr{\text{tr}}%

\global\long\def\Tr{\text{Tr}}%

\global\long\def\pd{\partial}%

\global\long\def\im{\text{Im}}%

\global\long\def\re{\text{Re}}%

\global\long\def\sgn{\text{sgn}}%

\global\long\def\Det{\text{Det}}%

\global\long\def\abs#1{\left|#1\right|}%

\global\long\def\up{\uparrow}%

\global\long\def\down{\downarrow}%

\global\long\def\vc#1{\mathbf{#1}}%

\global\long\def\bs#1{\boldsymbol{#1}}%

\global\long\def\t#1{\text{#1}}%

\allowdisplaybreaks

\begin{document}
\title{\titlePaper}

\author{Miguel Gonçalves}
\thanks{These authors contributed equally.}
\affiliation{Princeton Center for Theoretical Science, Princeton University, Princeton NJ 08544, USA}

\author{Jonah Herzog-Arbeitman}
\thanks{These authors contributed equally.}
\affiliation{Department of Physics, Princeton University, Princeton, New Jersey 08544, USA}

\author{Juan Felipe Mendez-Valderrama}
\affiliation{Department of Physics, Princeton University, Princeton, New Jersey 08544, USA}

\author{Andreas Feuerpfeil}
\affiliation{Institut f\"ur Theoretische Physik und Astrophysik and W\"urzburg-Dresden Cluster of Excellence ctd.qmat, Julius-Maximilians-Universit\"at W\"urzburg, Am Hubland, Campus S\"ud, W\"urzburg 97074, Germany}

\author{B. Andrei Bernevig}
\affiliation{Department of Physics, Princeton University, Princeton, New Jersey 08544, USA}
\affiliation{Donostia International Physics Center, P. Manuel de Lardizabal 4, 20018 Donostia-San Sebastian, Spain}
\affiliation{IKERBASQUE, Basque Foundation for Science, Bilbao, Spain}

\author{Nicolas Regnault}
\affiliation{Center for Computational Quantum Physics, Flatiron Institute, 162 5th Avenue, New York, NY 10010, USA}
\affiliation{Laboratoire de Physique de l'Ecole normale sup\'{e}rieure, ENS, Universit\'{e} PSL, CNRS, Sorbonne Universit\'{e}, Universit\'{e} Paris-Diderot, Sorbonne Paris Cit\'{e}, 75005 Paris, France}


\let\oldaddcontentsline\addcontentsline
\begin{abstract}
Moiré materials host integer and fractional Chern insulators (FCIs) where time-reversal is broken spontaneously by flat band ferromagnetism. In twisted MoTe$_{2}$, unlike in quantum Hall, opposite spins carry opposite Chern numbers and realize Ising ferromagnetism with gapped magnon excitations. Clarifying the nature of these collective modes and the origin of the magnon gap is key to understanding the stability of the Chern magnets in these materials. To do so, we develop a generalized single-mode approximation, which we call the symmetric mode ansatz (SyMA), to describe the low-energy magnon spectrum. We find it to be in exceptional agreement with numerical calculations at integer filling, and qualitatively accurate across fractional fillings including FCIs. To explain the accuracy of the SyMA, we present two families of Hubbard models (including the Haldane model and continuum ideal bands) whose collective modes are analytically solvable and for which the SyMA is either exact or an accurate approximation. For a class of these models where SyMA is exact, we prove two-sided bounds on the zero-momentum magnon gap in terms of the Chern number, Berry curvature, and quantum metric, revealing that uniform quantum geometry is sufficient for stabilizing magnetism in Chern bands.
\end{abstract}

\maketitle

\noindent\begin{minipage}[t]{1\columnwidth}%
\global\long\def\ket#1{\left| #1\right\rangle }%

\global\long\def\bra#1{\left\langle #1 \right|}%

\global\long\def\kket#1{\left\Vert #1\right\rangle }%

\global\long\def\bbra#1{\left\langle #1\right\Vert }%

\global\long\def\braket#1#2{\left\langle #1\right. \left| #2 \right\rangle }%

\global\long\def\bbrakket#1#2{\left\langle #1\right. \left\Vert #2\right\rangle }%

\global\long\def\av#1{\left\langle #1 \right\rangle }%

\global\long\def\tr{\text{tr}}%

\global\long\def\Tr{\text{Tr}}%

\global\long\def\dag{\dagger}%

\global\long\def\pd{\partial}%

\global\long\def\im{\text{Im}}%

\global\long\def\re{\text{Re}}%

\global\long\def\sgn{\text{sgn}}%

\global\long\def\Det{\text{Det}}%

\global\long\def\abs#1{\left|#1\right|}%

\global\long\def\up{\uparrow}%

\global\long\def\down{\downarrow}%

\global\long\def\la{\lambda}%

\global\long\def\al{\alpha}%

\global\long\def\be{\beta}%

\global\long\def\vc#1{\mathbf{#1}}%

\global\long\def\bs#1{\boldsymbol{#1}}%

\global\long\def\t#1{\text{#1}}%

\global\long\def\mbf#1{{\bf \mathbf{#1}}}%

\global\long\def\pmb#1{{\bf \mathbf{#1}}}%

\global\long\def\bsl#1{{\bf \mathbf{\boldsymbol{{\bf #1}}}}}%
\end{minipage}

\section{Introduction}
\label{sec:intro}

Interacting Chern insulators (CI) and fractional Chern insulators (FCI)
can arise from the interplay between band topology and electron interactions
without requiring an external magnetic field$\,$\cite{PhysRevLett.61.2015,doi:10.1126/science.1187485,PhysRevLett.106.236804,Sheng2011,PhysRevX.1.021014,tang2011}.
Moiré materials provide a versatile platform where such phases have been observed, most notably in homobilayer twisted MoTe$_2$, which hosts robust CIs over a broad range of twist angles, $\theta \approx 2.1^\circ$–$3.9^\circ$, at integer fillings $\nu=-1$ and $\nu=-3$$\,$\cite{Cai2023,Zeng2023,Park2023,PhysRevX.13.031037,Ji2024,Redekop2024,park2025observationhightemperaturedissipationlessfractional,li2025universalmagneticphasestwisted,xu2025signaturesunconventionalsuperconductivitynear},
as well as FCIs$\,$\cite{Cai2023,Zeng2023,Park2023,PhysRevX.13.031037,Ji2024,Redekop2024,park2025observationhightemperaturedissipationlessfractional,xu2025signaturesunconventionalsuperconductivitynear}. Both states spontaneously break time-reversal symmetry, leading to a finite spin/valley polarization$\,$\cite{PhysRevResearch.3.L032070,PhysRevLett.131.136501,PhysRevLett.131.136502,PhysRevB.108.245159,PhysRevB.108.085117,PhysRevB.107.L201109,MFCI_0,FTI_Yves,PhysRevLett.132.036501,PhysRevB.109.L121107,shen2024stabilizingfractionalcherninsulators,zaklama2025structurefactortopologicalbound,doi:10.1073/pnas.2316749121,PhysRevLett.133.066601,PhysRevLett.133.186602,PhysRevLett.132.236502,chen2025fractionalcherninsulatorquantum}.
In a conventional SU(2) ferromagnet, spontaneous spin polarization leads to gapless magnons by Goldstone's theorem, famously exemplified by Kallin and Halperin's calculation of the spin stiffness in quantum Hall ferromagnets$\,$\cite{PhysRevB.30.5655}. In contrast, if only spin-U(1) symmetry is preserved (for instance due to spin-orbit coupling), spontaneous spin polarization leads to an Ising ferromagnet whose excitations are \emph{gapped} magnons$\,$\cite{PhysRevLett.108.046806}.
Describing the nature of these gapped excitations is crucial for understanding the stability of magnetism in interacting Chern insulators. Theoretically, the simplest realization of a system with these symmetries is a toy model of time-reversal conjugate Landau levels serving as Chern $C = \pm 1$ bands. The magnon spectrum of this model at integer filling was solved in Ref.$\,$\cite{PhysRevLett.126.137601}. The electron and hole that bind to form the magnon occupy Landau levels
experiencing opposite magnetic fields and the two-body magnon problem separates into relative and center-of-mass guiding-center degrees of freedom. The relative angular momentum of the
electron-hole pair labels distinct magnon energy levels, while the level degeneracy
is determined by the center-of-mass guiding-center problem. The
center-of-mass problem is equivalent to a Landau-level problem for an effective
charge $-2|e|$. Translational invariance makes the energy independent of the center-of-mass guiding-center label, so each relative-motion magnon level is macroscopically degenerate and non-dispersive with twice the guiding-center degeneracy of a
single-particle Landau level. Because the model lacks SU(2) spin-rotation symmetry, no Goldstone theorem forces these energies to vanish. Hence the magnon spectrum consists of flat levels which are generically separated from the ferromagnetic ground state by a finite gap. This contrasts with the gapless and dispersive magnons of the SU(2) symmetric Landau level. Related studies have been conducted in the context of twisted bilayer graphene$\,$\cite{PhysRevLett.124.046403,PhysRevLett.124.166601,khalaf2020softmodesmagicangle,PhysRevLett.126.137601}, in toy models featuring time-reversed pairs of Chern bands supplemented with Hubbard interactions$\,$\cite{PhysRevB.92.245124,Gu2021}, in excitons formed in topological bands$\,$\cite{UchoaPNAS} and in magnetic superconductors where time-reversal symmetry (TRS) is broken$\,$\cite{PhysRevLett.90.057002,PhysRevLett.120.067003}.

Twisted MoTe$_{2}$ is an experimental platform realizing Ising
Chern ferromagnets due to spin and valley locking and thus a $U(1)$ symmetry. Being an Ising ferromagnet, the magnon excitations are generically gapped. These modes
were previously studied in Refs.$\,$\cite{JianXinTopoMagnonsMoTe2,FengchengTopoMagnonsMoTe2,chen2025quantumgeometricdipoletopologicalboost,xie2026collectivespinexcitationscorrelated} numerically, while Ref.$\,$\cite{chen2025quantumgeometricdipoletopologicalboost}
proposed the interpretation of the zero-momentum magnon gap in terms of
a quantum geometric dipole and suggested non-trivial electron topology
as a means to stabilize it.

Despite these advances, a general and rigorous understanding of the low-energy magnon spectrum - and most importantly what band geometric properties control its gap - remains elusive. In this work, we provide a significant step forward in answering these questions by proposing a generalization of the single-mode approximation (SMA) to construct magnon wave functions based on symmetry and locality -- which we call the Symmetric Mode Ansatz (SyMA). We prove that this ansatz is exact or an accurate approximation in two families of flat band Hubbard models, and then apply our theory at $\nu=-1$ to accurate models of MoTe$_{2}$ obtained from ab-initio for different twist-angles ranging from $\theta=2.13^\circ-3.89^\circ$ \cite{zhang2024universalmoiremodelbuildingmethodfitting,MFCI_0}, finding excellent agreement between the energies of the SyMA and the exact numerical results.

Within the flat band Hubbard models, we obtain quantum geometric upper and lower bounds for the magnon gap at zero-momentum:
\bea
 \label{eq:magnonbound}
\frac{\pi}{3} \frac{U \, |C|}{ \max_\mbf{k}|F(\mbf{k})|\Omega_{BZ}}\leq \Delta({\bf p}=0) \leq 4\pi \frac{ U G}{\Omega_{BZ} |\mbf{a}|^2} \,,
\eea
where $U$ is the Hubbard interaction, $F(\mbf{k})$ is the Berry curvature, $C$ is the Chern number, $G \geq |C|$ is the integrated Fubini-Study metric, $\Omega_{BZ}$ is the area of the Brillouin zone (BZ), and $\mbf{a}$ is the shortest lattice vector. The bounds in Eq.$\,$\ref{eq:magnonbound} quantify the relation between the magnon gap at ${\bf p}=0$,  the distribution of the Berry curvature and the integrated quantum metric. In particular, the lower bound in Eq.$\,$\ref{eq:magnonbound} implies that a uniform Berry curvature is sufficient to stabilize a zero-momentum magnon gap of order $U$.
We further apply our theory to capture magnon excitations at fractional fillings in MoTe$_2$. We observe that although our theory overestimates the magnon gap size, it reproduces the trends of the magnon gap and magnon dispersion over a range of densities including FCI fillings. These results show that our construction, derived rigorously at integer filling, can also provide faithful ansatz wavefunctions for the magnetic excitations over an extended doping regime of partial filling of the parent Chern band. Overall, our results provide a unified picture of magnetism in time-reversed Chern bands.

The remainder of the manuscript is organized as follows. In Sec.$\,$\ref{sec:moire_magnons} we provide numerical results for the magnon spectrum in a simple model of twisted MoTe$_2$, while in Sec.$\,$\ref{sec:SyMA} we describe our SyMA theory and illustrate its application in this model, describing how it can be used to capture the low-energy magnon modes and understand their energetic hierarchy. In Sec.$\,$\ref{sec:Solvable_Models}, we study solvable models of Ising ferromagnets, for which several analytical results can be derived for the magnon bound modes: (1) a family of flat band Hubbard lattice models on uniform lattices, and (2) a continuum ideal-band Bernevig-Hughes-Zhang (BHZ) model \cite{bernevig2013topological}. These models allow us to analytically understand the magnon excitations in different cases: (i) the case of a flat particle-hole continuum (for flat band Hubbard models), where the SyMA is exact and the quantum-geometric bounds can be proved; (ii) the case of a dispersive continuum (for ideal-band BHZ continuum models), where the SyMA is still an accurate approximation, but not exact, allowing us to obtain a simple approximate expression for the zero-momentum magnon gap in terms of the Chern number and Berry curvature of the electronic band.
Finally, in Sec.$\,$\ref{sec:connection_experiment}, we explore in detail different experimentally relevant regimes in twisted MoTe$_2$, focusing on using the SyMA theory to acquire a detailed understanding and make predictions. We first point out key differences between the magnon spectrum in the topological and trivial ferromagnetic phases at $\theta=3.7^{\circ}$. We then explore more realistic models of twisted MoTe$_2$, valid over a broad range of angles from $\theta=2.13^{\circ}$ to $\theta=3.89^{\circ}$, and study the evolution of the magnon spectrum with twist angle. Finally, we extend the application of our theory to the FCI at fractional filling $\nu=-2/3$, at $\theta=3.7^{\circ}$, showing that it provides a simple description of its low-energy magnon excitations.\\

\section{Moiré magnons}

\label{sec:moire_magnons}

We start by providing numerical results for the magnon spectrum in the first-harmonic (FH) model of twisted MoTe$_2$ - a simple continuum model of AA-stacked
twisted MoTe$_2$$\,$\cite{PhysRevLett.122.086402, MFCI_I}. Because the FH model we study in this section was obtained
by fitting the DFT band structure at $\theta=3.89^{\circ}$, it is expected to accurately describe twisted MoTe$_2$ only close to this angle. However, in this
section, we use the FH model outside its fitted range, over $1^\circ<\theta<5^\circ$, only as a controlled toy model to illustrate our theory, but not as a quantitatively accurate description of twisted MoTe$_2$. We leave the exploration of more sophisticated models that remain
accurate with respect to ab-initio calculations also in the small angle regime ($\theta\sim 2^\circ$) \cite{zhang2024universalmoiremodelbuildingmethodfitting}, to Sec.$\,$\ref{subsec:Realistic_MoTe2}.

\subsection{First-Harmonic model}

\label{subsec:FH_model}

Following the convention of Ref.$\,$\cite{MFCI_I}, the single-particle Hamiltonian for the FH model
of twisted MoTe$_2$ can be written as

\begin{widetext}

\begin{equation}
H_{\eta}^{0}=-\int d^{2}r\,\left(c_{\eta,-1,{\bf r}}^{\dagger}\,c_{\eta,1,{\bf r}}^{\dagger}\right)\left(\begin{array}{cc}
\frac{\hbar^{2}\nabla^{2}}{2m^{*}}+V_{\eta,-1}({\bf r})+\frac{V_{z}}{2} & t_{\eta}({\bf r})\\
t_{\eta}^{*}({\bf r}) & \frac{\hbar^{2}\nabla^{2}}{2m^{*}}+V_{\eta,1}({\bf r})-\frac{V_{z}}{2}
\end{array}\right)\left(\begin{array}{c}
c_{\eta,-1,{\bf r}}\\
c_{\eta,1,{\bf r}}
\end{array}\right)\,,\label{eq:c_eta_l_r_FH}
\end{equation}
\end{widetext}where the coarse-grained continuum operator $c_{\eta,l,{\bf r}}^{\dagger}$
creates a hole at position ${\bf r}$, layer $l=\pm1$ ($l=1$ for
top and $l=-1$ for bottom layer) and valley $\eta=\pm1$ (respectively
the single-layer $\pm{\bf K}_{SL}$ valleys) corresponding also to spin $\up(+)/\down(-)$ due to spin-valley locking, $V_{z}$ denotes the
energy difference between layers induced by a displacement field and we use the lowest-order Fourier expansion in terms of the moiré scattering momentum for the intralayer and interlayer couplings, respectively $V_{\eta,l}({\bf r})$ and $t_{\eta}({\bf r})$ (see App. \ref{sec:detalis_MoTe2_FH} for a detailed discussion).
We choose the model parameters extracted for the first-harmonic model
in Ref.$\,$\cite{MFCI_I} from fitting the DFT band structure at
$\theta=3.89^{\circ}$ (see Refs.$\,$\cite{PhysRevB.108.085117,PhysRevLett.132.036501} for alternative parameter choices). The full Hamiltonian respects time-reversal symmetry $\mathcal{T}$ which relates $H^0_{\up}$ and $H^0_{\down}$.

\begin{figure}
\begin{centering}
\includegraphics[width=\columnwidth]{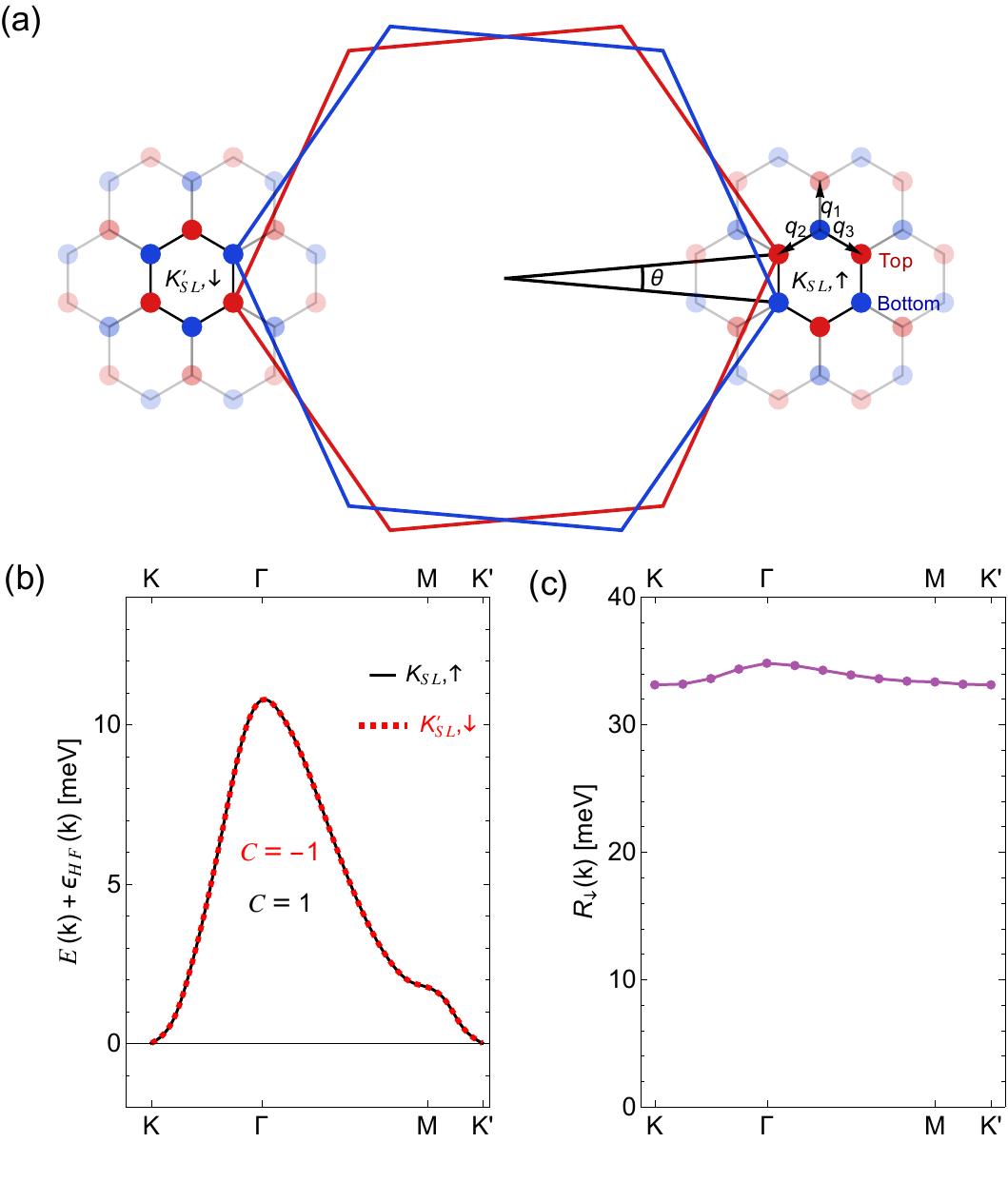}
\par\end{centering}
\caption{Twisted Brillouin zones of the TMD bilayer, with opposite spin-polarized low-energy states at the single-layer Dirac points $K_{SL}$ and $K'_{SL}$ related by time-reversal symmetry $\mathcal{T}$. A relative twist creates a moiré Brillouin zone shown in the figure.
Blue/red points represent moiré reciprocal lattice points in the
bottom/top layers. (b) Single-particle dispersion in Eq.$\,$\ref{eq:H_full_band_basis} for the lowest-energy hole band at $K_{SL}$ and $K'_{SL}$, for the FH model of MoTe$_2$ at $\theta=3.7^{\circ}$. The Chern number for this band in each valley is also indicated. Note that the dispersion of $\epsilon_{HF,\eta}({\bf k})$ is only 5$\%$ of the full dispersion. (c) Charge excitation energies computed using Eq.$\,$\ref{eq:Rmat} for the FH model of MoTe$_2$ at $\theta=3.7^{\circ}$.
\label{fig:1}
}
\end{figure}

We now introduce the plane-wave basis $c_{{\bf k},{\bf Q},\eta}^{\dagger}$
as
\begin{equation}
c_{\eta,l,{\bf r}}^{\dagger}=\frac{1}{\sqrt{\Omega_{tot}}}\sum_{{\bf Q}\in\mathcal{Q}_{l}^{\eta}}\sum_{{\bf k}\in{\rm MBZ}}e^{-i({\bf k}-{\bf Q})\cdot{\bf r}}c_{{\bf k},{\bf Q},\eta}^{\dagger}\,,
\label{eq:c_kQ_def}
\end{equation}
where ${\bf k}$ is defined in the moiré Brillouin zone (MBZ), $\Omega_{\textrm{tot}}$ is the total area of the sample and $\mathcal{Q}_{l}^{\eta}=\{{\bf G}+\eta(-1)^{(l+1)/2}{\bf q}_{1}\}$, with ${\bf G}$ the moiré reciprocal lattice vectors (see Fig.$\,$\ref{fig:1}(a) for illustration of the momentum-space lattice vectors ${\bf Q}$ and definition of $ {\bf q}_1 $).

Throughout this work, we consider the projected Hamiltonian in the lowest-energy hole band (equivalently, the topmost valence electron band). Working in the hole basis, this corresponds to the lowest energy hole band with dispersion $E_{\eta}({\bf k})$. We thus introduce the hole creation operators for this band as follows:
\begin{equation}
\gamma_{{\bf k},\eta}^{\dagger}=\sum_{{\bf Q}}[U_{\eta,{\bf k}}]_{{\bf Q}}c_{\eta,{\bf k},{\bf Q}}^{\dagger}\,,
\label{eq:gamma_def}
\end{equation}
where $[U_{\eta,{\bf k}}]_{{\bf Q}}$ is the topmost valence band eigenvector in the plane-wave basis and the sum over ${\bf Q}$ runs over $\mathcal{Q}=\{{\bf G}+{\bf q}_{1}\}\cup\{{\bf G}-{\bf q}_{1}\}$, unless otherwise specified. We choose a periodic embedding for $U_{\eta,{\bf k}}$ so that $\gamma_{{\bf k},\eta}^{\dagger} = \gamma_{{\bf k}+\mbf{G},\eta}^{\dagger}$. The form factors $M_{\eta}({\bf k},{\bf q}+{\bf G})$ computed from these eigenvectors are given by
\begin{equation}
M_{\eta}({\bf k},{\bf q}+{\bf G})=\sum_{{\bf Q}}[U_{\eta,{\bf k}+{\bf q}}^{*}]_{{\bf Q}-{\bf G}}[U_{\eta,{\bf k}}]_{{\bf Q}}\,.\label{eq:form_factor_1band}
\end{equation}
With these definitions, we now project the electron-electron interactions into the topmost valence bands for each valley. We will study the excitations around filling $\nu=-1$ (i.e. half-filling of the $C=\pm1$ bands). It is therefore convenient to write the conventional projected interaction $:H_I:$, normal-ordered with respect to the hole vacuum $\ket{0}$ defined by $\gamma_{{\bf k},\eta}\ket{0}=0$, in terms of the positive-semi-definite Hamiltonian $H_I$ normal ordered with respect to $\nu=-1$:

\bea
\label{eq:1bdymain}
:H_{I}: \,\,= H_{I} + \sum_{\mbf{k} \eta} \epsilon_{HF,\eta}({\bf k}) \gamma^\dag_{\mbf{k},\eta}\gamma_{\mbf{k},\eta} \ ,
\eea
where $\epsilon_{HF,\eta}({\bf k})$ is a Hartree-Fock-like term arising from the different normal orderings (see  Eq.$\,$\ref{eq:e_HF} in App. \ref{sec:detalis_MoTe2_FH} for explicit expression) and
\bea
H_{I}&=\frac{1}{2\Omega_{\textrm{tot}}}\sum_{{\bf G}}\sum_{{\bf q}\in\textrm{MBZ}} V(\bf{q}+\bf{G}) \delta \rho_{\bf{q}+\bf{G}} \delta \rho_{-\bf{q}-\bf{G}} \ , \\
\delta \rho_{\bf{q}+\bf{G}}&=\sum_{{\bf k},\eta} M^{(\eta)}({\bf k},{\bf q}+{\bf G})\big(\gamma_{{\bf k}+{\bf q},\eta}^{\dagger}\gamma_{{\bf k},\eta}-\frac{1}{2}\delta_{{\bf q},{\bf 0}}\big).
\label{eq:H_V_general}
\eea
Here, $H_I$ includes the Fourier-transformed interaction, $V(\bf{q})\geq0$. For all numerical results for MoTe$_2$, we take $V(\bf{q})$ as the double-gated Coulomb interaction with a dielectric constant of $\epsilon=10/0.9$ and a screening length of $20\textrm{nm}$, unless otherwise specified (see Eq.$\,$\ref{eq:dual_gate} in  App. \ref{sec:detalis_MoTe2_FH} for detailed expression). Overall, the full projected Hamiltonian normal-ordered with respect with the hole vacuum $\ket{0}$ can be written as
\bea
H&=H_{I}+\sum_{{\bf k},\eta}[E_{\eta}({\bf k})+\epsilon_{HF,\eta}({\bf k})]\gamma_{{\bf k},\eta}^{\dagger}\gamma_{{\bf k},\eta}. \label{eq:H_full_band_basis}
\eea
The single-particle dispersion in Eq.$\,$\ref{eq:H_full_band_basis} is shown in Fig.$\,$\ref{fig:1}(b) for $\theta=3.7^{\circ}$.

The key to our construction is that $H_{I}$ is a positive-semi-definite interaction which has the exact ferromagnetic ground-state as a zero-energy eigenstate \cite{PhysRevLett.125.257602,TBGIII,TBGIV,TBGV}
\bea
\ket{\eta}&=\prod_{{\bf k}}\gamma_{\mbf{k},\eta}^{\dagger}\ket{0} \,,
\eea
because $\delta \rho_{\mbf{q}+\mbf{G}} \ket{\eta} = 0$, i.e. $H_{I}$ is frustration-free. If $H$ possessed spin SU(2) symmetry, then $\ket{\eta}$ would be the maximally polarized state in a spin $N/2$ irrep. In our case, $H$ only retains a spin U(1) symmetry generated by $S_z$ and time-reversal $\mathcal{T}$. Hence, the ground-state multiplet is $\ket{\uparrow},\ket{\downarrow}$, and each state spontaneously breaks $\mathcal{T}$, giving it a non-zero Chern number. Note that our results thus far have not depended on the precise form of $V(\mbf{q}) \geq 0$.

In fact, since $\ket{\eta}$ is the Slater determinant of a filled band, it is also an eigenstate of Eq.$\,$\ref{eq:H_full_band_basis}, but in this case it is not guaranteed to be the ground-state. Nonetheless, our numerical calculations in App.$\,$\ref{subsec:App_Additional_MoTe2_Exact} show that the lowest-energy magnon modes are weakly affected by the single-particle term in Eq.$\,$\ref{eq:H_full_band_basis} and that $\ket{\eta}$ remains the ground state even when a realistic dispersion is included in the one-band projected limit (see Fig.$\,$\ref{fig:App_MoTe2_polarization_checks} of Appendix$\,$\ref{subsec:App_Additional_MoTe2_Exact}). As an example, we show in Fig.$\,$\ref{fig:App_MoTe2_variableDispersion} of Appendix$\,$\ref{subsec:App_Additional_MoTe2_Exact} that the Ising gap only closes when the single-particle bandwidth is artificially increased to five times its value at $\theta=3.7^\circ$. Because of this, in what follows we will neglect the single-particle term in Eq.$\,$\ref{eq:H_full_band_basis}. The calculations for the full Hamiltonian in Eq.$\,$\ref{eq:H_full_band_basis} can be found in App.$\,$\ref{subsec:App_Additional_MoTe2_Exact}, where it is explicitly shown that the low-energy magnon modes are only weakly affected by the single-particle terms.

\subsection{Exact magnon excitations}
\label{sec:exact_mag_excitations_Smatrix}

\begin{figure*}
\begin{centering}
\includegraphics[width=\textwidth]{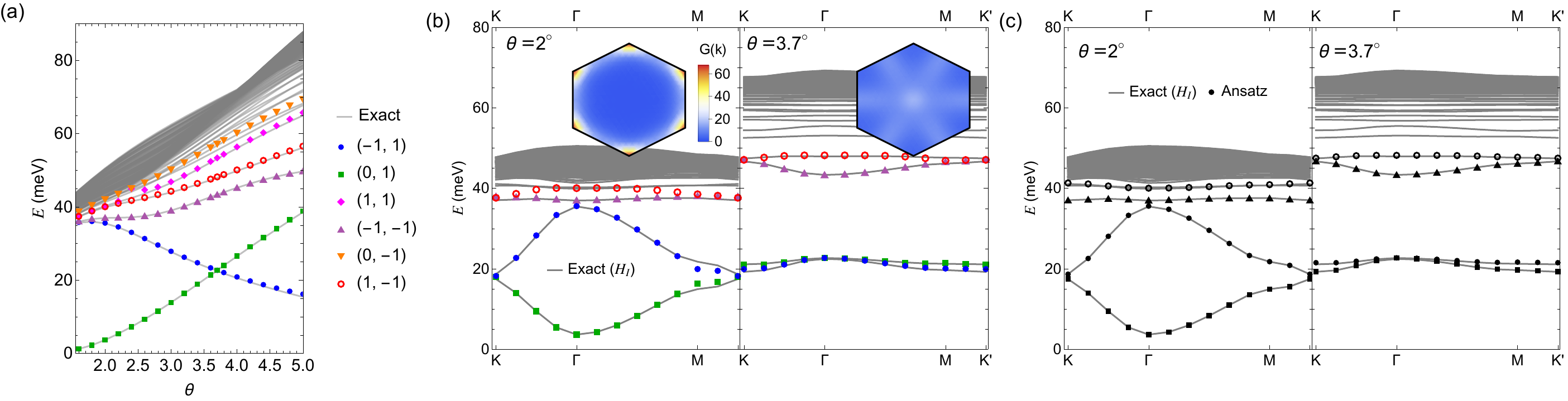}
\par\end{centering}
\caption{
Exact magnon spectrum (gray) vs. SyMA (other colors) results for the FH model of MoTe$_2$, using the flat band Hamiltonian $H_{I}$ in Eq.$\,$\ref{eq:H_V_general} where $V(\bf{q})$ corresponds to the double-gated Coulomb interaction.
(a) Comparison between exact (gray)
magnon spectrum at ${\bf p}=0$ with the SyMA (other colors) results of the individual
modes within the different symmetry sectors $(\kappa,\zeta)$. (b) Comparison between exact and SyMA magnon spectrum of the individual modes computed through Eq.$\,$\ref{eq:Ising_gap_1BPV} for variable $\bf{p}$, for the angles indicated in the figure. We exclude the modes $(\kappa,\zeta)=(1,1);(0,-1)$, that are in the continuum (plots for all the modes and different angles are provided in Fig.$\,$\ref{fig:App_MoTe2_SMA_energies} of
Appendix$\,$\ref{subsec:App_Additional_MoTe2_SMA}). The insets show the Berry curvature.
(c)
Comparison between the exact magnon spectrum and the SyMA magnon spectrum obtained by diagonalizing $H_I$ in the six-dimensional subspace spanned by all the individual SyMA modes $(\kappa,\zeta)$, with $\kappa=-1,0,1$ and $\zeta=-1,1$, as described at the end of Sec.$\,$\ref{subsec:syma_energies}. We show only the four lowest-energy magnon modes obtained from this subspace diagonalization, which are separated by a finite gap from the continuum. Nearly perfect agreement with the corresponding exact magnon modes is obtained. Note that, although we use the same symbols, the magnon modes obtained by the subspace diagonalization are different from the individual SyMA modes shown in (b) away from $\Gamma$.
\label{fig:2}
}
\end{figure*}

In addition to the ground-state, it is also possible to solve $H_I$ in Eq.$\,$\ref{eq:H_V_general} for exact excitations. The fermionic charge excitations $\gamma^\dag_{\mbf{k},\eta}\ket{-\eta}$ and $\gamma_{\mbf{k},\eta}\ket{\eta}$ are exact eigenstates of $H_I$ and have energies$\,$\cite{TBGV}
\begin{equation}
R^{\eta}({\bf k})
=
\frac{1}{2\Omega_{\textrm{tot}}}
\sum_{{\bf q},{\bf G}}
V({\bf q}+{\bf G})
|M^{(\eta)}({\bf k},{\bf q}+{\bf G})|^{2} \,,\label{eq:Rmat}
\end{equation}
which are non-negative for $V( {\bf q}) \geq 0$. For the interactions considered here they are strictly positive, since the form factors do not vanish for all ${\bf q},{\bf G}$ at fixed ${\bf k}$. Hence the ferromagnetic state $\ket{\eta}$ has a charge gap and is an insulator. An example calculation of the charge-1 energies is given in Fig.$\,$\ref{fig:1}(c) for $\theta=3.7^{\circ}$. Our focus will be on the magnon excitations. They can be obtained exactly by using the fact that the action of $H_I$ closes on states that contain a particle-hole excitation on top of the ferromagnetic ground-state $|\eta\rangle $$\,$\cite{TBGV}, so the exact magnon excitations obey
\bea
[H_{I},\gamma_{{\bf k}+{\bf p},-\eta}^{\dagger}\gamma_{{\bf k},\eta}]\ket{\eta} &= \sum_{\mbf{k}'} \gamma_{{\bf k'}+{\bf p},-\eta}^{\dagger}\gamma_{{\bf k'},\eta}\ket{\eta} [S^{(\eta)}(\mbf{p})]_{\mbf{k}',\mbf{k}}\,,
\label{eq:S_matrix_commutator}
\eea
where the effective magnon Hamiltonian is
\bea
[&S^{(\eta)}({\bf p})]_{{\bf k}+{\bf q},{\bf k}}
=
\delta_{{\bf q},{\bf 0}}
[R^{(\eta)}({\bf k})+R^{(-\eta)}({\bf k}+{\bf p})]
\\ &- 
\frac{1}{\Omega_{\textrm{tot}}}
\sum_{\bf G}
V({\bf q}+{\bf G})
M^\eta({\bf k},{\bf q}+{\bf G})^*
M^{-\eta}({\bf k}+{\bf p},{\bf q}+{\bf G}).
\label{eq:S_matrix_expression}
\eea
Here, Eq.$\,$\ref{eq:S_matrix_expression} consists of the charge excitation energy for an independent electron and hole in opposite valleys in the first line, and their scattering matrix in the second line. The exact results for the magnon excitation spectrum at $\theta=2^{\circ}$
and $\theta=3.7{}^{\circ}$ obtained by diagonalizing Eq.$\,$\ref{eq:S_matrix_expression}, using the flat band Hamiltonian $H_I$, are shown in Fig.$\,$\ref{fig:2}. Fig.$\,$\ref{fig:2}(a) shows the exact results at ${\bf p}=0$ in gray, while Fig.$\,$\ref{fig:2}(b) shows the exact results for $\theta=2^{\circ}$ and $\theta=3.7^{\circ}$ as function of ${\bf p}$. The colored symbols in Fig.$\,$\ref{fig:2}(a) and (b) compare the exact results with the results of the SyMA theory, that will be later introduced in Sec.$\,$\ref{sec:SyMA}. The calculation can also be performed for the full Hamiltonian in Eq.$\,$\ref{eq:H_full_band_basis}. At ${\bf p}=0$, we can prove using symmetry that the corrections to Eq.$\,$\ref{eq:S_matrix_expression} due to additional single-particle terms in Eq.$\,$\ref{eq:H_full_band_basis} vanish exactly for the FH model with $V_z=0$ (see App. \ref{subsec:1body_contribution}). At ${\bf p}\neq 0$ or $V_z \neq 0$ we observe numerically that the lowest-energy magnon modes are only weakly dependent on the additional single-particle terms in Eq.$\,$\ref{eq:H_full_band_basis} (see Fig.$\,$\ref{fig:App_MoTe2_H_vs_HI} in App. \ref{subsec:App_Additional_MoTe2_Exact}).

In Fig.$\,$\ref{fig:2}(b), we can identify four low-lying bound magnon modes for both $\theta=2^{\circ}$ and $\theta=3.7^{\circ}$, with the two lowest modes clearly separated from the continuum. The upper two of these four modes remain nearly flat at both angles, whereas the two lowest modes evolve from strongly dispersive at $\theta=2^{\circ}$ to nearly flat at $\theta=3.7^{\circ}$, where their bandwidth becomes approximately $5\,\textrm{meV}$. The remaining results in Figs.\ref{fig:2}(a) and (b), as well as in Fig.\ref{fig:2}(c), will be later discussed in Sec.$\,$\ref{subsec:syma_energies}. Understanding the spectrum of low-lying magnon modes  requires an analytical theory which we develop in the next section.

\section{Symmetric Mode Ansatz}

\label{sec:SyMA}

In this section, we describe our SyMA theory and illustrate its application to the FH model of twisted MoTe$_2$.

We begin in Sec.$\,$\ref{sec:sma_main} with a brief overview of the SMA, and of the complementary ingredients in our SyMA theory.
We then construct the SyMA for the FH model of MoTe$_2$ in Sec.$\,$\ref{sec:syma_fhmodel}. In Sec.$\,$\ref{subsec:syma_energies}, we derive an expression for the SyMA energies and compare the results with exact calculations in the FH model of MoTe$_2$. We next discuss some properties of the SyMA eigenfunctions in Sec.$\,$\ref{subsec:MoTe2_SMA_eigenfunctions}. In particular, based on these properties, we will propose a simple heuristic principle to understand the energy hierarchy of the magnon modes. Finally, in Sec.$\,$\ref{subsec:syma_constraints_node}, we will use symmetry arguments to verify the applicability of the node principle in the FH model of MoTe$_2$.

\subsection{From SMA to SyMA}
\label{sec:sma_main}

Feynman first introduced the single-mode-approximation to study the neutral excitations (sound modes) of bosonic fluids, proposing the variational ansatz $\rho_\mbf{p} \ket{GS}$, where $\rho_\mbf{p}$ is the Fourier transform of the local density operator \cite{PhysRev.94.262}. For all $\mbf{p} \neq 0$, Feynman's ansatz is orthogonal to the ground-state because it has different momentum, but reproduces the ground-state at $\mbf{p} \to 0$. Indeed, $\rho_\mbf{p} \ket{GS}$ is a good variational ansatz for the gapless neutral excitation and becomes exact as $\mbf{p} \to 0$. Later, Girvin, Platzman, and MacDonald introduced the SMA to the (fractional) quantum Hall context by projecting the density mode into the lowest Landau level \cite{PhysRevB.33.2481},
\bea
\rho_\mbf{p} = \int d^2r \, e^{i \mbf{p} \cdot \mbf{r}} \bar{c}^\dag(\mbf{r}) \bar{c}(\mbf{r}) \,,
\eea
where the projected operators obey $\{\bar{c}(\mbf{r}),\bar{c}^\dag(\mbf{r}')\} = \langle c^\dag(\mbf{r}')c(\mbf{r}) \rangle$, the correlator in the fully filled lowest Landau level. Girvin, Platzman, and MacDonald showed that this ansatz remains gapped as $\mbf{p} \to 0$ due to the non-commutativity of translations
in this Hilbert space, providing an understanding of incompressibility of the liquid. The gap predicted by the ansatz achieves a good agreement with ED calculations in the lowest Landau level, with a relative error of $20\%$ or below for $\nu=1/3$ with Coulomb interaction \cite{PhysRevB.95.075201} and for the bosonic Laughlin fractional quantum Hall state at $\nu=1/2$ with repulsive delta interaction (both using the torus geometry). 

In this work, we generalize the SMA to magnons in topological bands. A key result, illustrated by our numerics above, is that spontaneous time-reversal symmetry breaking in $\mathcal{T}$-related Chern bands can lead to \emph{multiple} low-energy modes, well separated from the particle-hole continuum (e.g. four modes shown in Fig.$\,$\ref{fig:2}(b)). Our SMA construction will allow for a simple understanding of their nature and energetic hierarchy.
Motivated by the conventional SMA, we will build an ansatz of magnon states by projecting local spin-flip bilinears into the active Chern band. As we will see in detail in Sec.$\,$\ref{sec:syma_fhmodel}, we can define an elementary plane-wave basis in terms of elementary projected spin-flip operators that contains contributions from all moiré reciprocal-lattice vectors ${\bf G}$. Because the Bloch eigenstates in the active band have the majority of their weight in the lowest reciprocal-lattice harmonics, our ansatz operators will be built by retaining only the smallest-$|{\bf G}|$ components. With this prescription, we finally use the symmetries of the Bloch bands to write magnon ansatz wave functions with distinct symmetry eigenvalues at the maximally symmetric momentum ${\bf p}= 0$. We will refer to this construction, which makes use of the Bloch band symmetries to build the distinct magnon modes, as the Symmetric Mode Ansatz (SyMA).

Before explicitly constructing the SyMA ansatz for the FH model of MoTe$_2$, we start by first summarizing the symmetries of this model in Sec.$\,$\ref{sec:symmetries_fhmodel}. These symmetries will be later used in Sec.$\,$\ref{sec:syma_fhmodel} to build distinct ansatz modes with distinct symmetry eigenvalues at ${\bf p}= 0$.

\subsection{Symmetries in the FH model of MoTe$_2$}

\label{sec:symmetries_fhmodel}

Before explicitly constructing the SyMA for the FH model of MoTe$_2$, we start by summarizing the symmetries of this model. These symmetries will be important later in the end of Sec.$\,$\ref{sec:syma_fhmodel} to build multiple ansatz modes with distinct symmetry eigenvalues at ${\bf p}= 0$.

For zero displacement field, $V_{z}=0$, the symmetry group of AA stacked moiré MoTe$_2$ is generated by the symmetries
\bea
T_\mbf{R}, C_{3z}, C_{2y}, \mathcal{I}, \mathcal{T}
\eea
which denote moiré lattice translations, rotational symmetries in- and out-of-plane, pseudo-inversion, and time-reversal symmetry (see Appendix
\ref{sec:MoTe2_symmetries_REP} for detailed definitions of the symmetry representations). The intra-valley symmetries are $C_{3z}, C_{2y} \mathcal{T}, \mathcal{I}$ generating the group $p6m'm'$, and opposite valleys are exchanged by time-reversal. Note that the symmetry $\mathcal{I}$ is approximate, and is only realized exactly in the FH model \cite{MFCI_0}. At $V_{z}\neq0$, both the $C_{2y}\mathcal{T}$ and $\mathcal{I}$
symmetries are broken. In this section, we only consider
the most symmetric $V_{z}=0$ case, while $V_{z}\neq0$ will be studied
in Sec.$\,$\ref{subsec:Trivial_vs_Topo_ferro}.

The interacting model preserves all the single-particle symmetries, but its symmetries are enhanced in the projected, flat band limit. $H_I$ in Eq.$\,$\ref{eq:H_V_general} also possesses a unitary
particle-hole symmetry $\mathcal{C}$ with representation $\mathcal{C}c_{{\bf k},{\bf Q},\eta}\mathcal{C}^{-1}=(-\eta)c_{-{\bf k},-{\bf Q},-\eta}^{\dagger}$ (see App.$\,$\ref{sec:MoTe2_symmetries_C}). This symmetry exists for generic projected Hamiltonians in the form of $H_I$ as long as the two valleys/spins are related by time-reversal symmetry, as we show in App.$\,$\ref{sec:MoTe2_symmetries_C}. No additional symmetry is required for the form factors. Additionally, $\mathcal C$ squares to $+1$ within the one-magnon Hilbert space spanned by the basis states $\gamma_{{\bf k}+{\bf p},-\eta}^{\dagger}\gamma_{{\bf k},\eta} \ket{\eta}$, as we show in detail in App.$\,$\ref{sec:MoTe2_symmetries_MAG}. This allows the magnon modes to be classified into $\mathcal C$-even and $\mathcal C$-odd sectors, as we explain below.

\subsection{SyMA in the FH model of MoTe$_2$}

\label{sec:syma_fhmodel}

We now build the SyMA explicitly in the FH model of MoTe$_2$. In Sec.$\,$\ref{subsec:syma_fhmodel_basis}, we first introduce the local
spin-flip bilinears that form the microscopic magnon basis of the construction, Fourier transform them, and project them into the active Chern band.
The resulting projected local basis contains components labeled by all moiré reciprocal-lattice vectors ${\bf G}$ and the moiré momentum ${\bf p}$ in the moiré Brillouin zone. We will then use this basis in Sec.$\,$\ref{subsec:syma_fhmodel_model} to build ansatz operators with distinct symmetry eigenvalues at ${\bf p}=0$. Crucially, to do so, we will employ a small $|{\bf G}|$ truncation. In particular, we will retain only the smallest-$|{\bf G}|$ basis elements from which a nonvanishing operator with the required symmetry eigenvalues can be constructed.

\subsubsection{SyMA basis of elementary magnons}

\label{subsec:syma_fhmodel_basis}

We start by defining the local magnon basis inspired by Feynman's SMA and constrained by the symmetries of the model. We introduce the real-space basis of ``elementary magnons'' given by
\bea
S_{\parallel,\mbf r,l,j}^{\dagger}=e^{-il{\bf q}_{j}\cdot\mbf r}c_{\down,l,\mbf r}^{\dag}c_{\up,l,\mbf r}, \quad S_{\perp,\mbf r,l}^{\dagger}=c_{\down,-l,\mbf r}^{\dag}c_{\up,l,\mbf r}\,,
\label{eq:elementary_magnons}
\eea
where we defined ${\bf q}_{j}=C_{3}^{j}{\bf q}_{1}$ and we distinguish between intralayer and interlayer operators respectively through the symbols ``$\parallel$'' and ``$\perp$''.
The operators $S^\dagger_{\parallel,\mathbf r,l,j}$ and $S^\dagger_{\perp,\mathbf r,l}$ create, respectively, intralayer and interlayer spin-flip excitations whose particle and hole are localized at the same real-space position $\mathbf r$. In direct analogy with the local projected-density operator used in the conventional SMA, we take these local spin-flip bilinears as the starting basis for our construction. More general spin-flip operators of the type $c_{\down,l,\mbf r}^{\dag}c_{\up,l',\mbf r'}$ with $\mbf{r} \neq \mbf{r}'$, will not be retained in our SyMA basis, as in the conventional SMA. The accuracy of this locality approximation will be assessed in the end of this section, as well as in Secs.$\,$\ref{subsec:syma_energies} and \ref{subsec:MoTe2_SMA_eigenfunctions} by comparing the SyMA ansatz built from these basis states with the exact magnon eigenstates. The restriction to local bilinears will be further motivated later in Sec.$\,$\ref{subsec:Flatband_Hubbard_models}, where we will show that the magnon eigenstates are exact superpositions of projected on-site spin flips in the flat band Hubbard models.
In creating the basis for the local bilinears in Eq.$\,$\ref{eq:elementary_magnons}, the moiré twist angle enters the phase factor $e^{-il{\bf q}_{j}\cdot\mbf r}$ in Eq.$\,$\ref{eq:elementary_magnons} since acting on $c_{\down,l,\mbf r}^{\dag}c_{\up,l,\mbf r}$ with the translation $T_\mbf{R}$ produces an extra phase factor $e^{-il{\bf q}_{j}\cdot\mbf r}$, implying that the intralayer operator $c_{\down,l,\mbf r}^{\dag}c_{\up,l,\mbf r}$ transforms projectively. This can be lifted to a conventional representation by including the factor $e^{-il{\bf q}_{j}\cdot\mbf r}$ in the definition of the intralayer operator, as done in Eq.$\,$\ref{eq:elementary_magnons}. Under the definitions in Eq.$\,$\ref{eq:elementary_magnons}, the elementary magnons transform conventionally:
\bea
T_{{\bf R}}S_{\parallel,\mbf r,l,j}^{\dagger}T_{{\bf R}}^{\dagger}&=S_{\parallel,\mbf r+{\bf R},l,j}^{\dagger}, \quad
T_{{\bf R}}S_{\perp,\mbf r,l}^{\dagger}T_{{\bf R}}^{\dagger} =S_{\perp,\mbf r+{\bf R},l}^{\dagger}\,.
\label{eq:interlayer_elementary_magnons_TR}
\eea
In the projected flat band limit, $H_I$ in Eq.$\,$\ref{eq:H_V_general} has the unitary particle-hole symmetry $\mathcal C$, so at ${\bf p}=0$ we can classify the ansatz states by their $\mathcal C$ eigenvalue. The action of the unitary symmetries $C_{3z},\mathcal{I}$ and $\mathcal{C}$ in the elementary magnons is given by the representations
\bea
C_{3}S_{\parallel,\mbf r,l,j}^{\dagger}C_{3}^{-1}&=e^{-\frac{2\pi}{3}i}S_{\parallel,C_{3}\mbf r,l,j+1}^{\dagger}\\
\mathcal{I}S_{\parallel,\mbf r,l,j}^{\dagger}\mathcal{I}^{-1}&=S_{\parallel,-\mbf r,-l,j}^{\dagger}\\
\mathcal{C}S_{\parallel,\mbf r,l,j}^{\dagger}\mathcal{C}^{-1}&=S_{\parallel,\mbf r,l,j}^{\dagger}
\,\,,
\label{eq:symmetries_Spar}
\eea
and
\bea
C_{3}S_{\perp,\mbf r,l}^{\dagger}C_{3}^{-1}&=e^{-\frac{2\pi}{3}i}S_{\perp,C_{3}\mbf r,l}^{\dagger}\\
\mathcal{I}S_{\perp,\mbf r,l}^{\dagger}\mathcal{I}^{-1}&=S_{\perp,-\mbf r,-l}^{\dagger}\\
\mathcal{C}S_{\perp,\mbf r,l}^{\dagger}\mathcal{C}^{-1}&=S_{\perp,\mbf r,-l}^{\dagger}\,\,.
\label{eq:symmetries_Sperp}
\eea

Eqs.$\,$\ref{eq:symmetries_Spar} and \ref{eq:symmetries_Sperp} show that the elementary magnon basis in Eq.$\,$\ref{eq:elementary_magnons} is closed under the symmetries of the model: each symmetry maps an elementary intralayer (interlayer) magnon operator into an intralayer (interlayer) magnon operator. More explicitly, for the intralayer operators, $C_{3z}$ cycles the index $j$ while $\mathcal I$ exchanges the two layers $l=\pm1$. Thus all six discrete choices $(l,j)$ are generated from a single representative intralayer operator. For the interlayer operators, $\mathcal I$ (or equivalently $\mathcal C$) exchanges the two choices $l=\pm1$. Thus, the intralayer and interlayer operators in Eq.$\,$\ref{eq:elementary_magnons} constitute the two symmetry-closed sets of local spin-flip bilinears used as our elementary magnon basis.

We now define the momentum-space elementary magnon basis by Fourier-transforming
the real-space elementary magnon operators in Eq.$\,$\ref{eq:elementary_magnons}. We will also project these operators to the lowest-energy hole band (per spin=valley) as in the conventional SMA \cite{PhysRevB.33.2481}. Taking $\mbf p\in MBZ$, we have
\bea
\bar{S}_{\parallel,\mbf p,{\bf G},l,j}^{\dagger}&=\int d^{2}{\bf r}\,\,e^{i({\bf p}+{\bf G})\cdot{\bf r}}\bar{S}_{\parallel,\mbf r,l,j}^{\dagger} \\
\bar{S}_{\perp,\mbf p,{\bf G},l}^{\dagger}&=\int d^{2}{\bf r}\,\,e^{i({\bf p}+{\bf G})\cdot{\bf r}}\bar{S}_{\perp,\mbf r,l}^{\dagger}\,,
\label{eq:eq:S_p,G,l_inter}
\eea
that are momentum ${\bf p}$ eigenstates. The over bar in the operators in Eq.$\,$\ref{eq:eq:S_p,G,l_inter} is used to indicate projection into the lowest-energy hole band and we will use this notation throughout the paper. Explicit expressions for the projected
elementary operators in Eq.$\,$\ref{eq:eq:S_p,G,l_inter} in terms of
the band operators $\gamma_{{\bf k},\eta}$ are given in
App.$\,$\ref{subsec:detailed_syma_bandbasis}. In principle, umklapp processes create magnon eigenmodes which are superpositions of basis elements in Eq.$\,$\ref{eq:eq:S_p,G,l_inter} with different moiré reciprocal lattice vectors $\mbf{G}$. The coefficients of this superposition cannot, in general, be determined analytically.
However, since we are projecting into the lowest-energy hole band, Bloch eigenstates in this band have most of their weight in basis elements with small ${|\bf G|}$ and so will the form factors entering the magnon Hamiltonian in Eq.$\,$\ref{eq:S_matrix_expression}). It is then natural to expect that the low-energy magnon modes also have most of their weight in basis elements with small ${|\bf G|}$.

This is explicitly verified in App.$\,$\ref{subsec:subspace_diagonalization_SyMA basis}, where we diagonalize the magnon Hamiltonian in Eq.$\,$\ref{eq:S_matrix_expression} in the subspace spanned by the elementary magnon operators in Eq.$\,$\ref{eq:eq:S_p,G,l_inter} (see fig.$\,$\ref{fig:subspace_diagonalization_SyMA basis}). Using the example of the FH model of MoTe2 $\theta=3.7^{\circ}$, we show that the (six) lowest-energy magnon modes are very well captured by restricting $|{\bf G}|\leq|{\bf b}_{M,1}|$, where ${\bf b}_{M,1}$ is a primitive reciprocal lattice vector. These results also show that including additional basis elements with larger $|{\bf G}|$ allows to capture well even higher-energy magnon modes.
It should however be noted that the projected elementary magnon operators in Eq.$\,$\ref{eq:eq:S_p,G,l_inter} associated with different ${\bf G}$ are not, in general, orthogonal or linearly independent. Therefore, basis elements with larger $|{\bf G}|$ need not have weight exclusively on higher-energy magnon modes, and can still have substantial overlap with the same low-energy modes already captured at smaller $|{\bf G}|$. Increasing the $|{\bf G}|$ cutoff instead systematically enlarges the variational subspace, allowing progressively higher-energy modes to be captured after diagonalizing the magnon Hamiltonian within this enlarged subspace.

The results mentioned in the previous paragraph directly verify that a small-$|{\bf G}|$ truncation of the elementary SyMA basis in Eq.$\,$\ref{eq:eq:S_p,G,l_inter} provides the low-rank subspace that spans the low-energy magnon modes of the FH model.
The low-energy magnon modes can then be accurately captured by diagonalizing the magnon Hamiltonian in Eq.$\,$\ref{eq:S_matrix_expression} in this low-rank subspace. A more ambitious question is whether it is possible to build simple analytical ansatz wave functions for the low-energy magnon modes by building on this low-rank construction. This question will be the focus of the next Sec.$\,$\ref{subsec:syma_fhmodel_model}. In particular, we will show that it is possible to analytically construct a minimal family of simple analytical ansatz wave functions with no free fitting parameters, by keeping only the set of basis elements in Eq.$\,$\ref{eq:eq:S_p,G,l_inter} with smallest possible $|{\bf G}|$ that allow to form modes with distinct symmetry eigenvalues at ${\bf p}=0$ (either ${\bf G}=0$ or $|{\bf G}|=|{\bf b}_{M,1}|$ depending on the symmetry sector). This is one of the major results of our work.

\subsubsection{Analytical SyMA modes}

\label{subsec:syma_fhmodel_model}

We now propose a family of analytical SyMA modes. In the one-band projected limit we focus on, the flat band magnon Hamiltonian in Eq.$\,$\ref{eq:S_matrix_expression} separates into $\mathcal C$-even and $\mathcal C$-odd sectors at each ${\bf p}$, as previously mentioned in Sec.$\,$\ref{sec:symmetries_fhmodel} and shown in App. \ref{sec:MoTe2_symmetries_MAG}. We denote the corresponding $\mathcal C$ eigenvalue by $\zeta=\pm1$. For the local SyMA operators, Eqs.$\,$\ref{eq:symmetries_Spar} and \ref{eq:symmetries_Sperp} show that the intralayer elementary magnons are $\mathcal C$ even, while $\mathcal C$ exchanges the two interlayer elementary magnons with opposite layer index. Thus $\mathcal C$-odd SyMA modes must be built from layer-antisymmetric combinations of interlayer operators.
At ${\bf p}=0$, the composition of $\mathcal C$ with the inversion symmetry $\mathcal{I}$ imposes a constraint on the magnon eigenstates: any magnon eigenstate in the one-band projected limit is invariant under the action of $\mathcal{I} \mathcal{C}$ (see App. \ref{sec:MoTe2_symmetries_MAG}). Therefore, the inversion and $\mathcal{C}$ eigenvalues of the magnon modes coincide, and we will therefore label this common eigenvalue by $\zeta=\pm 1$.
We start by building \emph{$\mathcal{C}$-even} ($\zeta=+1$) modes using exclusively the \emph{intralayer} magnons in Eq.$\,$\ref{eq:elementary_magnons}, which are $\mathcal{C}$-even by construction. All the intralayer eigenmodes with different $C_{3}$ eigenvalues at ${\bf p}=0$ can be built using ${\bf G}=0$ as

\begin{equation}
\bar{S}_{\mbf p}^{\dagger(\kappa,+)}=\frac{1}{3}\sum_{l=\pm1}\sum_{j=-1}^{1}e^{-\frac{2\pi i}{3}(\kappa+1)j}\bar{S}_{\parallel,\mbf p,{\bf 0},l,j}^{\dagger} \,, \label{eq:intralayer_Sp}
\end{equation}
\noindent having $C_{3z}$ and $\mathcal{I}/\mathcal{C}$ eigenvalues $e^{\frac{2\pi i}{3}\kappa}$ and $+1$ at ${\bf p}=0$.

To build the \emph{$\mathcal{C}$-odd} modes ($\zeta=-1$), \emph{interlayer} operators are required. The $\kappa=-1$ mode ($C_3$ eigenvalue $e^{-\frac{2\pi i}{3}}$)
can be obtained by setting ${\bf G}=0$ and is given by

\begin{equation}
\bar{S}_{\mbf p}^{\dagger(\kappa=-1,-)}=\sum_{l=\pm1}(-1)^{\frac{l+1}{2}}\bar{S}_{\perp,\mbf p,{\bf 0},l}^{\dagger}\,\,.
\label{eq:interlayer_Sp_k_m1}
\end{equation}

It can be readily checked that $\bar{S}_{\mbf p}^{\dagger(\kappa=-1,-)}$ has $C_{3z}$ and $\mathcal{I}/\mathcal{C}$ eigenvalues $e^{-\frac{2\pi i}{3}}$ and $-1$ at ${\bf p}=0$. The remaining modes (with $\kappa=0$ and $\kappa=1$)
cannot be captured with only ${\bf G}=0$,
which can be easily shown by Fourier transforming the $C_{3z}$ transformation in
Eq.$\,$\ref{eq:symmetries_Sperp}, that yields
\bea
C_{3z}S_{\perp,{\bf p},{\bf G},l}^{\dagger}C_{3z}^{-1}
=
e^{-2\pi i/3}
S_{\perp,C_{3z}{\bf p},C_{3z}{\bf G},l}^{\dagger}.
\eea
At ${\bf p}=0$ and ${\bf G}=0$, each interlayer operator therefore necessarily
has $C_{3z}$ eigenvalue $e^{-2\pi i/3}$.

However,
we can still propose a simple ansatz for the remaining interlayer modes with $\kappa=0$ and $\kappa=1$ by taking the lowest possible
shell given by ${\bf G}_{1,j}=C_{6}^{j}{\bf b}_{M,1}$. In particular,
 we can define

\begin{equation}
\bar{S}_{\mbf p}^{\dagger(\kappa,-)}=\frac{1}{6}\sum_{l=\pm1}\sum_{j=-2}^{3}e^{\frac{2\pi i}{3}(\kappa-2)j}(-1)^{\frac{l+1}{2}}\bar{S}_{\perp,\mbf p,{\bf G}_{1,j},l}^{\dagger}\,,
\label{eq:interlayer_Sp_k_0_1}
\end{equation}
where it can again be checked that $\bar{S}_{\mbf p}^{\dagger(\kappa)}$
has $C_{3z}$ and $\mathcal{I}/\mathcal{C}$ eigenvalues $e^{\frac{2\pi i}{3}\kappa}$ and $-1$.

It is important to note that an underlying approximation in the definition
of the modes in Eqs.$\,$\ref{eq:intralayer_Sp},\ref{eq:interlayer_Sp_k_m1} and \ref{eq:interlayer_Sp_k_0_1} is that the coefficients
are ${\bf p}$-independent, which is not true in general. However,
there is no a-priori generic prescription to introduce a ${\bf p}$-dependence
in the coefficients and we therefore utilize the simplest construction
where the coefficients derived by symmetry at ${\bf p}=0$ are also
used at ${\bf p}\neq0$. The analytical construction in this section
should therefore be the most accurate at ${\bf p}=0$.
Additionally, it should also be noted that by neglecting higher shells, moiré periodicity under ${\bf p}\rightarrow{\bf p}+{\bf G}$ is no longer guaranteed. Nevertheless, as we will verify in Sec.$\,$\ref{subsec:syma_energies}, these simplifications still allow our ansatz to achieve a good quantitative match even at $\bf{p}\neq 0$.

This finishes the construction of all the analytical SyMA modes $\bar{S}_{\mbf p}^{\dagger(\kappa,\zeta)}$ that can be distinguished by symmetry and by the requirement that only the lowest possible $|{\bf G}|$ appears. In App. \ref{subsec:detailed_syma_bandbasis}, we provide detailed expressions for all the projected magnon operators defined in this section written in the band basis.

\subsection{SyMA energies}

\label{subsec:syma_energies}

The analytical SyMA in Eqs.$\,$\ref{eq:intralayer_Sp},\ref{eq:interlayer_Sp_k_m1} and \ref{eq:interlayer_Sp_k_0_1}, built from superpositions of the \textit{local} elementary magnon operators in Eq.$\,$\ref{eq:elementary_magnons}, can all be
written in the projected plane-wave and band bases as

\bea
\bar{S}_{{\bf p}}^{(\kappa,\zeta)\dagger}&=\sum_{{\bf k}}\sum_{{\bf Q},{\bf Q}'}\bar{c}_{\down,{\bf k}+{\bf p},{\bf Q}'}^{\dagger}D^{(\kappa,\zeta)}_{{\bf Q}',{\bf Q}}\bar{c}_{\up,{\bf k},{\bf Q}}\\
&=\sum_{{\bf k}}\gamma_{{\bf k}+{\bf p},\down}^{\dagger}M^{(\down,\up)}_{(\kappa,\zeta)}({\bf k},{\bf p})\gamma_{{\bf k},\up}\,,
\label{eq:S_eta1_eta2}
\eea
where we defined the SyMA wavefunction

\begin{equation}
 M^{(\down,\up)}_{(\kappa,\zeta)}({\bf k},{\bf p})=U_{\down,{\bf k}+{\bf p}}^{\dagger}D^{(\kappa,\zeta)} U_{\up,{\bf k}}\,.
 \label{eq:Mdnup_general_def}
\end{equation}

Here, $D^{(\kappa,\zeta)}$ is a matrix that arises from writing the ansatz characterized by the symmetry labels $(\kappa,\zeta)$ (corresponding to $C_3$ and $\mathcal{C}$ eigenvalues $e^{\frac{2\pi i}{3}\kappa}$ and $\zeta$, see Eqs.$\,$\ref{eq:intralayer_Sp},\ref{eq:interlayer_Sp_k_m1} and \ref{eq:interlayer_Sp_k_0_1}) in the plane-wave basis (see App.$\,$\ref{subsec:detailed_syma_bandbasis}, Eqs.$\,$\ref{eq:D_par_FH} and \ref{eq:D_perp_FH} for detailed expressions).
The ${\bf k}$-independence of matrix $D^{(\kappa,\zeta)}$ follows from restricting the basis for the SyMA to the local spin-flip bilinears Eq.$\,$\ref{eq:elementary_magnons}. The ${\bf p}$-independence follows from the prescription introduced in the previous section in which the coefficients fixed by symmetry at ${\bf p}=0$ are used unchanged at ${\bf p}\neq0$. In the remaining expressions of this section, we will drop the $(\kappa,\zeta)$ labels in $D^{(\kappa,\zeta)}$, $\bar{S}_{{\bf p}}^{(\kappa,\zeta)\dagger}$ and $M^{(\down,\up)}_{(\kappa,\zeta)}({\bf k},{\bf p})$ for notational simplicity. The expressions we provide below therefore apply to each SyMA mode.

We adopt a simple normalization convention for the SyMA modes in Eqs.$\,$\ref{eq:intralayer_Sp},
\ref{eq:interlayer_Sp_k_m1} and \ref{eq:interlayer_Sp_k_0_1}, where we perform an average over the elementary magnon operators, weighted by phase factors that project onto the desired symmetry sector. The resulting magnon operators are however not necessarily normalized with this choice, but the corresponding matrices always satisfy
$\|D\|_2=\sqrt{\lambda_{\textrm{max}}(D^{\dagger}D)}\leq1$,
where $\lambda_{\textrm{max}}(D^{\dagger}D)$ is the maximum eigenvalue of matrix $D^{\dagger}D$. These definitions  establish an
upper bound $|M^{(\down,\up)}({\bf k},{\bf p})| \leq 1$ for all ${\bf k},{\bf p}$. 

Using that $\bar{S}_{{\bf p}}\ket{\up}=0$, the energy of the SyMA state $\bar{S}_{{\bf p}}^{\dagger}\ket{\up}$ can be computed through a double commutator expression given by

\begin{equation}
 \Delta^{(\down,\up)}({\bf p}) =\frac{\bra{\up}[\bar{S}_{{\bf p}},[H_{I},\bar{S}_{{\bf p}}^{\dagger}]]\ket{\up}}{\bra{\up}\bar{S}_{{\bf p}}\bar{S}_{{\bf p}}^{\dagger}\ket{\up}}\,.
\label{eq:Ising_gap_double_commutator}
\end{equation}

\,

Using Eq.$\,$\ref{eq:S_eta1_eta2}, this yields the following closed-form expression for any SyMA ansatz,
\begin{widetext}
\bea
\Delta^{\down\up}({\bf p})&=\frac{\frac{1}{2\Omega_{\textrm{tot}}}\sum_{{\bf k},{\bf q},{\bf G}}V({\bf q}+{\bf G})|M^{(\down)}({\bf k}+{\bf p},{\bf q}+{\bf G})M^{(\down,\up)}({\bf k},{\bf p})-M^{(\down,\up)}({\bf k}+{\bf q}+{\bf G},{\bf p})M^{(\up)}({\bf k},{\bf q}+{\bf G})|^{2}}{\sum_{{\bf k}}|M^{(\down,\up)}({\bf k},{\bf p})|^{2}} \\
&=\Big(\sum_{{\bf k}}\Tr[P^{\down}({\bf k}+{\bf p})DP^{\up}({\bf k})D^{\dagger}]\Big)^{-1}\times \frac{1}{2\Omega_{\textrm{tot}}}\sum_{{\bf k},{\bf q},{\bf G}}V({\bf q}+{\bf G})\Tr\Big[P^{\up}({\bf k})\Big(D^{\dagger}P^{\down}({\bf k}+{\bf p})-P^{\up}({\bf k}+{\bf q}+{\bf G})D^{\dagger}\Big)\\
&P^{\down}({\bf k}+{\bf q}+{\bf G}+{\bf p})\Big(P^{\down}({\bf k}+{\bf p})D-DP^{\up}({\bf k}+{\bf q}+{\bf G})\Big)\Big]
\,,\label{eq:Ising_gap_1BPV}
\eea
\end{widetext}
where we defined the gauge-invariant projectors $ P^{\eta}({\bf k})=U_{\eta,{\bf k}}U^{\dagger}_{\eta,{\bf k}}$, with the Bloch eigenvectors extended outside the MBZ using the periodic embedding convention introduced below Eq.$\,$\ref{eq:gamma_def}.

 In what follows, we benchmark the expression in Eq.$\,$\ref{eq:Ising_gap_1BPV} numerically, then we explain its quantum geometric meaning before turning to solvable models where more physically interpretable expressions can be obtained, and a simple elementary local SyMA basis analogous to Eq.$\,$\ref{eq:eq:S_p,G,l_inter} can be proven to be exact for all ${\bf p}$.

\begin{figure*}
\begin{centering}
\includegraphics[width=\textwidth]{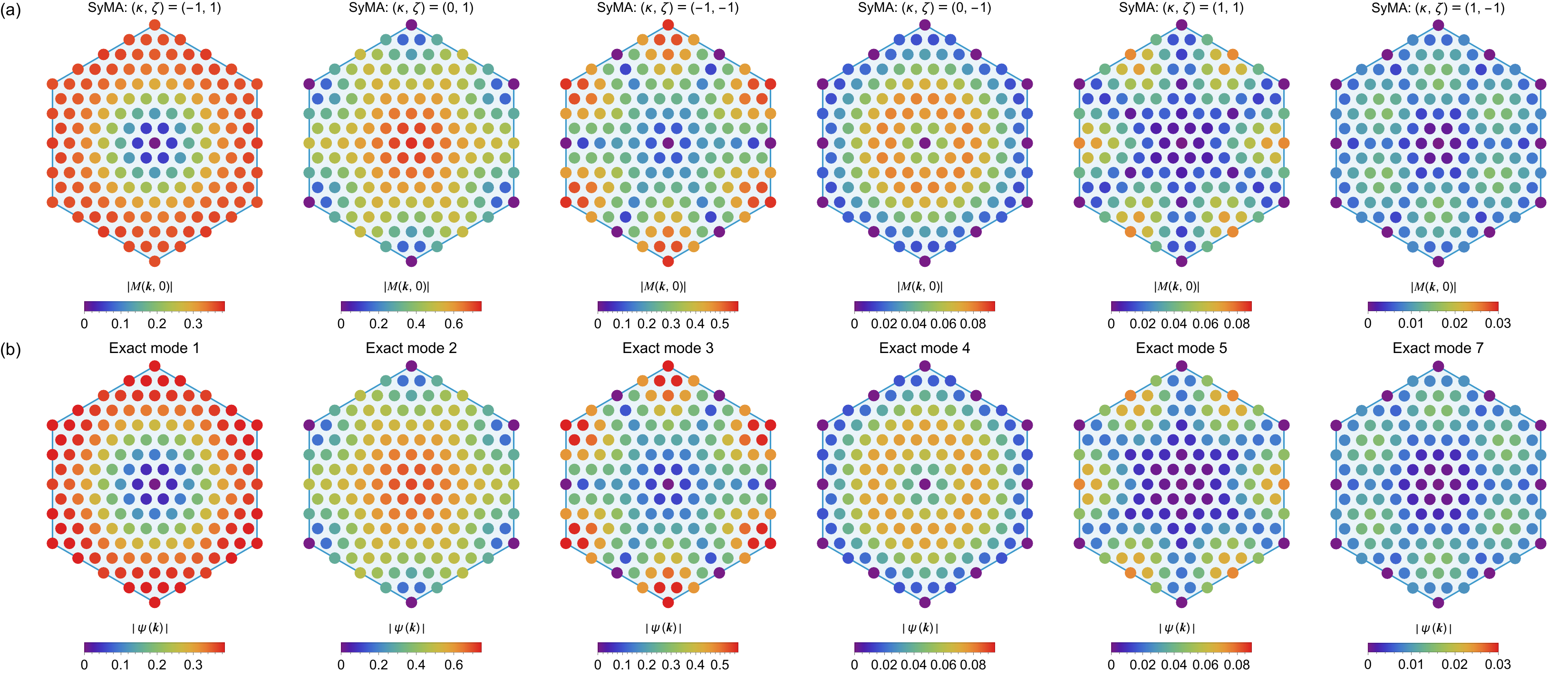}
\par\end{centering}
\caption{Comparison between absolute values of exact and SyMA magnon eigenstates at ${\bf p}=0$ for the FH model of MoTe$_2$ at $\theta=3.7^{\circ}$. The exact modes are ordered by increasing energy from left to right. The three lowest energy states have the smallest sum of the node degrees $\mathcal{N}$: 2 (double node), 2 and 4 respectively. All the higher-energy modes have $\mathcal N\geq 6$.}
\label{fig:3}
\end{figure*}

At ${\bf p}=0$, the analytical SyMA modes with different symmetry labels $(\kappa,\zeta)$ given in Eqs.$\,$\ref{eq:intralayer_Sp},\ref{eq:interlayer_Sp_k_m1} and \ref{eq:interlayer_Sp_k_0_1} are orthogonal by symmetry and are therefore expected to show the best agreement with the exact magnon modes. The results for the mode energies at ${\bf p}=0$ obtained using Eq.$\,$\ref{eq:Ising_gap_1BPV} are shown in Fig.$\,$\ref{fig:2}(a), showing remarkable agreement with the exact numerical results. At generic ${\bf p}\neq0$, $\kappa$ no longer labels a $C_{3z}$ eigenvalue, and modes with the same $\mathcal{C}$ eigenvalue need not remain orthogonal (note, however, that modes with opposite $\mathcal{C}$ eigenvalues remain orthogonal for all ${\bf p}$, since this symmetry holds at all $\bf p$ as shown in App.$\,$\ref{sec:MoTe2_symmetries_MAG}).
This is merely an artifact of our simplified analytical construction where the coefficients of the superposition between different basis elements derived using symmetry at ${\bf p}=0$ were used unchanged also at ${\bf p}\neq0$. Interestingly, very good agreement is also achieved for ${\bf p}\neq0$, with appreciable deviations only occurring near the edges of the moiré Brillouin zone, as shown in Fig.$\,$\ref{fig:2}(b) (see App. \ref{subsec:App_Additional_MoTe2_SMA} for additional results at other angles).

A more rigorous procedure is to diagonalize the magnon Hamiltonian in Eq.$\,$\ref{eq:S_matrix_expression} in the basis of all the SyMA modes $\bar{S}_{\mbf p}^{\dagger(\kappa,\zeta)}$ defined in Sec.$\,$\ref{sec:sma_main}. This is a direct test of whether the subspace spanned by the simple analytical SyMA modes defined in this section captures well the low-energy magnon subspace at every ${\bf p}$, and amounts to solve the generalized eigenvalue problem,

\bea
\sum_{\kappa',\zeta'} &
\bra{\uparrow}
\bar{S}_{\mbf p}^{(\kappa,\zeta)}
\,
{\bf S}^{(\uparrow)}({\bf p})
\,
\bar{S}_{\mbf p}^{\dagger(\kappa',\zeta')}
\ket{\uparrow}
v_{\kappa',\zeta'}({\bf p}) \\
&=
E({\bf p})
\sum_{\kappa',\zeta'}
\bra{\uparrow}
\bar{S}_{\mbf p}^{(\kappa,\zeta)}
\bar{S}_{\mbf p}^{\dagger(\kappa',\zeta')}
\ket{\uparrow}
v_{\kappa',\zeta'}({\bf p}),
\label{eq:SyMA_subspace_gevp}
\eea
where we defined $[{\bf S}^{(\eta)}({\bf p})]_{{\bf k}',{\bf k}}\equiv S^{(\eta)}({\bf k}',{\bf k};{\bf p})$, $\kappa,\kappa'=-1,0,1$ and $\zeta,\zeta'=-1,1$ run over
all the analytical SyMA modes defined in this section in Eqs.$\,$\ref{eq:intralayer_Sp},\ref{eq:interlayer_Sp_k_m1} and \ref{eq:interlayer_Sp_k_0_1}, and $v_{\kappa',\zeta'}$ are the eigenvector components in the subspace of SyMA modes.
This procedure yields remarkable agreement for the lowest-energy modes at all ${\bf p}$, as
shown in Fig.$\,$\ref{fig:2}(c). This result is consistent with the systematic analysis of App.$\,$\ref{subsec:subspace_diagonalization_SyMA basis}, where it can be clearly seen that the low-$|{\bf G}|$ elementary local-magnon basis in Eq.$\,$\ref{eq:eq:S_p,G,l_inter} spans the low-energy magnon subspace. Our analytical SyMA modes provide the minimal low-energy analytical basis constructed from the lowest possible $|{\bf G}|$ components in each symmetry sector.

\subsection{SyMA eigenfunctions and the Node Principle}
\label{subsec:MoTe2_SMA_eigenfunctions}

The success of our SyMA lies in the accuracy of the ansatz wave functions. At ${\bf p}=0$, they show remarkable agreement with the exact eigenfunctions as
shown in Fig.$\,$\ref{fig:3}, with an overlap above $97\%$ for all modes for $\theta=3.7^{\circ}$ (see also App. \ref{subsec:App_Additional_MoTe2_SMA}
for additional results at $\theta=2^{\circ}$). At ${\bf p}={\bf K}$, the overlap is above $75\%$ for the three lowest SyMA modes, and below $50\%$ for higher-energy modes. When the Hamiltonian is instead diagonalized in the subspace of all the SyMA modes, as described in the end of Sec.$\,$\ref{subsec:syma_energies} the overlaps at ${\bf p}={\bf K}$ are above $99\%$ for the four lowest-energy modes, falling below $90\%$ only for higher-energy modes.

An important feature of the exact magnon eigenfunctions, which is
successfully captured by our SyMA, is the presence of nodes in the magnon eigenstates, i.e. where $M^{(\down,\up)}(\mbf{k},\mbf{p}) = 0$. As an example, Fig.$\,$\ref{fig:3} shows results for $M^{(\down,\up)}(\mbf{k},\mbf{p}=0)$ and exact nodes can be seen either at $\mbf{k}=\Gamma$ for the $(\kappa,\zeta)=(-1,+1)$ mode or at ${\bf K}$ and ${\bf K}'$ for the $(\kappa,\zeta)=(0,+1)$ mode, which correspond to the two lowest modes.
The presence of nodes in the low-energy Ising magnons contrasts with the
SU(2)-symmetric case. For an SU(2)-invariant Hamiltonian, the global
spin-lowering operator, $S^-=\sum_{{\bf k},{\bf Q}}
c_{{\bf k},{\bf Q},\downarrow}^{\dagger}
c_{{\bf k},{\bf Q},\uparrow}$,
commutes with the Hamiltonian. Therefore, if $\ket{\uparrow}$ is the fully
polarized ground state with energy $E_0$, then
\begin{equation}
H S^-\ket{\uparrow}
=
S^-H\ket{\uparrow}
=
E_0 S^-\ket{\uparrow}.
\end{equation}
Thus, $S^-\ket{\uparrow}$ is an exact zero-energy state at ${\bf p}=0$, corresponding to the zero-momentum Goldstone mode.
Its wave function in the projected one-band Hilbert space is
\begin{equation}
\overline{S}^-=
\sum_{\bf k}
\gamma_{{\bf k},\downarrow}^{\dagger}
\left(
U_{\downarrow,{\bf k}}^{\dagger}
U_{\uparrow,{\bf k}}
\right)
\gamma_{{\bf k},\uparrow} = \sum_{\bf k}
\gamma_{{\bf k},\downarrow}^{\dagger}
\gamma_{{\bf k},\uparrow}.
\end{equation}
where we used that in the SU(2)-symmetric case, the two spin components of
the active band can be chosen to have the same Bloch eigenvector, and therefore $U_{\downarrow,{\bf k}}=U_{\uparrow,{\bf k}}.$
Consequently, the magnon wave function of the zero-momentum Goldstone state is $M^{(\downarrow,\uparrow)}({\bf k},{\bf 0})
=1$ for all ${\bf k}$ in the SU(2)-symmetric case. The zero-momentum Goldstone mode is therefore
nodeless, in contrast with the case of the Ising magnons.
In fact, the presence of nodes in the Ising class is known to be topological$\,$\cite{PhysRevLett.90.057002,PhysRevLett.120.067003,PhysRevLett.124.166601,PhysRevB.107.L201106,2023PhRvB.107t1106Y,UchoaPNAS,kwan2024texturedexcitoninsulators}. Letting $q_{j}$ denote the vorticity/winding number of the $j$-th node and considering the generalized case where the Chern number of the up (down) electron is $C_{\up}=C$ ($C_{\down}=-C$), it can be shown that the vorticities of all nodes of $M^{(\downarrow,\uparrow)}({\bf k},{\bf p})$ that can be found as ${\bf k}$ varies over the Brillouin zone for all fixed ${\bf p}$, satisfy
\begin{equation}
\sum_{j=1}q_{j}=C_{\down}-C_{\up}=-2C\,, \forall \, {\bf p}\,.\label{eq:qj_C}
\end{equation}
The result in Eq.$\,$\ref{eq:qj_C} applies for any one-band-projected magnon eigenstate, which we show in App. \ref{subsec:App_chern_vorticity_derivation}.
This immediately implies that the sum of the degrees of all nodes for $\bf k$ running in the moiré Brillouin zone, that we define as $\mathcal{N}$, is at least $2|C|$. Unlike in the SU(2) case, where opposite spins carry the same Chern number, the nodeless gapless mode cannot therefore be realized in the Ising U(1) case. We now provide a heuristic criterion to determine the energetic hierarchy of the modes
from $\mathcal{N}$ and the quantum geometry of the filled band.

The nodes correspond to the points where the norm of the magnon wave function, $\rho({\bf k},{\bf p})\equiv|M^{(\down,\up)}({\bf k},{\bf p})|$, vanishes. All else being equal, increasing the total node degree, $\mathcal{N}$, suppresses the denominator of Eq.$\,$\ref{eq:Ising_gap_1BPV}, and therefore tends to increase the energy of the magnon and thus the magnon gap. Because the magnon wave function also enters the numerator, however, the resulting dependence of the magnon gap on $\mathcal{N}$ can be more complex in general. We return to this point at the end of Sec.$\,$\ref{subsec:syma_constraints_node}. Additionally, the suppression of the denominator based on the existence of nodes assumes that $\rho(\mbf{k},\mbf{p})$ cannot grow too quickly away from the node, which can be proved explicitly for ${\bf p}=0$ (see App. \ref{subsec:App_GL_Isinggap}). Concretely, we can show that
\begin{equation}
|\nabla\rho({\bf k},{\bf 0})|^2 \leq 2\big(G^{\up}({\bf k})+G^{\down}({\bf k})\big)\,,\label{eq:grad_rho_bound_Gk}
\end{equation}
where $G^{\eta}({\bf k})$ is the trace of the quantum metric for the spin-$\eta$ band. The gradient bound implies that the growth rate of the magnon wave function norm is bounded by the quantum metric. Therefore, if the nodes are placed in regions with small quantum metric in a significant region around the node, $\rho({\bf k}, {\bf 0})$ also grows little away from the nodes and the total norm will be constrained to be small, tending to give rise to a large zero-momentum magnon gap (i.e. a magnon energy close to the particle-hole continuum). This is no longer necessarily the case if the nodes are placed in the regions with larger quantum metric, where $\rho({\bf k}, {\bf 0})$ can grow faster away from the nodes. In fact, although Eq.$\,$\ref{eq:grad_rho_bound_Gk} provides only an upper bound, we numerically verify in the FH model that modes whose nodes coincide with peaks of the quantum metric grow more rapidly away from the nodes and have a smaller Ising gap, as we will exemplify below. This behavior is similar to that of the charge-excitation spectrum, whose energy is larger in regions with small or vanishing quantum geometry and smaller in regions with large quantum geometry, see e.g. the example of twisted bilayer graphene$\,$\cite{TBGV}.
Fig.$\,$\ref{fig:2}(b) for $\theta=2^{\circ}$ shows an explicit example where different magnon modes can have lower or higher energy at ${\bf p}=0$ depending on whether the quantum geometry is peaked at the location of the node or away from it (see App. \ref{subsec:App_nodes_MoTe2}). In particular, $G({\bf k})$ is peaked at ${\bf k}={\bf K},{\bf K}'$
(see inset in Fig.$\,$\ref{fig:2}(b) for $\theta=2^{\circ}$). As a result, the $(\kappa,\zeta)=(0,+1)$ mode, whose nodes occur at
${\bf k}={\bf K}$ and ${\bf k}={\bf K}'$, has a significantly smaller energy at
${\bf p}=0$ than the $(\kappa,\zeta)=(-1,+1)$ mode, whose node occurs
at ${\bf k}={\bf \Gamma}$.

\subsection{Symmetry constraints on node degrees}

\label{subsec:syma_constraints_node}

At ${\bf p}=0$, the SyMA magnon modes defined in
Eqs.$\,$\ref{eq:intralayer_Sp}, \ref{eq:interlayer_Sp_k_m1}, and
\ref{eq:interlayer_Sp_k_0_1} have symmetry eigenvalues
$e^{i2\pi\kappa/3}$ ($C_{3z}$) and $\zeta$ ($\mathcal{C}$). We now establish an elegant relation between
these magnon symmetry eigenvalues to those of the Chern band occupied
in the symmetry-broken ground state (which we take to be $\ket{\up}$, without loss of generality). We denote
 the occupied band (spinful) $C_{3z}$ eigenvalues by $\xi_{\mbf{k}^*}$ for the $C_{3z}$-symmetric points $\mbf{k}^* = \Gamma, \mbf{K},\mbf{K}'$. For the intralayer/$\mathcal{C}$-even modes, we find the relation (see App.$\,$\ref{subsec:App_nodes_MoTe2})

\begin{equation}
e^{- i\frac{2\pi}{3}\kappa} M^{(\down,\up)}_{(\kappa,+1)}
({\bf k}^{*},{\bf 0})= \xi_{\mbf{k}^*} \xi_{-\mbf{k}^*} M^{(\down,\up)}_{(\kappa,+1)}({\bf k}^{*},{\bf 0}) \ . \label{eq:node_condition_intra}
\end{equation}
where $M^{(\down,\up)}_{(\kappa,+1)}$ is the intralayer magnon wave function (see Eq.$\,$\ref{eq:Mdnup_general_def} and App.$\,$\ref{subsec:detailed_syma_bandbasis} for explicit expressions).
Eq.$\,$\ref{eq:node_condition_intra} shows that a mismatch of the magnon and electron eigenvalues enforces a node at ${\bf k}={\bf k}^{*}$.
We now show that Eq.$\,$\ref{eq:node_condition_intra} also implies that in a topological phase with $C\neq0\pmod 3$, all SyMA modes necessarily have nodes.
If
$M^{(\downarrow,\uparrow)}_{(\kappa,+1)}({\bf k}^{*},{\bf 0})\neq 0$
at all three points ${\bf k}^{*}=\Gamma,{\bf K},{\bf K}'$, then
Eq.~\ref{eq:node_condition_intra} requires $e^{-i2\pi\kappa/3}
=
\xi_{{\bf k}^{*}}\xi_{-{\bf k}^{*}}$
at each point. Multiplying the three relations and using
$-\Gamma=\Gamma$, $-{\bf K}={\bf K}'$, and
$-{\bf K}'={\bf K}$ modulo reciprocal lattice vectors yields
\begin{equation}
\left(\xi_{\Gamma}\xi_{\bf K}\xi_{{\bf K}'}\right)^{2}
=
e^{-i2\pi\kappa}
=
1
=
e^{i4\pi C/3}
,
\end{equation}
where in the last step we used the identity
$e^{2\pi i C/3}=-\xi_{\Gamma}\xi_{\bf K}\xi_{{\bf K}'}$ \cite{PhysRevB.86.115112}.
For integer $C$, this condition holds only when
$C=0\pmod 3$. Therefore, when $C\neq0\pmod 3$,
Eq.~\ref{eq:node_condition_intra} enforces a node at least one
of the three $C_{3z}$-invariant momenta.
This is no longer true in a trivial phase, which is consistent with Eq.$\,$\ref{eq:qj_C} for $C=0$ - when $C=0$, the sum of the node vorticities vanishes, which is compatible with a nodeless magnon wave function.

Combining the vorticity sum rule in Eq.$\,$\ref{eq:qj_C} with the symmetry eigenvalue constraints inferred from Eq.$\,$\ref{eq:node_condition_intra} leads to additional constraints for the node degree, $\mathcal{N}$, in each magnon mode wave function. 
We now derive the minimal number of nodes and minimal total node degrees we can have for the intralayer modes of the FH model of MoTe$_2$ defined in Eq.$\,$\ref{eq:intralayer_Sp}. These have nodes only at the high-symmetry points $\mbf{k}^* = \Gamma, \mbf{K},\mbf{K}'$, so we will proceed with the working assumption that there are no nodes away from these points. 
Note also that any node away from the $C_{3z}$-invariant momenta must be accompanied by its two $C_{3z}$-related images. Thus, a generic node of this type occurs in a triplet and increases the total node degree $\mathcal N$ by at least $3$. According to the node principle discussed
in Sec.$\,$\ref{subsec:MoTe2_SMA_eigenfunctions}, modes containing such additional node triplets are therefore expected to lie at higher energy.
We first note that for the fully filled band with $\eta=\up, C=1$, we have $\xi_{\Gamma}=-1$ and $\xi_{{\bf K}}=\xi_{{\bf K}'}=e^{i \pi/3}$.
For $\kappa = -1$, Eq.$\,$\ref{eq:node_condition_intra} is compatible at $\mbf{K}$ and $\mbf{K}'$ but incompatible at $\Gamma$. This implies a node at $\Gamma$ and, by the vorticity sum rule in Eq.$\,$\ref{eq:qj_C}, this node must have degree $2$. Next, for $\kappa = 0$, the $\Gamma$ point is compatible but there must be nodes at $\mbf{K},\mbf{K}'$, and each can be degree $1$ to satisfy the sum rule. Hence both the $\kappa = -1,0$ modes have $\mathcal{N}=2$, the minimal sum of the node degrees, and are expected to have the lower energies. 
In contrast, the $\kappa=+1$ mode is incompatible at all three high-symmetry points and therefore has nodes at $\Gamma,{\bf K},{\bf K}'$. The $C_3$ symmetry further constrains their vorticities modulo $3$ (see App.$\,$\ref{subsec:App_nodes_MoTe2}).
Combining these constraints with the total-vorticity sum rule gives a minimum total node degree $\mathcal{N}=\sum_j |q_j|=6$, where we restricted ourselves, as above, to the minimal situation
with no additional nodes away from the high-symmetry points. The configuration realized by the $\kappa=+1$ intralayer mode shown in Fig.$\,$\ref{fig:3} has $q_\Gamma=2$ and $q_{\bf K}=q_{{\bf K}'}=-2$.
This analysis clarifies our findings in Fig.$\,$\ref{fig:2}(a), in particular, it explains the observation that the $\mathcal{C}$-even mode $\kappa = +1$ has higher-energy than the $\mathcal{C}$-even $\kappa = -1$ and $\kappa = 0$ modes.

We next consider the interlayer modes. First, we can prove the following compatibility condition (see App. \ref{subsec:App_nodes_MoTe2})
\bea
M^{(\down,\up)}_{(\kappa,-1)}({\bf k}^*,{\bf p})&=- M^{(\down,\up)}_{(\kappa,-1)}(-{\bf k}^*-{\bf p},{\bf p}) \,,
\eea
where $M^{(\down,\up)}_{(\kappa,-1)}({\bf k},{\bf p})$ are the interlayer magnon eigenstates (see App. \ref{subsec:detailed_syma_bandbasis} for explicit expressions),
 which implies that for the $\mathcal{C}$-odd modes ($\zeta=-1$) we are guaranteed to have nodes at $2 \mbf{k}^* = -\mbf{p} \mod \mbf{G}$ which has four solutions in the Brillouin zone. Hence, we have $\mathcal{N}\geq4$. For instance at ${\bf p}=0$,
the nodes occur at the four time-reversal invariant points. Based on the larger $\mathcal{N}$, it is expected that the $\mathcal{C}$-odd modes lie at higher energy than the lowest $\mathcal{C}$-even modes with $\mathcal{N}=2$. Indeed, the lowest $\mathcal{C}$-odd mode has $\mathcal{N}=4$ (see Fig.$\,$\ref{fig:3}) and corresponds to the third-lowest-energy mode at ${\bf p}=0$. Throughout the range $2^\circ\leq\theta\leq5^\circ$, this mode lies consistently closer to the particle-hole continuum than the two lowest modes with $\mathcal{N}=2$, as shown in Fig.$\,$\ref{fig:2}(a). All remaining higher-energy bound modes have $\mathcal{N}\geq6$.

So far, we have not considered the dependence of
$M^{(\down,\up)}({\bf k},{\bf p})$ in the numerator of
Eq.$\,$\ref{eq:Ising_gap_1BPV}. Nevertheless we can achieve a good heuristic understanding of the hierarchy of nodes using only symmetry and quantum geometry.
To better understand the full Eq.$\,$\ref{eq:Ising_gap_1BPV}, in
App.$\,$\ref{subsec:App_GL_Isinggap} we rewrite it as an approximate
Ginzburg-Landau functional.
This approximation is carried out for a flat band Hamiltonian with Hubbard interaction - defined by a ${\bf q}$-independent $V({\bf q})$ in Eq.$\,$\ref{eq:H_V_general} - and is valid when the band form
factors $M^{(\eta)}({\bf k},{\bf q}+{\bf G})$ are strongly dominated by small momentum transfers, i.e. small $|{\bf q}+{\bf G}|$ in Eq.$\,$\ref{eq:Ising_gap_1BPV}, see App.$\,$\ref{subsec:GL_derivation_first}. This approximation provides a qualitative understanding of how the nodes affect the numerator of Eq.$\,$\ref{eq:Ising_gap_1BPV}, as discussed in detail in App.$\,$\ref{subsec:Interpretation_IsingGap_Approximation}, allowing to explicitly show that the magnon gap increases with the number of nodes and node degree in some simple limiting cases discussed in App.$\,$\ref{subsec:simple_limiting_cases}.
For the FH model of MoTe$_2$, this
approximation captures the qualitative trends of the different magnon
energies of the two lowest SyMA modes, although it is not quantitatively accurate in general, see Fig.\ref{fig:Ising_gap_approximation_test} in App.$\,$\ref{subsec:GL_derivation_first}. To acquire a more rigorous and deeper understanding, we now turn to a new class of solvable models.

\section{Solvable models}

\label{sec:Solvable_Models}

In this section, we discuss different solvable models for which it is possible to obtain analytically tractable expressions for the Ising magnons. In Sec.$\,$\ref{subsec:Flatband_Hubbard_models}, we will introduce a family of solvable flat band Hubbard models \cite{2022arXiv220900007H} with $S_z$ and $\mathcal{T}$ which have a flat particle-hole continuum. We will prove that the SyMA is exact in this family, obtain new counting rules, and derive properties of the binding energies. The flat band Haldane-Hubbard model falls in this family of models and reproduces the number, ordering and dispersion of the lowest-energy magnon branches in the FH model at small angles $\theta\approx2^{\circ}$ (outside the range of applicability to MoTe$_2$), as we will show in this section. At larger twist angles near $\theta\approx 3.89^\circ$ where the quantum geometry is more homogeneous, the FH model is closer to the time-reversal symmetric Landau level limit discussed in Sec. \ref{sec:intro}, as shown in Fig.$\,$\ref{fig:4}$(a,b)$. We refer the reader to App. \ref{secapp:LLL} for a detailed discussion of this limit.

In Sec.$\,$\ref{subsec:Ideal_band_BHZ_model}, we consider a family of continuum BHZ-like models with ideal $|C| \geq 1$ flat bands. These ideal-band BHZ models have full rotational symmetry, allowing us to study the SyMA analytically beyond the limit of flat charge excitations.

\subsection{Flat band Hubbard models}

\label{subsec:Flatband_Hubbard_models}

\begin{figure}
\begin{centering}
\includegraphics[width=1\columnwidth]{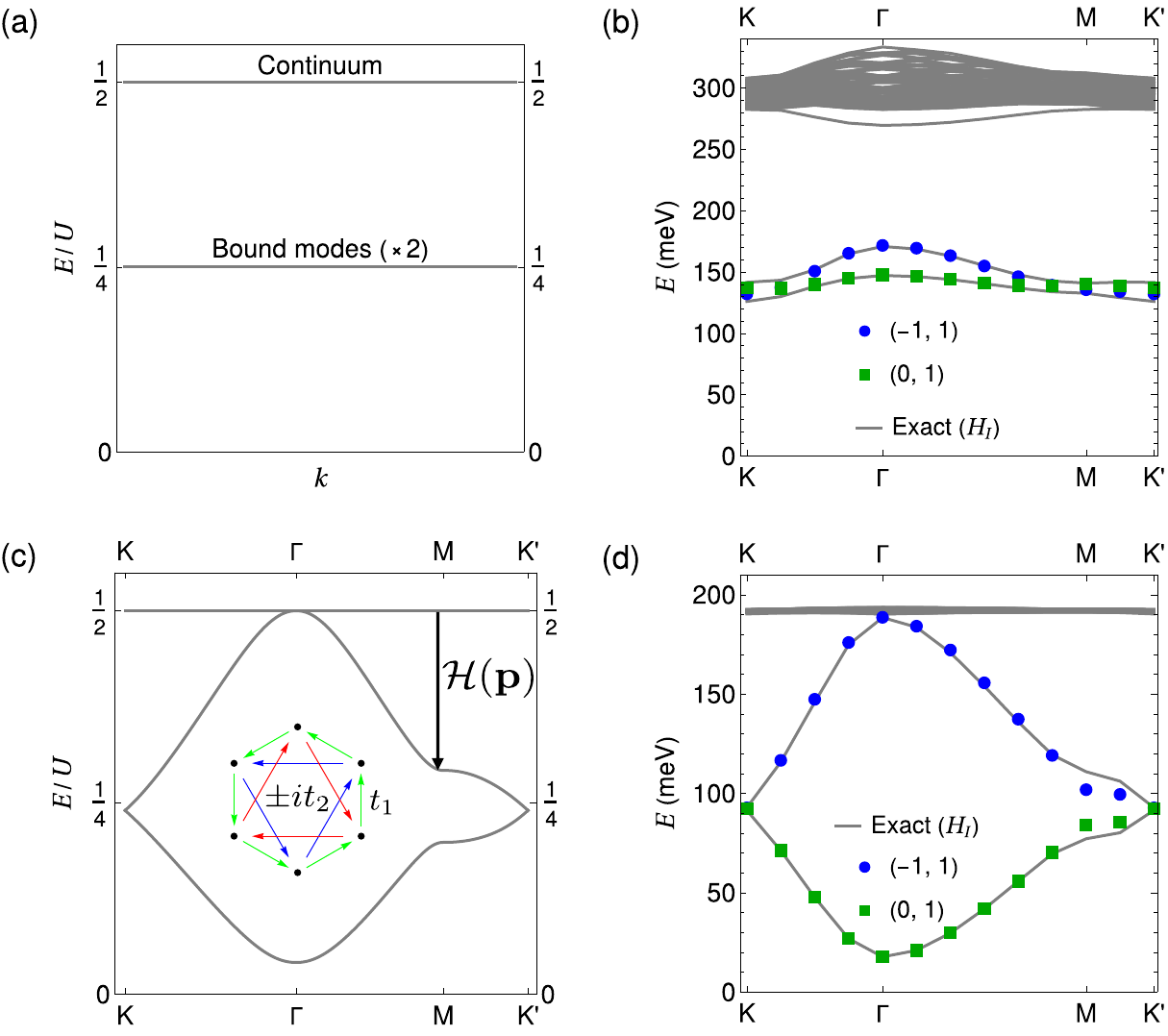}
\par\end{centering}
\caption{ (a) Magnon spectrum for the time-reversal
symmetric Landau level limit with $V_{0}$ pseudopotential (see Ref.$\,$\cite{PhysRevLett.126.137601} or App. \ref{secapp:LLL} for details). (b) Exact magnon spectrum obtained for the flat band Hamiltonian $H_{I}$
in Eq.$\,$\ref{eq:H_V_general} (full gray lines) using the Hubbard interaction
defined by setting $V({\bf q})=V({\bf q}=0)$ in Eq.$\,$\ref{eq:H_V_general}, for $\theta=3.7^{\circ}$.
The blue and green points are the results for the SyMA modes defined in Sec.$\,$\ref{sec:sma_main} with symmetry eigenvalues $(\kappa,\zeta)$ specified in the figure.
(c) Magnon spectrum of the flat band Haldane model, for ratio between next-nearest-neighbor and nearest-neighbor hopping strength (see inset) of $t_{2}/t_1=1/10$.
(d) Same as (b), for $\theta=2^{\circ}$.\label{fig:4}}
\end{figure}

\label{subsec:Flatband_Hubbard}

We now exhibit a large family of exactly solvable Hubbard models for flat band ferromagnetism, where the charged fermionic excitations and magnons are exactly solvable. This family of models can be defined on any lattice with Bravais lattice $\mbf{R}$ with $N$ unit cells and orbitals $\alpha = 1,\dots,N_{orb}$ per unit cell at position $\mbf{r}_\al$. The Hamiltonian reads
\bea
\label{eq:flatbandhubbard}
H &= \sum_{\mbf{k} \al \be \eta} c^\dag_{\mbf{k},\al \eta} h_{\al \be}^\eta(\mbf{k}) c_{\mbf{k},\be \eta} + \sum_{\mbf{R} \al} V_\al n_{\mbf{R},\al, \uparrow} n_{\mbf{R},\al, \downarrow} \,,
\eea
where $h^\eta(\mbf{k})$ is the single-particle Hamiltonian. We will focus on the case where $h^\eta(\mbf{k})$ is $U(1)$ symmetric with $S_z$ conservation and assume time-reversal symmetry $\mathcal{T}$ with the representation $h^\uparrow(\mbf{k}) = h^\downarrow(-\mbf{k})^*$. The solvability condition we require is that $h^\eta(\mbf{k})$ has an exactly flat band at $E=0$ and that $V_\al > 0$. This is possible for generic gapped bands at the cost of introducing exponentially decaying hoppings$\,$\cite{Chen_2014,PhysRevLett.106.236804}. One can immediately see that the Slater determinant state
\bea
\ket{\eta} &= \prod_{\mbf{k}} \gamma^\dag_{\mbf{k},\eta} \ket{0}, \quad \gamma^\dag_{\mbf{k},\eta} = \sum_\al c^\dag_{\mbf{k},\al \eta} U_{\eta,\al,\mbf{k}} \,,
\eea
where $U_{\eta,\al,\mbf{k}}$ is the Bloch wave function of the flat band,
is an exact ground-state since it is annihilated by $H$, which is positive semi-definite. Unlike in the SU(2) case where the ground-states form a ferromagnetic irrep$\,$\cite{1999JPhA...32.8411M,1993CMaPh.158..341M,2003CMaPh.242..445T,tasaki2020manybody,2016PhRvB..94x5149T,kang2024quantum}, in the Ising case, $\ket{\eta}$ are the only two ground-states at integer filling and spontaneously break $\mathcal{T}$. Going forward, we will project to the flat band Hilbert space assuming that the gap is much larger than $V_\al$.

The simplest exactly solvable case is obtained for \emph{uniform} lattices, where every orbital $\al$ is related by a space group element \cite{2022arXiv220900007H,2016PhRvB..94x5149T} and $V_\al = U$ is identical for every orbital by symmetry. As an example, we will later consider the Haldane honeycomb model with two $s$ orbitals related by $C_6$. We show in App. \ref{App:FlatBandHubbard_details} that the more general case is solvable when $V_\al$ is chosen so that $V_\al \bra{\eta}n_{\mbf{R},\al}\ket{\eta}$ is constant\cite{2024PhRvR...6a3256T}.

We start by obtaining the exact charge excitations. The hole excitations $\gamma_{\mbf{k},\eta} \ket{\eta}$ have exactly zero energy, and we show that the electron excitations $\gamma^\dag_{\mbf{k},-\eta} \ket{\eta}$ are eigenstates with energy (see App. \ref{app:hubbardcharge})
\bea
E_{-\eta}(\mbf{k}) = \frac{U}{N_{orb}} \,,
\eea
and are also exactly flat. This result illustrates two points: the charge gap is nonzero, and the excitations remain exactly flat \emph{independent} of all wave function properties in the Hubbard limit on uniform lattices (although for generic interactions this is not the case). These properties, however, are extremely important in the magnon excitations.

Because the charge excitations are exactly flat and the interaction is low-rank, the bound states can be obtained analytically. Following the techniques introduced in Ref.$\,$\cite{2022arXiv220900007H}, we find that the magnon binding energies (the energy below the flat continuum at $U/N_{orb}$) are described by the effective single-particle Hamiltonian
\bea
\label{eq:bindingenergies}
\mathcal{W}_{\al \be}(\mbf{p}) &= \frac{U}{N} \sum_{\mbf{k}} P^{\downarrow}_{\al \be}(\mbf{k}+\mbf{p}) P^{\uparrow}_{\be \al}(\mbf{k}) \,, \\
\eea
where $P^\eta(\mbf{k}) = U^{\eta}(\mbf{k})U^{\eta \dag}(\mbf{k})$ is the single-particle projector on the flat band (without loss of generality, we assumed magnon excitations on top of the $\ket{\up}$ ground-state). Our first observation is that the magnon energies are determined completely by the flat band wave functions, encoded in the gauge-invariant projector, and that the maximum possible number of bound states is $N_{orb}$, \emph{not} the number of flat bands. The presence of multiple bound states is thus a feature of non-trivial flat bands, which require multi-orbital models. Fig.$\,$\ref{fig:4}(c) shows an example of the magnon dispersion in the Haldane model$\,$\cite{PhysRevLett.61.2015}, with a ratio between next-nearest-neighbor and nearest-neighbor hopping strength (see inset) of $t_{2}/t_1=1/10$. A discussion of its general features from symmetry properties may be found in App. \ref{App:FlatBandHubbard_details}. A proof of a gapless, quadratic Goldstone mode in the $SU(2)$ symmetric case may be found in Ref.$\,$\cite{2022arXiv220900007H}.

Second, the presence of an effective single-particle Hamiltonian exactly describing the magnons suggests that they can be thought of as tightly-bound bosons hopping on the lattice. This picture is substantiated by the magnon wave functions, which are (see App.$\,$\ref{sec:magnoneigenstates})
\bea
\ket{S_{\mbf{p},\mu}} &= \sum_{\mbf{k}} M^{\downarrow \uparrow}_\mu(\mbf{k},\mbf{p})\gamma^\dag_{\mbf{k}+\mbf{p},\downarrow} \gamma_{\mbf{k},\uparrow} \ket{\uparrow} \,, \\
M^{\downarrow \uparrow}_\mu(\mbf{k},\mbf{p}) &= \sum_\al U^{*}_{\downarrow,\al,\mbf{k}+\mbf{p}} u^\mu_\al(\mbf{p}) U_{\uparrow,\al,\mbf{k}} \,, \\
\eea
in exact analogy to Eq.$\,$\ref{eq:Mdnup_general_def}. Here $u^\mu(\mbf{p})$ is the $\mu$-th eigenvector of the boson Hamiltonian $\mathcal{W}(\mbf{p})$ in Eq.$\,$\ref{eq:bindingenergies}. It is enlightening to rewrite these states in position space (bar again denotes projection to the flat band):
\bea
\ket{S_{\mbf{p},\mu}} &= \sum_{\mbf{R} \al} e^{i \mbf{p} \cdot (\mbf{R}+\mbf{r}_\al)} u^\mu_\al(\mbf{p}) \bar{c}^\dag_{\mbf{R},\al,\downarrow} \bar{c}_{\mbf{R},\al,\uparrow} \ket{\uparrow} \,,
\eea
from which we see that the projected local spin-flip SyMA basis is a complete basis for the exact magnon eigenstates, in direct analogy with the elementary basis in Eq.$\,$\ref{eq:elementary_magnons}. This proves the exactness of the SyMA in flat band Hubbard models with uniform lattices. The exactness of the SyMA in the more general case with constant $V_\al \bra{\eta}n_{\mbf{R},\al}\ket{\eta}$ is proven in App.$\,$\ref{sec:magnoneigenstates}.

In general, $u^\mu(\mbf{p})$ must be determined from the boson Hamiltonian $\mathcal{W}(\mbf{p})$. However, in the SU(2) symmetric case where $P_\uparrow = P_\downarrow$, the lowest energy magnon is a Goldstone mode with $u_\al(0) = 1/\sqrt{N_{orb}}$ (see App. \ref{app:magnonandSU2}). If we take this eigenvector as a SyMA ansatz in the Ising case, we obtain
\bea
\Delta(\mbf{p}) \leq \frac{U}{N_{orb}} \left( 1 - \int \frac{d^2k}{\Omega_{BZ}}|M^{\uparrow \downarrow}(\mbf{k},\mbf{p})|^2 \right) \,,
\eea
which substantiates the intuition in Sec. \ref{subsec:MoTe2_SMA_eigenfunctions}: topologically-protected nodes in
$M^{\uparrow\downarrow}({\bf k},{\bf p})$ suppress the norm of this
trial wave function and thereby increase the corresponding variational
estimate of the magnon energy. We will pursue explicit bounds in Sec. \ref{sec:boundApp}.

Our results in this family of models relied on the exactly flat particle-hole continuum. To continue our analytical study, in the next section we introduce a simple model with arbitrary Chern number whose particle-hole continuum dispersion can be controlled. We will show that the SyMA remains highly accurate.

\subsection{Ideal-band BHZ model}

\label{subsec:Ideal_band_BHZ_model}

In this section, we turn to a solvable model beyond the flat-charge-excitation limit, where the SyMA is no longer exact. The main purpose of this discussion is to understand why the SyMA can be very accurate even away from the flat continuum limit. To this end, we consider a modified continuum version of the BHZ model$\,$\cite{doi:10.1126/science.1133734} described by the Hamiltonian
\bea
H_{\eta}^{0}=\sum_{{\bf k}}c_{\eta,{\bf k}}^{\dagger}
\left(\begin{array}{cc}
V^{2n} & -(\eta ke^{\eta i\theta}V)^{n}\\
-(\eta ke^{-\eta i\theta}V)^{n} & k^{2n}
\end{array}\right)
c_{\eta,{\bf k}} \,,\\
\label{eq:ideal_band_BHZ_H0}
\eea
where $c_{\eta,{\bf k}}^{\dagger}=(c_{0,\eta,{\bf k}}^{\dagger}\,,c_{1,\eta,{\bf k}}^{\dagger})$,
with $0$ and $1$ denoting different sublattice degrees of freedom, $\eta=\pm1$ denotes the valley/spin, and we denote the momentum in polar and cartesian coordinates as $ke^{\pm i\theta}=k_{x}\pm ik_{y}$. The parameter $V$ tunes the single-particle gap between an exact flat band at $E=0$ and a higher-energy band. The wave function of the flat band is given by,
\begin{equation}
 \psi_{n,\eta}(k/V,\theta)=\big(1+(k/V)^{2n}\big)^{-1/2}\big[(\eta e^{i\eta\theta}k/V)^{n}\,,1\big]^T \,,
 \label{eq:psi_n_eta_BHZ}
\end{equation}
For $\eta=+1$ ($-1$), the flat band wave function in Eq.$\,$\ref{eq:psi_n_eta_BHZ} is holomorphic (anti-holomorphic), and the band has Chern number $C=-\eta n$. The flat band is therefore ideal \cite{Wang2021ExactLandauLevel,WangLiu2022IdealFlatbands,Ledwith2022IdealChernFlatbands}. The parameter $V$ also controls the concentration of the flat band's Berry curvature (the maximum of the Berry curvature in this band is proportional to $V^{-2}$, see Appendix$\,$\ref{subsec:BHZ_berry_curvature}). In the remainder of this section, we will drop the index "$n$" from the flat band wave function in Eq.$\,$\ref{eq:psi_n_eta_BHZ} and other quantities that depend on it, for notational simplicity.

We now consider the Hubbard interaction projected into the flat band, with
\begin{equation}
H_{I}=\frac{1}{2N}\sum_{{\bf q}}\overline{\delta\rho}_{{\bf q}}\overline{\delta\rho}_{-{\bf q}}\,,\label{eq:H_nI_BHZ}
\end{equation}
where $\gamma_{{\bf k},\eta}^{\dagger}$ creates an electron with valley/spin $\eta$ in the flat band.
We take $N=L^2$ and discretize the momentum mesh in units of $2\pi/L$ on a $L\times L$ system. Note that the choice of interaction strength in Eq.$\,$\ref{eq:H_nI_BHZ} is the equivalent of taking $V({\bf q})=1$ in Eq.$\,$\ref{eq:H_V_general}. The projected density operator is
\bea
\overline{\delta\rho}_{{\bf q}}&=\sum_{{\bf k},\eta}M^{\eta}({\bf k},{\bf q})\big(\gamma_{{\bf k}+{\bf q},\eta}^{\dagger}\gamma_{{\bf k},\eta}-\frac{1}{2}\delta_{{\bf q},{\bf 0}}\big)\,\,\,,\\
& M^{\eta}({\bf k},{\bf q})=\psi_{\eta,{\bf k}+{\bf q}}^{\dagger}\psi_{\eta,\bf k} \ .
\eea
Because the model is defined in the continuum, we introduce a cutoff $\Lambda$ in all calculations with the prescription $\gamma_{|{\bf k}| > \Lambda,\eta} = 0$.

We can compute the single charged excitation energies via Eq.$\,$\ref{eq:Rmat} to find (see App. \ref{subsec:BHZ_charge_gap})
\begin{equation}
\begin{array}{c}
R^{\eta}(k_{v})=\frac{(-1)^{-1/n}V^2}{8 \pi n(1+k_{v}^{2})} \Big(\mathcal{B}_{-x^{-n}}(\frac{1}{n},0)-k_{v}^{2}\mathcal{B}_{-x^{-n}}(\frac{n+1}{n},0)\Big) \,,
\end{array}\label{eq:Rmat_BHZ_general}
\end{equation}
where we defined $k_{v}=k/V, x=(V/\Lambda)^{2}$ and the Beta function $\mathcal{B}_{z}(a,b)\equiv\int_{0}^{z}t^{a-1}(1-t)^{b-1}\,dt$. To leading order in $x$, we
can show that the charge-1 gap is minimal at $k=0$ and we have
\begin{equation}
R^{\eta}(k_{v}=0) = \,\frac{V^2}{8n} \begin{cases}
\frac{1}{\pi}\log(\frac{1}{x}), & n=1\\
\csc(\pi/n), & n>1
\end{cases} \quad (x\rightarrow0)\,.
\label{eq:BHZ_R_smallx}
\end{equation}
so that the charge gap is increasing in $V^2$.

Next, employing a straightforward adaptation of Eq.$\,$\ref{eq:S_matrix_commutator}, we can obtain the magnon spectrum. Example numerical results are shown in Figs.$\,$\ref{fig:BHZ}(a,b) for varying $n$ and $V$, where we observe at most three bound modes separated from the continuum. We will now derive these modes analytically and explain them via the SyMA.

We again exploit the low-rank structure of the interaction. We write the magnon Hamiltonian as a matrix in $\mbf{k},\mbf{k}'$ in the form
\begin{equation}
S(\mbf{p}) = \mathcal{R}({\bf p})-\frac{1}{L^2} \mathcal{U}{(\bf p})\mathcal{U}^{\dagger}({\bf p}) \ ,
\label{eq:Smatrix_BHZ}
\end{equation}
where we defined the diagonal matrix $[\mathcal{R}(\mbf{p})]_{{\bf k},{\bf k}'}=[R^{\down}({\bf k}+{\bf p})+R^{\up}({\bf k})]\delta_{{\bf k},{\bf k}'}$, and the rectangular matrix
$
\mathcal U_{{\bf k},(\alpha,\beta)}({\bf p})
=
\psi^{\alpha*}_{\down,{\bf k}+{\bf p}}\psi^{\beta}_{\up,{\bf k}}$,
with row indices ${\bf k}$ and column indices $(\alpha,\beta)$.
For a bound-state energy $\lambda$ lying outside the spectrum of $\mathcal{R}({\bf p})$, we have $\det[\mathcal{R}({\bf p})-\lambda]\neq0$, and the eigenvalue equation $\det[S({\bf p})-\lambda\mathbb{1}]=0$ can be reduced to
\begin{equation}
\det\big(\mathcal{R}({\bf p})-\lambda\big)\det\Big(1-\frac{1}{L^2(\mathcal{R}({\bf p})-\lambda)}\mathcal{U}({\bf p})\mathcal{U}^{\dagger}({\bf p})\Big)=0\,.
\end{equation}
We will further use the matrix determinant lemma $\det(\mathbb{1}+AB)=\det(\mathbb{1}+BA)$, valid for any rectangular
matrices $A,B$. Taking $A=\big(\mathcal{R}(\bf p) - \lambda \big)^{-1} \mathcal{U}({\bf p})$ and $B=\mathcal{U}^{\dagger}({\bf p})$, the bound-states can be found by solving $\det \big( \mathbb{1}-T_{\lambda}({\bf p}) \big) = 0$ for $\lambda$, where
\bea
T_{\lambda}({\bf p})&=\frac{1}{L^2}[\mathcal{U}^{\dagger}({\bf p})\frac{1}{\mathcal{R}({\bf p})-\lambda}\mathcal{U}({\bf p})] \\
&=\int \frac{d^2 k}{(2\pi)^2}\,\frac{P^{\down}({\bf k}+{\bf p})\otimes P^{\up}({\bf k})^T}{R^{\down}({\bf k}+{\bf p})+R^{\up}({\bf k})-\lambda} \,,
\label{eq:Tp_kernel}
\eea
\noindent where $P^{\eta}({\bf k}) = \psi_{\eta,{\bf k}} \psi^\dagger_{\eta,{\bf k}}$, and hence we have reduced the eigenvalue problem to a $4\times 4$ matrix $T_{\lambda}({\bf p})$.
It is possible to obtain a closed-form expression for $T_{\lambda}({\bf p})$ at $\mbf{p}=0$, that we provide in App. \ref{subsec:BHZ_exact_magnon_energies}. In particular, it can be shown that $T_{\lambda}({\bf p}=0)$ has rank 3, so there can be at most three magnon bound states. These bound modes can be obtained as solutions of an implicit equation, as shown in detail in App. \ref{subsec:BHZ_exact_magnon_energies}.
The exact bound modes as a function of $V$ are shown
in Figs.$\,$\ref{fig:BHZ}(c,d), together with the particle-hole continuum that
can also be obtained analytically using Eq.$\,$\ref{eq:Rmat_BHZ_general}. The model has a special point where the continuum
becomes exactly flat, in which case the simpler approach in Sec.$\,$\ref{subsec:Flatband_Hubbard_models}
can be employed.

More generally, the reduction in Eq.$\,$\ref{eq:Tp_kernel} applies to
models with a finite number of orbitals and local interactions of the type in Eq.$\,$\ref{eq:H_nI_BHZ}, for
which the interaction kernel has finite rank even when the
particle-hole continuum is dispersive (see for instance Ref. \cite{2025arXiv250821127H}).

\subsection{SyMA in the Ideal-band BHZ model}

\label{subsec:idealbandBHZ_SyMA}

Having obtained the exact expression in Eq.$\,$\ref{eq:Tp_kernel}, that allows for the exact calculation of the magnon bound modes, we will now formulate the SyMA and compare it with these exact
results. We start by considering a general expression for the SyMA mode given by
\bea
S_{D}^{\dagger}({\bf p})&=\sum_{\alpha,\alpha'}\int d^{2}{\bf r}\,\,e^{i{\bf p}\cdot{\bf r}}\bar{c}_{\alpha,\down,{\bf r}}^{\dagger}D_{\alpha\alpha'}\bar{c}_{\alpha',\up,{\bf r}}\\
&=\sum_{{\bf k}}M_{D}^{(\down,\up)}({\bf k},{\bf p})\gamma_{{\bf k}+{\bf p},\down}^{\dagger}\gamma_{{\bf k},\up}\,\,,
\eea
where $\bar{c}_{\alpha,\eta,{\bf r}}^{\dagger}=N^{-1/2}\sum_{{\bf k}}\,e^{-i{\bf k}\cdot{\bf r}}\bar{c}_{\alpha,\eta,{\bf k}}^{\dagger}$ and once again the over bar denotes projection onto the flat band.
The SyMA magnon wave function is
\bea
M_{D}^{(\down,\up)}({\bf k},{\bf p})&=\psi_{\down,{\bf k}+{\bf p}}^{\dagger}D\psi_{\up,{\bf k}} \,.
\label{eq:MD_idealband}
\eea
Just like in previous sections, we will define different ansatze through
different choices of $D$ that respect the symmetries of the model
at ${\bf p}=0$.

At ${\bf p}=0$, the model has a SO(2) rotational symmetry.
As shown in App.$\,$\ref{subsec:BHZ_magnon_gaps_SyMA}, the Fock-space representation of SO(2) rotations in this model is
\begin{equation}
W(\phi)=e^{-i\phi J_z},
\qquad
J_z=
\sum_{\eta,{\bf k}}
c_{\eta,{\bf k}}^\dagger
\left(
-i\partial_\theta-\frac{n\eta}{2}\sigma_z
\right)
c_{\eta,{\bf k}},
\label{eq:BHZ_Jz_main}
\end{equation}
where $J_z$ is the total angular-momentum operator. We define the angular-momentum eigenvalue $w$ of a generic
zero-momentum magnon mode through
\begin{equation}
W(\phi)S_{D}^\dagger{({\bf 0})} W^\dagger(\phi)
=
e^{-iw\phi}S_{D}^\dagger{({\bf 0})}.
\label{eq:BHZ_w_definition_main}
\end{equation}

In addition, the interaction Hamiltonian in
Eq.$\,$\ref{eq:H_nI_BHZ} has the same form as in Eq.$\,$\ref{eq:H_V_general} and therefore has the same unitary particle-hole symmetry
$\mathcal C$ introduced previously in Sec.$\,$\ref{sec:symmetries_fhmodel}. This symmetry is
defined for every magnon momentum ${\bf p}$ and acts on the local
spin-flip operators as
\begin{equation}
\mathcal C S_D^\dagger({\bf p})\mathcal C^{-1}
=
S_{D^T}^\dagger({\bf p}).
\label{eq:BHZ_C_SD_main}
\end{equation}
Thus, a mode with $\mathcal C$-eigenvalue $\zeta=\pm1$ satisfies
\begin{equation}
\mathcal C S_D^\dagger({\bf p})\mathcal C^{-1}
=
\zeta S_D^\dagger({\bf p}),
\, \textrm{with }
D^T=\zeta D.
\label{eq:BHZ_C_D_main}
\end{equation}

We now use these symmetries to construct the SyMA modes by specifying the corresponding $D$ matrices. Starting
with an arbitrary complex $2\times2$ $D$ matrix,
\begin{equation}
D=
\begin{pmatrix}
d_{11} & d_{12}\\
d_{21} & d_{22}
\end{pmatrix}.
\end{equation}
and using the flat band wave function in
Eq.$\,$\ref{eq:psi_n_eta_BHZ}, the corresponding magnon wave function
at ${\bf p}=0$ is
\begin{equation}
\begin{aligned}
M_D^{(\down,\up)}({\bf k},{\bf 0})
={}&\frac{1}{1+k_v^{2n}}
\Big[
(-1)^n d_{11}e^{2in\theta}k_v^{2n}
\\
&\quad
+\big((-1)^n d_{12}+d_{21}\big)e^{in\theta}k_v^n
+d_{22}
\Big].
\end{aligned}
\label{eq:general_D_BHZ}
\end{equation}
The three terms in Eq.$\,$\ref{eq:general_D_BHZ} transform with
angular-momentum eigenvalues $w=2n$, $w=n$, and $w=0$.
Therefore, the coefficients of the $D$ matrix must be chosen in order to select only one of these sectors.

The $w=2n$ and $w=0$ sectors are generated
by
\begin{equation}
D=\frac{1+\sigma_z}{2}
\qquad\text{and}\qquad
D=\frac{1-\sigma_z}{2},
\end{equation}
respectively. Both matrices are symmetric and hence have
$\mathcal C$ eigenvalue $\zeta=1$.

The $w=n$ sector is generated by the two off-diagonal matrix elements.
Imposing a definite particle-hole eigenvalue,
$D^T=\zeta D$, fixes their relative coefficient. For $\zeta=1$ one
obtains $D=\sigma_x$, whereas for $\zeta=-1$ one obtains
$D=\sigma_y$, up to an overall normalization. Consequently, the four
simultaneous symmetry eigenmatrices and corresponding projected magnon wave functions are
\bea
D=\frac{1+\sigma_{z}}{2}: & \quad &
M_{D}^{(\down,\up)}({\bf k},{\bf 0})
=
\frac{(-1)^{n}e^{2in\theta}k_{v}^{2n}}
{1+k_{v}^{2n}},
\\
D=\sigma_{y}: & \quad &
M_{D}^{(\down,\up)}({\bf k},{\bf 0})
=
\frac{i[1-(-1)^{n}]e^{in\theta}k_{v}^{n}}
{1+k_{v}^{2n}},
\\
D=\sigma_{x}: & \quad &
M_{D}^{(\down,\up)}({\bf k},{\bf 0})
=
\frac{[1+(-1)^{n}]e^{in\theta}k_{v}^{n}}
{1+k_{v}^{2n}},
\\
D=\frac{1-\sigma_{z}}{2}: & \quad &
M_{D}^{(\down,\up)}({\bf k},{\bf 0})
=
\frac{1}{1+k_{v}^{2n}}.
\label{eq:Mk_different_modes_nmodel}
\eea
Equation$\,$\ref{eq:Mk_different_modes_nmodel} further shows that the
$\mathcal C$-even $w=n$ mode generated by $D=\sigma_x$ vanishes
identically for odd $n$, whereas the $\mathcal C$-odd $w=n$ mode
generated by $D=\sigma_y$ vanishes identically for even $n$.
Therefore, for any fixed $n$, only three of the four symmetry-adapted
local operators produce nonvanishing SyMA modes. Their symmetry
eigenvalues are summarized in Table$\,$\ref{tab:n_model_winding_table}.

\begin{table}
\begin{centering}
\begin{tabular}{|c|c|c|}
\hline
$D$ & $w$ & $\zeta$\tabularnewline
\hline
\hline
$(1+\sigma_{z})/2$ & $2n$ & $1$\tabularnewline
\hline
$\sigma_{y}$ (odd $n$ only) & $n$ & $-1$\tabularnewline
\hline
$\sigma_{x}$ (even $n$ only) & $n$ & $1$\tabularnewline
\hline
$(1-\sigma_{z})/2$ & $0$ & $1$\tabularnewline
\hline
\end{tabular}
\par\end{centering}
\caption{$\mathcal{C}$ eigenvalue $\zeta$ and total angular momentum eigenvalue $w$ of all the modes
in Eq.$\,$\ref{eq:Mk_different_modes_nmodel} for the Chern number $n$ model.
\label{tab:n_model_winding_table}}
\end{table}

The choices of $D$ in Eq.$\,$\ref{eq:Mk_different_modes_nmodel}
also coincide with the three nonzero ${\bf p}=0$ eigenvectors of the reduced kernel
$T_\lambda({\bf p}=0)$ in Eq.$\,$\ref{eq:Tp_kernel}, which can be read as $D_{\alpha'\alpha}$,
combining $(\alpha',\alpha)$ into a single index (see App. \ref{subsec:BHZ_exact_magnon_energies}).
The SyMA wave functions are, however, not exact
eigenstates of the full magnon Hamiltonian in general. At the special point where
the particle-hole continuum is flat, we have
$\mathcal{R}({\bf 0})=R_0\mathbb{1}$, and the S-matrix in Eq.$\,$\ref{eq:Smatrix_BHZ} at ${\bf p}=0$ becomes
\begin{equation}
S({\bf 0})
=
R_0\mathbb{1}
-
\frac{1}{L^2}\mathcal{U}({\bf 0})\mathcal{U}^{\dagger}({\bf 0}).
\end{equation}
Therefore, if $D$ is an eigenvector of
$L^{-2}\mathcal{U}^{\dagger}({\bf 0})\mathcal{U}({\bf 0})$, then
$\sum_{\alpha,\alpha'}\mathcal{U}_{{\bf k},(\alpha',\alpha)}D_{\alpha'\alpha}=M_{D}^{(\down,\up)}({\bf k},{\bf 0})$ is an exact
eigenstate of $S({\bf 0})$. Hence, the SyMA is exact at this point.
Away from the flat-continuum limit,
$[\mathcal{R}({\bf 0})]_{{\bf k},{\bf k}'}=[R^{\down}({\bf k})+R^{\up}({\bf k})]\delta_{{\bf k},{\bf k}'}$ is
${\bf k}$-dependent, which implies that
the SyMA is no longer an exact eigenstate of the full magnon Hamiltonian, although  the corresponding $D$ matrix remains an eigenvector of the reduced kernel
 in Eq.$\,$\ref{eq:Tp_kernel}.

The rank-3 structure of $T_\lambda({\bf 0})$ mentioned at the end of Sec.$\,$\ref{subsec:Ideal_band_BHZ_model} also follows directly from Eq.$\,$\ref{eq:Mk_different_modes_nmodel}. For odd $n$, the $D=\sigma_x$ mode vanishes identically after projection, whereas for even $n$ the $D=\sigma_y$ mode vanishes. The remaining three projected wave functions have distinct angular momenta $w=0,n,2n$ and are therefore linearly independent. Hence $\mathcal U({\bf 0})$ has rank three, and consequently so does $T_\lambda({\bf 0})$ for $\lambda$ outside the particle-hole continuum.

\begin{figure}
\begin{centering}
\includegraphics[width=1\columnwidth]{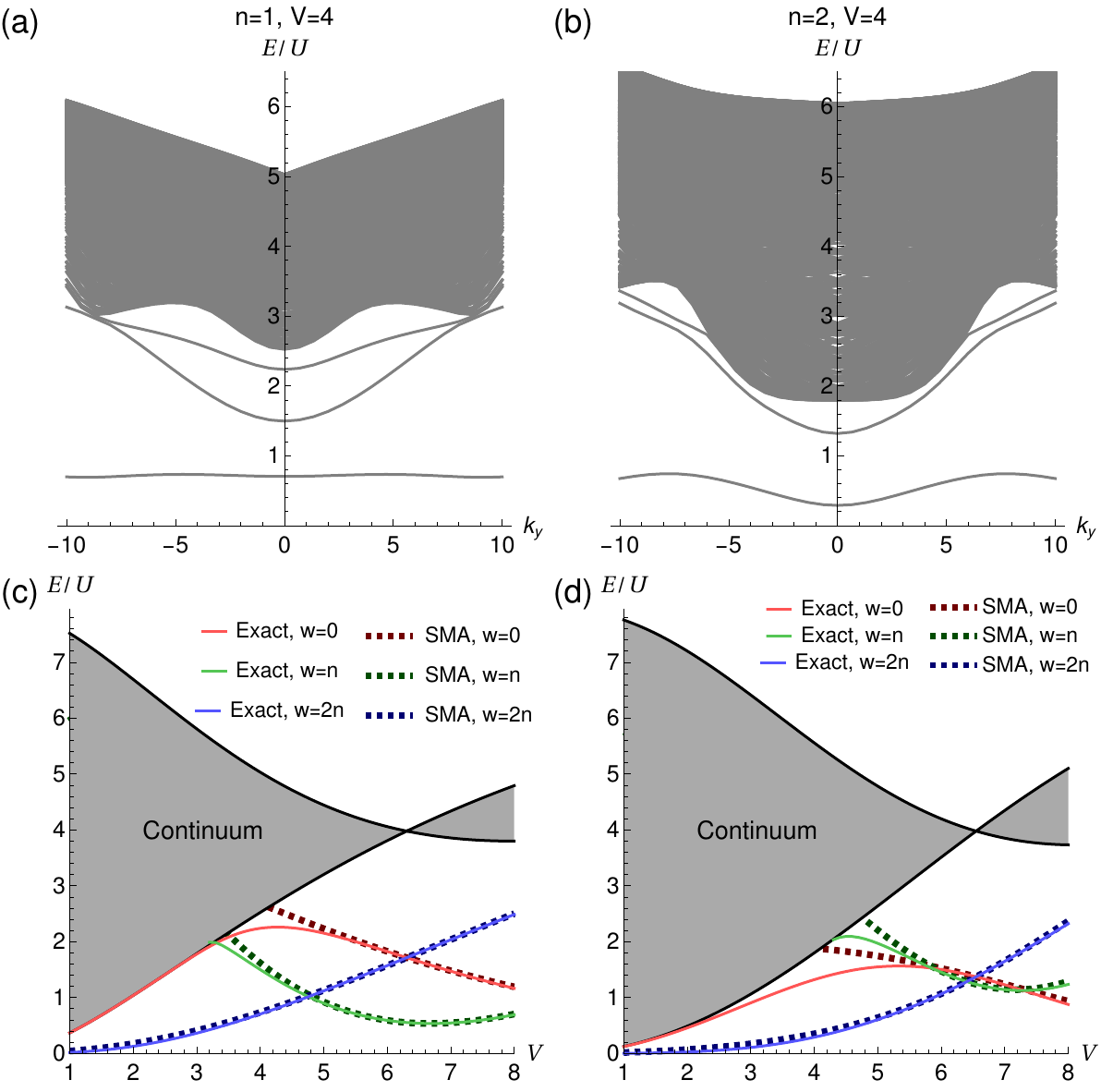}
\par\end{centering}
\caption{(a,b) Exact magnon mode spectra obtained numerically for $V=4$ and $\Lambda=10$, with $n=1$
(a) and $n=2$ (b). Three bound states are seen. (c,d) Exact and analytical SyMA results for the magnon bound modes
at ${\bf p}=0$ for $n=1$ (c) and $n=2$ (d). \label{fig:BHZ}}

\end{figure}

The SyMA magnon gaps at ${\bf p}=0$ can be evaluated analytically in
closed form, as shown in App.$\,$\ref{subsec:BHZ_magnon_gaps_SyMA}.
In Figs.$\,$\ref{fig:BHZ}(c) and (d), we compare these analytical results
with the exact bound-state energies and find very good agreement.

\section{Topological Bound on the Magnon Gap}

Using the SyMA in MoTe$_2$ allowed us to identify topologically protected momentum-space nodes as playing a key role in controlling the zero-momentum magnon gap. In this section, we make this connection quantitative by proving a bound when the SyMA is exact in the flat band Hubbard models. We will then derive a related approximate expression in the ideal-band BHZ model, where the SyMA is accurate but not exact.

\label{subsec:analytical_maxF_relations}
\subsection{Exact bound in Flat Band Hubbard models}
\label{sec:boundApp}

We now return to the flat band Hubbard models introduced in Sec. \ref{subsec:Flatband_Hubbard_models}. To formulate our bound, we will restrict ourselves to 2-band models (per spin) with (spin-less) $C_2$ symmetry exchanging the orbitals. In this case, the representation of $C_2$ can be chosen as $\sigma_x$ without loss of generality). This class of Hamiltonians includes the Haldane model, and supports bands of arbitrary Chern number. We define the lowest energy magnon branch as $\Delta(\mbf{p})$ (which is simply the momentum-resolved magnon gap). Our full result is a two-sided quantum geometric bound for $\Delta({\mbf 0})$,
\bea
\label{eq:2sidebound}
\frac{\pi}{3} U \frac{|C|}{ \max_\mbf{k}|F(\mbf{k})| \Omega_{BZ}} \leq \Delta({\mbf 0}) \leq 4\pi U \frac{G}{\Omega_{BZ}|\mbf{a}|^2} \,,
\eea
where $G \geq |C|$ is the integrated Fubini-Study metric and $\mbf{a}$ is the smallest lattice vector. For square lattices, $\Omega_{BZ}|\mbf{a}|^2 = (2\pi)^2$ and for triangular lattices, $\Omega_{BZ}|\mbf{a}|^2 = \frac{2}{\sqrt{3}} (2\pi)^2$.
Eq.$\,$\ref{eq:2sidebound} shows that the band topology, quantum geometry, and distribution of Berry curvature constrain the zero-momentum magnon gap in the flat band limit.
In particular, Chern bands support a \emph{nonzero} lower bound, which is higher when the Berry curvature is smoothly distributed. Hot spots reduce this bound, consistent with tuning the band through a topological phase transition, where $F(\mbf{k})$ will diverge at the band inversion point. Similarly, trivial atomic bands where $G$ can be taken zero have an upper bounded magnon gap that can be taken to zero, implying weak magnetism.

We now present our proof. First, since Eq.$\,$\ref{eq:bindingenergies} yields the binding energies below the continuum at $U/2$, we see that the magnon gaps above the ground-state are the eigenvalues of the Hamiltonian
\bea
\mathcal{H}_{\al \be}(\mbf{p}) &= \frac{U}{2} \delta_{\al \be} - \mathcal{W}_{\al \be}(\mbf{p}) \\
&= \frac{U}{N} \sum_{\mbf{k}} Q^{\downarrow}_{\al \be}(\mbf{k}+\mbf{p}) P^{\uparrow}_{\be \al}(\mbf{k}) \\
&= \frac{U}{N} \sum_{\mbf{k}} Q^*_{\al \be}(-\mbf{k}-\mbf{p}) P_{\be \al}(\mbf{k}) \,, \\
\eea
where $P = 1- Q$ is the spin-$\uparrow$ projector and we used the identity $\frac{1}{N} \sum_\mbf{k} P^{\uparrow}_{\al \al}(\mbf{k}) = \frac{1}{2}$$\,$\cite{2022arXiv220900007H,2022PhRvB.106a4518H}. Next, we use the symmetry representation $\sigma_x \mathcal{H}(\mbf{p})\sigma_x = \mathcal{H}(-\mbf{p})$, which means that at $\mbf{p}=0$, the eigenvectors of $\mathcal{H}$ are $u_\pm = (1,\pm 1)^T/\sqrt{2}$. We must determine which eigenvector has lower-energy. In fact, $u_-$ has exactly zero binding energy, and hence is in the continuum as is shown in Fig.$\,$\ref{fig:4}(c). This is because of a particle-hole symmetry of the two-band model with $C_2 = \sigma_x$: $\sigma_z P(\mbf{k})^* \sigma_z = Q(-\mbf{k})$. With this identity, we find
\bea
u^\dag_- \mathcal{H}(\mbf{0}) u_- &= \frac{U}{2N} \sum_{\mbf{k}} \Tr \big[ \, \sigma_z Q^*(-\mbf{k}) \sigma_z P(\mbf{k}) \big] = \frac{U}{2} \,, \\
\eea
which is equal to the energy of the particle-hole continuum. Thus the lowest energy excitation at $\mbf{p}=0$ is the $s$-wave magnon $u_+$ with gap
\bea
\Delta({\mbf 0}) = u^\dag_+ \mathcal{H}(\mbf{0}) u_+ &= \frac{U}{2N} \sum_{\mbf{k}} \Tr [ Q^*(-\mbf{k}) P(\mbf{k}) ] \, \ . \\
\eea
We see that $\Delta({\mbf 0}) = 0$ if and only if $P(\mbf{k})=P(-\mbf{k})^*$, giving this expression a simple interpretation: it is the amount of spinless time-reversal breaking in the wave function. This suggests that a nonzero Chern number, which prevents spinless time-reversal symmetry, should provide a nonzero lower bound. To quantify this, we use the expression $P(\mbf{k}) = \frac{1}{2}(1+ \hat{\mbf{d}}(\mbf{k}) \cdot \pmb{\sigma})$ where the flat band Hamiltonian is written $h(\mbf{k}) = \mbf{d}(\mbf{k}) \cdot \pmb{\sigma}$. We obtain
\bea
\Delta({\mbf 0}) &= \frac{U}{2N} \sum_{\mbf{k},ij} \frac{1}{4} \Tr (\mathbb{1} - \hat{d}_i(-\mbf{k}) \hat{d}_j(\mbf{k}) \sigma^*_i \sigma_j) \\
&= \frac{U}{2N} \sum_{\mbf{k}} \frac{1}{2} (1 - \hat{d}_x(\mbf{k})^2 - \hat{d}_y(\mbf{k})^2 + \hat{d}_z(\mbf{k})^2) \\
&= \frac{U}{2N} \sum_{\mbf{k}} \hat{d}_z(\mbf{k})^2 = \frac{U}{2} \int \frac{d^2k}{\Omega_{BZ}} \hat{d}_z(\mbf{k})^2 \,,
\label{eq:Ising_gap_Haldane_exact}
\eea
using the fact that $d_y$ and $d_z$ are odd in $\mbf{k}$ with the representation $C_2 = \sigma_x$, and that $|\hat{\mbf{d}}|^2 = 1$. In the last line, we took the thermodynamic limit. We will now change variables from $\hat{\mbf{d}}$ to angles on the Bloch sphere. To be explicit, we write
\bea
\hat{d}_z(\mbf{k}) &= \cos \theta (\mbf{k}), \ \hat{d}_x(\mbf{k}) \pm i \hat{d}_y(\mbf{k}) = e^{i \phi(\mbf{k})} \sin \theta (\mbf{k}) \,, \\
\eea
from which one derives the well-known identity for the Jacobian $J(\mbf{k}) = \det \frac{\del (\theta,\phi)}{\del \mbf{k}}$:
\bea
\sin \theta \, J(\mbf{k}) = \hat{d} \cdot (\del_{k_x} \hat{d} \times \del_{k_y} \hat{d}) = 2 F(\mbf{k}) \,,
\eea
where $F(\mbf{k})$ is the Berry curvature. To introduce the Jacobian, we write $F_{max} = \max_\mbf{k} |F(\mbf{k})|$ and
\bea
\Delta({\mbf 0}) &\geq \frac{U}{2} \int \frac{d^2k}{\Omega_{BZ}} \hat{d}_z(\mbf{k})^2 \frac{|F(\mbf{k})|}{F_{max}} \\
&= \frac{U}{4 F_{max}} \int \frac{d^2k}{\Omega_{BZ}} \sin \theta \, |J(\mbf{k})| \cos^2 \theta(\mbf{k}) \\
&\geq \frac{U|C|}{4 F_{max}} \int \frac{\sin \theta d\theta d\phi}{\Omega_{BZ}} \cos^2 \theta \\
&= \frac{U|C|}{4 F_{max}} \frac{4\pi}{3\Omega_{BZ}} \\
&= \frac{\pi}{3} U \frac{|C|}{F_{max} \Omega_{BZ}} \,, \\
\label{eq:Ising_gap_Haldane_bound}
\eea The third line requires a generalized change-of-variables integration formula which does not assume that the map between the two domains (the BZ and Bloch sphere) is bijective, since multiple $\mbf{k}$ point may map to the same angles $\pmb{\theta}$ on the Bloch sphere. This requires the more general formula $\int f(\mbf{k}) |J(\mbf{k})| d^2k = \int f(\pmb{\theta}(\mbf{k})) \#(\pmb{\theta}) d\Omega$ where $\#(\pmb{\theta}) \in \mathbb{N}$ is the number of $\mbf{k}$-points mapped to $\pmb{\theta}$. If the Chern number is $C$, then $\hat{d}(\mbf{k})$ winds the Bloch sphere $|C|$ times and hence $\#(\pmb{\theta}) \geq |C|$. This completes the proof.

A few comments are in order. First, we see that $F_{max}$ appears in the denominator of Eq. \ref{eq:Ising_gap_Haldane_bound}, capturing the fact that if the Berry curvature is highly peaked, the magnon gap can be small even if $C \neq 0$. Second, we consider whether our bound can be saturated. The first inequality in the proof of Eq. \ref{eq:Ising_gap_Haldane_bound} is an equality if $|F(\mbf{k})| = F_{max}$ is constant and Berry curvature is uniform. While it is possible to construct such models using exponentially decaying hoppings (see Ref. \cite{2022ScPP...12..118V}), there is an obstruction in the two-band case relevant to our proof. This is because the Berry curvature of a two-band model is forced to have a zero\cite{PhysRevB.104.045103,2021PhRvB.104d5104M}, and thus cannot be made flat if $|C| \neq 0$.

In the Haldane model with only two hoppings, the ratio between the lower bound and the exact magnon gap peaks at $66\%$  [Fig.$\,$\ref{fig:Hubbard_BHZ_maxF}(b)], near the point where the band is closest to ideal, as indicated by the minimum of $G-C$ in Fig.$\,$\ref{fig:Hubbard_BHZ_maxF}(a).
For comparison, in Fig.$\,$\ref{fig:Hubbard_BHZ_maxF}(c) we also evaluate the same lower-bound expression for the FH model of MoTe$_2$ with a Hubbard interaction. Although the bound has not been proven for this model, the resulting estimate can also be very tight in the regime where the Berry curvature is flatter.

We remark that although the magnon gap is always at ${\bf p}=0$ for the Haldane model example, this is not necessarily the case in general. For instance, for the continuum models of MoTe$_2$ (both the FH model and the more realistic models that will be considered in Sec.$\,$\ref{subsec:Realistic_MoTe2}), the magnon gap can also occur at ${\bf p}={\bf K},{\bf K'}$, in which case we also observe that the lowest-energy magnon branch is weakly dispersive with ${\bf p}$ (see Fig.$\,$\ref{fig:App_MoTe2_SMA_energies} in App. \ref{subsec:App_Additional_MoTe2_SMA} and Fig.$\,$\ref{fig:MoTe2_realistic}(c) below). The possible existence of a more general bound for all ${\bf p}$ remains an open question and a fruitful direction for future work.

Finally, we prove the upper bound of Eq. \ref{eq:2sidebound}. We first need a lemma known as Poincare's inequality (see also Ref. \cite{2022PhRvL.128h7002H} for other uses of this bound in quantum geometry). For any periodic function $d(\mbf{k}) = \sum_\mbf{R} e^{i \mbf{k} \cdot \mbf{R}} d_\mbf{R}$, if $d_0 = 0$, then
\bea
\int \frac{d^2k}{\Omega_{BZ}}|\nabla d(\mbf{k})|^2 &= \sum_\mbf{R} |\mbf{R}|^2 |d_\mbf{R}|^2 \geq |\mbf{a}|^2 \int \frac{d^2k}{\Omega_{BZ}}|d(\mbf{k})|^2 \ .
\eea
where $\mbf{a}$ is the smallest lattice vector. Since $d_z(\mbf{k})$ is odd, it obeys the conditions of the inequality and hence Eq. \ref{eq:Ising_gap_Haldane_exact} yields
\bea
\Delta({\mbf 0}) &= \frac{U}{2} \int \frac{d^2k}{\Omega_{BZ}} \hat{d}_z(\mbf{k})^2 \\
&\leq \frac{U}{2|\mbf{a}|^2}\int \frac{d^2k}{\Omega_{BZ}} |\nabla \hat{d}_z(\mbf{k})|^2 \\
&\leq \frac{U}{2|\mbf{a}|^2}\int \frac{d^2k}{\Omega_{BZ}} |\nabla \hat{d}(\mbf{k})|^2 \\
&= \frac{2U}{|\mbf{a}|^2}\int \frac{d^2k}{\Omega_{BZ}} \Tr \, g(\mbf{k}) \\
&\equiv \frac{4\pi U}{\Omega_{BZ} |\mbf{a}|^2} G \,, \\
\label{eq:upper_bound_Hubbard}
\eea
where we used the identity $\frac{1}{4}\del_i \hat{d} \cdot \del_j \hat{d} = \frac{1}{2} \Tr \, \del_i P \del_j P = g_{ij}(\mbf{k})$ for the Fubini-Study metric. This bound is shown in the Haldane model in Fig. \ref{fig:Hubbard_BHZ_maxF}(b), in cyan. It is minimized when the band is most ideal and $G = 1.07 > C=1$ and reaches $\sim 50\%$ saturation. Our bound shows that quantum geometry sets an upper limit on the magnon gap, and goes to zero (where it is a tight inequality) when the band is in the trivial atomic limit (and hence the sites are totally decoupled in the flat band limit). In a topological band, this bound cannot be saturated. This is because saturation would require $|\nabla \hat{d}_z|^2 = |\nabla \hat{d}|^2$ which is only possible if $|\nabla \hat{d}_x|^2 = |\nabla \hat{d}_y|^2 = 0$. But such a band cannot have a Chern number since it cannot wind the Bloch sphere.

\subsection{Approximate analytical expression in ideal-band BHZ model}
\label{subsec:Ideal_band_BHZ_analytical_expression}

For the ideal-band BHZ model in Sec.$\,$\ref{subsec:Ideal_band_BHZ_model}
it is also possible to obtain a simple expression relating the zero-momentum SyMA
magnon gap with $\max_{{\bf k}}|F({\bf k})|$ and $C$ for
small $x=(V/\Lambda)^{2}$. In this limit, we can show that the SyMA modes with angular-momentum eigenvalues $w=0$ and $w=n$ merge into the particle-hole continuum, while the mode with
$w=2n$ is always below (see App.$\,$\ref{subsec:BHZ_magnon_gaps_SyMA}).

Therefore, in this regime, the SyMA magnon gap is determined by the
mode with $w=2n$ and it is given by

\begin{equation}
(V/\Lambda)\rightarrow0:\,
\Delta_{w=2n}^{\textrm{SyMA}}({\bf 0})
=
\frac{f_{n}}{\max_{{\bf k}}|F({\bf k})|}\,,
\label{eq:BHZ_SMA_asymptotic}
\end{equation}
with
\begin{equation}
\begin{aligned}
f_{1}&=\frac{1}{4\pi}, && n=1,\\
f_{n}&=\frac{(n-1)^{2-1/n}(n+1)^{1+1/n}\pi\csc(\pi/n)}
{16 \pi n^{3}}, && n>1,\\
f_n&=\frac{n-1}{16\pi}+\mathcal{O}(1/n), && n\gg1.
\end{aligned}
\label{eq:nmodel_Isinggap_largen_simple}
\end{equation}

\begin{figure}

\begin{centering}
\includegraphics[width=1\columnwidth]{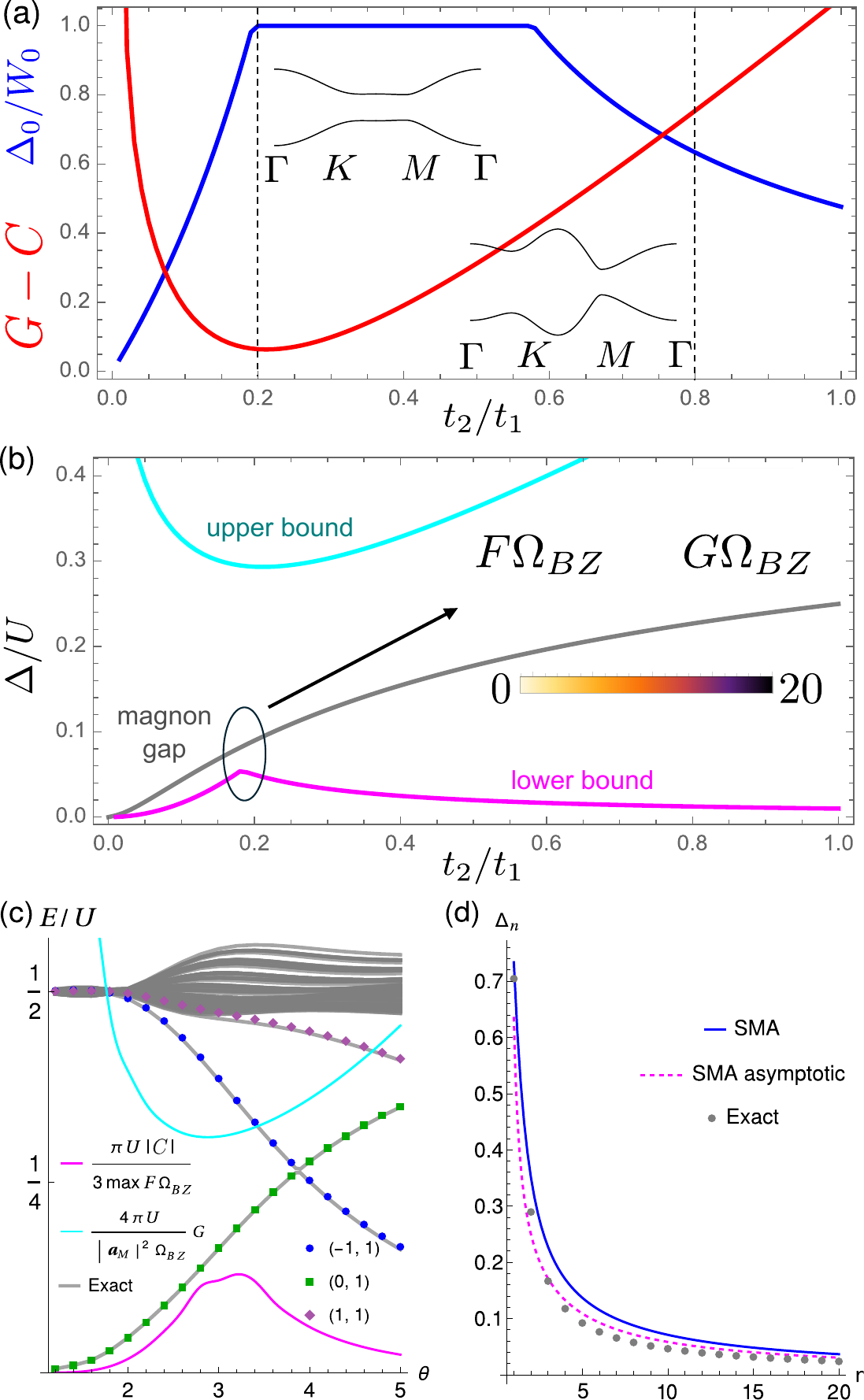}\caption{
(a) Ratio between single-particle gap $\Delta_0$ and bandwidth $W_0$ (blue) and ideal geometry violation (red), the difference between integrated quantum metric and Chern number, for the Haldane model as a function of the ratio between next-nearest-neighbor and nearest-neighbor hoppings $t_2/t_1$. Band structures are shown for $t_2/t_1 = 0.2, 0.8$. (b) Comparison between exact magnon gap obtained for the flat band Haldane model using Eq.$\,$\ref{eq:Ising_gap_Haldane_exact} (gray), lower bound in Eq.$\,$\ref{eq:Ising_gap_Haldane_bound} (pink) and upper bound in Eq.$\,$\ref{eq:upper_bound_Hubbard} (cyan). The inset shows the momentum-resolved Berry curvature and quantum geometry at $t_2/t_1 = 0.2$ where the two bounds are tightest. This also corresponds to the most ideal point.
(c) Comparison between the magnon spectrum at ${\bf p}=0$ and the bounds in Eqs.$\,$\ref{eq:Ising_gap_Haldane_bound} and \ref{eq:upper_bound_Hubbard} computed for the FH model of MoTe$_2$ with a Hubbard interaction, defined by choosing a constant (${\bf q}$-independent) $V({\bf q})$ in Eq.$\,$\ref{eq:H_V_general}. Since the particle-hole continuum of the two-orbital flat band Hubbard model lies at energy $U/2$, we define an effective interaction strength for the FH model by setting $U/2$ equal to the average energy of the spin-flip modes in the particle-hole continuum at ${\bf p}=0$.
(d) Comparison between exact and SyMA results for the magnon gap at
${\bf p}=0$, for the ideal-band BHZ model in Sec.$\,$\ref{subsec:Ideal_band_BHZ_model},
for $V=4$ and $\Lambda=10$. The full and dashed lines correspond
respectively to the full SyMA result (see App. \ref{subsec:BHZ_magnon_gaps_SyMA}
for complete expression) and the simplified asymptotic expression
in Eq.$\,$\ref{eq:BHZ_SMA_asymptotic}. \label{fig:Hubbard_BHZ_maxF}}
\par\end{centering}
\end{figure}

Eq.$\,$\ref{eq:nmodel_Isinggap_largen_simple} shows that, in this small $x$ limit, the SyMA gap takes a simple form controlled by $n=|C|$ and $\max_{\bf k}|F({\bf k})|$.
It is however important to remark that at small $x = (V/\Lambda)^2$, the asymptotic
behavior of the exact gap is different from the SyMA result. Taking
$n=1$ as an example, it can be shown that $\Delta_{n=1}^{\textrm{exact}}({\bf 0})\approx V^{2} x\log^{2}(x)/(4\pi)$
(while $\Delta_{n}^{\textrm{SyMA}}({\bf 0})\approx V^2$). The reason is that, as can be seen in Figs.$\,$\ref{fig:BHZ}(c) and (d), when $V\rightarrow0$, (i) the particle-hole continuum is strongly dispersive - in contrast with the intermediate $V$ regime where the continuum is weakly dispersive, becoming flat at a special point; and (ii) the charge gap is asymptotically small. The SyMA is therefore
expected to fail in this small $V$ regime.
Nonetheless, the simple Eq.$\,$\ref{eq:BHZ_SMA_asymptotic}
still provides an accurate approximate result for intermediate
$x$, as shown in Fig.$\,$\ref{fig:Hubbard_BHZ_maxF}(d).

\section{Connection with experiments}

\label{sec:connection_experiment}

In this Section we employ the SyMA theory introduced in Sec.$\,$\ref{sec:SyMA} to understand some aspects
of experiments on twisted AA-stacked MoTe$_2$.
We will first study the magnon spectrum at $\nu=-1$, comparing the topological and trivial ferromagnetic phases in Sec.$\,$\ref{subsec:Trivial_vs_Topo_ferro} and examining more realistic models of twisted MoTe$_2$ - valid also at smaller twist angles - in Sec.$\,$\ref{subsec:Realistic_MoTe2}. We will then extend the SyMA construction to fractional fillings in Sec.$\,$\ref{subsec:Fractional_fillings}.

\subsection{Magnon spectrum for topological and trivial ferromagnets}

\label{subsec:Trivial_vs_Topo_ferro}

In Ref.$\,$\cite{Park2023}, it was experimentally observed that
the displacement field required to suppress the $\nu=-1$ ferromagnetic state
at $\theta=3.7^{\circ}$ is larger than the one needed to suppress
the quantum anomalous Hall phase, implying the existence of a trivial ferromagnetic phase.
In this section, we will compare the magnon spectra of the topological and trivial fully polarized ferromagnetic phases of the FH model of twisted MoTe$_2$ at $\theta=3.7^{\circ}$, obtained by tuning the displacement field $V_z$. We assume a fully polarized ferromagnetic ground-state,
in which case the tools in Sec.$\,$\ref{sec:moire_magnons} can be
applied.

\begin{figure}
\begin{centering}
\includegraphics[width=1\columnwidth]{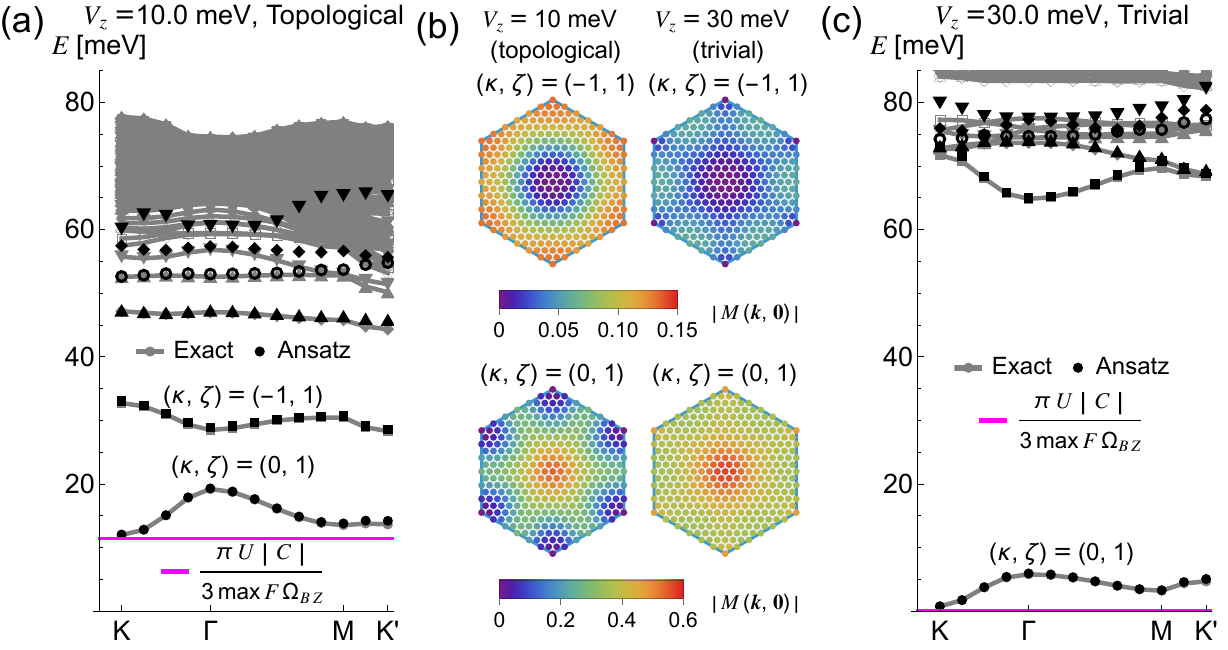}\caption{(a) Exact spectrum of magnon excitations of a fully polarized ferromagnetic
ground-state, obtained for the flat band Hamiltonian $H_{I}$ in Eq.$\,$\ref{eq:H_V_general}
(full gray lines), for $\theta=3.7^{\circ}$ and $V_{z}=10\,\textrm{meV}$
(topological ground-state). The black points are the results obtained
by diagonalizing $H_{I}$ in the subspace of SyMA modes $\bar{S}_{\mbf p}^{\dagger(\kappa,\zeta)}$ defined in Sec.$\,$\ref{sec:sma_main}. In magenta, we also plot the ${\bf p}=0$ lower bound in Eq.$\,$\ref{eq:Ising_gap_Haldane_bound} (note that this
lower bound vanishes in the trivial phase because $C=0$). Like in Fig.$\,$\ref{fig:Hubbard_BHZ_maxF}, we also choose $U/2$ to be equal to the average energy of the spin-flip modes in particle-hole continuum at ${\bf p}=0$.
(b) SyMA eigenstates at ${\bf p}=0$ for the two lowest magnon modes,
for $V_{z}=10\,\textrm{meV}$ (left column) and $V_{z}=30\,\textrm{meV}$
(right column). (c) Same as (a), for $V_{z}=30\,\textrm{meV}$ (trivial
ground-state).\label{fig:MoTe2_topo_vs_trivial}}
\par\end{centering}
\end{figure}

The results are shown in Fig.$\,$\ref{fig:MoTe2_topo_vs_trivial}.
A crucial difference between the topological and trivial phases is that, while there are
two low-energy magnon modes deep inside the topological phase, there
is only one deep inside the trivial phase. To understand this change,
we show in Fig.$\,$\ref{fig:MoTe2_topo_vs_trivial}(b) the
zero-momentum SyMA wave functions of the two lowest magnon modes in
both phases. The symmetry-enforced node locations follow directly from
Eq.$\,$\ref{eq:node_condition_intra} and the $C_3$ symmetry eigenvalues of
the occupied band. Because these symmetry eigenvalues change across
the topological transition, the $(\kappa,\zeta)=(0,+1)$ mode changes
from having nodes at ${\bf K}$ and ${\bf K}'$ to being nodeless, while
the $(\kappa,\zeta)=(-1,+1)$ mode changes from having a total node degree of
$\mathcal{N}=2$ to $\mathcal{N}=4$, as shown in
Fig.$\,$\ref{fig:MoTe2_topo_vs_trivial}(b).

In the topological phase, the two lowest modes both have
$\mathcal{N}=2$, so their relative energies cannot be understood from
the total node degree alone. Instead, their ordering depends on the
location of the nodes relative to the quantum geometry. For
$V_z=10\,\textrm{meV}$, the quantum metric is peaked near ${\bf K}$ for
$\eta=\up$ and near ${\bf K}'$ for $\eta=\down$. Consistently with the node principle discussed in
Sec.$\,$\ref{subsec:MoTe2_SMA_eigenfunctions}, the
$(\kappa,\zeta)=(0,+1)$ magnon mode whose nodes occur at ${\bf K}$ and
${\bf K}'$ has lower energy than the
$(\kappa,\zeta)=(-1,+1)$ mode, whose node occurs at $\Gamma$.

In the trivial phase, the lowest $(\kappa,\zeta)=(0,+1)$ mode is
nodeless, with $\mathcal{N}=0$, whereas the next
$(\kappa,\zeta)=(-1,+1)$ mode has $\mathcal{N}=4$. This large difference in the total node degree is consistent with the strong
energy separation between these modes in
Fig.$\,$\ref{fig:MoTe2_topo_vs_trivial}(c), leaving only the nodeless
mode well below the particle-hole continuum.

For completeness, we also evaluate the zero-momentum bounds derived in
Sec.$\,$\ref{sec:boundApp}, where, as in Fig.$\,$\ref{fig:Hubbard_BHZ_maxF}, we set $U/2$ equal to the average energy of the spin-flip modes in particle-hole continuum at ${\bf p}=0$. Recall that the rigorous bounds were proven only for the flat band Hubbard models in Sec.$\,$\ref{sec:boundApp}, and should therefore be understood as estimates for MoTe$_2$. The lower-bound estimates are shown in
magenta in Figs.$\,$\ref{fig:MoTe2_topo_vs_trivial}(a,c). The
corresponding upper-bound estimates lie significantly above the particle-hole continuum, being $349.5\,\textrm{meV}$ and
$422.4\,\textrm{meV}$ for $V_z=10\,\textrm{meV}$ and
$V_z=30\,\textrm{meV}$, respectively, and are therefore not shown in the figure.

In App.$\,$\ref{subsec:App_Additional_MoTe2_SMA}, we also show the magnon spectrum obtained using the full projected Hamiltonian in Eq.$\,$\ref{eq:H_full_band_basis}, and compare it with the results in Figs.$\,$\ref{fig:MoTe2_topo_vs_trivial}(a,c).
Because $\mathcal{I}$ is broken in the presence of a displacement field $V_z\neq0$, there is no symmetry guarantee that the flat band Hamiltonian $H_I$ in Eq.$\,$\ref{eq:H_V_general} and the full projected Hamiltonian in Eq.$\,$\ref{eq:H_full_band_basis} yield similar low-energy magnon spectra even at ${\bf p}\approx0$.
However, Fig.$\,$\ref{fig:App_MoTe2_H_vs_HI} in App. \ref{subsec:App_Additional_MoTe2_SMA} shows that this is
the case for the lowest-energy magnon modes, implying that the
flat band Hamiltonian can still be reliably used to understand the
low-energy magnon physics in this regime.

\subsection{Realistic models of MoTe$_2$ at variable angle}

\label{subsec:Realistic_MoTe2}

As we discussed in Sec.$\,$\ref{sec:moire_magnons}, the FH model
explored there can only be reliably used for twist angles in the vicinity
of $\theta=3.89^{\circ}$ for an accurate description of the single-particle
band structure of twisted MoTe$_2$. To accurately capture the effects of lattice relaxation and provide a more realistic description of MoTe$_2$ at small twist angles, we adopt the models proposed in Ref.$\,$\cite{zhang2024universalmoiremodelbuildingmethodfitting}, which are constructed directly from large-scale ab-initio calculations without parameter fitting. We will
refer to these models as the no-fitting models (see also Ref.$\,$\cite{ZhangDiXiao2025} for an alternative continuum model that also accurately matches the DFT band structure at smaller twist angles).

\begin{figure*}
\includegraphics[width=1\textwidth]{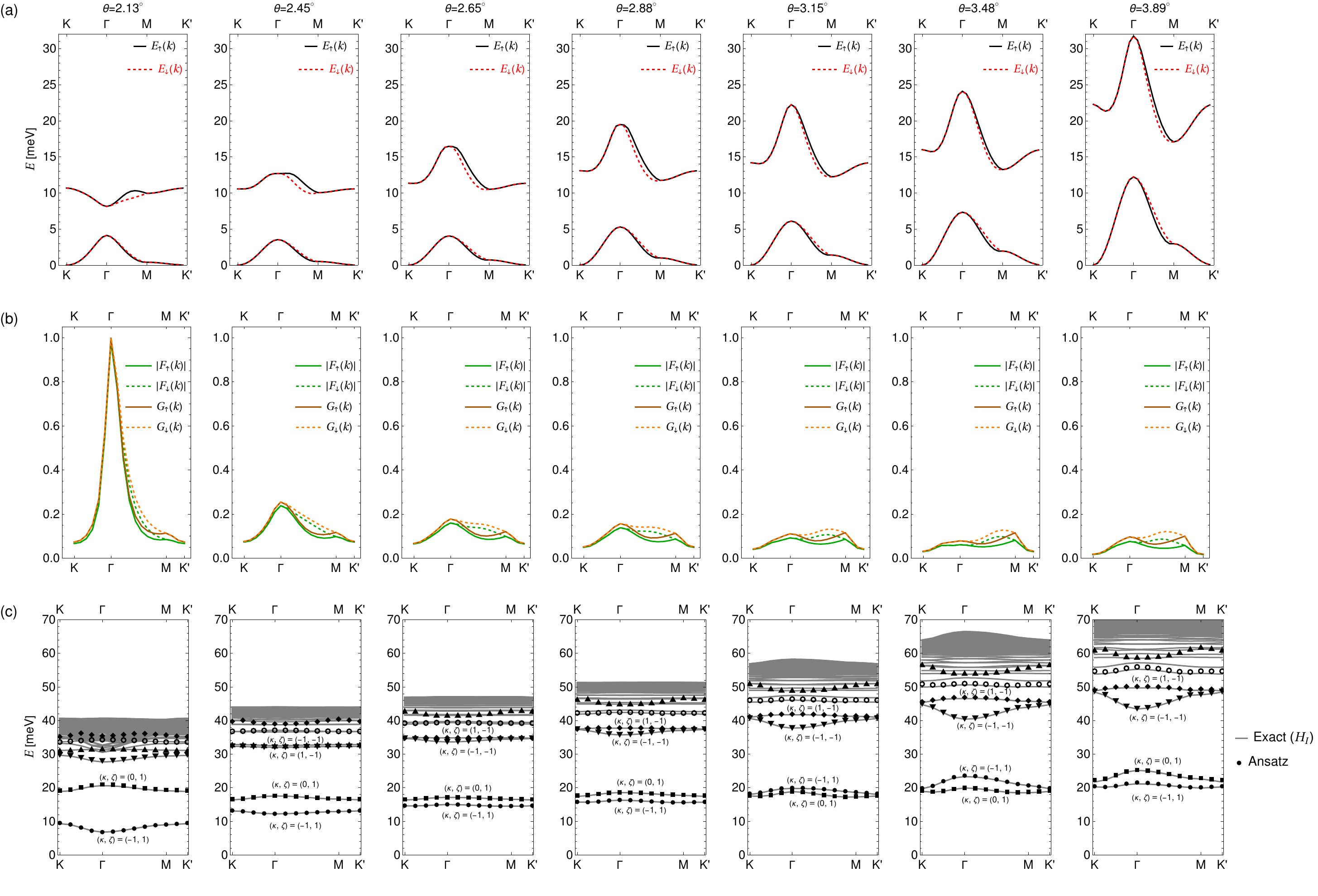}

\caption{Results for the no-fitting models introduced in Ref.$\,$\cite{zhang2024universalmoiremodelbuildingmethodfitting}
using the standard basis, for variable $\theta$ indicated on top
of each column. (a) Band structure for the two lowest-energy hole bands.
(b) Berry curvature $F_{\eta}({\bf k})$ and trace of quantum
metric $G_{\eta}({\bf k})$. All quantities are normalized to the maximum of $G_{\eta}({\bf k})$, obtained for the smallest twist angle. (c) Exact spectrum of magnon excitations
of a fully polarized ferromagnetic ground-state, obtained for the
flat band Hamiltonian $H_{I}$ in Eq.$\,$\ref{eq:H_V_general} (see Fig.$\,$\ref{fig:App_MoTe2_H_vs_HI_nofitting} for comparison with the full projected Hamiltonian in Eq.$\,$\ref{eq:H_full_band_basis}). The black points
are the results obtained by diagonalizing $H_{I}$ in the subspace
of all SyMA modes $\bar{S}_{\mbf p}^{\dagger(\kappa,\zeta)}$ defined in Sec.$\,$\ref{sec:sma_main}. \label{fig:MoTe2_realistic}}

\end{figure*}

The results are shown in Fig.$\,$\ref{fig:MoTe2_realistic}. Figure$\,$\ref{fig:MoTe2_realistic}(a) shows the single-particle hole-band dispersions obtained for the no-fitting models introduced in the previous paragraph. The lowest-energy hole band, which corresponds to the topmost valence electron band, remains separated from the second-lowest-energy hole band for every angle considered. The Chern number of this lowest-energy hole band is $C=\eta$ throughout the angle range shown.

The Berry curvature and quantum metric of the lowest-energy hole band remain relatively uniform for $\theta=2.65^{\circ}$-$3.89^{\circ}$, as shown in Fig.$\,$\ref{fig:MoTe2_realistic}(b). At the smallest angle, $\theta=2.13^{\circ}$, the direct gap between the two lowest-energy hole bands is strongly reduced near $\Gamma$, and the Berry curvature and quantum metric become sharply concentrated around this momentum.

Figure$\,$\ref{fig:MoTe2_realistic}(c) shows the exact magnon spectra obtained using the flat band interaction Hamiltonian $H_I$ in Eq.$\,$\ref{eq:H_V_general}. For $\theta=2.65^{\circ}$-$3.89^{\circ}$, the two lowest magnon branches are weakly dispersive and close in energy. This resembles the time-reversal-symmetric Landau-level model with the $V_0$ pseudopotential shown in Fig.$\,$\ref{fig:4}(a), where the two bound magnon modes are exactly flat and degenerate (see App.$\,$\ref{secapp:LLL}). The Landau-level limit therefore provides a useful qualitative description of the two lowest-energy magnon modes in this angle range.

At $\theta=2.13^{\circ}$, the two lowest branches split more strongly, and the magnon gap at ${\bf p}=0$ is set by the $(\kappa,\zeta)=(-1,+1)$ mode. The node of this magnon mode occurs at $\Gamma$, where the quantum geometry is strongly peaked, while the $(\kappa,\zeta)=(0,+1)$ mode has nodes at ${\bf K}$ and ${\bf K}'$ and has significantly higher energy at ${\bf p}=0$. The node locations remain unchanged throughout the angle range considered because the Chern number and symmetry eigenvalues of the active band (the lowest-energy hole band) do not change.

The ordering of the two lowest-energy magnon modes for the no-fitting model at $\theta=2.13^{\circ}$ studied in this section, differs from that obtained using the FH model at $\theta=2^{\circ}$ in Fig.$\,$\ref{fig:2}(b). In the FH model, the lowest-energy magnon mode at ${\bf p}=0$ is the $(\kappa,\zeta)=(0,+1)$ mode, in contrast with the no-fitting model at $\theta=2.13^{\circ}$ considered here. This opposite ordering can be understood from the different small-angle quantum geometries predicted by the two models. In the FH model, the quantum geometry of the lowest-energy hole band is peaked near ${\bf K}$ and ${\bf K}'$, energetically favoring the $(\kappa,\zeta)=(0,+1)$ mode, whose nodes occur at these momenta. In contrast, in the no-fitting model at $\theta=2.13^{\circ}$, the quantum geometry is peaked near $\Gamma$, favoring the $(\kappa,\zeta)=(-1,+1)$ mode, whose node occurs at this momentum. We again emphasize that the FH model is not reliable to describe small angles like $\theta\approx2^{\circ}$. Instead, the no-fitting models analyzed in this section describe well the DFT band structure in this regime \cite{zhang2024universalmoiremodelbuildingmethodfitting}.

The evolution of the magnon gap in Fig.$\,$\ref{fig:MoTe2_realistic}(c) is qualitatively consistent with the measurements of the Curie temperature in Ref.$\,$\cite{li2025universalmagneticphasestwisted}, which was found to decrease with decreasing $\theta$. Although the results shown in Fig.$\,$\ref{fig:MoTe2_realistic}(c) were obtained using the flat band interaction Hamiltonian $H_I$ in Eq.$\,$\ref{eq:H_V_general}, we show in Fig.$\,$\ref{fig:App_MoTe2_H_vs_HI_nofitting} of App.$\,$\ref{sec:no_fitting_flatband_vs_exact} that including the single-particle terms in the full one band projected Hamiltonian in Eq.$\,$\ref{eq:H_full_band_basis} only weakly affects the lowest-energy magnon modes. For more quantitative predictions, however, accounting for band-mixing effects is crucial$\,$\cite{MFCI_0,FTI_Yves,goncalves2025spinlessspinfulchargeexcitations}, which is beyond the scope of this work.

\subsection{Fractional fillings}

\label{subsec:Fractional_fillings}

Until this point, we focused on the magnon excitation spectrum at integer filling. In this final section, we extend our analysis to fractional fillings, focusing primarily on twisted MoTe$_2$ at $\nu=-2/3$. The main purpose of this section is to understand the origin of the low-energy magnon modes at fractional filling using the SyMA theory.

\begin{figure}
\begin{centering}
\includegraphics[width=1\columnwidth]{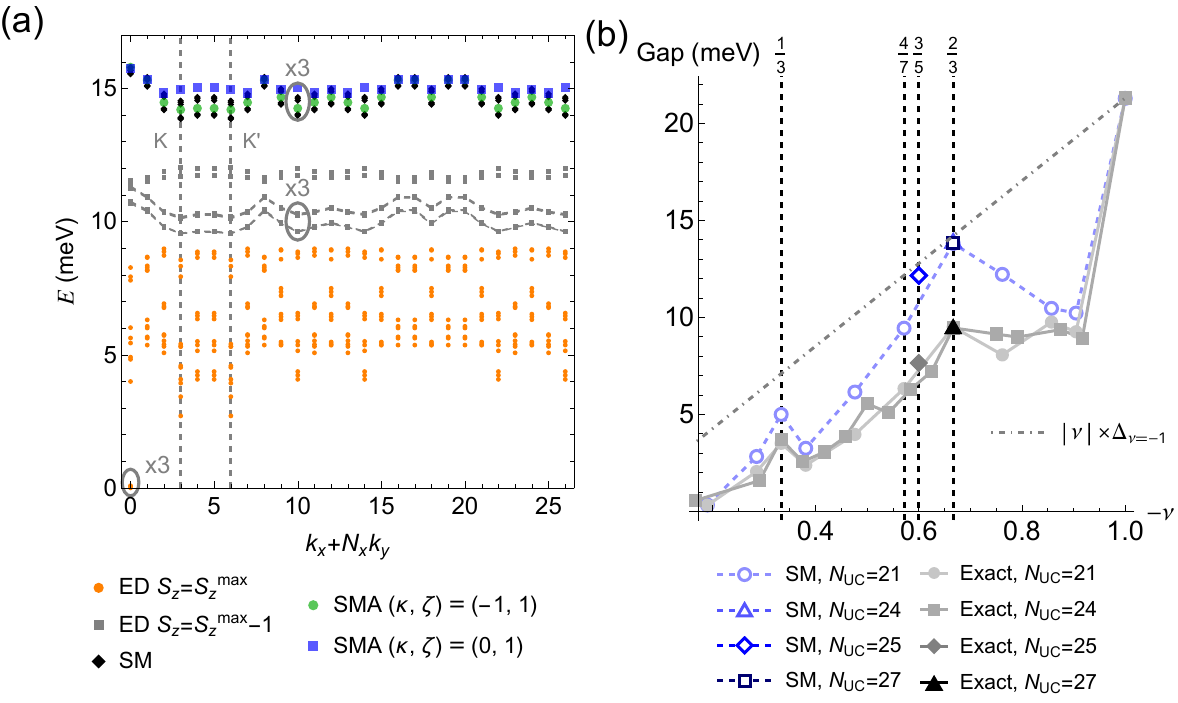}\caption{
(a) Exact-diagonalization results at filling $\nu=-2/3$ and $\theta=3.7^{\circ}$ for the FH model of twisted MoTe$_2$ in the no-dispersion limit, where the noninteracting part of the Hamiltonian in Eq.$\,$\ref{eq:H_full_band_basis} is neglected. The results were obtained using the double-gated screened Coulomb interaction with dielectric constant $\epsilon=10$ on a 27-unit-cell $C_6$-symmetric
momentum grid. In orange, we show the 12 lowest-energy states in the fully polarized sector $S_z=S_z^{\textrm{max}}$. The three quasi-degenerate FCI ground states occur at ${\bf k}=0$. In gray, we
show the eight lowest-energy states in the magnon sector $S_z=S_z^{\textrm{max}}-1$. At each momentum, the six lowest gray states form two magnon branches, each containing three quasi-degenerate states. We use dashed gray lines to highlight these two branches and distinguish them from the magnon modes immediately above in energy.
In black, we show the two lowest-energy branches obtained by diagonalizing the Hamiltonian in the bilinear subspace generated by acting with all operators $\gamma_{{\bf k}+{\bf p},\down}^{\dagger}
\gamma_{{\bf k},\up}$ on each of the three FCI ground states, to which we refer to as S-matrix (SM) results. This bilinear-subspace diagonalization overestimates the exact magnon gap but approximately
reproduces the dispersion of the two lowest exact branches.
In green and blue, we show the energy expectation values obtained by applying the SyMA operators in Eq.$\,$\ref{eq:intralayer_Sp}, with $\kappa=-1$ and $\kappa=0$, respectively, to the lowest-energy FCI
ground state. These SyMA energies closely follow the S-matrix results for the two lowest-energy branches obtained from diagonalization in the full bilinear subspace.
Only the SyMA states constructed from the lowest energy FCI ground state are displayed. Applying each of the two SyMA operators to all three FCI ground states produces three quasi-degenerate states in each of the two low-energy branches.
(b) Comparison between the exact magnon gap and the
bilinear-subspace, or S-matrix (SM), result for the same model parameters as in (a) and variable filling. The SM results were obtained by diagonalizing the Hamiltonian in the subspace generated by applying the magnon bilinears to the selected low-energy fully polarized states indicated in Fig.$\,$\ref{fig:exact_vs_sm_variable_filling_21site_selection}. Besides the 27-unit-cell lattice used in (a), we considered systems with $N_{UC}=21$, $24$, and $25$. The corresponding momentum grids
are shown in Fig.$\,$\ref{fig:App_momentum_meshes} of
App.$\,$\ref{sec:Additional_results_Fractional}.
\label{fig:Fractional_Fillings}}
\par\end{centering}
\end{figure}

Some results for the magnon spectrum at $\nu=-2/3$ have been previously obtained in Ref.$\,$\cite{goncalves2025spinlessspinfulchargeexcitations} using exact diagonalization (ED), where low-lying magnon modes were identified.
Here we show the exact results for the low-energy magnon spectrum in Fig.$\,$\ref{fig:Fractional_Fillings}(a) in the no-dispersion limit, where the non-interacting part of the Hamiltonian in Eq.$\,$\ref{eq:H_full_band_basis} is neglected. In this limit, it is possible to numerically obtain very good quality FCI ground-states, showing a very small finite-size energy splitting - only $3\%$ of the FCI gap.
We can clearly identify in Fig.$\,$\ref{fig:Fractional_Fillings}(a) two lowest-energy magnon branches, each containing three quasi-degenerate states. We highlighted these branches in Fig.$\,$\ref{fig:Fractional_Fillings}(a) with dashed gray lines. The additional exact magnon states, also shown in gray, immediately above these two branches also have low energy. Thus, unlike at $\nu=-1$, the two lowest-energy branches are not separated from the rest of the exact magnon spectrum by a sizable energy gap.

At $\nu=-1$, the full Hilbert space in the magnon sector can be constructed by applying the bilinears $\gamma_{{\bf k}+{\bf p},\down}^{\dagger}\gamma_{{\bf k},\up}$ to the fully polarized ground state $\ket{\uparrow}\equiv\prod_{\bf k}\gamma_{{\bf k},\up}^{\dagger}\ket{0}$. The magnon spectrum can therefore be obtained exactly by diagonalizing the Hamiltonian in this subspace. At fractional fillings, applying these bilinears to the ground state(s) no longer generates the full magnon Hilbert space. In fact, the fraction of the Hilbert space generated in this way vanishes with increasing system size.

Nevertheless, we can construct an approximate magnon subspace by acting with all the bilinears $\gamma_{{\bf k}+{\bf p},\down}^{\dagger}\gamma_{{\bf k},\up}$ on all the quasi-degenerate FCI ground states and diagonalizing the Hamiltonian in the resulting subspace. This is the fractional-filling generalization of the numerical S-matrix approach described in Sec.$\,$\ref{sec:exact_mag_excitations_Smatrix} and used throughout the manuscript at $\nu=-1$. The resulting spectrum is shown in black in Fig.$\,$\ref{fig:Fractional_Fillings}(a). Although this subspace diagonalization overestimates the magnon gap by approximately $45\%$, it reproduces the dispersion of the two lowest-energy magnon branches. Moreover, as shown in Table$\,$\ref{tab:overlaps_gaps_2o3} of App.$\,$\ref{sec:Additional_results_Fractional}, the squared projection of the exact lowest-energy magnon eigenstates onto the bilinear subspace is relatively high, ranging from $0.76$ to $0.81$ for the system sizes considered.

Finally, each of the two lowest-energy branches contains three quasi-degenerate states because the bilinear operators can act on each of the three quasi-degenerate FCI ground states, giving a total of six low-energy magnon states.

In App.$\,$\ref{sec:Additional_results_Fractional}, we provide a detailed analysis supporting this interpretation. We also show results at $\nu=-3/5$, where each magnon branch instead contains five quasi-degenerate states associated with the five quasi-degenerate FCI ground states. The bilinear subspace does not, however, capture the additional exact magnon states lying immediately above the two lowest-energy branches in Fig.$\,$\ref{fig:Fractional_Fillings}(a). Instead, the subspace diagonalization predicts a sizable gap above these branches, similar to the spectrum at $\nu=-1$.

A natural follow-up question is whether the lowest-energy states obtained within the full bilinear subspace can be approximated by applying the lowest-energy SyMA operators in Eq.$\,$\ref{eq:intralayer_Sp} to the FCI ground states. The corresponding expectation values of the Hamiltonian in the SyMA states are shown in green and blue in Fig.$\,$\ref{fig:Fractional_Fillings}(a), and closely follow the lowest-energy branches obtained by diagonalizing the Hamiltonian in the full bilinear subspace, shown in black.

We finish this section by showing results for the magnon gap at different filling fractions in Fig.$\,$\ref{fig:Fractional_Fillings}(b), comparing with the results obtained by diagonalizing the Hamiltonian in the bilinear subspace. Note that because the spectrum is gapless at non-commensurate fillings, we selected a subspace of low-energy modes in the fully-polarized sector to apply the bilinears to, which we show explicitly in Fig.$\,$\ref{fig:exact_vs_sm_variable_filling_21site_selection} in App.$\,$\ref{sec:Additional_results_Fractional}. The bilinear diagonalization results closely track the exact results, with the largest deviations interestingly occurring for the FCI fillings. These results show that although the bilinear subspace only captures the magnon spectrum exactly at $\nu=-1$, it can still capture reasonably well the lowest-energy magnon modes at general fractional fillings.

\section{Discussion}

In this work, we developed a theory to understand the low-energy magnon excitations in Ising ferromagnetic Chern insulators.
Building on the single-mode approximation, we introduced the Symmetric Mode Ansatz (SyMA), an ansatz for the low-energy magnon modes built using locality and the symmetries of the Bloch bands. By considering several solvable models, we explained the success of the SyMA in twisted MoTe$_2$ by identifying a class of flat band Hubbard models in which it is exact, as well as an ideal-band continuum model in which it is not exact but remains highly accurate.

Our SyMA theory identifies quantum geometry and the nodal structure of the magnon wave function as important ingredients governing the magnon spectrum. In particular, while the lowest-energy magnon modes have the smallest possible sum of node degrees in their wave function $\mathcal{N}$, this number $\mathcal{N}$ is lower bounded by $2|C|$ in Chern bands. Additionally, the lower-energy magnons place their nodes in regions where the quantum geometry of the band is most strongly peaked. In the family of two-band Hubbard models, we proved rigorous lower and upper bounds for the zero-momentum magnon gap in terms of elementary band geometric quantities like the Chern number $C$ and the quantum geometry. Using the maximum value of the quantum metric $G_{\text{max}}$, our bounds can alternatively be stated as
\bea
 \frac{\pi}{3} U \frac{|C|}{G_{\text{max}}\Omega_{BZ}} \leq \Delta({\bf p}= 0) \leq \frac{4\pi U}{\Omega_{BZ} |\mbf{a}|^2} G \ .
\eea

Hence, we identify global topology and local quantum geometry as fundamental quantities constraining magnetic stability in flat bands with local Hubbard interactions. Our bounds suggest two natural extensions for future work. In particular, whether a lower bound be formulated in terms of other topological invariants \cite{2016PhRvB..94x5149T,PhysRevLett.124.167002,2022PhRvL.128h7002H,2025PhRvL.135h6401Y,4k6g-hy6y,mxk6-mmym,PhysRevB.90.165139,PhysRevB.104.045103,PhysRevB.110.075135} can be derived, and whether bounds for all $\mbf{p}$ can be obtained.
Ref.$\,$\cite{chen2025quantumgeometricdipoletopologicalboost} introduced the quantum-geometric dipole as a gauge-invariant measure of the separation between the particle and hole forming a magnon, and showed that a larger separation weakens their attraction and increases the magnon energy. In Ref.$\,$\cite{chen2025quantumgeometricdipoletopologicalboost}, a single-mode approximation was also employed, but with a particular single ansatz corresponding to the choice $D=\mathbb{1}$ in our notation, which therefore does not necessarily capture the lowest-energy magnon mode. Within this local-mode approximation and a small-${\bf q}$ expansion, Ref.$\,$\cite{chen2025quantumgeometricdipoletopologicalboost} derived the approximate lower bound $\Delta({\bf 0})\gtrsim
|C_{\up}-C_{\down}|\,U(\bar{g})$
where $\bar{g}$ is the quantum metric averaged over spin/valley and the Brillouin zone, and $U(\bar{g})$ is an effective interaction scale that depends on it.
Our SyMA generalizes this construction by including multiple projected local spin-flip modes belonging to distinct symmetry sectors, thereby capturing several low-lying magnon branches rather than a single trial state. Moreover, for the class of two-band flat band Hubbard models considered here, the SyMA is exact and our result provides rigorous lower and upper bounds on the exact zero-momentum magnon gap.
The lower bound is nonzero for $C\neq0$ and increases as the Berry-curvature distribution becomes more uniform, while the integrated quantum metric independently constrains the gap from above.

One crucial prediction relevant to experiment is that there is a change in the $C_3$ character at ${\bf p}=0$ of the lowest-energy mode around $\theta=3.7^{\circ}$ in twisted MoTe$_2$. This should be possible to capture through optical measurements such as circularly polarized RILS$\,$\cite{PhysRevResearch.3.023040,PhysRevLett.123.146801,Liang2024}. 
Finally, the success of our SyMA theory - which describes the magnon modes as local bound states - demonstrates that these excitations are best thought of as tightly-bound bosons. As such, we speculate that the behavior of the finite-temperature phase is likely to be well-described as a dilute gas of magnons. Residual interactions among the magnons may lead to interacting bosonic phases \cite{PhysRevB.71.045339,PhysRevB.92.245124,PhysRevB.104.155129}.

A further important question, beyond the minimal setting studied in this paper, is the effect of band mixing, which was recently shown to substantially affect spinful excitations$\,$\cite{goncalves2025spinlessspinfulchargeexcitations}, in particular decreasing the magnon gap at $\nu=-2/3$.

We finally highlight the extension of the SyMA construction to fractional filling. At $\nu=-2/3$, the exact low-energy magnon states have substantial projection onto the subspace generated by acting with spin-flip bilinears on the quasi-degenerate FCI ground states instead of the single Slater determinant ground-state at $\nu=-1$. Within this bilinear subspace, the lowest-energy states are in turn well approximated by applying the SyMA operators to the FCI ground states. These results suggest that the SyMA may also be extended to describe reasonably well the low-energy magnon modes away from integer filling.
It is important to note, however, that while the agreement with the exact results for the lowest-energy modes is good, these are close in energy to an additional set of low-lying excitations. This contrasts with the integer case, where the lowest-energy magnon modes are typically well separated in energy from higher-energy excitations. It will be interesting to understand the nature of these extra excitations in the future and whether they can also be described by a simple ansatz, most likely involving an ansatz built from quasi-holes, as in Jain's theory of the SMA$\,$\cite{PhysRevB.61.13064}.

\begin{acknowledgments}
We thank Yves Kwan, Jiabin Yu, Zhaoyu Han, Daniel Muñoz-Segovia, Eslam Khalaf, Di Xiao, Xiaodong Xu, Nicolás Morales-Durán and Jennifer Cano for fruitful discussions. M.G. is supported by a postdoctoral research fellowship
at the Princeton Center for Theoretical Science. B.A.B. was supported by the Gordon and Betty Moore Foundation through Grant No.~GBMF8685 towards the Princeton theory program, the Gordon and Betty Moore Foundation’s EPiQS Initiative (Grant No.~GBMF11070), the Global Collaborative Network Grant at Princeton University, the Simons Investigator Grant No.~404513, the NSF-MERSEC (Grant No.~MERSEC DMR 2011750), the Simons Collaboration on New Frontiers in Superconductivity (SFI-MPS-NFS-00006741-01), the Schmidt Foundation at the Princeton University, and the Princeton Catalysis Initiative. This project was partially supported by the European Research Council (ERC) under the European Union’s Horizon 2020 Research and Innovation Programme (Grant Agreement No. 101020833). The Flatiron Institute is a division of the Simons Foundation. J.H.-A. is supported by a Hertz Fellowship. A.F. was supported by the Deutsche Forschungsgemeinschaft (DFG, German Research Foundation) through Project-ID 258499086 - SFB 1170 and through the W\"urzburg-Dresden Cluster of Excellence on Complexity, Topology and Dynamics in Quantum Matter - ctd.qmat Project-ID 390858490 - EXC 2147.

Part of the simulations presented in this article were performed on computational resources managed and supported by Princeton Research Computing, a consortium of groups including the Princeton Institute for Computational Science and Engineering (PICSciE) and Research Computing at Princeton University.

\end{acknowledgments}
\clearpage

\appendix

\section{Details on the first-harmonic model of MoTe$_2$}
\label{sec:detalis_MoTe2_FH}
In this Appendix, we provide additional details on the FH model of MoTe$_2$ studied in Sec.$\,$\ref{subsec:FH_model}.
Following Ref.$\,$\cite{MFCI_I}, the intralayer and interlayer couplings $V_{\eta,l}({\bf r})$ and $t_{\eta}({\bf r})$ in Eq.$\,$\ref{eq:c_eta_l_r_FH} are given by

\begin{equation}
V_{\eta,l}({\bf r})=2V\sum_{j=1}^{3}\cos({\bf g}_{j}\cdot{\bf r}-(-1)^{\frac{1-l}{2}}\psi)\,,\label{eq:Vr_FH}
\end{equation}
\begin{equation}
t_{\eta}({\bf r})=w\sum_{j=1}^{3}e^{-{\rm i}\eta{\bf q}_{j}\cdot{\bf r}}\,,\label{eq:tr_FH}
\end{equation}

\noindent where ${\bf g}_{j}=C_{3}^{j-1}{\bf b}_{M,1}$, with $C_{n}$ denoting an in-plane rotation by an angle $2\pi/n$. Here,
${\bf b}_{M,1}=4\pi(\sqrt{3}a_{M})^{-1}(1,0)^{T}$ and
${\bf b}_{M,2}=C_{6}{\bf b}_{M,1}$ are the moiré reciprocal-lattice vectors,
$a_{M}=a_{0}/[2\sin(\theta/2)]$ is the moiré lattice constant,
$a_{0}=3.52\,\text{\AA}$ is the monolayer lattice constant, and $\theta$ is the twist angle. Furthermore,
${\bf q}_{1}=8\pi(3a_{0})^{-1}\sin(\theta/2)(0,1)^{T}$ and
${\bf q}_{j+1}=C_{3}^{j}{\bf q}_{1}$ for $j=1,2$.
The parameters $V$, $\psi$, and $w$ denote, respectively, the amplitude and phase of the intralayer moiré potential and the interlayer tunneling amplitude.

We work with the Hamiltonian projected into the topmost valence electron band with eigenvectors $U_{\eta,{\bf k}}$, and introduce the periodic convention $U_{\eta,{\bf k}+{\bf q}+{\bf G}}=V_{{\bf G}}U_{\eta,{\bf k}+{\bf q}}$,
with $(V_{{\bf G}})_{{\bf Q}{\bf Q}'}=\delta_{{\bf {\bf Q}},{\bf Q}'+{\bf G}}$
the embedding matrix, and ${\bf Q} \in \{{\bf G}+{\bf q}_{1}\}\cup\{{\bf G}-{\bf q}_{1}\}$, as introduced in Sec.$\,$\ref{subsec:FH_model} (see Fig.$\,$\ref{fig:1}(a) for illustration of the momentum-space lattice vectors ${\bf Q}$). The form factors defined in Eq.$\,$\ref{eq:form_factor_1band} from these eigenvectors
satisfy hermiticity, $M^{(\eta)*}({\bf k}-{\bf q},{\bf q+{\bf G}})=M^{(\eta)}({\bf k},-{\bf q-{\bf G}})$. Throughout the paper, we picked the time-reversal gauge

\begin{equation}
\begin{array}{c}
U_{\up,{\bf k}}=U_{{\bf k}}\,\,,\\
U_{\down,{\bf k}}=D_{\mathcal{I}}\,U_{-{\bf k}}^{*}\,\,,
\end{array}\label{eq:TR_gauge}
\end{equation}
where $[D_{\mathcal{I}}]_{{\bf Q},{\bf Q}'}=\delta_{{\bf Q},-{\bf Q}'}$,
in which case we have $M^{(\down)*}({\bf k},{\bf q}+{\bf G})=M^{(\up)}(-{\bf k},-{\bf q}-{\bf G})$.

Using Eqs.$\,$\ref{eq:c_kQ_def} and \ref{eq:gamma_def}, the full Hamiltonian projected to the lowest energy hole band can be written as
\begin{equation}
H
=
H_I
+
\sum_{{\bf k}\in{\rm MBZ},\eta}
\left[
E_{\eta}({\bf k})
+
\epsilon_{HF,\eta}({\bf k})
\right]
\gamma_{{\bf k},\eta}^{\dagger}
\gamma_{{\bf k},\eta} \,,
\label{eq:H_full_band_basis_2}
\end{equation}
where $H_I$ is the positive-semidefinite interaction Hamiltonian defined in Eq.$\,$\ref{eq:H_V_general}, $E_{\eta}({\bf k})$ is the bare hole energy dispersion, and $\epsilon_{HF,\eta}({\bf k})$ is the one-body term generated when the interaction, initially normal ordered with respect to the hole vacuum, is rewritten in a form normal ordered with respect to $\nu=-1$. Explicitly,
\begin{equation}
\label{eq:e_HF}
\epsilon_{HF,\eta}({\bf k})
=
\sum_{\eta',{\bf k}'}
V_{\eta\eta'}({\bf k},{\bf k}',{\bf 0})
-
\sum_{{\bf q}}
V_{\eta\eta}({\bf k}-{\bf q},{\bf k},{\bf q}),
\end{equation}
with
\begin{equation}
V_{\eta\eta'}({\bf k},{\bf k}',{\bf q})
=
\sum_{{\bf G}}
V({\bf q}+{\bf G})
M^{(\eta)}({\bf k},{\bf q}+{\bf G})
M^{(\eta')}({\bf k}',-{\bf q}-{\bf G}).
\label{eq:V_band_basis}
\end{equation}

Throughout the paper, we mostly use the double-gated Coulomb potential defined in Ref.$\,$\cite{MFCI_0}, given by

\begin{equation}\label{eq:dual_gate}
 V({\bf q})= V_C({\bf q})=\frac{e^{2}\pi\xi}{4\pi\epsilon\epsilon_{0}}\tanh(\xi|{\bf q}|/2)/(\xi|{\bf q}|/2)
\end{equation}
with gate distance $\xi=20\,\textrm{nm}$. We parameterize the interaction strength by $u=10/\epsilon$ and take $u=0.9$ unless otherwise specified, where $\epsilon$ is the dielectric constant.
In Figs.$\,$\ref{fig:4}(b,d) and \ref{fig:Hubbard_BHZ_maxF}(c),
we also considered the simplified Hubbard interaction $V({\bf q})=V_{C}({\bf q}={\bf 0})$.

In most of the paper, we focused on the flat band Hamiltonian $H_I$. We will show below in App.$\,$\ref{subsec:App_Additional_MoTe2_Exact} that this choice is well justified to characterize the lowest-energy magnon modes. In particular, we will show that the spectra of low-energy magnon modes obtained using the flat band Hamiltonian $H_I$ and the full Hamiltonian in Eq.$\,$\ref{eq:H_full_band_basis_2} are very similar. For the FH model at zero displacement field ($V_z=0$, see Eq.$\,$\ref{eq:c_eta_l_r_FH} for definition in the FH model Hamiltonian), the contribution of the additional one-body term in Eq.$\,$\ref{eq:H_full_band_basis_2} to the magnon spectrum will be shown in App.$\,$\ref{subsec:1body_contribution} to vanish exactly at ${\bf p}={\bf 0}$ due to the existing pseudo-inversion symmetry. Away from ${\bf p}={\bf 0}$, we will check numerically in App.$\,$\ref{subsec:App_Additional_MoTe2_Exact} (see in particular Fig.$\,$\ref{fig:App_MoTe2_H_vs_HI}) that including the renormalized dispersion $E_{\eta}({\bf k})+\epsilon_{HF,\eta}({\bf k})$ in Eq.$\,$\ref{eq:H_full_band_basis_2} produces only small changes in the lowest-energy magnon modes for the parameters studied, although it modifies the higher-energy continuum.

\section{Symmetries of AA-stacked moiré MoTe$_2$}

\label{sec:MoTe2_symmetries}

In this section, we will detail the action of the different symmetries
on the single-particle and magnon operators, for the FH model of MoTe$_2$.

\subsection{Symmetry representations}

\label{sec:MoTe2_symmetries_REP}

We use $T_{{\bf R}}$ for
the moiré translation operator, with ${\bf R}$ being a moiré lattice
vector, $C_{3}$ for an in-plane $120^{\circ}$ rotation, $C_{2y}$
for a $180^{\circ}$ rotation about the $y$ axis. $\mathcal{I}$
for pseudo-inversion, $\mathcal{T}$ for time-reversal, and $\mathcal{C}$ for unitary particle-hole
symmetry. $C_{2y} \mathcal{T}$ and $\mathcal{I}$ are only present for $V_{z}=0$.
$\mathcal{I}$ only holds for the FH model and breaks down when higher-order
harmonics are used for expanding the terms in Eqs.$\,$\ref{eq:Vr_FH}
and \ref{eq:tr_FH}.

We will employ the following representation for the action of symmetries
on the continuum hole operators $c_{\eta,l,{\bf r}}^{\dagger}$ introduced
in Eq.$\,$\ref{eq:c_eta_l_r_FH}$\,$\cite{MFCI_I}:

\begin{equation}
\begin{array}{c}
T_{\mbf R}c_{\eta,l,\mbf r}^{\dag}T_{\mbf R}^{\dag}=e^{-i\eta l{\bf q}_{1}\cdot\mbf R}c_{\eta,l,\mbf r+\mbf R}^{\dag}\,\,\,\,(\textrm{moiré translation})\\
C_{3}c_{\eta,l,{\bf r}}^{\dagger}C_{3}^{-1}=c_{\eta,l,C_{3}{\bf r}}^{\dagger}e^{i\eta\frac{\pi}{3}}\\
(C_{2y}\mathcal{T})c_{\eta,l,{\bf r}}^{\dagger}(C_{2y}\mathcal{T})^{-1}=c_{\eta,-l,C_{2y}{\bf r}}^{\dagger}\\
\mathcal{I}c_{\eta,l,{\bf r}}^{\dagger}\mathcal{I}^{-1}=c_{\eta,-l,-{\bf r}}^{\dagger}\,\,.\\
\mathcal{T}c_{\eta,l,{\bf r}}^{\dagger}\mathcal{T}^{-1}=(-\eta)c_{-\eta,l,{\bf r}}^{\dagger}\\
\end{array}\label{eq:representations_cr_MoTe2}
\end{equation}

The relations in Eq.$\,$\ref{eq:representations_cr_MoTe2} also fix
the algebra of the symmetry representations on the microscopic fermion operators. With these conventions, we have
\begin{equation}
C_{3}^{3}=\mathcal{T}^{2}=(-1)^{F},
\qquad
(C_{2y}\mathcal{T})^{2}=\mathcal{I}^{2}=1,
\label{eq:symmetry_algebra_MoTe2}
\end{equation}
where $(-1)^{F}$ is the fermion-parity operator. Thus $C_{3}^{3}$ and
$\mathcal{T}^{2}$ yield a minus sign when acting
on a single fermion operator. On neutral bilinears, such as density
operators or spin-flip operators, this fermion-parity sign cancels.

Under the definitions in Eq.$\,$\ref{eq:elementary_magnons}, the elementary magnons transform as

\begin{equation}
\begin{array}{c}
T_{{\bf R}}S_{\parallel,\mbf r,l,j}^{\dagger}T_{{\bf R}}^{\dagger}=S_{\parallel,\mbf r+{\bf R},l,j}^{\dagger}\\
C_{3}S_{\parallel,\mbf r,l,j}^{\dagger}C_{3}^{-1}=e^{-\frac{2\pi}{3}i}S_{\parallel,C_{3}\mbf r,l,j+1}^{\dagger}\\
(C_{2y}\mathcal{T})S_{\parallel,\mbf r,l,j}^{\dagger}(C_{2y}\mathcal{T})^{-1}=S_{\parallel,C_{2y}\mbf r,-l,C_{2y}j}^{\dagger}\\
\mathcal{I}S_{\parallel,\mbf r,l,j}^{\dagger}\mathcal{I}^{-1}=S_{\parallel,-\mbf r,-l,j}^{\dagger}
\,\,,
\end{array}
\end{equation}

\begin{equation}
\begin{array}{c}
T_{{\bf R}}S_{\perp,\mbf r,l}^{\dagger}T_{{\bf R}}^{\dagger}=S_{\perp,\mbf r+{\bf R},l}^{\dagger}\\
C_{3}S_{\perp,\mbf r,l}^{\dagger}C_{3}^{-1}=e^{-\frac{2\pi}{3}i}S_{\perp,C_{3}\mbf r,l}^{\dagger}\\
(C_{2y}\mathcal{T})S_{\perp,\mbf r,l}^{\dagger}(C_{2y}\mathcal{T})^{-1}=S_{\perp,C_{2y}\mbf r,-l}^{\dagger}\\
\mathcal{I}S_{\perp,\mbf r,l}^{\dagger}\mathcal{I}^{-1}=S_{\perp,-\mbf r,-l}^{\dagger}
\,\,,
\end{array}
\end{equation}

\noindent where we simply have $C_{2y}j=-j$ (which follows from $C_{2y}{\bf q}_{j}={\bf q}_{-j}$).

We will also define the action of symmetries in the plane-wave and
band operators. We define the action of the symmetry operation $g$
in the plane-wave basis introduced in Eq.$\,$\ref{eq:c_kQ_def} in terms of a representation matrix $D[g]$
as

\begin{equation}
gc_{{\bf k},{\bf Q},\eta}^{\dagger}g^{\dagger}=\sum_{{\bf Q}'\eta',}c_{g{\bf k},{\bf Q}',\eta'}^{\dagger}D_{{\bf Q}'\eta',{\bf Q}\eta}[g]\,,
\end{equation}

The symmetry relation $gH_{0}g^{\dagger}=H_{0}$ implies that

\bea
D[g]{\bf h}({\bf k})D^{\dagger}[g]={\bf h}(g{\bf k})\\
\leftrightarrow D[g]{\bf h}({\bf k})={\bf h}(g{\bf k})D[g]\,,
\label{eq:Dhk_hgk_D}
\eea
where we defined the single-particle Hamiltonian in the plane-wave
basis $[{\bf h}({\bf k})]_{{\bf Q}\eta,{\bf Q}'\eta'}=\delta_{\eta\eta'}h_{{\bf Q}{\bf Q}'}^{\eta}({\bf k})$.
$D[g]U_{\eta,m,{\bf k}}$ is an eigenstate of ${\bf h}(g{\bf k})$
with energy $E_{\eta,m}({\bf k})$,

\bea
&{\bf h}(g{\bf k})(D[g]U_{\eta,m,{\bf k}})=D[g]{\bf h}({\bf k})U_{\eta,m,{\bf k}} \\ &=E_{\eta,m}({\bf k})(D[g]U_{\eta,m,{\bf k}})\,,
\eea
which implies that we can write $D[g]U_{\eta,m,{\bf k}}$ as a linear
combination of the degenerate eigenstates $\{U_{\eta',m',g{\bf k}}\}$
with the same energy $E_{\eta',m'}({\bf k})=E_{\eta,m}({\bf k})$,
i.e.,

\begin{equation}
\begin{array}{c}
D[g]U_{\eta,m,{\bf k}}=\sum_{\eta',m'}[B^{g}({\bf k})]_{\eta m,\eta'm'}U_{\eta',m',g{\bf k}}\end{array}\,,\label{eq:symmetry_relation_MoTe2}
\end{equation}
where we defined the sewing matrix $B^{g}({\bf k})$, that can be
computed through

\begin{equation}
[B^{g}({\bf k})]_{\eta m,\eta'm'}=U_{\eta',m',g{\bf k}}^{\dagger}D[g]U_{\eta,m,{\bf k}}\,.\label{eq:sewing_relation}
\end{equation}

The representation matrices for all symmetries are then given by

\begin{equation}
\begin{array}{c}
D_{{\bf Q}\eta,{\bf Q}'\eta'}[C_{3}]=e^{i\eta\frac{\pi}{3}}\delta_{\eta,\eta'}\delta_{{\bf Q},C_{3}{\bf Q}'}\\
D_{{\bf Q}\eta,{\bf Q}'\eta'}[C_{2y}\mathcal{T}]=\delta_{\eta,\eta'}\delta_{{\bf Q},-C_{2y}{\bf Q}'}\mathcal{K}\\
D_{{\bf Q}\eta,{\bf Q}'\eta'}[\mathcal{I}]=\delta_{\eta,\eta'}\delta_{{\bf Q},-{\bf Q}'}\\
D_{{\bf Q}\eta,{\bf Q}'\eta'}[\mathcal{T}]=(-\eta')\delta_{\eta,-\eta'}\delta_{{\bf Q},-{\bf Q}'}\mathcal{K}\,,
\end{array}\label{eq:D_representations_MoTe2}
\end{equation}
where $C_{2y}\mathcal{T}(Q_{x},Q_{y})=(Q_{x},-Q_{y})$. Using Eq.$\,$\ref{eq:sewing_relation} and the unitarity of the representation
matrices,
this implies that the sewing matrices for non-degenerate eigenstates
at momentum ${\bf k}$ are given by

\begin{equation}
\begin{array}{c}
[B^{C_{3}}({\bf k})]_{\eta m,\eta'm'}=\delta_{m,m'}\delta_{\eta,\eta'}e^{i\varphi_{m\eta}^{C_{3}}({\bf k})}\\{}
[B^{C_{2y}\mathcal{T}}({\bf k})]_{\eta m,\eta'm'}=\delta_{m,m'}\delta_{\eta,\eta'}e^{i\varphi_{m\eta}^{C_{2y}\mathcal{T}}({\bf k})}\\{}
[B^{\mathcal{I}}({\bf k})]_{\eta m,\eta'm'}=\delta_{m,m'}\delta_{\eta,\eta'}e^{i\varphi_{m\eta}^{\mathcal{I}}({\bf k})}\\{}
[B^{\mathcal{T}}({\bf k})]_{\eta m,\eta'm'}=(-\eta)\delta_{m,m'}\delta_{\eta,-\eta'}e^{i\varphi_{m\eta}^{\mathcal{T}}({\bf k})}\,,
\end{array}\label{eq:sewing_matrices_MoTe2}
\end{equation} 
where $\varphi_{m\eta}^{g}({\bf k})$ are sewing phases. The action
of $g$ on the band fermion operators $\gamma_{{\bf k},m,\eta}^{\dagger}=\sum_{{\bf Q}}[U_{\eta,m,{\bf k}}]_{{\bf Q}}c_{\eta,{\bf k},{\bf Q}}^{\dagger}$
is then given by

\begin{equation}
g\gamma_{{\bf k},m,\eta}^{\dagger}g^{-1}=\sum_{m'\eta'}[B^{g}({\bf k})]_{\eta m,\eta'm'}\gamma_{g{\bf k},m',\eta'}^{\dagger}
\end{equation}

For $\mathcal{T}$, the symmetry relation in Eq.$\,$\ref{eq:symmetry_relation_MoTe2}
implies that

\bea
[U_{\up,m,{\bf k}}^{*}]_{-{\bf Q}}=e^{i\varphi_{m\up}^{\mathcal{T}}({\bf k})}[U_{\down,m,-{\bf k}}]_{{\bf Q}} \\ \leftrightarrow[U_{\up,m,{\bf k}}]_{{\bf Q}}=e^{-i\varphi_{m\up}^{\mathcal{T}}({\bf k})}[U_{\down,m,-{\bf k}}^{*}]_{-{\bf Q}}
\eea

We will fix the gauge with $\varphi_{m\eta}^{\mathcal{T}}({\bf k})=0$
by choosing

\bea
[U_{\up,m,{\bf k}}]_{{\bf Q}}=[U_{m,{\bf k}}]_{{\bf Q}} \\
[U_{\down,m,{\bf k}}]_{{\bf Q}}=[U_{m,-{\bf k}}^{*}]_{-{\bf Q}}\,,
\eea
from which we can derive

\begin{equation}
 M^{(\up)}_{mn}(-{\bf k},-{\bf q}-{\bf G}) = M^{(\down)*}_{mn}({\bf k},{\bf q}+{\bf G}) \,,
 \label{eq:ff_TR_multiband}
\end{equation}
where $M^{(\eta)}_{mn}({\bf k},{\bf q}+{\bf G})=U^{\dagger}_{\eta,m,{\bf k}+{\bf q}}V^{\dagger}_{{\bf G}}U_{\eta,n,{\bf k}}$ are the multiband form factors.

Further using Eq.$\,$\ref{eq:c_kQ_def}
and Eq.$\,$\ref{eq:representations_cr_MoTe2}, we have $\mathcal{T}c_{{\bf k},{\bf Q},\eta}^{\dagger}\mathcal{T}^{-1}=(-\eta)c_{-{\bf k},-{\bf Q},-\eta}^{\dagger}$,
which therefore implies

\begin{equation}
\mathcal{T}\gamma_{{\bf k},m,\eta}^{\dagger}\mathcal{T}^{-1}=(-\eta)\gamma_{-{\bf k},m,-\eta}^{\dagger}\,\,(\textrm{gauge fixed})\,.\label{eq:TR_band_operators}
\end{equation}

\subsection{Particle-hole symmetry $\mathcal{C}$}

\label{sec:MoTe2_symmetries_C}

In this section, we provide details on the unitary particle-hole (or charge-conjugation) symmetry $\mathcal{C}$ introduced in Sec.$\,$\ref{sec:symmetries_fhmodel}. Unlike the other symmetries in the previous section, $\mathcal{C}$ is only a symmetry of the flat band Hamiltonian $H_I$ given in Eq.$\,$\ref{eq:H_V_general}, and not of the full Hamiltonian in Eq.$\,$\ref{eq:H_full_band_basis}. Additionally, as we will see below, this symmetry is general for density-density Hamiltonians of the form of $H_I$ (i.e. beyond the FH model here considered), as long as the opposite valleys are time-reversal partners, that is, the form factors satisfy Eq.$\,$\ref{eq:ff_TR_multiband}.

The representation of $\mathcal{C}$ in the continuum operators $c_{\eta,l,{\bf r}}^{\dagger}$ is given by

\bea
\mathcal{C}c_{\eta,l,{\bf r}}^{\dagger}\mathcal{C}^{-1}=(-\eta)c_{-\eta,l,{\bf r}} \,,
\label{eq:representations_cr_MoTe2_C}
\eea

Applying Eq.$\,$\ref{eq:representations_cr_MoTe2_C} twice gives
$\mathcal{C}^{2}c_{\eta,l,{\bf r}}^{\dagger}\mathcal{C}^{-2}
=-c_{\eta,l,{\bf r}}^{\dagger}$ (and analogously for
$c_{\eta,l,{\bf r}}$), and therefore
$\mathcal{C}^{2}=(-1)^{F}$.
From Eqs.$\,$\ref{eq:representations_cr_MoTe2} and \ref{eq:representations_cr_MoTe2_C}, it follows that $\mathcal{C}$ commutes with
the symmetry operations in Eq.$\,$\ref{eq:representations_cr_MoTe2},
\begin{equation}
[\mathcal{C},T_{\bf R}]
=
[\mathcal{C},C_{3}]
=
[\mathcal{C},C_{2y}\mathcal{T}]
=
[\mathcal{C},\mathcal{I}]
=
[\mathcal{C},\mathcal{T}]
=
0.
\label{eq:C_commutation_relations_MoTe2}
\end{equation}

Additionally, the elementary magnon operators transform under $\mathcal{C}$ as

\begin{equation}
\begin{array}{c}
\mathcal{C}S_{\parallel,\mbf r,l,j}^{\dagger}\mathcal{C}^{-1}=S_{\parallel,\mbf r,l,j}^{\dagger} \\
\mathcal{C}S_{\perp,\mbf r,l}^{\dagger}\mathcal{C}^{-1}=S_{\perp,\mbf r,-l}^{\dagger}
\,\,.
\end{array}
\end{equation}

The action of $\mathcal{C}$ on the plane-wave operators $c_{{\bf k},{\bf Q},\eta}$ and band operators $\gamma_{{\bf k},m,\eta}$ can also be derived using Eq.$\,$\ref{eq:representations_cr_MoTe2_C} to be

\bea
\mathcal{C}c_{{\bf k},{\bf Q},\eta}\mathcal{C}^{-1}=(-\eta)c_{-{\bf k},-{\bf Q},-\eta}^{\dagger} \\
\mathcal{C}\gamma_{{\bf k},m,\eta}^{\dagger}\mathcal{C}^{-1}=(-\eta)\gamma_{-{\bf k},m,-\eta}.\label{eq:C_c_gamma}
\eea

Using Eq.$\,$\ref{eq:C_c_gamma}, we can show explicitly that
$\mathcal C$ is a symmetry of the flat band projected interaction.
We start from the multiband projected density fluctuation,
\begin{align}
\delta\rho_{{\bf q}+{\bf G}}
&=
\sum_{{\bf k},\eta}\sum_{m,n}
M^{(\eta)}_{mn}({\bf k},{\bf q}+{\bf G})
\Big(
\gamma_{{\bf k}+{\bf q},m,\eta}^{\dagger}
\gamma_{{\bf k},n,\eta}
\nonumber\\
&\hspace{3.0cm}
-\frac{1}{2}\delta_{{\bf q},{\bf 0}}\delta_{mn}
\Big).
\label{eq:deltarho_multiband_appendix}
\end{align}
The bilinear operator appearing in this expression transforms as
\begin{align}
&\mathcal C
\Big(
\gamma_{{\bf k}+{\bf q},m,\eta}^{\dagger}
\gamma_{{\bf k},n,\eta}
-\frac{1}{2}\delta_{{\bf q},{\bf 0}}\delta_{mn}
\Big)
\mathcal C^{-1}
\nonumber\\
&\quad =
\gamma_{-{\bf k}-{\bf q},m,-\eta}
\gamma_{-{\bf k},n,-\eta}^{\dagger}
-\frac{1}{2}\delta_{{\bf q},{\bf 0}}\delta_{mn}
\nonumber\\
&\quad =
-\gamma_{-{\bf k},n,-\eta}^{\dagger}
\gamma_{-{\bf k}-{\bf q},m,-\eta}
+\frac{1}{2}\delta_{{\bf q},{\bf 0}}\delta_{mn}
\nonumber\\
&\quad =
-\Big(
\gamma_{-{\bf k},n,-\eta}^{\dagger}
\gamma_{-{\bf k}-{\bf q},m,-\eta}
-\frac{1}{2}\delta_{{\bf q},{\bf 0}}\delta_{mn}
\Big).
\label{eq:C_density_bilinear_action}
\end{align}
Therefore,
\begin{align}
&\mathcal C\delta\rho_{{\bf q}+{\bf G}}\mathcal C^{-1}
\nonumber\\
&=
-\sum_{{\bf k},\eta}\sum_{m,n}
M^{(\eta)}_{mn}({\bf k},{\bf q}+{\bf G})
\Big(
\gamma_{-{\bf k},n,-\eta}^{\dagger}
\gamma_{-{\bf k}-{\bf q},m,-\eta}
\nonumber\\
&\hspace{3.1cm}
-\frac{1}{2}\delta_{{\bf q},{\bf 0}}\delta_{mn}
\Big).
\label{eq:C_deltarho_firstline}
\end{align}
We now relabel ${\bf k}'=-{\bf k}-{\bf q}$ and $\eta'=-\eta$.
Then
\begin{align}
&\mathcal C\delta\rho_{{\bf q}+{\bf G}}\mathcal C^{-1}
\nonumber\\
&=
-\sum_{{\bf k}',\eta'}\sum_{m,n}
M^{(-\eta')}_{mn}(-{\bf k}'-{\bf q},{\bf q}+{\bf G})
\Big(
\gamma_{{\bf k}'+{\bf q},n,\eta'}^{\dagger}
\gamma_{{\bf k}',m,\eta'}
\nonumber\\
&\hspace{3.1cm}
-\frac{1}{2}\delta_{{\bf q},{\bf 0}}\delta_{mn}
\Big)
\nonumber\\
&=
-\sum_{{\bf k}',\eta'}\sum_{m,n}
M^{(-\eta')}_{nm}(-{\bf k}'-{\bf q},{\bf q}+{\bf G})
\Big(
\gamma_{{\bf k}'+{\bf q},m,\eta'}^{\dagger}
\gamma_{{\bf k}',n,\eta'}
\nonumber\\
&\hspace{3.1cm}
-\frac{1}{2}\delta_{{\bf q},{\bf 0}}\delta_{mn}
\Big),
\label{eq:C_deltarho_before_ff}
\end{align}
where in the last equality we exchanged the dummy band indices $m$ and
$n$. Comparing with Eq.$\,$\ref{eq:deltarho_multiband_appendix}, we see
that
\begin{equation}
\mathcal C\delta\rho_{{\bf q}+{\bf G}}\mathcal C^{-1}
=
-\delta\rho_{{\bf q}+{\bf G}}
\label{eq:C_deltarho}
\end{equation}
provided
\begin{equation}
M^{(-\eta)}_{nm}(-{\bf k}-{\bf q},{\bf q}+{\bf G})
=
M^{(\eta)}_{mn}({\bf k},{\bf q}+{\bf G}).
\label{eq:C_form_factor_condition}
\end{equation}
This condition follows from the time-reversal relation between
opposite-valley form factors in Eq.$\,$\ref{eq:ff_TR_multiband},
together with form-factor hermiticity condition,
\begin{equation}
M^{(\eta)*}_{mn}({\bf k}-{\bf q},{\bf q}+{\bf G})
=
M^{(\eta)}_{nm}({\bf k},-{\bf q}-{\bf G}).
\end{equation}
Eq.$\,$\ref{eq:C_deltarho} then immediately implies that $H_I$ in Eq.$\,$\ref{eq:H_V_general} is invariant under the action of $\mathcal{C}$,
\begin{equation}
\mathcal C H_I\mathcal C^{-1}=H_I.
\end{equation}

The derivation above also clarifies the physical role of the valley exchange in $\mathcal C$. After band projection, the
density operator contains the band form factors. Since the particle-hole transformation defined in Eq.$\,$\ref{eq:representations_cr_MoTe2_C} sends ${\bf k}$ to $-{\bf k}$, the form factors must satisfy an additional relation for $H_I$ to remain invariant. The valley exchange in $\mathcal{C}$ guarantees that this additional symmetry is time-reversal, which exists in the class of models we are describing. A valley-preserving particle-hole transformation would require an analogous additional symmetry in the
form-factors, which is not present in general.
For the general case of $N_b$ retained bands per valley, $\mathcal C$ is defined relative to half filling of the two-valley projected Hilbert space, namely $N_b$ holes per moiré unit cell. In the one-band projected problem used in the main text, this is the filling $\nu=-1$ that we considered.

\subsection{Additional details on the symmetry properties of the magnon eigenstates}

\label{sec:MoTe2_symmetries_MAG}

In this section, we show that the flat band magnon Hamiltonian defined in
Eq.$\,$\ref{eq:S_matrix_expression} separates into $\mathcal C$-even and
$\mathcal C$-odd sectors for every total momentum ${\bf p}$ in the
one-band projected limit. We define the magnon bilinears
\begin{equation}
\mathcal{O}_{\bf k}^{\dagger}({\bf p})
=
\gamma_{{\bf k}+{\bf p},\downarrow}^{\dagger}
\gamma_{{\bf k},\uparrow}\,,
\label{eq:Okp_def}
\end{equation}
which, acting on the fully polarized state $\ket{\uparrow}$, generate a
complete basis for the one-magnon Hilbert space at fixed ${\bf p}$.

Using Eq.$\,$\ref{eq:C_c_gamma} and fermionic anticommutation, we obtain
\begin{align}
\mathcal C
\mathcal{O}_{\bf k}^{\dagger}({\bf p})
\mathcal C^{-1}
&=
\gamma_{-{\bf k}-{\bf p},\uparrow}
\left(-\gamma_{-{\bf k},\downarrow}^{\dagger}\right)
\nonumber\\
&=
\mathcal{O}_{-{\bf k}-{\bf p}}^{\dagger}({\bf p})\,.
\label{eq:C_action_Okp}
\end{align}
Thus, $\mathcal C$ preserves the one-magnon Hilbert space at each fixed
${\bf p}$. The fully polarized state $\ket{\uparrow}$ is also mapped to
itself by $\mathcal C$ up to an overall phase. This follows because
$\mathcal C$ exchanges occupied spin-$\uparrow$ states with holes in the
spin-$\downarrow$ sector (the signs arising from reordering the fermionic
operators contribute only to this overall phase).

Defining
$\ket{{\bf k};{\bf p}}
\equiv
\mathcal O_{\bf k}^{\dagger}({\bf p})\ket{\uparrow}$,
and factoring out the common phase acquired by $\ket{\uparrow}$, the
induced action of $\mathcal C$ in the one-magnon Hilbert space is therefore
\begin{equation}
\ket{{\bf k};{\bf p}}
\longrightarrow
\ket{-{\bf k}-{\bf p};{\bf p}}\,,
\end{equation}
or, equivalently,
\begin{equation}
[U_{\mathcal C}({\bf p})]_{{\bf k}',{\bf k}}
=
\delta_{{\bf k}',-{\bf k}-{\bf p}}\,.
\label{eq:UC_matrix}
\end{equation}
Since applying this transformation twice returns ${\bf k}$ to itself,
$[U_{\mathcal C}({\bf p})]^2=\mathbb{1}$, and its eigenvalues are
$\zeta=\pm1$.

Finally, because $\mathcal C$ is a symmetry of the flat-band Hamiltonian,
$\mathcal C H_I\mathcal C^{-1}=H_I$, and the one-magnon Hilbert space at
fixed ${\bf p}$ is invariant under $\mathcal C$, the restriction of
$H_I$ to this subspace must commute with the induced representation
$U_{\mathcal C}({\bf p})$. Therefore,
\begin{equation}
[S^{(\uparrow)}({\bf p}),U_{\mathcal C}({\bf p})]=0\,.
\label{eq:S_commutes_UC}
\end{equation}
The magnon Hamiltonian consequently separates into $\mathcal C$-even
and $\mathcal C$-odd sectors at every ${\bf p}$, and eigenstates with
opposite $\mathcal C$ eigenvalues are orthogonal.

We finish this Appendix by showing that, in the one-band projected limit, the combined symmetry $\mathcal{I}\mathcal{C}$ acts trivially on the one-magnon Hilbert space at ${\bf p}=0$. From Eq.$\,$\ref{eq:C_action_Okp}, the magnon bilinears defined in Eq.$\,$\ref{eq:Okp_def} transform under $\mathcal{C}$ as
\begin{equation}
\mathcal{C}\mathcal{O}_{\bf k}^{\dagger}({\bf 0})\mathcal{C}^{-1}
=
\mathcal{O}_{-{\bf k}}^{\dagger}({\bf 0}).
\end{equation}
Using the inversion sewing phases defined in Eq.$\,$\ref{eq:sewing_matrices_MoTe2}, we then obtain
\begin{align}
\mathcal{I}\mathcal{C}
\mathcal{O}_{\bf k}^{\dagger}({\bf 0})
\mathcal{C}^{-1}\mathcal{I}^{-1}
&=
\mathcal{I}
\mathcal{O}_{-{\bf k}}^{\dagger}({\bf 0})
\mathcal{I}^{-1}
\nonumber\\
&=
e^{i[\varphi_{\down}^{\mathcal{I}}(-{\bf k})
-\varphi_{\up}^{\mathcal{I}}(-{\bf k})]}
\mathcal{O}_{\bf k}^{\dagger}({\bf 0}).
\label{eq:I_C_equation}
\end{align}
It therefore remains to show that
\begin{equation}
\varphi_{\up}^{\mathcal{I}}({\bf k})
=
-\varphi_{\up}^{\mathcal{I}}(-{\bf k})
=
\varphi_{\down}^{\mathcal{I}}({\bf k}).
\label{eq:varphi_I_relations}
\end{equation}
Using the sewing relations in Eqs.$\,$\ref{eq:sewing_matrices_MoTe2} and \ref{eq:TR_band_operators}, we have
\begin{align}
\mathcal{I}\mathcal{T}
\gamma_{{\bf k},\eta}^{\dagger}
\mathcal{T}^{-1}\mathcal{I}^{-1}
&=
e^{i\varphi_{-\eta}^{\mathcal{I}}(-{\bf k})}
(-\eta)\gamma_{{\bf k},-\eta}^{\dagger},
\\
\mathcal{T}\mathcal{I}
\gamma_{{\bf k},\eta}^{\dagger}
\mathcal{I}^{-1}\mathcal{T}^{-1}
&=
e^{-i\varphi_{\eta}^{\mathcal{I}}({\bf k})}
(-\eta)\gamma_{{\bf k},-\eta}^{\dagger}.
\end{align}
Since $[\mathcal{I},\mathcal{T}]=0$, this yields
\begin{equation}
\varphi_{-\eta}^{\mathcal{I}}(-{\bf k})
=
-\varphi_{\eta}^{\mathcal{I}}({\bf k})
\qquad (\mathrm{mod}\ 2\pi).
\label{eq:varphi_I_aux}
\end{equation}
In addition, $\mathcal{I}^{2}=1$ implies
\begin{equation}
\varphi_{\eta}^{\mathcal{I}}({\bf k})
+
\varphi_{\eta}^{\mathcal{I}}(-{\bf k})
=
0
\qquad (\mathrm{mod}\ 2\pi),
\end{equation}
which, together with Eq.$\,$\ref{eq:varphi_I_aux}, gives Eq.$\,$\ref{eq:varphi_I_relations}. Hence the phase factor in Eq.$\,$\ref{eq:I_C_equation} is unity, and
\begin{equation}
\mathcal{I}\mathcal{C}
\mathcal{O}_{\bf k}^{\dagger}({\bf 0})
\mathcal{C}^{-1}\mathcal{I}^{-1}
=
\mathcal{O}_{\bf k}^{\dagger}({\bf 0})
\end{equation}
for every ${\bf k}$. Thus, $\mathcal{I}\mathcal{C}$ acts trivially on the ${\bf p}=0$ one-magnon Hilbert space, and every magnon eigenstate at ${\bf p}=0$ is invariant under $\mathcal{I}\mathcal{C}$.

\section{Additional details and results on magnon excitations for twisted MoTe$_2$}

\label{sec:App_Additional_MoTe2}

\subsection{Details on the numerical calculation}

\label{sec:sec_details_magnon_exact}

In this section we provide additional details on the exact numerical calculation of the magnon excitations introduced in Sec.$\,$\ref{sec:exact_mag_excitations_Smatrix}.

At each momentum ${\bf p}$, the magnon excitation spectrum can
be numerically obtained from Eq.$\,$\ref{eq:S_matrix_commutator} by diagonalizing the matrix
$[S^{(\eta)}({\bf p})]_{{\bf k}',{\bf k}}$ using a finite mesh of ${\bf k}$-points. We consider momentum meshes
defined by the two reciprocal lattice vectors $\bsl f_{1}$ and $\bsl f_{2}$
given by

\begin{equation}
\begin{array}{cc}
\bsl f_{1} & =n_{11}\bsl b_{M,1}+n_{12}\bsl b_{M,2}\\
\bsl f_{2} & =n_{21}\bsl b_{M,1}+n_{22}\bsl b_{M,2}\,.
\end{array}
\end{equation}
where $\bsl b_{M,1}$ and $\bsl b_{M,2}$ are the moiré reciprocal
lattice vectors and $n_{11},n_{12},n_{21},n_{22}$ are integers satisfying
$|n_{11}n_{22}-n_{12}n_{21}|=1$. This latter condition ensures that the transformation from $(\bsl b_{M,1},\bsl b_{M,2})$ to $(\bsl f_1,\bsl f_2)$ is invertible using integer coefficients. Consequently, every vector of the original moiré reciprocal lattice can also be written as a linear combination of $\bsl f_1$ and $\bsl f_2$ with integer coefficients, so the $\bsl f_1$ and $\bsl f_2$ vectors generate the full moiré reciprocal lattice rather than only a subset of its points. We take momentum meshes of $N_{x},N_{y}$
momenta along the directions $\bsl f_{1}$ and $\bsl f_{2}$ respectively,
that is,

\begin{equation}
\bsl k=\frac{k_{x}}{N_{x}}\bsl f_{1}+\frac{k_{y}}{N_{y}}\bsl f_{2}\,,
\end{equation}
where $k_{x}=0,\cdots,N_{x}-1$ and $k_{y}=0,\cdots,N_{y}-1$, corresponding
to a system size of $N_{UC}=N_{x}N_{y}$ unit cells. In most of the calculations in the main text, we used the momentum mesh defined by $\ensuremath{(N_{x},N_{y},n_{11},n_{12},n_{21},n_{22})=(12,12,1,0,0,1)}$.

\subsection{Contribution of one-body terms to S-matrix}

\label{subsec:1body_contribution}

In the main text, we showed the explicit S-matrix expression for the magnon excitations of the flat band Hamiltonian $H_I$ in Eq.$\,$\ref{eq:S_matrix_expression}. The same procedure can be applied to the Hamiltonian in Eq.$\,$\ref{eq:H_full_band_basis}.
The single-particle terms in this Hamiltonian can be accounted for by simply re-defining the one-body contributions in Eq.$\,$\ref{eq:S_matrix_expression}
as
\bea
R^{(\pm\eta)}({\bf k})
\rightarrow
R^{(\pm\eta)}({\bf k})
\mp
\left[
E_{\pm\eta}({\bf k})
+\epsilon_{HF,\pm\eta}({\bf k})
\right].
\label{eq:R_one_body_replacement}
\eea
As stated in Sec.$\,$\ref{sec:exact_mag_excitations_Smatrix} of the main text, the additional contributions from the renormalized dispersion
$E_{\eta}({\bf k})+\epsilon_{HF,\eta}({\bf k})$
cancel exactly in the magnon Hamiltonian at ${\bf p}=0$ in the presence of time reversal $\mathcal T$
and pseudo-inversion $\mathcal I$ symmetries. We now prove this result.

Setting $g=\mathcal T$ in
Eq.$\,$\ref{eq:Dhk_hgk_D}, we obtain
\begin{equation}
{\bf h}(-{\bf k})
=
D[\mathcal T]{\bf h}({\bf k})D[\mathcal T]^\dagger.
\end{equation}
Using the representation in
Eq.$\,$\ref{eq:D_representations_MoTe2}, the operator $\mathcal K$
complex conjugates the Hamiltonian matrix, while the Kronecker deltas
exchange $\eta\rightarrow-\eta$ and
${\bf Q}\rightarrow-{\bf Q}$. Therefore,
\begin{equation}
h^\eta_{{\bf Q},{\bf Q}'}(-{\bf k})
=
\left[
h^{-\eta}_{-{\bf Q},-{\bf Q}'}({\bf k})
\right]^*.
\label{eq:h_TR_relation}
\end{equation}
Similarly, pseudo-inversion symmetry gives, by using $g=\mathcal I$ in
Eq.$\,$\ref{eq:Dhk_hgk_D},
\begin{equation}
h^\eta_{{\bf Q},{\bf Q}'}(-{\bf k})
=
h^\eta_{-{\bf Q},-{\bf Q}'}({\bf k}).
\label{eq:h_I_relation}
\end{equation}
Combining Eq.$\,$\ref{eq:h_TR_relation} and Eq.$\,$\ref{eq:h_I_relation} then yields
\begin{equation}
h^{-\eta}_{{\bf Q},{\bf Q}'}({\bf k})
=
\left[
h^\eta_{{\bf Q},{\bf Q}'}({\bf k})
\right]^*,
\end{equation}
or equivalently, $h^{-\eta}({\bf k})=h^\eta({\bf k})^*$.
The two Hamiltonian matrices consequently have identical spectra,
implying that $E_{-\eta}({\bf k})=E_\eta({\bf k})$.

We can also
use $D[\mathcal{I}]U_{\eta,{\bf k}}=e^{i\varphi_{\eta}^{\mathcal{I}}({\bf k})}U_{\eta,-{\bf k}}$ (recall Eq.$\,$\ref{eq:gamma_def} for definition of $U_{\eta,{\bf k}}$, as well as Eqs.$\,$\ref{eq:symmetry_relation_MoTe2}-\ref{eq:sewing_matrices_MoTe2} for this equality)
to prove that

\begin{equation}
\begin{array}{c}
M_{\eta}({\bf k},{\bf {\bf G}})=U_{\eta,{\bf k}}^{\dagger}D^{\dagger}[\mathcal{I}]D[\mathcal{I}]V_{{\bf G}}^{\dagger}D^{\dagger}[\mathcal{I}]D[\mathcal{I}]U_{\eta,{\bf k}}\\
=U_{\eta,-{\bf k}}^{\dagger}V_{-{\bf G}}^{\dagger}U_{\eta,-{\bf k}}=M_{\eta}(-{\bf k},-{\bf {\bf G}})\,,
\end{array}
\end{equation}

Similarly, under $\mathcal{I}$ we also have $|M_{\eta}({\bf k},{\bf q+{\bf G}})|=|M_{\eta}(-{\bf k},-{\bf q-{\bf G}})|$.
Using these properties, we can readily show that $\epsilon_{HF,\up}({\bf k})=\epsilon_{HF,\down}({\bf k})$. Therefore, at ${\bf p}=0$, the particle and hole contributions from the renormalized dispersion entering the $S$-matrix in Eq.$\,$\ref{eq:S_matrix_expression} with the substitution in Eq.$\,$\ref{eq:R_one_body_replacement} cancel exactly.

\begin{figure}[h!]
\begin{centering}
\includegraphics[width=1\columnwidth]{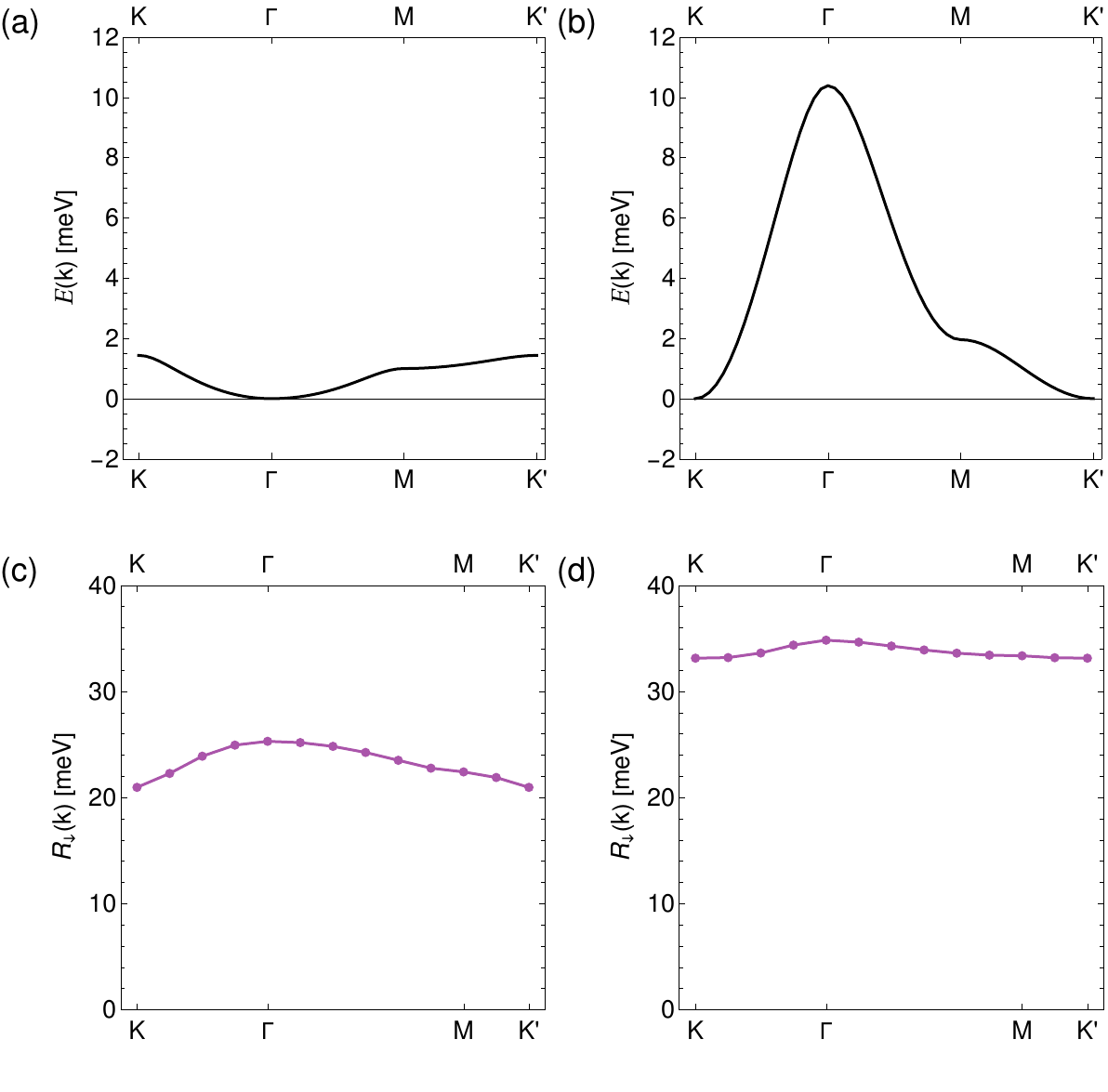}\caption{Band structure and charge-1 gap for the FH model of twisted MoTe$_2$.
(a,b) Band structure of the two lowest-energy hole bands for $\theta=2^{\circ}$
(a) and $\theta=3.7^{\circ}$ (b). (c,d) Charge-1 gaps computed using
Eq.$\,$\ref{eq:Rmat} for $\theta=2^{\circ}$ (c) and $\theta=3.7^{\circ}$
(d). \label{fig:App_MoTe2_BSvsR}}
\par\end{centering}
\end{figure}

\subsection{Additional results for FH model of MoTe$_2$}

\label{subsec:App_Additional_MoTe2_Exact}

In this section, we provide additional results for the FH
model of MoTe$_2$, to complement the results in Sec.$\,$\ref{sec:moire_magnons}.

The hole band structure for the two lowest-energy hole bands is shown
in Fig.$\,$\ref{fig:App_MoTe2_BSvsR}, together with the exact charge-1
gap $R_{\down}({\bf k})$.

\begin{figure}[h!]
\includegraphics[width=1\columnwidth]{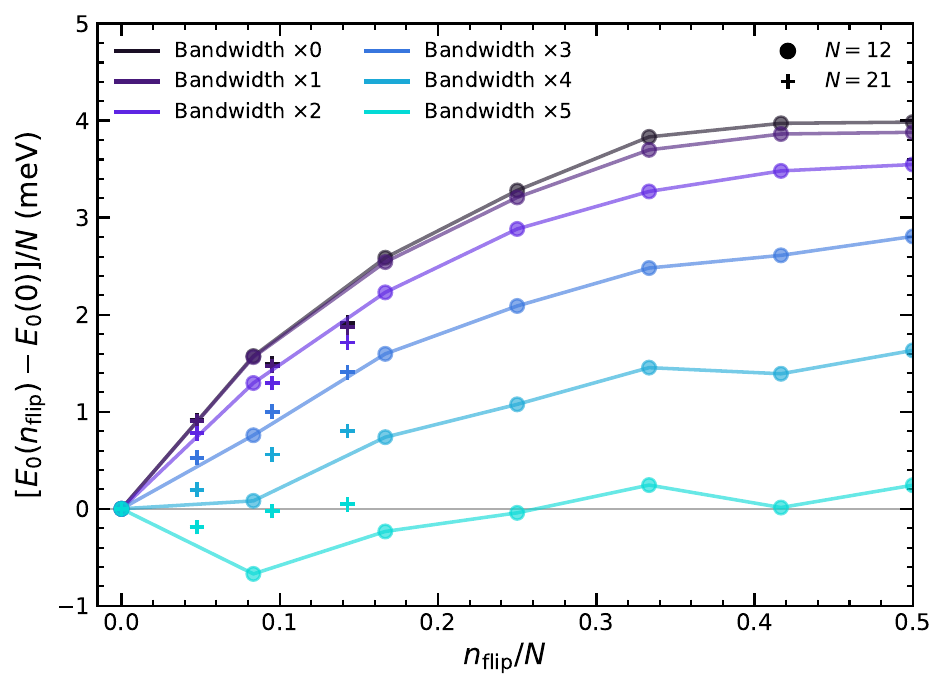}
\caption{Energy per particle for the FH model of MoTe$_2$ at
$\theta=3.7^\circ$, with an artificial increase of the single-particle
bandwidth by the factor shown on top of each figure, as a function of
the number of flipped spins $n_{\rm flip}$ relative to the fully
polarized state. The spin sectors are therefore defined by $S_z=(N_\uparrow-N_\downarrow)/2$, where
$N_\uparrow=N-n_{\rm flip}$ and $N_\downarrow=n_{\rm flip}$.
We show results for a 12-site system (circles), characterized by the
$k$-mesh
$(N_x,N_y,n_{11},n_{12},n_{21},n_{22})=(6,2,1,1,0,1)$,
and a 21-site system (crosses), characterized by
$(N_x,N_y,n_{11},n_{12},n_{21},n_{22})=(21,1,1,-5,0,1)$.
\label{fig:App_MoTe2_polarization_checks}}
\end{figure}

\begin{figure}[h!]
\includegraphics[width=1\columnwidth]{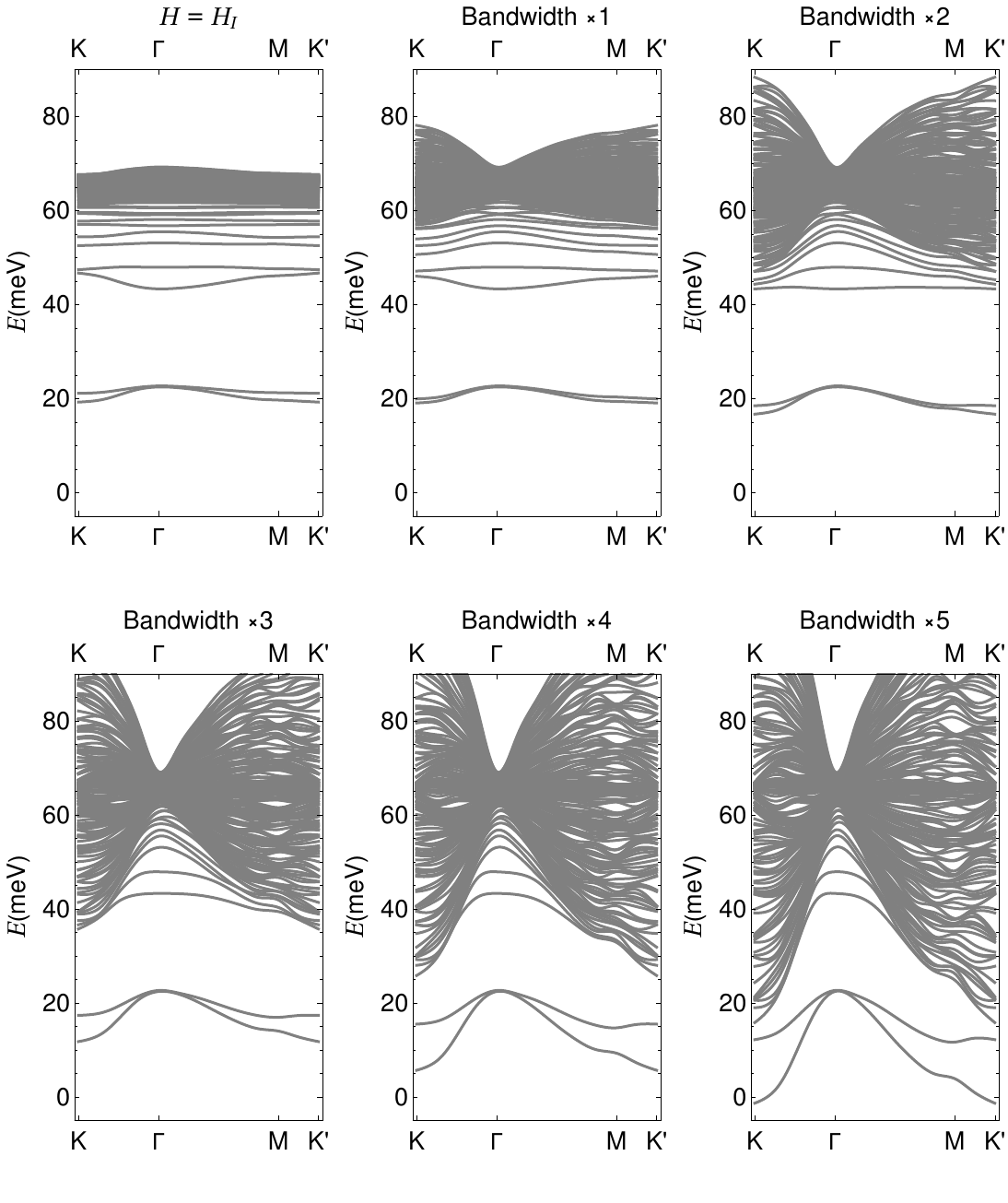}\caption{Magnon spectrum for the FH model for MoTe$_2$ for $\theta=3.7\text{°}$,
with an artificial increase of the single-particle bandwidth by the
factor shown on top of each figure. The first figure shows the results
for the simplified flat band Hamiltonian $H_{I}$ in Eq.$\,$\ref{eq:H_V_general}.\label{fig:App_MoTe2_variableDispersion}}
\end{figure}

In the main text, in Sec.$\,$\ref{subsec:FH_model}, we claimed that the fully-polarized state $\ket{\eta}$ is the ground-state even when the single-particle dispersion in Eq.$\,$\ref{eq:H_full_band_basis} is considered. We now present numerical results to support this claim (note however, that we do not have a proof for it, just numerical evidence on finite system sizes). First, in Fig.~\ref{fig:App_MoTe2_polarization_checks}, we show, for each spin sector, the minimum energy over all many-body momentum sectors using exact diagonalization of both a 12-site and a 21-site lattice. We label the spin sectors by $n_{\rm flip}$, the number of spins flipped relative to the fully polarized state, such that $N_\uparrow = N - n_{\rm flip}$, and
$N_\downarrow = n_{\rm flip}$. To probe the stability of the fully polarized state against an increased kinetic-energy scale, we artificially enhance the single-particle bandwidth by a factor $f_{\rm band}$ while keeping the interaction strength fixed. We find that the fully polarized state remains the ground state up to $f_{\rm band}=4$. Upon further increasing the bandwidth, the first instability occurs in the $n_{\rm flip}=1$ sector for both the 12-site and 21-site lattices (as pointed out by the negative energies with respect to the fully polarized state), indicating that the lowest-energy state is obtained by a single spin flip on top of the fully polarized state, i.e. a magnon. To characterize this softening in more detail, in Fig.~\ref{fig:App_MoTe2_variableDispersion} we show the exact single-spin-flip spectrum obtained from Eq.~\ref{eq:S_matrix_commutator}, computed by diagonalizing the matrix $[S^{(\eta)}({\bf p})]_{{\bf k}',{\bf k}}$ on a finite mesh of ${\bf k}$-points, and by including dispersion through the substitution in Eq.$\,$\ref{eq:R_one_body_replacement}. At ${\bf p}=0$, pseudo-inversion symmetry guarantees that the results are always exactly the same as for the flat band limit. Away from ${\bf p}=0$, substantial
changes in the low-lying modes are only observed when the bandwidth
is increased by a factor larger than $f_{\rm band}=3$, with the magnon gap collapsing
at ${\bf p}={\bf K},{\bf K}'$ when the factor is approximately $f_{\rm band}=5$ which is consistent with the exact diagonalization results in Fig.~\ref{fig:App_MoTe2_polarization_checks}.

\begin{figure*}
\includegraphics[width=1\textwidth]{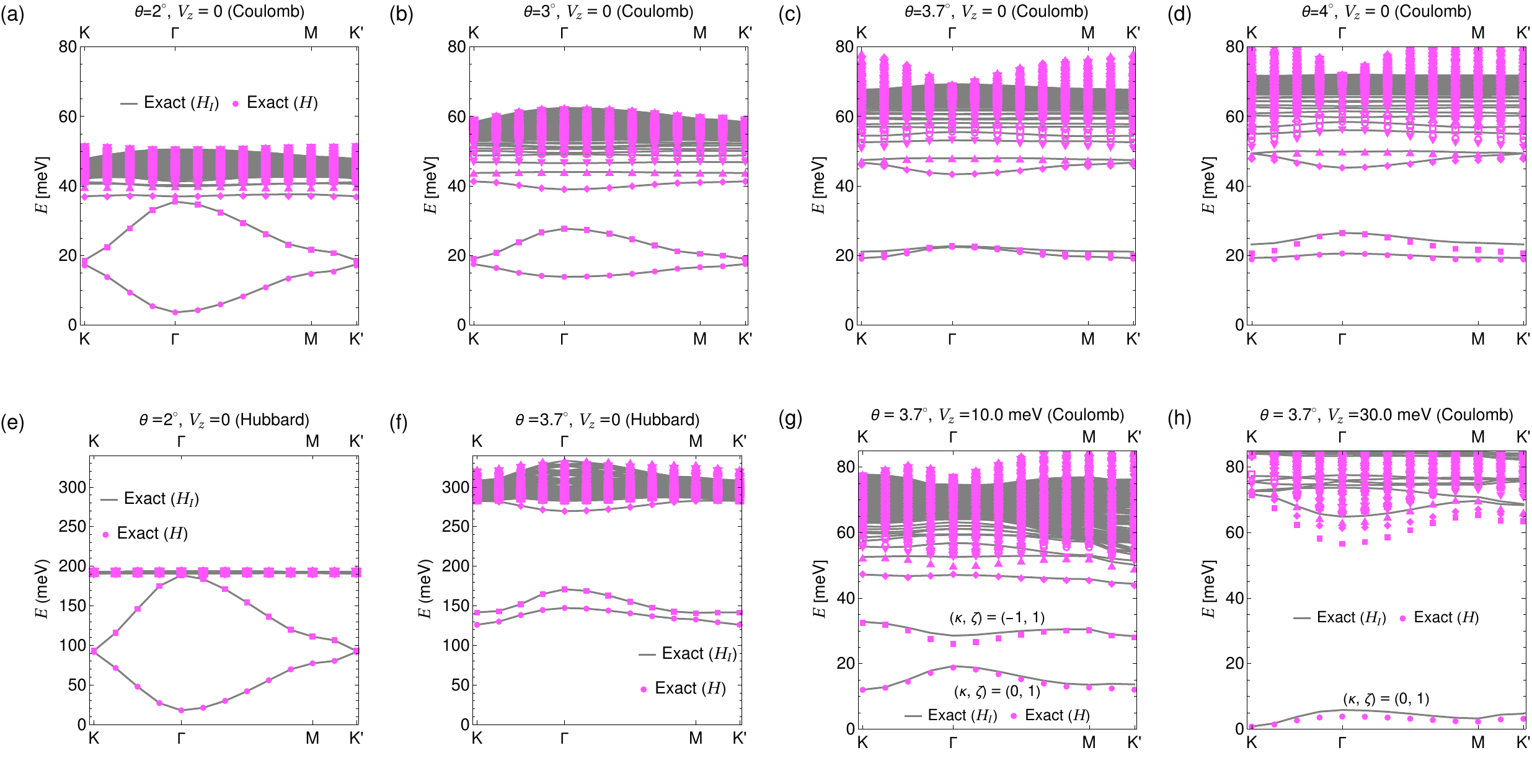}\caption{Comparison of magnon spectrum results for full Hamiltonian in Eq.$\,$\ref{eq:H_full_band_basis}
(pink points) and the flat band Hamiltonian in Eq.$\,$\ref{eq:H_V_general}
(full gray lines), for the FH model of MoTe$_2$ with different parameters
specified on top of each figure. A momentum mesh defined by $\ensuremath{(N_{x},N_{y},n_{11},n_{12},n_{21},n_{22})=(12,12,1,0,0,1)}$
was used in all the results. \label{fig:App_MoTe2_H_vs_HI}}
\end{figure*}

In the main text, in Secs.$\,$\ref{subsec:FH_model} and \ref{sec:exact_mag_excitations_Smatrix}, we also mentioned that the results for the lowest-energy magnon
modes are very similar no matter whether we consider the full Hamiltonian
in Eq.$\,$\ref{eq:H_full_band_basis}, or the simplified flat band
Hamiltonian $H_{I}$. We now present results supporting this claim.
In Fig.$\,$\ref{fig:App_MoTe2_H_vs_HI}, we compare the magnon spectra
obtained using the full Hamiltonian in Eq.$\,$\ref{eq:H_full_band_basis}
and the simplified flat band Hamiltonian in Eq.$\,$\ref{eq:H_V_general},
where an excellent agreement can be observed for the lowest-energy
magnon modes, while significant deviations arise in the continuum.
The results in Fig.$\,$\ref{fig:App_MoTe2_H_vs_HI} include all the
parameters used in Figs.$\,$\ref{fig:2}(b,c), Figs.$\,$\ref{fig:4}(b,d)
and Figs.$\,$\ref{fig:MoTe2_topo_vs_trivial}(a,c) of the main text.

\subsection{Diagonalization of the magnon Hamiltonian in the elementary SyMA basis}
\label{subsec:subspace_diagonalization_SyMA basis}

In Sec.$\,$\ref{sec:syma_fhmodel} we introduced the elementary
projected magnon operators
\begin{equation}
\bar{S}_{\parallel,\mbf p,{\bf G},l,j}^{\dagger}
\quad\text{and}\quad
\bar{S}_{\perp,\mbf p,{\bf G},l}^{\dagger}
\label{eq:app_elementary_SyMA_ops}
\end{equation}
defined in Eq.$\,$\ref{eq:eq:S_p,G,l_inter}, which form a
local basis of elementary intralayer and interlayer spin-flip bilinears. The analytical SyMA modes
of Sec.$\,$\ref{subsec:syma_fhmodel_model} are particular linear
combinations of a small subset of these operators, selected by symmetry
and by retaining only the smallest $|{\bf G}|$ compatible with each
symmetry sector. Here we instead diagonalize the magnon Hamiltonian
directly in the subspace spanned by \emph{all} elementary operators in
Eq.$\,$\ref{eq:app_elementary_SyMA_ops} with $|{\bf G}|$ below a
tunable cutoff. This provides a controlled numerical test of the
small-$|{\bf G}|$ truncation underlying the SyMA construction, without
imposing any additional symmetry-adapted coefficients.

Concretely, for a fixed total magnon momentum ${\bf p}$ and cutoff
$G_{\max}$, we collect all the operators in
Eq.$\,$\ref{eq:eq:S_p,G,l_inter} into the normalized states
\begin{equation}
\ket{\mathcal{O}_a({\bf p})}
=
\frac{
\bar{S}_a^{\dagger}({\bf p})\ket{\uparrow}
}{
\|\bar{S}_a^{\dagger}({\bf p})\ket{\uparrow}\|
},
\qquad
a=1,\dots,N_B(G_{\max}),
\end{equation}
where $\bar{S}_a^{\dagger}({\bf p})$ denotes the operators in
Eq.$\,$\ref{eq:app_elementary_SyMA_ops}, with the index $a$ running over $a=(\parallel,{\bf G},l=\pm 1,j=\{1,2,3\})$ for the 
intralayer operators and $a=(\perp,{\bf G},l=\pm 1)$ for the interlayer operators, with reciprocals ${\bf G}$ satisfying  in both cases
\begin{equation}
|{\bf G}|\le G_{\max}|{\bf b}_{M,1}|.
\end{equation}
Here $N_B(G_{\max})$ is the number of retained elementary operators.

Because these states are generally nonorthogonal and may become
linearly dependent as $G_{\max}$ is increased, the variational magnon
energies are obtained from the generalized eigenvalue problem
\bea
\sum_{b=1}^{N_B}
\mathcal{S}^{\rm proj}_{ab}({\bf p}){\bf v}_b
=
E({\bf p})
\sum_{b=1}^{N_B}
O_{ab}({\bf p}){\bf v}_b.
\label{eq:app_SyMA_gevp}
\eea
Here
\bea
\mathcal{S}^{\rm proj}_{ab}({\bf p})
&=
\bra{\mathcal{O}_a({\bf p})}
S^{(\uparrow)}({\bf p})
\ket{\mathcal{O}_b({\bf p})},
\\
O_{ab}({\bf p})
&=
\braket{\mathcal{O}_a({\bf p})}{\mathcal{O}_b({\bf p})},
\eea
are, respectively, the projected magnon matrix and overlap matrix, and
${\bf v}\in\mathbb{C}^{N_B}$ is the vector of expansion coefficients.
Eq.$\,$\ref{eq:app_SyMA_gevp} is the Rayleigh--Ritz projection of
$S^{(\uparrow)}({\bf p})$ onto the subspace generated by the elementary
projected operators.

To handle the possible near-null space of the overlap matrix
$O({\bf p})$, we first compute its eigendecomposition
\begin{equation}
O({\bf p})=W\Lambda W^\dagger,
\end{equation}
where $\Lambda=\mathrm{diag}(\lambda_1,\dots,\lambda_{N_B})$ and $W$
is the unitary matrix whose columns are the corresponding orthonormal
eigenvectors. We retain only those eigenvectors satisfying
$\lambda_i>\texttt{tol}\,\lambda_{\max}$, with
$\texttt{tol}=10^{-8}$, collecting them in $W_{\mathrm{keep}}$ with
eigenvalues $\Lambda_{\mathrm{keep}}$, and define
\begin{equation}
\widetilde{W}
=
W_{\mathrm{keep}}\Lambda_{\mathrm{keep}}^{-1/2}.
\end{equation}
Substituting ${\bf v}=\widetilde{W}\tilde{\bf v}$ into
Eq.$\,$\ref{eq:app_SyMA_gevp} reduces it to the standard eigenvalue
problem
\begin{equation}
\widetilde{\mathcal{S}}({\bf p})\tilde{\bf v}
=
E({\bf p})\tilde{\bf v},
\qquad
\widetilde{\mathcal{S}}({\bf p})
=
\widetilde{W}^{\dagger}
\mathcal{S}^{\rm proj}({\bf p})
\widetilde{W}.
\end{equation}
The resulting eigenvalues $E({\bf p})$ are the variational magnon
energies in the retained subspace. The number of retained eigenvectors
defines an effective basis dimension
$N_B^{\mathrm{eff}}({\bf p})\leq N_B$, which can vary with ${\bf p}$.

\begin{figure*}
\begin{centering}
\includegraphics[width=\textwidth]{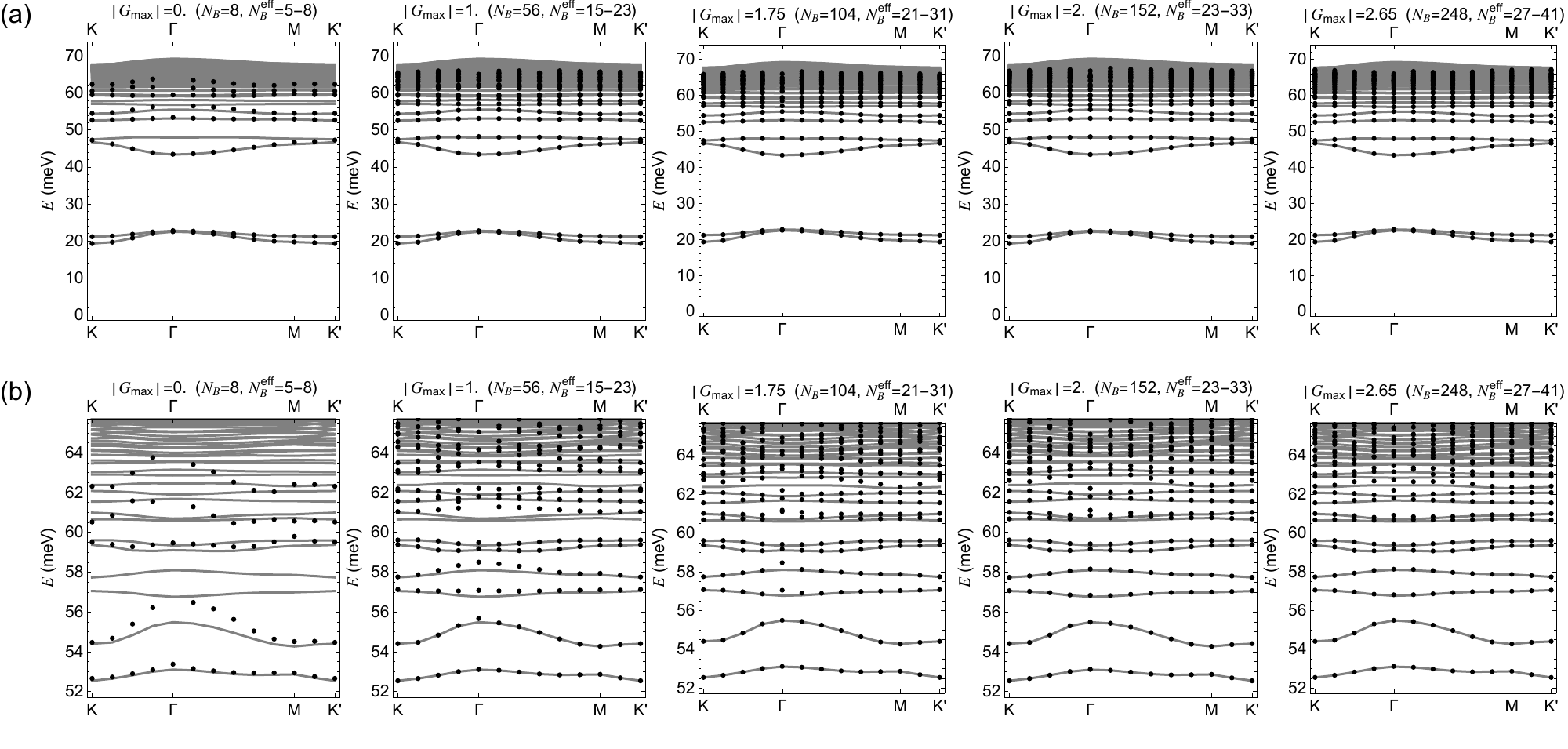}
\par\end{centering}
\caption{Exact magnon spectrum (gray) versus diagonalization of
the magnon Hamiltonian $S^{(\uparrow)}({\bf p})$ in the subspace
spanned by the elementary projected SyMA operators of
Eq.$\,$\ref{eq:eq:S_p,G,l_inter} with
$|{\bf G}|\le G_{\max}|{\bf b}_{M,1}|$ (black points), for the FH
model of MoTe$_2$ at $\theta=3.7^{\circ}$ and $V_z=0$. All results were
obtained for the flat band Hamiltonian $H_I$ in
Eq.$\,$\ref{eq:H_V_general} with the double-gated Coulomb interaction,
on the momentum mesh
$(N_x,N_y,n_{11},n_{12},n_{21},n_{22})=(12,12,1,0,0,1)$.
Each column corresponds to increasing $G_{\max}$. The panel titles
indicate the number of retained elementary operators $N_B$ and the
range of effective (linearly independent) dimensions
$N_B^{\mathrm{eff}}$ along the plotted path (see text for details). (a) Full spectrum.
(b) Zoom-in on the higher-energy window $E\in[52-65]\, \textrm{meV}$.
\label{fig:subspace_diagonalization_SyMA basis}}
\end{figure*}

Figure$\,$\ref{fig:subspace_diagonalization_SyMA basis} shows the
resulting spectrum for the FH model of MoTe$_2$ at $\theta=3.7^{\circ}$
and $V_z=0$, using the flat band Hamiltonian $H_I$ in
Eq.$\,$\ref{eq:H_V_general} with the double-gated Coulomb interaction.
Each column corresponds to a successively larger cutoff $G_{\max}$,
with the raw operator count $N_B$ and the range of retained dimensions
$N_B^{\mathrm{eff}}$ along the plotted path in the moiré Brillouin zone.

Some comments on these results are in order. First, already at
$G_{\max}=0$ - corresponding to the eight elementary operators with
${\bf G}={\bf 0}$ (six intralayer and two interlayer) - the two
lowest-energy magnon branches are captured very accurately
throughout the moir\'e Brillouin zone. This provides direct numerical
support for the analytical construction of Sec.$\,$\ref{subsec:syma_fhmodel_model},
in which the two lowest $\mathcal{C}$-even modes are built exclusively
from ${\bf G}={\bf 0}$ intralayer operators. Second, increasing
$G_{\max}$ allows to capture an increasing portion of higher-energy
magnon modes. Interestingly, restricting
$|{\bf G}|\le|{\bf b}_{M,1}|$ ($G_{\max}=1$) already accounts for the
six lowest-energy magnon modes with high accuracy, while larger cutoffs
continue to improve the description of states closer to the dense
particle-hole continuum.
Third, as $G_{\max}$ is increased, $N_B^{\mathrm{eff}}$ becomes significantly smaller than the raw number of included basis elements $N_B$, reflecting that a significant number of basis elements with larger $|{\bf G}|$ become linearly
dependent and do not enlarge the physical subspace.

These results therefore establish that the elementary SyMA basis of
Eq.$\,$\ref{eq:app_elementary_SyMA_ops}, truncated to a small number of
reciprocal-lattice shells, captures well the low-energy magnon subspace of the
FH model.
The analytical SyMA modes of
Sec.$\,$\ref{subsec:syma_fhmodel_model} are particular symmetry-adapted linear combinations within this low-rank subspace.

\subsection{Detailed SyMA expressions in the band basis}

\label{subsec:detailed_syma_bandbasis}

In this Appendix we provide explicit expressions in the band basis for all the magnon modes defined in Sec.$\,$\ref{sec:sma_main}.
There, we defined the different SyMA modes in Eqs.$\,$\ref{eq:intralayer_Sp} and \ref{eq:interlayer_Sp_k_0_1}. Using the hole band creation/annihilation operators, these SyMA modes can be written as

\begin{align}
 \bar{S}_{\mbf p}^{\dagger(\kappa,+)}&=\sum_{{\bf k}}M^{(\down,\up)}_{(\kappa,+1)}({\bf k},{\bf p})\gamma_{{\bf k}+{\bf p},\down}^{\dagger}\gamma_{{\bf k},\up} \label{eq:Sp_par_band}\\
\bar{S}_{\mbf p}^{\dagger(\kappa,-)}
&=\sum_{{\bf k}}M^{(\down,\up)}_{(\kappa,-1)}({\bf k},{\bf p})
\gamma_{{\bf k}+{\bf p},\down}^{\dagger}\gamma_{{\bf k},\up}.
\end{align}

with the magnon wave functions defined by
\bea
M^{(\down,\up)}_{(\kappa,+1)}({\bf k},{\bf p})&=U_{\down,{\bf k}+{\bf p}}^{\dagger}D_{(\kappa)}^{\parallel}U_{\up,{\bf k}} \ , \\
D_{(\kappa)}^{\parallel} &= \frac{1}{3}\sum_{j=0}^{2}e^{-\frac{2\pi i}{3}(\kappa+1)j}\theta_{j}
\label{eq:D_par_FH}
\eea
for the intra-layer modes, and
\bea
M^{(\down,\up)}_{(\kappa,-1)}({\bf k},{\bf p})=U_{\down,{\bf k}+{\bf p}}^{\dagger}D_{(\kappa)}^{\perp}U_{\up,{\bf k}}\\
D_{(\kappa)}^{\perp}=\begin{cases}
\tau_{-1} &,\kappa=-1\\
\frac{1}{6}\sum_{j=-2}^{3}e^{\frac{2\pi i}{3}(\kappa-2)j}\tau_{-1}\tilde{\theta}_{j} &,\kappa=0,1
\end{cases}
\label{eq:D_perp_FH}
\eea
for the inter-layer modes, and we defined the matrices

\bea
[\theta_{j}]_{{\bf Q},{\bf Q}'}=\delta_{{\bf {\bf Q}}',{\bf Q}+(-1)^{{\bf Q}}C_{3}^{j}{\bf q}_{1}}
\label{eq:thetaj}
\eea

\bea
[\tilde{\theta}_{j}]_{{\bf Q},{\bf Q}'}=\delta_{{\bf {\bf Q}}',{\bf Q}-C_{6}^{j}{\bf b}_{1}}
\label{eq:thetajtilde}
\eea

\bea
[\tau_{-1}]_{{\bf Q},{\bf Q}'}=-(-1)^{{\bf Q}}\delta_{{\bf Q}',{\bf Q}}
\label{eq:tauminus1}
\eea

and introduced the notation
\begin{equation}
 (-1)^{{\bf Q}}=\pm1 \textrm{ for } {\bf Q}\in\{{\bf G}\pm{\bf q}_{1}\} \,.
\end{equation}

We will carry out the detailed derivation for the intralayer modes, for clarity.
The projected version of the basis elements defined in Eq.$\,$\ref{eq:eq:S_p,G,l_inter} is given by

\begin{widetext}
\bea
\bar{S}_{\parallel,\mbf p,{\bf G},l,j}^{\dagger}&=\int d^{2}{\bf r}\,\,e^{i({\bf p}+{\bf G})\cdot{\bf r}}e^{-il{\bf q}_{j}\cdot\mbf r}\bar{c}_{\down,l,\mbf r}^{\dag}\bar{c}_{\up,l,\mbf r}\\
&=\frac{1}{\Omega_{tot}} \sum_{{\bf k},{\bf k}'}\sum_{{\bf Q}'\in\mathcal{Q}_{l}^{\down},{\bf Q}\in\mathcal{Q}_{l}^{\up}}\int d^{2}{\bf r}\,\,e^{i({\bf p}+{\bf G}-{\bf k}'+{\bf Q}'+{\bf k}-{\bf Q})\cdot{\bf r}}e^{-il{\bf q}_{j}\cdot\mbf r}\bar{c}_{{\bf k}',{\bf Q}',\down}^{\dagger}\bar{c}_{{\bf k},{\bf Q},\up}\\
&=\frac{1}{\Omega_{tot}}\sum_{{\bf k},{\bf k}'}\sum_{{\bf Q}'\in\mathcal{Q}_{l}^{\down},{\bf Q}\in\mathcal{Q}_{l}^{\up}}\int d^{2}{\bf r}\,\,e^{i({\bf p}-{\bf k}'+{\bf k}+{\bf Q}'-{\bf Q}+{\bf G}-l{\bf q}_{j})\cdot{\bf r}}[U_{\down,{\bf k}'}^{*}]_{{\bf Q}'}[U_{\up,{\bf k}}]_{{\bf Q}}\gamma_{{\bf k}',\down}^{\dagger}\gamma_{{\bf k},\up}\\
&=\sum_{{\bf k}}\sum_{{\bf Q}'\in\mathcal{Q}_{l}^{\down},{\bf Q}\in\mathcal{Q}_{l}^{\up}}[U_{\down,{\bf k}+{\bf p}}^{*}]_{{\bf Q}'}\delta_{{\bf Q}',{\bf Q}-{\bf G}+l{\bf q}_{j}}[U_{\up,{\bf k}}]_{{\bf Q}}\gamma_{{\bf k}+{\bf p},\down}^{\dagger}\gamma_{{\bf k},\up}\,\,.
\label{eq:S_intra_band_basis_derivation}
\eea
\end{widetext}
where $\mathcal{Q}_{l}^{\eta}$ were defined below Eq.$\,$\ref{eq:c_kQ_def} ($\mathcal{Q}_{l}^{\eta}=\{{\bf G}+\eta(-1)^{(l+1)/2}{\bf q}_{1}\}$) and we used Eqs.$\,$\ref{eq:c_kQ_def},\ref{eq:gamma_def} and the Fourier identity
\bea
\int d^{2}{\bf r}\,\,e^{i({\bf p}-{\bf k}'+{\bf k}+{\bf Q}'-{\bf Q}+{\bf G}-l{\bf q}_{j})\cdot{\bf r}} =\Omega_{tot}\delta_{{\bf k}',{\bf k}+{\bf p}}\delta_{{\bf Q}',{\bf Q}-{\bf G}+l{\bf q}_{j}} \ .
\eea
For completeness, we also add below the expression for the interlayer basis elements defined in Eq.$\,$\ref{eq:eq:S_p,G,l_inter}, which is given by

\begin{equation}
\bar{S}_{\perp,\mbf p,{\bf G},l}^{\dagger}
=
\sum_{{\bf k}}
\sum_{{\bf Q}'\in\mathcal{Q}_{-l}^{\down},
      {\bf Q}\in\mathcal{Q}_{l}^{\up}}
[U_{\down,{\bf k}+{\bf p}}^{*}]_{{\bf Q}'}
\delta_{{\bf Q}',{\bf Q}-{\bf G}}
[U_{\up,{\bf k}}]_{{\bf Q}}
\gamma_{{\bf k}+{\bf p},\down}^{\dagger}
\gamma_{{\bf k},\up}.
\label{eq:S_inter_band_basis}
\end{equation}

The proof for the intralayer modes is completed by plugging Eq.$\,$\ref{eq:S_intra_band_basis_derivation} into Eq.$\,$\ref{eq:intralayer_Sp}, which yields

\begin{widetext}
\bea
\bar{S}_{\mbf p}^{\dagger(\kappa,+)}&=\frac{1}{3}\sum_{l=\pm1}\sum_{j=-1}^{1}e^{-\frac{2\pi i}{3}(\kappa+1)j}\bar{S}_{\parallel,\mbf p,{\bf 0},l,j}^{\dagger}\\
&=\frac{1}{3}\sum_{l=\pm1}\sum_{\bf k}\sum_{{\bf Q}'\in\mathcal{Q}_{l}^{\down},{\bf Q}\in\mathcal{Q}_{l}^{\up}}[U_{\down,{\bf k}+{\bf p}}^{*}]_{{\bf Q}'}\sum_{j=-1}^{1}\delta_{{\bf Q}',{\bf Q}+l{\bf q}_{j}}e^{-\frac{2\pi i}{3}(\kappa+1)j}[U_{\up,{\bf k}}]_{{\bf Q}}\gamma_{{\bf k}+{\bf p},\down}^{\dagger}\gamma_{{\bf k},\up}\\
&=\sum_{\bf k}\sum_{{\bf Q}',{\bf Q}\in\mathcal{Q}}[U_{\down,{\bf k}+{\bf p}}^{*}]_{{\bf Q}'}\Big(\frac{1}{3} \sum_{j=-1}^{1}\delta_{{\bf Q}',{\bf Q}+(-1)^{{\bf Q}}{\bf q}_{j}}e^{-\frac{2\pi i}{3}(\kappa+1)j}\Big)[U_{\up,{\bf k}}]_{{\bf Q}}\gamma_{{\bf k}+{\bf p},\down}^{\dagger}\gamma_{{\bf k},\up}\,,
\eea
\label{eq:S_ppar_proof_final}
\end{widetext}
where we recall that $\mathcal{Q}=\{{\bf G}+{\bf q}_{1}\}\cup\{{\bf G}-{\bf q}_{1}\}$. To go from the second to third lines, we used that $l=(-1)^{\bf Q}$ when ${\bf Q}\in \mathcal{Q}_l^\uparrow$. Eq.$\,$\ref{eq:D_par_FH} directly follows from the last line of Eq.$\,$\ref{eq:S_ppar_proof_final}.

\subsection{SyMA results for FH model}

\label{subsec:App_Additional_MoTe2_SMA}

In this section, we show additional results comparing the SyMA spectrum
 and eigenfunctions with the exact results. We will start by showing
the results obtained by calculating the magnon energies using Eq.$\,$\ref{eq:Ising_gap_1BPV}
for the individual SyMA modes.

\begin{figure*}

\includegraphics[width=1\textwidth]{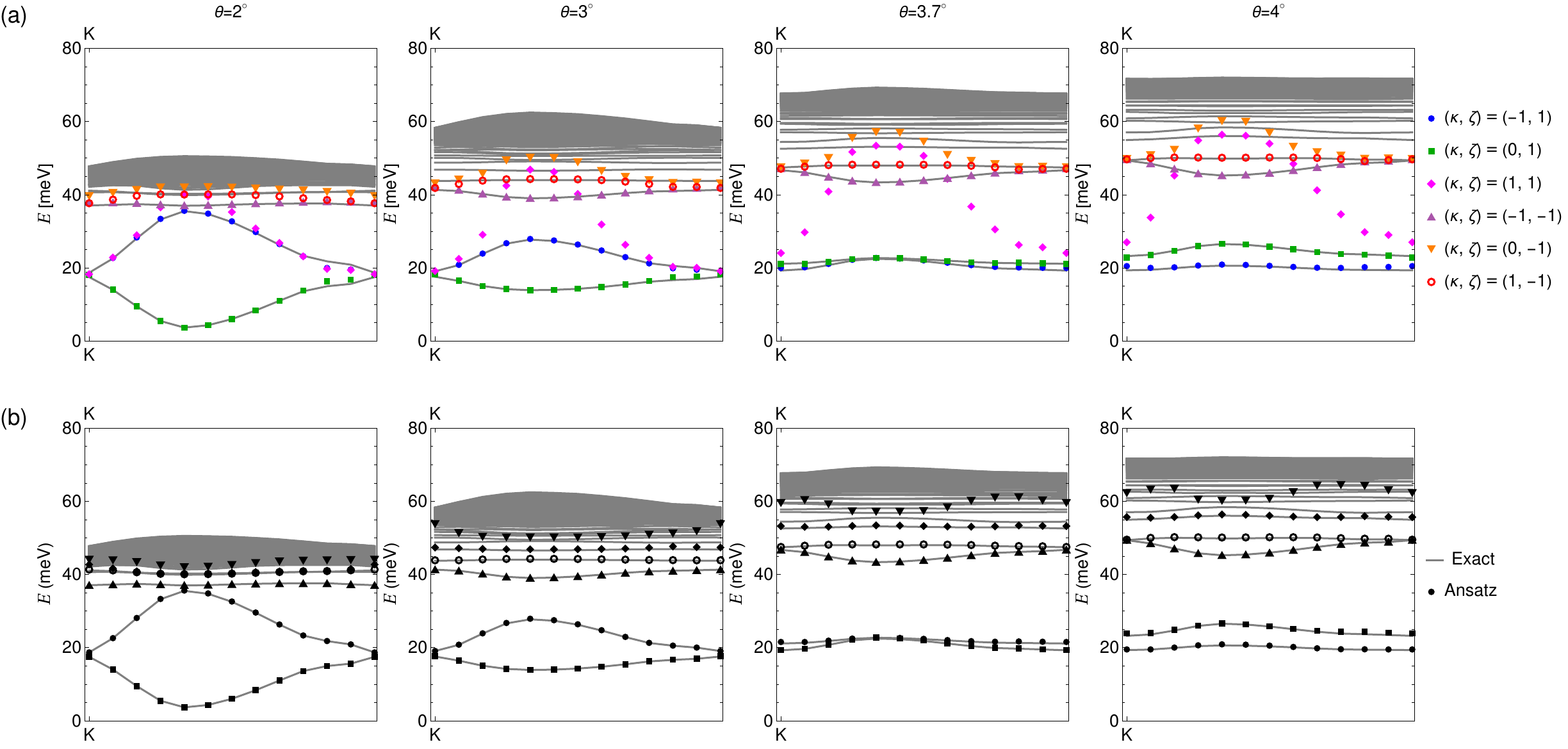}\caption{Exact (gray) vs. SyMA (other colors) results for the FH model of MoTe$_2$,
using the flat band interaction $H_{I}$ in Eq.$\,$\ref{eq:H_V_general},
for different angles. (a) All 6 SyMA modes $\bar{S}_{\mbf p}^{\dagger(\kappa,\zeta)}$ defined in Sec.$\,$\ref{sec:sma_main}.
(b) Diagonalization of the Hamiltonian in the subspace of the 6 modes
in (a). A momentum mesh defined by $\ensuremath{(N_{x},N_{y},n_{11},n_{12},n_{21},n_{22})=(12,12,1,0,0,1)}$
was used in all the results. \label{fig:App_MoTe2_SMA_energies}}

\end{figure*}

The results are shown in Fig.$\,$\ref{fig:App_MoTe2_SMA_energies}(a)
for different $\theta$. Although the simplified expressions for the
SyMA modes in Eq.$\,$\ref{eq:D_par_FH} and \ref{eq:D_perp_FH} only
guarantee that they are orthogonal at ${\bf p}=0$, as discussed in Sec.$\,$\ref{subsec:syma_energies} of the main text, the
lowest-energy magnon mode energies are reasonably well captured even for ${\bf p}\neq0$, for the whole range of explored angles.
In Fig.$\,$\ref{fig:App_MoTe2_SMA_energies}(b) we show the results
obtained by diagonalizing $H_{I}$ in the subspace of all the SyMA modes defined in Sec.$\,$\ref{sec:sma_main}, which yields excellent agreement
with the exact results for the five lowest energy modes.

\begin{figure*}
\includegraphics[width=1\textwidth]{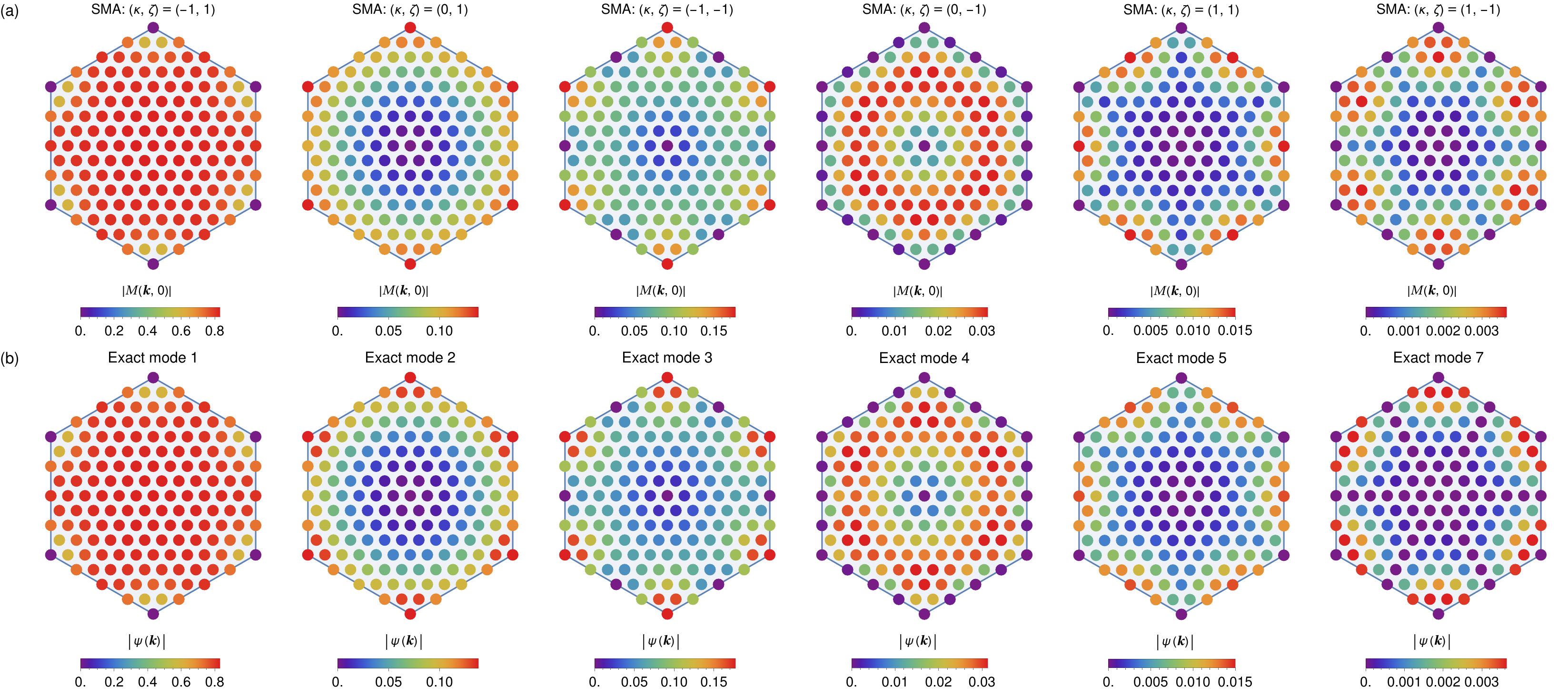}\caption{SyMA (a) vs. exact (b) lowest 5 eigenmodes at ${\bf p}=0$ for the
FH model for MoTe$_2$, using the flat band interaction $H_{I}$ in Eq.$\,$\ref{eq:H_V_general},
for $\theta=2^{\circ}$. The normalization of the exact modes was
chosen in order to match the normalization of each SyMA mode - where
$|M({\bf k},0)|\protect\leq1\,$, for any ${\bf k}$, by construction. The modes are ordered by increasing energy from left to right.\label{fig:App_MoTe2_eigenfuncs}}
\end{figure*}

In Fig.$\,$\ref{fig:App_MoTe2_eigenfuncs}, we complement the results in Fig.$\,$\ref{fig:3} of the main text by showing the comparison of the exact results obtained for $H_I$ by numerically diagonalizing Eq.$\,$\ref{eq:S_matrix_expression} and SyMA eigenstate amplitudes at $\bf{p}=0$ for a different angle $\theta=2^{\circ}$.

\subsection{Results for no-fitting models using the full Hamiltonian
in Eq.$\,$\ref{eq:H_full_band_basis}}

\label{sec:no_fitting_flatband_vs_exact}

\begin{figure*}
\includegraphics[width=1\textwidth]{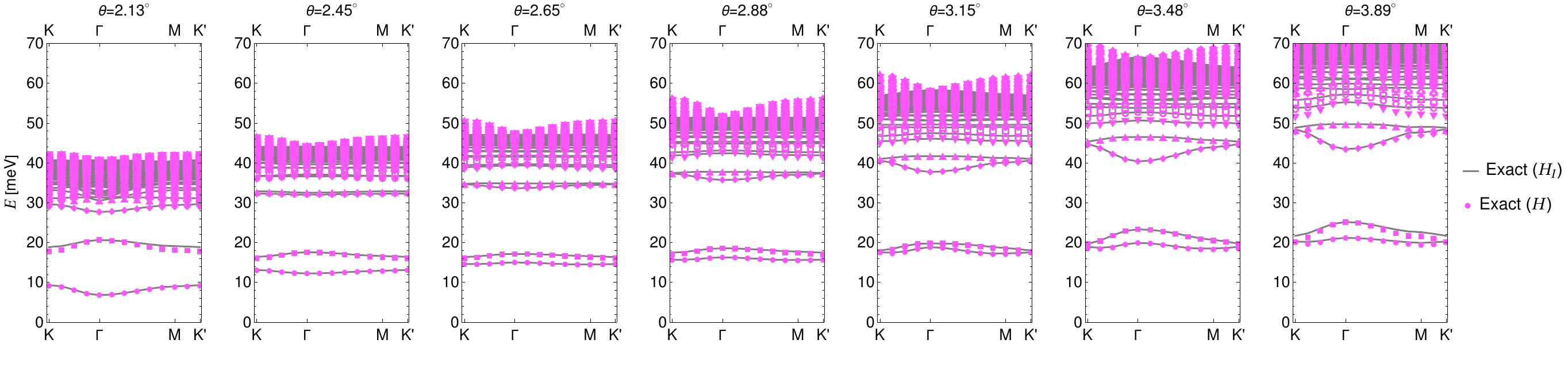}\caption{Comparison of exact magnon spectrum results for full Hamiltonian in Eq.$\,$\ref{eq:H_full_band_basis}
(pink points) and the flat band Hamiltonian in Eq.$\,$\ref{eq:H_V_general}
(full gray lines), for the no-fitting model of MoTe$_2$, with different parameters
specified on top of each figure. A momentum mesh defined by $\ensuremath{(N_{x},N_{y},n_{11},n_{12},n_{21},n_{22})=(12,12,1,0,0,1)}$
was used in all the results. \label{fig:App_MoTe2_H_vs_HI_nofitting}}
\end{figure*}

We finish this Appendix by comparing the exact magnon spectra obtained using the full Hamiltonian in Eq.$\,$\ref{eq:H_full_band_basis} and the flat band Hamiltonian $H_I$ in Eq.$\,$\ref{eq:H_V_general} for the no-fitting models introduced in Ref.$\,$\cite{zhang2024universalmoiremodelbuildingmethodfitting} and studied in Sec.$\,$\ref{subsec:Realistic_MoTe2}.
The results are shown in Fig.$\,$\ref{fig:App_MoTe2_H_vs_HI_nofitting},
where an excellent agreement can be seen, similarly to the case of
the FH model. Since the single-particle terms in Eq.$\,$\ref{eq:H_full_band_basis} containing the renormalized band dispersion only weakly affect the exact magnon spectrum, the SyMA results shown in Fig.$\,$\ref{fig:MoTe2_realistic} still faithfully capture the low-energy modes.

\section{Additional details on magnon eigenstates}

\subsection{Chern number and nodes}

\label{subsec:App_chern_vorticity_derivation}

In this section, we will prove Eq.$\,$\ref{eq:qj_C}. This expression
was obtained multiple times before in different references$\,$\cite{PhysRevLett.90.057002,PhysRevLett.120.067003,PhysRevLett.124.166601,UchoaPNAS,kwan2024texturedexcitoninsulators},
but we re-derive it here to make the paper self-contained.

We start by writing a generic magnon eigenstate as
\begin{equation}
\ket{\mathcal{M}_{\bf p}}=\sum_{{\bf k}}M^{(\down,\up)}({\bf k},{\bf p})\gamma_{{\bf k}+{\bf p},\down}^{\dagger}\gamma_{{\bf k},\up}\ket{\up}\,,
\end{equation}
where $\ket{\up}=\prod_{{\bf k}}\gamma_{{\bf k},\up}^{\dagger}\ket 0$.
Using the magnon wave function $M^{(\down,\up)}({\bf k},{\bf p})$, we define the gauge-invariant vector field given by

\begin{equation}
{\bf v}_{{\bf p}}({\bf k})={\bf A}_{M}^{{\bf p}}({\bf k})+{\bf A}^{\up}({\bf k})-{\bf A}^{\down}({\bf k}+{\bf p})\,,\label{eq:vp_k}
\end{equation}
where we defined
${\bf A}_{M}^{{\bf p}}({\bf k})
=\nabla_{{\bf k}}\arg M^{(\down,\up)}({\bf k},{\bf p})$
and
${\bf A}^{\eta}({\bf k})
=iU_{\eta,{\bf k}}^{\dagger}\nabla_{{\bf k}}U_{\eta,{\bf k}}$
as the magnon-phase and Berry connections, respectively.

The precise form of ${\bf v}_{\bf p}({\bf k})$
is fixed by gauge invariance. Consider independent gauge transformations
of the Bloch eigenvectors,
\begin{equation}
U_{\eta,{\bf k}}
\rightarrow
e^{i\chi_{\eta}({\bf k})}U_{\eta,{\bf k}}.
\label{eq:Bloch_gauge_transformation_nodes}
\end{equation}
Since
\begin{equation}
\gamma_{{\bf k},\eta}^{\dagger}
\rightarrow
e^{i\chi_{\eta}({\bf k})}
\gamma_{{\bf k},\eta}^{\dagger},
\end{equation}
the magnon wave function must transform as
\begin{equation}
M^{(\down,\up)}({\bf k},{\bf p})
\rightarrow
e^{i\left[
\chi_{\up}({\bf k})
-
\chi_{\down}({\bf k}+{\bf p})
\right]}
M^{(\down,\up)}({\bf k},{\bf p}),
\label{eq:magnon_wavefunction_gauge_transformation}
\end{equation}
so that the physical magnon state is unchanged, up to the overall gauge
phase of the fully polarized ground state. It follows that
\begin{equation}
{\bf A}_{M}^{\bf p}({\bf k})
\rightarrow
{\bf A}_{M}^{\bf p}({\bf k})
+
\nabla_{\bf k}\chi_{\up}({\bf k})
-
\nabla_{\bf k}\chi_{\down}({\bf k}+{\bf p}),
\end{equation}
whereas the Berry connections transform as
\begin{equation}
{\bf A}^{\eta}({\bf k})
\rightarrow
{\bf A}^{\eta}({\bf k})
-
\nabla_{\bf k}\chi_{\eta}({\bf k}).
\end{equation}
Therefore, the gauge-dependent terms cancel exactly in the combination in Eq.$\,$\ref{eq:vp_k}, that defines ${\bf v}_{\bf p}({\bf k})$.

Considering
a domain $\mathcal{D}_{\epsilon}$ containing the whole Brillouin
zone, but excluding tiny disks of radius $\epsilon$ around the nodes
of $M^{(\down,\up)}({\bf k},{\bf p})$, we integrate the curl of ${\bf v}_{{\bf p}}({\bf k})$
over this domain, getting
\bea
I_{v}&=\frac{1}{2\pi}\int_{\mathcal{D}_{\epsilon}}d^{2}{\bf k}\,\nabla\times{\bf v}_{{\bf p}}({\bf k})\\
&=\frac{1}{2\pi}\int_{\mathcal{D}_{\epsilon}}d^{2}{\bf k}\,[\nabla\times{\bf A}^{\up}({\bf k})-\nabla\times{\bf A}^{\down}({\bf k}+{\bf p})]\\
&=C_{\up}-C_{\down}\,\,(\epsilon\rightarrow0)\,.
\label{eq:Iv_C}
\eea

On the other hand, ${\bf v}_{{\bf p}}({\bf k})$ is gauge invariant and
single-valued on the punctured domain $\mathcal{D}_{\epsilon}$, so we
can apply Stokes' theorem to obtain
\bea
I_{v}&=-\frac{1}{2\pi}\sum_{j}\int_{(\pd\mathcal{D}_{\epsilon})^{j}}\,d{\bf k}\cdot{\bf v}_{{\bf p}}({\bf k})\\
&=-\frac{1}{2\pi}\sum_{j}\int_{(\pd\mathcal{D}_{\epsilon})^{j}}\,d{\bf k}\cdot\nabla_{{\bf k}}\arg M^{(\down,\up)}({\bf k},{\bf p})\\
&=-\sum_{j}q_{j}
\label{eq:Iv_q}
\eea
where $(\pd\mathcal{D}_{\epsilon})^{j}$ denotes the boundary
of the tiny disk of radius $\epsilon$ around node $j$,
\bea
q_j
=
\frac{1}{2\pi}
\oint_{(\partial\mathcal{D}_{\epsilon})^j}
d{\bf k}\cdot
\nabla_{\bf k}\arg M^{(\down,\up)}({\bf k},{\bf p})
\eea
is the vorticity of node $j$, and we used that

\bea
& |\int_{(\pd\mathcal{D}_{\epsilon})^{j}}\,d{\bf k}\cdot[{\bf A}^{\up}({\bf k})-{\bf A}^{\down}({\bf k}+{\bf p})]|\leq \\ & 2\pi\epsilon\sup_{(\pd\mathcal{D}_{\epsilon})^{j}}|{\bf A}^{\up}({\bf k})-{\bf A}^{\down}({\bf k}+{\bf p})|\rightarrow0
\eea

This is because we can choose a smooth gauge for the Berry connections on the disk boundaries. Equating Eqs.$\,$\ref{eq:Iv_C} and \ref{eq:Iv_q}, we obtain Eq.$\,$\ref{eq:qj_C}.

\subsection{Magnon gap as approximate Ginzburg-Landau-like stiffness}

\label{subsec:App_GL_Isinggap}

In this section, we derive an approximation of
Eq.$\,$\ref{eq:Ising_gap_1BPV} to identify the main contributions
controlling the zero-momentum magnon gap. The aim of this section
is not to obtain rigorous proofs on the magnon gap, but instead to identify some of the important contributions, complementing the analysis in the main text.

We note that a similar set of approximations that will be made in this section was carried out in Ref.$\,$\cite{chen2025quantumgeometricdipoletopologicalboost}.
However, in this reference, the authors
introduced a local-mode approximation in which a single local spin-flip
ansatz is projected into the active band. In our notation this corresponds to
the particular SyMA choice $D=\mathbb{1}$ in
Eq.$\,$\ref{eq:Mdnup_general_def}.
Consequently, the local-mode approximation of
Ref.$\,$\cite{chen2025quantumgeometricdipoletopologicalboost}
constructs one trial magnon state at each momentum and uses its
variational energy as an approximation to the lowest magnon branch.
This state is not guaranteed in general to approximate the lowest-energy
mode, although its energy still provides a variational upper bound on the
lowest one-magnon energy. By contrast, our more general SyMA allows us to construct and compare different possible local
magnon modes belonging to different symmetry sectors.
Additionally, in Ref.$\,$\cite{chen2025quantumgeometricdipoletopologicalboost}
the magnon gap was decomposed in terms of a quantum geometric dipole.
Here we will show that it can be written as a Ginzburg-Landau-like
stiffness in momentum-space.

\subsubsection{Approximate simplification of the Ising gap expression in Eq.$\,$\ref{eq:Ising_gap_1BPV}}

\label{subsec:GL_derivation_first}

We start by assuming a Hubbard interaction for simplicity, so that
$\Omega_{\textrm{tot}}^{-1}V({\bf q}+{\bf G})=N^{-1}V_{0}$
where $N$ is the number of unit cells and $V_{0}$ is the
interaction strength that has units of energy. We will also take ${\bf p}=0$
throughout this section. Identifying 
\bea
M({\bf k})=M^{(\down,\up)}({\bf k},{\bf 0})=\rho({\bf k})e^{i\theta_{{\bf k}}} \, ,
\label{eq:Mk_def}
\eea
we can write Eq.$\,$\ref{eq:Ising_gap_1BPV} as

\begin{widetext}

\begin{equation}
\begin{array}{c}
\Delta^{\down\up}({\bf 0})=\frac{\frac{V_{0}}{2N}\sum_{{\bf k},{\bf q},{\bf G}}(|M^{(\down)}({\bf k},{\bf q}+{\bf G})|\rho({\bf k})-\rho({\bf k}+{\bf q})|M^{(\up)}({\bf k},{\bf q}+{\bf G})|)^{2}+2\rho({\bf k})\rho({\bf k}+{\bf q})|M^{(\down)}({\bf k},{\bf q}+{\bf G})||M^{(\up)}({\bf k},{\bf q}+{\bf G})|[1-\cos(\nu_{{\bf k},{\bf q}+{\bf G}})]}{\sum_{{\bf k}}\rho^{2}({\bf k})}\end{array}\label{eq:Ising_gap_SMA_1BPV_p=00003D0}
\end{equation}
\end{widetext}where we defined the gauge-invariant phase

\begin{equation}
\nu_{{\bf k},{\bf q}}=\theta_{{\bf k}+{\bf q}}-\theta_{{\bf k}}+\alpha_{{\bf k},{\bf q}}^{\up}-\alpha_{{\bf k},{\bf q}}^{\down}\,,\label{eq:=00005Cnu_kq}
\end{equation}
with $\alpha_{{\bf k},{\bf q}}^{\eta}$ the phases of the form factors
defined as $M^{(\eta)}({\bf k},{\bf q})=|M^{(\eta)}({\bf k},{\bf q})|e^{i\alpha_{{\bf k},{\bf q}}^{\eta}}$.

In order to make analytical progress, we assume that the density form
factors are strongly peaked around zero momentum transfer and decay
sufficiently rapidly as $|{\bf q}|$ increases. Under this assumption,
the momentum sum in Eq.$\,$\ref{eq:Ising_gap_SMA_1BPV_p=00003D0} is dominated by the small-${\bf q}$ contributions, where
the form factors can be expressed in terms of the quantum metric. In
particular, we approximate their magnitude by
\begin{equation}
|M^{(\eta)}({\bf k},{\bf q})|
\approx
\exp\left[
-\frac{1}{2}q_i g_{ij}^{\eta}({\bf k})q_j
\right].
\label{eq:form_factor_gaussian_approx}
\end{equation}
This expression is exact in the lowest Landau level, where
$g_{ij}^{\eta}({\bf k})$ is also ${\bf k}$ independent
\cite{PhysRevB.104.045103}. For a generic Chern band, only its
small-${\bf q}$ expansion is universal and given by
\begin{equation}
|M^{(\eta)}({\bf k},{\bf q})|
=
1-\frac{1}{2}q_iq_jg_{ij}^{\eta}({\bf k})
+\mathcal{O}(q^3).
\end{equation}
Therefore, Eq.$\,$\ref{eq:form_factor_gaussian_approx} should be understood as a gaussian continuation of the universal small-${\bf q}$ behavior to finite momentum transfer. The assumption that this continuation accurately captures the decay of the form factors, and
therefore suppresses contributions from outside the small-${\bf q}$ region, is exact in the lowest Landau level but is an additional approximation for a generic Chern band. We will test the validity of this approximation in the end of this section for the FH model of MoTe$_2$.

We will further assume that we have inversion (or pseudo-inversion like for the FH model of MoTe$_2$) and time-reversal symmetries. These symmetries imply
\begin{equation}
g_{ij}^{\eta}({\bf k})
=
g_{ij}^{\eta}(-{\bf k})
\end{equation}
and
\begin{equation}
g_{ij}^{\up}({\bf k})
=
g_{ij}^{\down}(-{\bf k}),
\end{equation}
respectively. In the presence of both symmetries, the quantum metrics
of the two valleys are therefore equal at the same momentum, and we
define
\begin{equation}
\Gamma_{ij}({\bf k})
\equiv
g_{ij}^{\up}({\bf k})
=
g_{ij}^{\down}({\bf k}).
\end{equation}

Finally, within the same approximation of rapidly decaying form
factors, terms with ${\bf G}\neq0$ involve momentum transfers
$|{\bf q}+{\bf G}|$ outside the region centered around zero momentum
transfer and are assumed to be strongly suppressed. We therefore
neglect these positive-semidefinite contributions. Under these
assumptions, Eq.$\,$\ref{eq:Ising_gap_SMA_1BPV_p=00003D0} becomes

\begin{widetext}

\begin{equation}
\begin{array}{c}
\Delta^{\down\up}({\bf 0})=\frac{\frac{V_{0}}{2N}\sum_{{\bf k},{\bf q}} \exp (- {\bf q}^T \Gamma({\bf k}) {\bf q})\big[\big(\rho({\bf k})-\rho({\bf k}+{\bf q})\big)^{2}+2\rho({\bf k})\rho({\bf k}+{\bf q})\big(1-\cos(\nu_{{\bf k},{\bf q}})\big)\big]}{\sum_{{\bf k}}\rho^{2}({\bf k})}\end{array}\label{eq:Ising_gap_SMA_1BPV_approx}
\end{equation}
\end{widetext}

Because of the gaussian integrating kernel in Eq.$\,$\ref{eq:Ising_gap_SMA_1BPV_approx}, only small-${\bf q}$ contributions from the term in the square brackets should be relevant. We will therefore approximate this term by its contributions to lowest-order in ${\bf q}$, using the following expansions

\begin{equation}
\begin{array}{c}
\rho({\bf k}+{\bf q})=\rho({\bf k})+q_{i}\pd_{i}\rho({\bf k})+\mathcal{O}(q^{2}) \\
\alpha_{{\bf k},{\bf q}}^{\eta}=q_{i}A_{i}^{\eta}({\bf k})+\mathcal{O}(q^{2}) \\
\nu_{{\bf k},{\bf q}}={\bf q}\cdot{\bf v}_{{\bf 0}}({\bf k})+\mathcal{O}(q^{2}) \,,
\end{array}
\end{equation}
where in the second line we used that the phase of the form factor $\alpha_{{\bf k},{\bf q}}^{\eta}$ can be written in terms of the Berry connection $A_{i}^{\eta}({\bf k})$, to first-order in ${\bf q}$, and replaced this result in Eq.$\,$\ref{eq:=00005Cnu_kq} to obtain the third line,
recognizing
\begin{equation}
{\bf v}_{{\bf 0}}({\bf k})=\nabla\theta_{{\bf k}}+A^{\up}({\bf k})-A^{\down}({\bf k}),
\label{eq:v0_def}
\end{equation}
as the gauge-invariant vector field defined in Eq.$\,$\ref{eq:vp_k},
evaluated at ${\bf p}=0$.

The numerator in Eq.$\,$\ref{eq:Ising_gap_SMA_1BPV_approx} then becomes

\bea
\mathcal{N}_{\Delta}\approx \frac{V_{0}}{2\Omega_{BZ}}\sum_{{\bf k}} & \Big([\nabla\rho({\bf k})]^{T}\bar{\Gamma}({\bf k})\nabla\rho({\bf k}) \\ & +\rho^{2}({\bf k}) {\bf v}_{{\bf 0}}^{T}({\bf k})\bar{\Gamma}({\bf k}){\bf v}_{{\bf 0}}({\bf k})\Big)
\label{eq:Ising_gap_SMA_1BPV_approx2}
\eea
where we defined
\begin{equation}
 \bar{\Gamma}({\bf k})=\int d^{2}{\bf q}\,\exp\big(-{\bf q}^{T}\Gamma({\bf k}){\bf q}\big)\,{\bf q}{\bf q}^{T}\,.
 \label{eq:gaussian_integral}
\end{equation}

We again remark that Eq.$\,$\ref{eq:Ising_gap_SMA_1BPV_approx2} assumes that the higher-order contributions in ${\bf q}$ of the term within square brackets in Eq.$\,$\ref{eq:Ising_gap_SMA_1BPV_approx} can be neglected due to the gaussian pre-factor that multiplies them. This allows us to extend the domain of the integral for any ${\bf q}$ and therefore perform the gaussian integral in Eq.$\,$\ref{eq:gaussian_integral} analytically.
Performing the gaussian integral yields
\begin{widetext}
\begin{equation}
\Delta^{\down\up}({\bf 0})\approx\frac{\pi V_{0}}{4\Omega_{BZ} \sum_{{\bf k}}\rho^{2}({\bf k})}  \sum_{{\bf k}} \Big[ \frac{[\nabla\rho({\bf k})]^{T}\Gamma({\bf k})^{-1}\nabla\rho({\bf k})}{\sqrt{\det\Gamma({\bf k})}}+\frac{\rho^{2}({\bf k}){\bf v}_{{\bf 0}}^{T}({\bf k})\Gamma({\bf k})^{-1}{\bf v}_{{\bf 0}}({\bf k})}{\sqrt{\det\Gamma({\bf k})}} \Big] \,.\label{eq:numerator_approximation_less_approx}
\end{equation}
\end{widetext}

Eq.$\,$\ref{eq:numerator_approximation_less_approx} already separates
the contributions to the magnon gap into an amplitude-stiffness term (first term)
and a phase-stiffness term (second term). It is useful to define the
momentum-dependent stiffness tensor,
\begin{equation}
K_{ij}({\bf k})
=
\frac{
\left[\Gamma({\bf k})^{-1}\right]_{ij}
}{
\sqrt{\det\Gamma({\bf k})}
}.
\label{eq:stiffness_tensor}
\end{equation}
In terms of this tensor,
Eq.$\,$\ref{eq:numerator_approximation_less_approx} takes the form
\begin{equation}
\begin{aligned}
\mathcal{N}_{\Delta}
\approx
\frac{\pi V_{0}}{4\Omega_{BZ}}
\sum_{\bf k}
\sum_{i,j=x,y}
\Big[
&
\partial_i\rho({\bf k})
K_{ij}({\bf k})
\partial_j\rho({\bf k})
\\
&
+
\rho^2({\bf k})
{\bf v}_{{\bf 0},i}({\bf k})
K_{ij}({\bf k})
{\bf v}_{{\bf 0},j}({\bf k})
\Big].
\end{aligned}
\label{eq:numerator_stiffness_tensor}
\end{equation}
where ${\bf v}_{{\bf 0},i}({\bf k})$ is the $i$-th component of the vector ${\bf v}_{{\bf 0}}({\bf k})$.
In general, $K_{ij}({\bf k})$ is anisotropic and depends on the full
quantum-metric tensor.
A simpler scalar expression can be obtained by using the additional
assumption that the quantum metric is locally isotropic. This assumption requires the metric to
have the same value along every direction, and therefore to be
proportional to the identity, that is,
\begin{equation}
\Gamma_{ij}({\bf k})
=
\frac{G({\bf k})}{2}\delta_{ij},
\qquad
G({\bf k})
=
\operatorname{Tr}\Gamma({\bf k}).
\label{eq:isotropic_metric_approximation}
\end{equation}

Under the locally isotropic approximation, we also have
\begin{equation}
\sqrt{\det\Gamma({\bf k})}
=
\frac{G({\bf k})}{2}.
\end{equation}
The stiffness tensor defined in
Eq.$\,$\ref{eq:stiffness_tensor} therefore becomes
\begin{equation}
K_{ij}({\bf k})
=
\frac{4}{G^{2}({\bf k})}\delta_{ij}.
\label{eq:isotropic_stiffness_tensor}
\end{equation}
Substituting Eq.$\,$\ref{eq:isotropic_stiffness_tensor} into
Eq.$\,$\ref{eq:numerator_stiffness_tensor}, we obtain
\begin{equation}
\Delta^{\down\up}({\bf 0})
\approx
\frac{\pi V_{0}}{\Omega_{BZ}}
\frac{
\displaystyle
\int d^{2}{\bf k}\,
\frac{
|\nabla\rho({\bf k})|^{2}
+
\rho^{2}({\bf k})
|{\bf v}_{{\bf 0}}({\bf k})|^{2}
}{
G^{2}({\bf k})
}
}{
\displaystyle
\int d^{2}{\bf k}\,
\rho^{2}({\bf k})
}.
\label{eq:approximation_Ising_gap_di=00005Crho}
\end{equation}
The simpler Eq.$\,$\ref{eq:approximation_Ising_gap_di=00005Crho} should be understood as the locally isotropic limit of the general
tensor expression in
Eq.$\,$\ref{eq:numerator_stiffness_tensor}.

\begin{figure}
\centering{}\includegraphics[width=1\columnwidth]{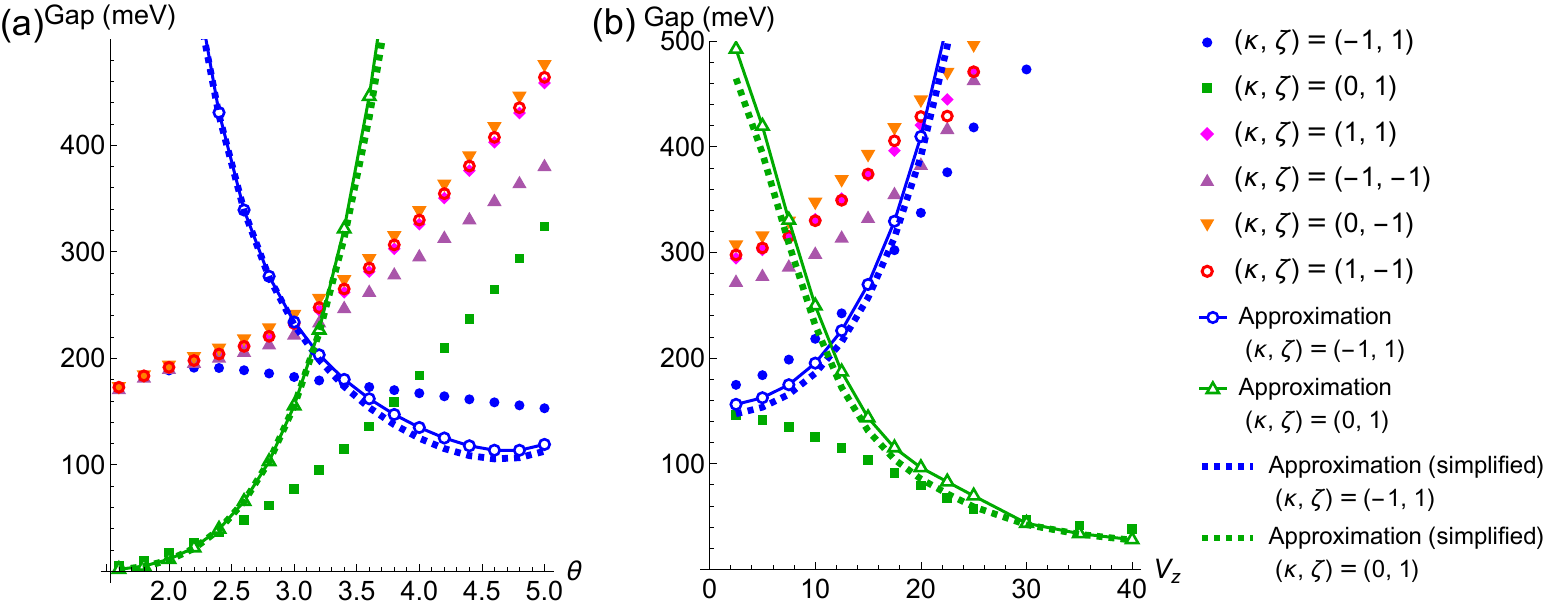}\caption{Magnon gap approximation in Eq.$\,$\ref{eq:approximation_Ising_gap_di=00005Crho}
for the FH model of MoTe$_2$ with Hubbard interaction (see definition
in the end of Appendix$\,$\ref{sec:detalis_MoTe2_FH}), for (a) $V_{z}=0$ and variable
$\theta$ and (b) $\theta=3.7^{\circ}$ for variable $V_{z}$. The
approximation in Eq.$\,$\ref{eq:numerator_approximation_less_approx} is shown with open plot markers, together with the simplified version in Eq.$\,$\ref{eq:approximation_Ising_gap_di=00005Crho} and with the
SyMA results in full markers. For the $V_z\neq0$ case in (b), because inversion symmetry is broken, we took $\Gamma_{ij}({\bf k})$ as the average between the spin $\up$ and $\down$ quantum geometries, $\Gamma_{ij}({\bf k})\rightarrow\frac{1}{2}[g_{ij}^{\up}({\bf k})+g_{ij}^{\down}({\bf k})]$. \label{fig:Ising_gap_approximation_test}}
\end{figure}

Eq.$\,$\ref{eq:numerator_approximation_less_approx} and its simplified version Eq.$\,$\ref{eq:approximation_Ising_gap_di=00005Crho} are applied
in Fig.$\,$\ref{fig:Ising_gap_approximation_test} to the FH model
of MoTe$_2$, for variable $\theta$ and variable $V_{z}$. The results of Eqs.$\,$\ref{eq:numerator_approximation_less_approx} and \ref{eq:approximation_Ising_gap_di=00005Crho} are very similar, justifying the simplifications leading to the latter, as see in Fig.$\,$\ref{fig:Ising_gap_approximation_test}. The quantitative agreement with the SyMA results is not good in general, but the qualitative evolution of the mode gaps is reasonably well captured. In particular, the approximation reproduces the crossing of the two lowest modes as a function of $\theta$ in Fig.$\,$\ref{fig:Ising_gap_approximation_test}(a), as well as their opposite trends with increasing $V_z$ in Fig.$\,$\ref{fig:Ising_gap_approximation_test}(b).

\subsubsection{Interpretation of the magnon gap as a Ginzburg-Landau-like stiffness}

\label{subsec:Interpretation_IsingGap_Approximation}

Treating the SyMA wave function as a complex order parameter, $M({\bf k})=\rho({\bf k})e^{i\theta_{\bf k}}$, as in Eq.$\,$\ref{eq:Mk_def}, 
and defining the relative Berry connection
\begin{equation}
a_i({\bf k})
=
A_i^{\up}({\bf k})-A_i^{\down}({\bf k}),
\end{equation}
we introduce the gauge-covariant derivative
\begin{equation}
\mathcal{D}_iM({\bf k})
=
\left[
\partial_i+ia_i({\bf k})
\right]M({\bf k}).
\end{equation}
Using
\begin{equation}
\mathcal{D}_iM({\bf k})
=
e^{i\theta_{\bf k}}
\left[
\partial_i\rho({\bf k})
+
i\rho({\bf k}){\bf v}_{0,i}({\bf k})
\right],
\end{equation}
where ${\bf v}_{0,i}({\bf k})
=
\partial_i\theta_{\bf k}+a_i({\bf k})$, as follows directly from Eq.$\,$\ref{eq:v0_def}.
Eq.$\,$\ref{eq:numerator_approximation_less_approx} can be written,
without making any additional assumption about the form of
$\Gamma({\bf k})$, as the Ginzburg--Landau-like functional
\begin{equation}
\Delta^{\down\up}({\bf 0})
\approx
\frac{\pi V_0}{4\Omega_{BZ}}
\frac{
\displaystyle
\sum_{i,j=x,y} \int d^2{\bf k}\,
\left[\mathcal{D}_iM({\bf k})\right]^*
K_{ij}({\bf k})
\left[\mathcal{D}_jM({\bf k})\right]
}{
\displaystyle
\int d^2{\bf k}\,
\rho^2({\bf k})
},
\label{eq:GL_Ising_gap_tensor}
\end{equation}
where
\begin{equation}
K_{ij}({\bf k})
=
\frac{
\left[\Gamma({\bf k})^{-1}\right]_{ij}
}{
\sqrt{\det\Gamma({\bf k})}
}
\label{eq:GL_stiffness_tensor}
\end{equation}
is a momentum-dependent stiffness tensor. Indeed, since
$K_{ij}({\bf k})$ is real and symmetric,
\begin{equation}
\begin{aligned}
&\left[\mathcal{D}_iM({\bf k})\right]^*
K_{ij}({\bf k})
\left[\mathcal{D}_jM({\bf k})\right]
= \\ &
= \partial_i\rho({\bf k})
K_{ij}({\bf k})
\partial_j\rho({\bf k})
+
\rho^2({\bf k})
{\bf v}_{0,i}({\bf k})
K_{ij}({\bf k})
{\bf v}_{0,j}({\bf k}).
\end{aligned}
\end{equation}
Thus, a Ginzburg-Landau interpretation is valid for a general
anisotropic quantum metric, in which case the stiffness depends on the
full quantum-metric tensor.

Eq.$\,$\ref{eq:GL_Ising_gap_tensor} can be further simplified by adopting the isotropic approximation introduced above in Eq.$\,$\ref{eq:isotropic_metric_approximation} to derive
Eq.$\,$\ref{eq:approximation_Ising_gap_di=00005Crho}.
This approximation replaces the full anisotropic stiffness tensor by a
scalar stiffness determined in terms of the trace of the quantum metric, $G({\bf k})$, in which case
Eq.$\,$\ref{eq:GL_Ising_gap_tensor} reduces precisely to
\bea
\Delta^{\down\up}({\bf 0})
&\approx
\frac{\pi V_{0}}{\Omega_{BZ}}
\frac{
\displaystyle
\int d^2{\bf k}\,
\frac{1}{2m^{*}({\bf k})}
\sum_i \left|\mathcal{D}_{i}M({\bf k})\right|^{2}
}{
\displaystyle
\int d^2{\bf k}\,
\rho^{2}({\bf k})
}
\\
&=
\frac{\pi V_{0}}{\Omega_{BZ}}
\frac{
\displaystyle
\int d^2{\bf k}\,
\frac{1}{2m^{*}({\bf k})}
\left|
\left(-i\nabla+{\bf a}({\bf k})\right)M({\bf k})
\right|^{2}
}{
\displaystyle
\int d^2{\bf k}\,
\rho^{2}({\bf k})
}.
\label{eq:GL_Ising_gap}
\eea
where $m^{*}({\bf k})=G^{2}({\bf k})/2$ is a ${\bf k}$-dependent
effective mass (inverse stiffness). The gauge field ${\bf a}({\bf k})$
felt by the magnon order parameter $M({\bf k})$ is simply the relative
Berry connection between the two valleys. The effective magnetic field
is the difference of the Berry curvatures given by $\nabla\times{\bf a}({\bf k})=F_{\up}({\bf k})-F_{\down}({\bf k})$.
We note that a similar expression as Eq.$\,$\ref{eq:GL_Ising_gap} was previously derived in the chiral
limit of twisted bilayer graphene to describe the energetics of inter-Chern
particle-hole excitations$\,$\cite{PhysRevResearch.3.013033,khalaf2020softmodesmagicangle}.

A number of important insights can then be drawn from Eqs.$\,$\ref{eq:numerator_approximation_less_approx} and \ref{eq:GL_Ising_gap_tensor}, and their isotropic versions in Eqs.$\,$\ref{eq:approximation_Ising_gap_di=00005Crho} 
and \ref{eq:GL_Ising_gap}:
\begin{itemize}
\item The proof in Sec.$\,$\ref{subsec:App_chern_vorticity_derivation}
implies that a constant $\rho({\bf k})$ (no nodes) is only possible
when $C=0$. A constant $\rho({\bf k})$ is however not sufficient
to ensure that the gap vanishes, in contrast with the $SU(2)$ case.
In fact, even for constant $\rho({\bf k})$, the phase stiffness term,
set by the mismatch of the Berry connections of the opposite valleys,
can still be finite.

For ${\bf v}_{0}({\bf k})$ to vanish at every ${\bf k}$, which is
necessary for the phase-stiffness contribution in
Eqs.$\,$\ref{eq:numerator_approximation_less_approx} and
\ref{eq:approximation_Ising_gap_di=00005Crho} to vanish, we require
$\nabla\theta_{\bf k}
+
{\bf A}^{\up}({\bf k})
-
{\bf A}^{\down}({\bf k})
=
0$.
For a nodeless mode, one can choose a smooth relative gauge in which
$\theta_{\bf k}=0$, in which case this condition becomes
${\bf A}^{\up}({\bf k})={\bf A}^{\down}({\bf k})$. Together with
time reversal, this implies
$F^{\eta}({\bf k})=-F^{\eta}(-{\bf k})$.
This condition, which is simply spinless time-reversal, can be violated even for a trivial parent band.
\item The nodes are vortices of the order parameter $M({\bf k})$.
They affect the amplitude-stiffness contribution by enforcing spatial
variation of $\rho({\bf k})$ and by reducing the norm in the
denominator. Their nonzero vorticity also constrains the phase texture
${\bf v}_{0}({\bf k})$ and can therefore contribute to the phase
stiffness. Eqs.$\,$\ref{eq:numerator_approximation_less_approx} and
\ref{eq:GL_Ising_gap} thus provide a simple understanding
of why modes with a larger number or degree of nodes can tend to have
higher energies.
\item The quantum metric also constrains the amplitude stiffness.
From Eq.$\,$\ref{eq:grad_rho_bound_Gk}, a nonzero amplitude gradient
$|\nabla\rho({\bf k})|$ requires a nonzero quantum metric at that
momentum. Moreover, the stiffness itself depends on the local quantum
metric through Eqs.$\,$\ref{eq:numerator_approximation_less_approx}
and \ref{eq:approximation_Ising_gap_di=00005Crho}. Consequently, the
energetic cost associated with a node depends on its position relative
to the quantum-metric profile.
This is in agreement with the simple picture described in the main text for the magnon wave function. We now explore this
in more detail in Sec.$\,$\ref{subsec:simple_limiting_cases} under this Ginzburg-Landau perspective.
\end{itemize}

\subsubsection{Simple limiting cases}

\label{subsec:simple_limiting_cases}

We now explore simple limiting cases to obtain a better understanding
of the different contributions to the magnon gap within the
approximation in Eq.$\,$\ref{eq:approximation_Ising_gap_di=00005Crho}.
We decompose the gap as
\begin{equation}
\Delta^{\down\up}({\bf 0})
\approx
\Delta_A^{\down\up}({\bf 0})
+
\Delta_P^{\down\up}({\bf 0}),
\end{equation}
where the amplitude-stiffness contribution is

\begin{equation}
\Delta_A^{\down\up}({\bf 0})
=
\frac{\pi V_{0}}{\Omega_{BZ}}
\frac{
\displaystyle
\int d^{2}{\bf k}\,
\frac{
|\nabla\rho({\bf k})|^{2}
}{
G^{2}({\bf k})
}
}{
\displaystyle
\int d^{2}{\bf k}\,
\rho^{2}({\bf k})
},
\label{eq:approximation_Ising_gap_di=00005Crho_Amplitude}
\end{equation}
and the phase-stiffness contribution is
\begin{equation}
\Delta_P^{\down\up}({\bf 0})
=
\frac{\pi V_{0}}{\Omega_{BZ}}
\frac{
\displaystyle
\int d^{2}{\bf k}\,
\frac{
\rho^{2}({\bf k})
|{\bf v}_{\bf 0}({\bf k})|^{2}
}{
G^{2}({\bf k})
}
}{
\displaystyle
\int d^{2}{\bf k}\,
\rho^{2}({\bf k})
}.
\label{eq:approximation_Ising_gap_di=00005Crho_Phase}
\end{equation}
Both terms are positive semidefinite. In particular, the phase
stiffness can produce a finite magnon gap even when the amplitude is
constant. For example, taking $\rho({\bf k})=\rho_0$ and
$G({\bf k})=G$ yields
\begin{equation}
\Delta_P^{\down\up}({\bf 0})
=
\frac{\pi V_{0}}{\Omega_{BZ}^{2}G^{2}}
\int d^{2}{\bf k}\,
|{\bf v}_{\bf 0}({\bf k})|^{2}.
\label{eq:phase_only_limiting_case}
\end{equation}

In the remainder of this subsection, we will focus on
$\Delta_A^{\down\up}({\bf 0})$ in order to isolate the energetic
contribution associated directly with the nodes of $\rho({\bf k})$.
The phase contribution depends on the full gauge-invariant and ${\bf k}$-resolved phase texture
${\bf v}_{\bf 0}({\bf k})$, and is therefore not fixed by the number
and amplitude order of the nodes alone.

We will now obtain simple expressions for $\Delta_A^{\down\up}({\bf 0})$ in Eq.$\,$\ref{eq:approximation_Ising_gap_di=00005Crho_Amplitude} in terms of the node number and degree.
We will consider two simple cases. We first assume $G({\bf k})\approx G\,,\forall_{{\bf k}}$ and
that there are $w$ non-overlapping nodes of order $m$. We also assume
perfect radial symmetry around each node located at ${\bf k}={\bf k}_{i}^{*}$
for simplicity and that the nodes behave identically i.e. that we
have
\begin{equation}
\rho(k)=\begin{cases}
\rho_{0}k^{m} &,k<k_{0}\\
\rho_{0}k_{0}^{m}=Z &,k\geq k_{0}
\end{cases} \,,
\end{equation}
with $k\equiv|{\bf k}-{\bf k}_{i}^{*}|$, which implies that
\begin{equation}
|\nabla\rho({\bf k})|=\begin{cases}
m\rho_{0}k^{m-1} &,k<k_{0}\\
0 &,k>k_{0} \,.
\end{cases}
\end{equation}

Note also that we necessarily have $Z\leq1$ from our definition of the magnon eigenfunction in Eq.$\,$\ref{eq:Mdnup_general_def}. These
assumptions imply that

\begin{equation}
\Delta_A^{\down\up}({\bf 0})\approx\frac{\pi V_{0}}{\Omega_{BZ}^{2}G^{2}}\frac{\pi wm}{1-\frac{w\pi k_{0}^{2}}{\Omega_{BZ}}\frac{m}{m+1}}.
\end{equation}

 $\Delta_A^{\down\up}({\bf 0})$ is a strictly increasing function
of $w$. Note also that the denominator is always positive, which
follows from $w\pi k_{0}^{2}\leq\Omega_{BZ}$ (since the $w$ nodes are assumed
to be non-overlapping, their total area cannot exceed the
Brillouin-zone area). This
shows a clear case where increasing the number of nodes $w$ and order
of the node $m$ increases the magnon gap. The intuition is that increasing
the number of nodes and their order will decrease the norm and also
increase the gradient term in the numerator by a factor $wm$: so
both contribute to increasing the magnon gap. If we further assume
the bound in Eq.$\,$\ref{eq:grad_rho_bound_Gk} is
tight at $k=k_{0}$, i.e. $|\nabla\rho(k_{0})|\approx2\sqrt{G}$,
we have $m\rho_{0}k_{0}^{m-1}=2\sqrt{G}\leftrightarrow mZ/k_{0}=2\sqrt{G}\leftrightarrow k_{0}=mZ/(2\sqrt{G})$
and therefore

\begin{equation}
\Delta_A^{\down\up}({\bf 0})\approx\frac{\pi V_{0}}{\Omega_{BZ}^{2}G^{2}}\frac{\pi wm}{1-\frac{Z^{2}w\pi}{4G\Omega_{BZ}}\frac{m^{3}}{m+1}}.
\end{equation}

We will now consider a second example, where we assume a very peaked quantum metric
around one point in the Brillouin zone (${\bf k}=0$ without loss
of generality), with
\begin{equation}
G({\bf k})\approx\begin{cases}
G &,k<k_{0}\\
G_{\textrm{out}} &,k\geq k_{0}
\end{cases}
\end{equation}
and $G_{\textrm{out}}\ll G$. We will consider two types of nodes:

(i) We first assume an order $m$ node exactly at ${\bf k}=0$,
with

\begin{equation}
\rho(k)=\begin{cases}
\rho_{0}k^{m} &,k<k_{0}\\
\rho_{0}k_{0}^{m} &,k>k_{0}
\label{eq:rho_k_case2}
\end{cases}
\end{equation}

Under these assumptions, we have
\begin{equation}
\Delta^{\down\up}_A({\bf 0})\approx\frac{\pi V_{0}}{\Omega_{BZ}G^{2}}\frac{\pi m}{\Omega_{BZ}-\frac{m}{m+1}\pi k_{0}^{2}}\,.
\end{equation}
If we further assume that $G({\bf k})$ closely tracks the Berry curvature, we have $\pi k_{0}^{2}G\approx |C|$, where $C$ is the Chern
number, implying
\begin{equation}
\Delta^{\down\up}_A({\bf 0})\approx\frac{\pi^{2}V_{0}m}{\Omega_{BZ}^{2}G^{2}}\frac{1}{1-\frac{m}{m+1}\frac{|C|}{\Omega_{BZ}G}}\,,
\end{equation}
This again implies that the magnon gap grows with the order of the
node, as in the first example.

(ii) We now consider an order-$m$ node located outside the region in
which the quantum metric is concentrated. We place the node at
${\bf k}={\bf k}_{n}$ and introduce a nodal-core radius $q_{0}$ such
that
\begin{equation}
|{\bf k}_{n}|>k_{0}+q_{0}.
\end{equation}
The nodal core therefore lies entirely in the region where
$G({\bf k})\approx G_{\mathrm{out}}\ll G$. Defining
\begin{equation}
k\equiv|{\bf k}|,
\qquad
q\equiv|{\bf k}-{\bf k}_{n}|,
\end{equation}
we approximate the amplitude by
\begin{equation}
\rho({\bf k})\approx
\begin{cases}
\rho_{\mathrm{out}}
\left(\dfrac{q}{q_{0}}\right)^{m},
& q<q_{0},
\\[8pt]
\rho_{\mathrm{out}}+\rho_{0}(k_{0}-k),
& k<k_{0},
\\[4pt]
\rho_{\mathrm{out}},
& \text{otherwise}.
\end{cases}
\label{eq:rho_node_outside_metric}
\end{equation}
The first region describes an order-$m$ node at
${\bf k}={\bf k}_{n}$, with the amplitude growing from zero to
$\rho_{\mathrm{out}}$ inside the nodal core. Since
$G_{\mathrm{out}}\ll G$, Eq.$\,$\ref{eq:grad_rho_bound_Gk} implies
that the amplitude varies only weakly outside the nodal core and
before reaching the high-metric region. We therefore approximate it
by the constant value $\rho_{\mathrm{out}}$ in this intermediate
region. Once the high-metric region $k<k_{0}$ is reached, the amplitude
can grow much more rapidly. We assume the largest possible linear
growth compatible with Eq.$\,$\ref{eq:grad_rho_bound_Gk}, such that
$\rho_{0}\approx2\sqrt{G}$.

The amplitude-stiffness numerator receives contributions from the
high-metric region and from the nodal core:
\begin{equation}
\begin{aligned}
\int d^{2}{\bf k}\,
\frac{|\nabla\rho({\bf k})|^{2}}{G^{2}({\bf k})}
={}&
\frac{2\pi\rho_{0}^{2}}{G^{2}}
\int_{0}^{k_{0}}k\,dk
\\
&+
\frac{2\pi m^{2}\rho_{\mathrm{out}}^{2}}
{G_{\mathrm{out}}^{2}q_{0}^{2m}}
\int_{0}^{q_{0}}q^{2m-1}\,dq
\\
={}&
\frac{\pi\rho_{0}^{2}k_{0}^{2}}{G^{2}}
+
\frac{\pi m\rho_{\mathrm{out}}^{2}}
{G_{\mathrm{out}}^{2}}.
\end{aligned}
\label{eq:amplitude_numerator_node_outside}
\end{equation}
The norm of the mode is
\begin{equation}
\begin{aligned}
\int d^{2}{\bf k}\,\rho^{2}({\bf k})
={}&
\Omega_{BZ}\rho_{\mathrm{out}}^{2}
+
\frac{2\pi}{3}\rho_{\mathrm{out}}\rho_{0}k_{0}^{3}
+
\frac{\pi}{6}\rho_{0}^{2}k_{0}^{4}
\\
&-
\frac{\pi m}{m+1}\rho_{\mathrm{out}}^{2}q_{0}^{2}.
\end{aligned}
\label{eq:amplitude_norm_node_outside}
\end{equation}
We therefore obtain
\begin{equation}
\Delta_A^{\down\up}({\bf 0})
\approx
\frac{
\displaystyle \frac{\pi V_{0}}{\Omega_{BZ}} \Big(
\frac{\pi\rho_{0}^{2}k_{0}^{2}}{G^{2}}
+
\frac{\pi m\rho_{\mathrm{out}}^{2}}
{G_{\mathrm{out}}^{2}}\Big)
}{
\displaystyle
\Omega_{BZ}\rho_{\mathrm{out}}^{2}
+
\frac{2\pi}{3}\rho_{\mathrm{out}}\rho_{0}k_{0}^{3}
+
\frac{\pi}{6}\rho_{0}^{2}k_{0}^{4}
-
\frac{\pi m}{m+1}\rho_{\mathrm{out}}^{2}q_{0}^{2}
}.
\label{eq:gap_node_outside_metric_general}
\end{equation}

We now consider the limit in which $\rho_{\mathrm{out}}$ is sufficiently
small that the terms proportional to $\rho_{\mathrm{out}}$ can be
neglected in both the numerator and denominator. More precisely, we
assume
\begin{equation}
\frac{m\rho_{\mathrm{out}}^{2}}{G_{\mathrm{out}}^{2}}
\ll
\frac{\rho_{0}^{2}k_{0}^{2}}{G^{2}},
\end{equation}
and
\begin{equation}
\Omega_{BZ}\rho_{\mathrm{out}}^{2},
\quad
\rho_{\mathrm{out}}\rho_{0}k_{0}^{3},
\quad
\rho_{\mathrm{out}}^{2}q_{0}^{2}
\ll
\rho_{0}^{2}k_{0}^{4}.
\end{equation}
Under these assumptions,
\begin{equation}
\begin{aligned}
\Delta_A^{\down\up}({\bf 0})
&\approx
\frac{\pi V_{0}}{\Omega_{BZ}}
\frac{
\pi\rho_{0}^{2}k_{0}^{2}/G^{2}
}{
\pi\rho_{0}^{2}k_{0}^{4}/6
}
\\
&=
\frac{6\pi V_{0}}
{\Omega_{BZ}G^{2}k_{0}^{2}}.
\end{aligned}
\end{equation}
Assuming again that the quantum metric closely tracks the Berry
curvature, such that
\begin{equation}
\pi k_{0}^{2}G\approx |C|,
\end{equation}
we obtain
\begin{equation}
\Delta_A^{\down\up}({\bf 0})
\approx
\frac{6\pi^{2}V_{0}}
{\Omega_{BZ}G |C|}.
\label{eq:gap_node_outside_metric_limit}
\end{equation}

The amplitude-stiffness contribution therefore scales as $G^{-1}$
when the node lies outside the region of concentrated quantum metric,
whereas it scales as $G^{-2}$ when the node lies inside this region,
as found in case (i). Within the present illustrative example, the
former is therefore parametrically larger in the large-$G$ limit.

This is a simple illustrative
example showing that placing the node of the magnon wave function in a region of
small quantum metric produces a larger amplitude-stiffness
contribution than placing it in a region with large quantum metric.
As discussed in the main text, this distinction between nodes located in regions of large and small quantum metric is realized in the
small-angle regime of the FH model and well as of the no-fitting models of
twisted MoTe$_2$ introduced in
Ref.$\,$\cite{zhang2024universalmoiremodelbuildingmethodfitting} and
studied in Sec.$\,$\ref{subsec:Realistic_MoTe2} and
App.$\,$\ref{sec:no_fitting_flatband_vs_exact}. In particular, one of the magnon modes
aligns its nodes with the quantum-geometry peak(s) and becomes the
lowest-energy mode, while the other has a node in a region of smaller
quantum geometry and is pushed to higher energy.

\subsection{Derivation of upper bound on $|\nabla\rho|$ in Eq.$\,$\ref{eq:grad_rho_bound_Gk}}

In this section, we will derive the upper bound established
in Eq.$\,$\ref{eq:grad_rho_bound_Gk}. We start by re-writing the general SyMA in Eq.$\,$\ref{eq:Mdnup_general_def}, in bra-ket notation for readability as
\begin{equation}
 M({\bf k})=\bra{u_{\down,{\bf k}}}D\ket{u_{\up,{\bf k}}}
\end{equation}
where $D$ is a matrix characterizing different SyMA ansatze, as introduced in Eq.$\,$\ref{eq:Mdnup_general_def}.

Rewriting $\ket{\pd_{i}u_{\eta,{\bf k}}}=Q_{\eta}({\bf k})\ket{\pd_{i}u_{\eta,{\bf k}}}+\braket{u_{\eta,{\bf k}}}{\pd_{i}u_{\eta,{\bf k}}}\ket{u_{\eta,{\bf k}}}$,
with $Q_{\eta}({\bf k})=\mathbb{1}-\ket{u_{\eta,{\bf k}}}\bra{u_{\eta,{\bf k}}}$,
we have

\begin{equation}
\begin{aligned}
\partial_i M({\bf k})
={}&
\bra{\partial_i u_{\down,{\bf k}}}
D
\ket{u_{\up,{\bf k}}}
+
\bra{u_{\down,{\bf k}}}
D
\ket{\partial_i u_{\up,{\bf k}}}
\\
={}&
\bra{\partial_i u_{\down,{\bf k}}}
Q_{\down}({\bf k})D
\ket{u_{\up,{\bf k}}}
+
\bra{u_{\down,{\bf k}}}
D Q_{\up}({\bf k})
\ket{\partial_i u_{\up,{\bf k}}}
\\
&-
iM({\bf k})
\left[
A_{\up,i}({\bf k})-A_{\down,i}({\bf k})
\right].
\end{aligned}
\label{eq:diM_intermediate-1}
\end{equation}

Using
\begin{equation}
\partial_i\rho^2({\bf k})
=
\partial_i\left[M^*({\bf k})M({\bf k})\right]
=
2\operatorname{Re}
\left[
M^*({\bf k})\partial_iM({\bf k})
\right],
\end{equation}
together with Eq.$\,$\ref{eq:diM_intermediate-1}, we obtain
\begin{equation}
\begin{aligned}
\partial_i\rho^2({\bf k})
=
2\operatorname{Re}\Big[
M^*({\bf k})
\Big(
&
\bra{\partial_i u_{\down,{\bf k}}}
Q_{\down}({\bf k})D
\ket{u_{\up,{\bf k}}}
\\
&+
\bra{u_{\down,{\bf k}}}
D Q_{\up}({\bf k})
\ket{\partial_i u_{\up,{\bf k}}}
\Big)
\Big].
\end{aligned}
\end{equation}
Since
$\partial_i\rho^2({\bf k})
=
2\rho({\bf k})\partial_i\rho({\bf k})$,
taking the absolute value of both sides and using
$|\operatorname{Re}z|\leq |z|$ together with the triangle inequality gives
\begin{equation}
\begin{aligned}
2\rho({\bf k})|\partial_i\rho({\bf k})|
\leq
2\rho({\bf k})
\Bigg[
&
\left|
\bra{\partial_i u_{\down,{\bf k}}}
Q_{\down}({\bf k})D
\ket{u_{\up,{\bf k}}}
\right|
\\
&+
\left|
\bra{u_{\down,{\bf k}}}
D Q_{\up}({\bf k})
\ket{\partial_i u_{\up,{\bf k}}}
\right|
\Bigg].
\end{aligned}
\end{equation}
and therefore, for $\rho({\bf k})\neq0$,
\begin{equation}
\begin{aligned}
|\partial_i\rho({\bf k})|
\leq
&
\left|
\bra{\partial_i u_{\down,{\bf k}}}
Q_{\down}({\bf k})D
\ket{u_{\up,{\bf k}}}
\right|
\\
&+
\left|
\bra{u_{\down,{\bf k}}}
D Q_{\up}({\bf k})
\ket{\partial_i u_{\up,{\bf k}}}
\right|.
\end{aligned}
\end{equation}
Finally, using the Cauchy-Schwarz inequality and the definition of
the spectral norm, we have
\begin{equation}
\begin{aligned}
|\partial_i\rho({\bf k})|
&\leq
\|D\|_2
\left[
\|
Q_{\down}({\bf k})
\ket{\partial_i u_{\down,{\bf k}}}
\|
+
\|
Q_{\up}({\bf k})
\ket{\partial_i u_{\up,{\bf k}}}
\|
\right]
\\
&=
\|D\|_2
\left[
\sqrt{g_{ii}^{\down}({\bf k})}
+
\sqrt{g_{ii}^{\up}({\bf k})}
\right].
\end{aligned}
\label{eq:di_=00005Crho_expression}
\end{equation}
where we used the definition of quantum metric $g_{ii}^{\eta}({\bf k})=\|Q_{\eta}\ket{\pd_{i}u_{\eta,{\bf k}}}\|^2$.

Using the definition of the spin-resolved trace of the quantum metric,
\begin{equation}
G^{\eta}({\bf k})
\equiv
\operatorname{Tr}g^{\eta}({\bf k})
=
\sum_i g_{ii}^{\eta}({\bf k}),
\end{equation}
we now sum the componentwise bound in
Eq.$\,$\ref{eq:di_=00005Crho_expression} over the momentum-space
directions,
\begin{equation}
\begin{aligned}
|\nabla\rho({\bf k})|^{2}
&=
\sum_i
|\partial_i\rho({\bf k})|^{2}
\\
&\leq
\|D\|_{2}^{2}
\sum_i
\left[
\sqrt{g_{ii}^{\down}({\bf k})}
+
\sqrt{g_{ii}^{\up}({\bf k})}
\right]^{2}.
\end{aligned}
\end{equation}
For each momentum component $i$, we use
\begin{equation}
\left[
\sqrt{g_{ii}^{\down}({\bf k})}
+
\sqrt{g_{ii}^{\up}({\bf k})}
\right]^{2}
\leq
2\left[
g_{ii}^{\down}({\bf k})
+
g_{ii}^{\up}({\bf k})
\right],
\end{equation}
which follows from the bound $(a+b)^{2}\leq2(a^{2}+b^{2})$,
valid for any real numbers $a$ and $b$, upon taking
$a=\sqrt{g_{ii}^{\down}({\bf k})}$ and
$b=\sqrt{g_{ii}^{\up}({\bf k})}$. Therefore,
\begin{equation}
\begin{aligned}
|\nabla\rho({\bf k})|^{2}
&\leq
2\|D\|_{2}^{2}
\sum_i
\left[
g_{ii}^{\down}({\bf k})
+
g_{ii}^{\up}({\bf k})
\right]
\\
&=
2\|D\|_{2}^{2}
\left[
G^{\down}({\bf k})
+
G^{\up}({\bf k})
\right].
\end{aligned}
\end{equation}
If the two spin-resolved trace metrics are equal,
\begin{equation}
G^{\up}({\bf k})
=
G^{\down}({\bf k})
\equiv
G({\bf k}),
\end{equation}
we finally obtain
\begin{equation}
|\nabla\rho({\bf k})|^{2}
\leq
4\|D\|_{2}^{2}G({\bf k}).
\end{equation}
For the SyMA matrices considered here, which satisfy
$\|D\|_{2}\leq1$, this gives the desired upper bound in
Eq.$\,$\ref{eq:grad_rho_bound_Gk}.

\subsection{Nodes in twisted MoTe$_2$ }

\label{subsec:App_nodes_MoTe2}

In this section we will derive the positions and properties of the
nodes in MoTe$_2$ discussed in Sec.$\,$\ref{subsec:MoTe2_SMA_eigenfunctions}
using symmetry.

\subsubsection{Nodes of intralayer modes}

We will start by deriving Eq.$\,$\ref{eq:node_condition_intra}.
Writing the $C_{3}$ representation matrix in Eq.$\,$\ref{eq:D_representations_MoTe2}
as
\begin{equation}
 \big[D_{\eta}[C_{3}]\big]_{{\bf Q},{\bf Q}'}\equiv D_{{\bf Q}\eta,{\bf Q}'\eta}[C_{3}]=e^{i\eta\frac{\pi}{3}}\,[D_{C_{3}}]_{{\bf Q},{\bf Q}'}\,,
 \label{eq:D_C3_spinless_spinful}
\end{equation}
where $[D_{C_{3}}]_{{\bf Q},{\bf Q}'}=\delta_{{\bf Q},C_{3}{\bf Q}'}$
(i.e. explicitly splitting the spinless and spinful representations),
we can show that

\begin{equation}
D_{C_{3}}\theta_{j}D_{C_{3}}^{\dagger}=\theta_{j+1}\,.
\end{equation}
where $[\theta_{j}]_{{\bf Q},{\bf Q}'}=\delta_{{\bf Q}',{\bf Q}+(-1)^{{\bf Q}}C_{3}^{j}{\bf q}_{1}}$ was defined in Eq.$\,$\ref{eq:thetaj}.

Eq.$\,$\ref{eq:symmetry_relation_MoTe2} implies that

\begin{equation}
 D_{\eta}[C_{3}]U_{\eta,{\bf k}}=e^{i\varphi_{\eta}({\bf k})}U_{\eta,C_{3}{\bf k}}
 \label{eq:C3_sewing}
\end{equation}
where $\varphi_{\eta}({\bf k})$ is the $C_3$ sewing phase in Eq.$\,$\ref{eq:D_representations_MoTe2} and we dropped the $C_3$ superscript for readability in the calculation that follows. Using these properties,
we have

\begin{widetext}

\bea
M^{(\down,\up)}_{(\kappa,+1)}({\bf k},{\bf 0})&=\frac{1}{3}\sum_{j=-1}^{1}e^{-\frac{2\pi i}{3}(\kappa+1)j}U_{\down,{\bf k}}^{\dagger}\theta_{j}U_{\up,{\bf k}}\\
&=\frac{1}{3}\sum_{j=-1}^{1}e^{-\frac{2\pi i}{3}(\kappa+1)j}U_{\down,{\bf k}}^{\dagger}D_{\down}^{\dagger}[C_{3}]D_{\down}[C_{3}]\theta_{j}D_{\up}^{\dagger}[C_{3}]D_{\up}[C_{3}]U_{\up,{\bf k}}\\
&=\frac{1}{3}\sum_{j=-1}^{1}e^{-\frac{2\pi i}{3}(\kappa+1)j}e^{i[\varphi_{\up}({\bf k})-\varphi_{\down}({\bf k})]}U_{\down,C_{3}{\bf k}}^{\dagger}(e^{-\frac{\pi}{3}i}D_{C_{3}})\theta_{j}(e^{-\frac{\pi}{3}i}D_{C_{3}}^{\dagger})U_{\up,C_{3}{\bf k}}\\
&=\frac{1}{3}\sum_{j=-1}^{1}e^{-\frac{2\pi i}{3}(\kappa+1)j}e^{i[\varphi_{\up}({\bf k})-\varphi_{\down}({\bf k})-\frac{2\pi}{3}]}U_{\down,C_{3}{\bf k}}^{\dagger}\theta_{j+1}U_{\up,C_{3}{\bf k}}\\
&=e^{i[\varphi({\bf k})+\varphi(-{\bf k})+\frac{2\pi}{3}\kappa]}M^{(\down,\up)}_{(\kappa,+1)}(C_{3}{\bf k},{\bf 0})\,.
\eea
\end{widetext}where from the second to third lines we used Eqs.$\,$\ref{eq:D_C3_spinless_spinful} and \ref{eq:C3_sewing}, and defined $\varphi({\bf k})\equiv\varphi_{\up}({\bf k})=-\varphi_{\down}(-{\bf k})$.

At $C_{3}$ symmetric points, ${\bf k}^{*}=C_{3}{\bf k}^{*}\textrm{ mod }{\bf G}$,
the phase $e^{i\varphi({\bf k}^{*})}$ is gauge-invariant and becomes
the spinful $C_{3}$ symmetry eigenvalue of the filled band (spin-$\up$)
eigenstate. $ $ Writing $\varphi({\bf k}^{*})=\frac{2\pi}{3}m_{{\bf k}^{*}}+\frac{\pi}{3}$,
where $m_{{\bf k}^{*}}$ is the spinless character and $\pi/3$ is
the additional spin-$\up$ angular momentum (see Eq.$\,$\ref{eq:representations_cr_MoTe2}),
and using $M^{(\down,\up)}_{(\kappa,+1)}(C_{3}{\bf k}^{*},{\bf 0})=M^{(\down,\up)}_{(\kappa,+1)}({\bf k}^{*},{\bf 0})$,
we arrive at Eq.$\,$\ref{eq:node_condition_intra}, that we rewrite here as

\begin{equation}
 [1-e^{i\frac{2\pi}{3}m_{{\bf k}^{*}}^{\kappa}}]M^{(\down,\up)}_{(\kappa,+1)}({\bf k}^{*},{\bf 0})=0 \,,
 \label{eq:node_condition_intra_SM}
\end{equation}
where we defined

\bea
 m_{{\bf k}^{*}}^{\kappa}&= e_{{\bf k}^{*}}+\kappa \,,
 \label{eq:m_kstar}
\eea

\bea
 e_{{\bf k}^{*}}&=m_{{\bf k}^{*}}+m_{-{\bf k}^{*}}+1 \,.
 \label{eq:e_kstar}
\eea

We will now use Eq.$\,$\ref{eq:node_condition_intra_SM} to verify that
within a topological phase with $C\neq0\pmod 3$, all modes
necessarily have nodes (in accordance with the general proof in Sec.$\,$\ref{subsec:App_chern_vorticity_derivation}).
The restriction of $C\neq0\pmod 3$ follows from the usage of $C_{3}$ symmetry
in the proof. However, we recall that we derived the most general proof in Appendix.$\,$\ref{subsec:App_chern_vorticity_derivation},
in which case the claim is valid for any $C\neq0 \, (\mod 3)$ and any one-band projected magnon mode. The proof that follows is complementary, but also adds further information for the intralayer SyMA modes, requiring that at least one of
the three $C_{3}$-symmetric points ${\bf k}=\Gamma,{\bf K},{\bf K}'$ to be a node at ${\bf p}=0$ for topological phases with $C\neq0\pmod 3$.

Defining the sum of characters at $C_{3}$
symmetric points as
\begin{equation}
 m_{C_{3}}=m_{\Gamma}+m_{{\bf K}}+m_{{\bf K}'}\,,
 \label{eq:m_C3}
\end{equation}
and using $\exp(\frac{2\pi i}{3}C)=\exp(\frac{2\pi}{3}i\sum_{{\bf k}=\Gamma,{\bf K},{\bf K}'}m_{{\bf k}})$
\cite{PhysRevB.86.115112}, we have

\begin{equation}
\begin{cases}
m_{C_{3}}\mod 3=0 &,\, \, C=0\pmod3\\
m_{C_{3}}\mod 3=\pm1 &,\textrm{ Topological with \ensuremath{C=\pm1}}\,.
\end{cases}
\end{equation}

In the trivial phase or in a topological phase with $C=0\pmod 3$, we have $m_{{\bf k}^{*}}^{\kappa}=2m_{\Gamma}+\kappa+1\textrm{ mod }\ensuremath{3}$, which can be shown by using

\begin{equation}
m_{\bf K}+m_{{\bf K}'}
\equiv
m_{C_3}-m_\Gamma
\equiv
m_{C_3}+2m_\Gamma
\pmod 3.
\label{eq:m_C3_relations}
\end{equation}
and Eqs.$\,$\ref{eq:m_kstar} and \ref{eq:m_C3}.
Therefore, in this case there always necessarily exists an intralayer SyMA mode with
no nodes at the $C_3$ symmetric points.

On the other hand, in a topological phase with $C\neq0\pmod 3$ we have using again Eq.$\,$\ref{eq:m_C3_relations},

\begin{equation}
\begin{array}{c}
m_{{\bf k}^{*}}^{\kappa}=\begin{cases}
2m_{\Gamma}+\kappa+1\textrm{ mod }\ensuremath{3}\,,{\bf k}^{*}=\Gamma\\
2m_{\Gamma}+m_{C_3}+\kappa+1\textrm{ mod }\ensuremath{3}\,,{\bf k}^{*}={\bf K},{\bf K}'
\end{cases},\end{array}
\end{equation}
which implies at least a node at $C_3$-symmetric points for all $\kappa$ because $m_{C_{3}}\neq0$ in this case.

\subsubsection{Vorticity of nodes and $C_{3}$ characters of intralayer modes}

We now relate the vorticity of the nodes to the electron and magnon
$C_{3}$ eigenvalues for the intralayer modes. For an isolated node of
the magnon wave function at a $C_{3}$-symmetric point ${\bf k}={\bf k}^{*}$, we define its
vorticity as the integer winding number of the phase of
$M^{(\down,\up)}_{(\kappa,+1)}({\bf k},{\bf 0})$ along a small
counterclockwise loop enclosing the node:
\begin{equation}
q_{{\bf k}^{*}}
\equiv
\frac{1}{2\pi}
\oint_{\mathcal C_{{\bf k}^{*}}}
d{\bf k}\cdot
\nabla_{\bf k}
\arg
M^{(\down,\up)}_{(\kappa,+1)}({\bf k},{\bf 0})
\in\mathbb Z,
\label{eq:vorticity_definition}
\end{equation}
where $\mathcal C_{{\bf k}^{*}}$ encloses ${\bf k}^{*}$ but no other
nodes.

We introduce the momentum difference from the $C_{3}$-symmetric point
${\bf k}^{*}$ as
$\delta{\bf k}={\bf k}-{\bf k}^{*}$, and the corresponding complex
coordinates
\begin{equation}
z=(k_x-k_x^{*})+i(k_y-k_y^{*}),
\qquad
\bar z=(k_x-k_x^{*})-i(k_y-k_y^{*}).
\end{equation}
We now consider a general expansion of the magnon wave function around ${\bf k}={\bf k}^*$, that we write as
\begin{equation}
 M^{(\down,\up)}_{(\kappa,+1)}({\bf k}^{*}+\delta{\bf k},{\bf 0})\equiv M^{(\down,\up)}_{(\kappa,+1)}({\bf k}^{*};z,\bar{z})=\sum_{m,n\geq0}c_{mn}z^{m}\bar{z}^{n}\,,
\end{equation}
for general expansion coefficients $c_{mn}$.
The action of $C_{3}$ makes $z\rightarrow\omega z$ and $\bar{z}\rightarrow\omega^{-1}\bar{z}$,
with $\omega=e^{2\pi i/3}$, which follows from ${\bf k}={\bf k}^{*}$ being a $C_3$ symmetric point. Then we have

\begin{equation}
\begin{aligned}
M^{(\down,\up)}_{(\kappa,+1)}
({\bf k}^{*}+\delta{\bf k},{\bf 0})
={}&
e^{i[
\varphi({\bf k}^{*}+\delta{\bf k})
+\varphi(-{\bf k}^{*}-\delta{\bf k})
+\frac{2\pi}{3}\kappa
]}
\\
&\times
M^{(\down,\up)}_{(\kappa,+1)}
({\bf k}^{*}+C_{3}\delta{\bf k},{\bf 0}).
\end{aligned}
\end{equation}
which implies that
\bea
\sum_{m,n\geq0}c_{mn}^{{\bf k}^{*}}z^{m}\bar{z}^{n}=\omega^{m_{{\bf k}^{*}}^{\kappa}+\mathcal{O}(\delta{\bf k})}\sum_{m,n\geq0}c_{mn}^{{\bf k}^{*}}\omega^{m-n}z^{m}\bar{z}^{n} \,
\label{eq:C3_polynomial_constraint}
\eea
where $m_{{\bf k}^{*}}^{\kappa}
=
m_{{\bf k}^{*}}
+
m_{-{\bf k}^{*}}
+
1
+
\kappa
=
e_{{\bf k}^{*}}
+
\kappa$
was defined in Eq.$\,$\ref{eq:m_kstar}.

Taking the lowest total polynomial of degree $s_0$, we define the corresponding
leading-order polynomial as
\begin{equation}
P_{s_0}(z,\bar z)
=
\sum_{m+n=s_0}
c_{mn}^{{\bf k}^{*}}
z^m\bar z^n.
\end{equation}
Keeping the terms of degree $s_0$ in
Eq.$\,$\ref{eq:C3_polynomial_constraint} gives
\begin{equation}
P_{s_0}(z,\bar z)
=
\omega^{e_{{\bf k}^{*}}+\kappa}
P_{s_0}(\omega z,\omega^{-1}\bar z).
\end{equation}
Since the monomials $z^m\bar z^n$ are linearly independent, every nonzero
coefficient $c_{mn}^{{\bf k}^{*}}$ appearing at this lowest order must
therefore satisfy
\begin{equation}
m-n
=
-\left(e_{{\bf k}^{*}}+\kappa\right)
\mod 3,
\end{equation}
where
$e_{{\bf k}^{*}}
=
m_{{\bf k}^{*}}
+
m_{-{\bf k}^{*}}
+
1$
was defined in Eq.$\,$\ref{eq:e_kstar}.

We now parameterize a small circle around ${\bf k}^{*}$ by
$z=\rho e^{i\varphi}$. Since all monomials contributing to
$P_{s_0}$ have the same value of $m-n$ modulo $3$, we can write
\begin{equation}
P_{s_0}(\rho,\varphi)
=
\rho^{s_0}
e^{-i(e_{{\bf k}^{*}}+\kappa)\varphi}
\sum_{m+n=s_0}
c_{mn}^{{\bf k}^{*}}
e^{i3\ell_{mn}\varphi},
\end{equation}
where $\ell_{mn}\in\mathbb{Z}$. The second factor can contribute only
an integer multiple of $3$ to the phase winding around the node.
Therefore, defining the vorticity as
\bea
q_{{\bf k}^{*}}
=
\lim_{\rho\rightarrow0}
\frac{1}{2\pi}
\int_{0}^{2\pi}
d\varphi\,
\partial_{\varphi}
\arg\left[
M^{(\down,\up)}_{(\kappa,+1)}
({\bf k}^{*}+\rho e^{i\varphi},{\bf 0})
\right],
\eea
we obtain
\begin{equation}
q_{{\bf k}^{*}}
=
-\left(e_{{\bf k}^{*}}+\kappa\right)
\mod 3.
\end{equation}

Although $C_{3}$ symmetry also allows nodes at generic momenta occurring
in $C_{3}$-related triplets, we observe numerically that the magnon wave functions
of all the intralayer SyMA modes (and corresponding exact modes) have nodes only at the
$C_{3}$-symmetric momenta in the FH model of MoTe$_2$, as shown in Figs.$\,$\ref{fig:3} and
\ref{fig:App_MoTe2_eigenfuncs}. We therefore focus below on the
vorticities of the nodes at $\Gamma$, ${\bf K}$, and ${\bf K}'$.
Using $e_{\Gamma}=3$ and $e_{{\bf K}}=e_{{\bf K}'}=1$,
which is true within the $V_{z}=0$ topological phase at $\theta=3.7^{\circ}$,
we have:

\begin{equation}
\begin{array}{cc}
\kappa=-1:\textrm{ single node at \ensuremath{\Gamma}}\rightarrow q_{\Gamma}=1\mod 3=-2\\
\kappa=0:\textrm{ nodes at {\bf K},{\bf K}'}\rightarrow q_{{\bf K}}=q_{{\bf K}'}=-1\mod 3=-1
\end{array}
\end{equation}
and
\begin{equation}
\begin{array}{cc}
\kappa=1:\textrm{ nodes at \ensuremath{\Gamma,{\bf K},{\bf K}'}} \ ,\\
q_{{\bf K}}=q_{{\bf K}'}=-2\mod 3, \quad q_{\Gamma}=-4\mod 3 \,.\\
\end{array}
\end{equation}

Combining these conditions with the total-vorticity sum rule in
Eq.$\,$\ref{eq:qj_C}, and restricting to the configurations with the
minimum possible total node degree $\mathcal N=\sum_j|q_j|$, we obtain

\bea
\begin{cases}
q_{{\bf K}}=q_{{\bf K}'}=-2\,,q_{\Gamma}=2\\
q_{{\bf K}}=q_{{\bf K}'}=1\,,q_{\Gamma}=-4\,,
\end{cases}
\eea
as discussed in the main text.

\subsubsection{Nodes of interlayer modes}

For the $\mathcal{C}$-odd interlayer mode with $\kappa=-1$, we write the spinless $\mathcal{I}$-representation
matrix $[D_{\mathcal{I}}]_{{\bf Q},{\bf Q}'}=\delta_{{\bf Q},-{\bf Q}'}$ (see Eq.$\,$\ref{eq:D_representations_MoTe2})
and use $\{D_{\mathcal{I}},\tau_{-1}\}=0$, where $[\tau_{-1}]_{{\bf Q},{\bf Q}'}=-(-1)^{{\bf Q}}\delta_{{\bf Q}',{\bf Q}}$ was defined in Eq.$\,$\ref{eq:tauminus1}, to obtain
\bea
M^{(\down,\up)}_{(-1,-1)}({\bf k},{\bf p})&=U_{-{\bf k}-{\bf p}}^{T}D_{\mathcal{I}}D_{(-1)}^{\perp}U_{{\bf k}}\\
&=U_{-{\bf k}-{\bf p}}^{T}D_{\mathcal{I}}\tau_{-1}U_{{\bf k}}\\
&=U_{{\bf k}}^{T}\tau_{-1}D_{\mathcal{I}}U_{-{\bf k}-{\bf p}}\\
&=-U_{{\bf k}}^{T}D_{\mathcal{I}}\tau_{-1}U_{-{\bf k}-{\bf p}}\\
&=-M^{(\down,\up)}_{(-1,-1)}(-{\bf k}-{\bf p},{\bf p}).
\label{eq:interlayer_modes_node_condition}
\eea
where from the second to the third lines we transposed the result (which is valid since $M^{(\down,\up)}_{(-1,-1)}({\bf k}, {\bf p})$ is a scalar). For the remaining $\mathcal{C}$-odd modes, the same can be shown using $D_{\mathcal{I}}\tilde{\theta}_{j}^{T}=\tilde{\theta}_{j}D_{\mathcal{I}}$
and $[\tilde{\theta}_{j},\tau_{-1}]=0$, where $[\tilde{\theta}_{j}]_{{\bf Q},{\bf Q}'}=\delta_{{\bf {\bf Q}}',{\bf Q}-C_{6}^{j}{\bf b}_{1}}$ was defined in Eq.$\,$\ref{eq:thetajtilde}. The conditions for the
nodes of the $\mathcal{C}$-odd interlayer modes described in Sec.$\,$\ref{subsec:MoTe2_SMA_eigenfunctions}
can readily be obtained from Eq.$\,$\ref{eq:interlayer_modes_node_condition}. Namely, the magnon wave function must vanish at the momenta satisfying
\begin{equation}
{\bf k}=-{\bf k}-{\bf p}\pmod{\bf G}.
\end{equation}

\section{Lowest Landau level with arbitrary interaction}
\label{secapp:LLL}

In the main text, we argued that the almost flat magnon modes in MoTe$_2$ can be understood from a simple toy model of time-reversal conjugate Landau levels with opposite Chern number. The magnon spectrum of this model at integer filling was first solved in Ref.$\,$\cite{PhysRevLett.126.137601}. In this Appendix, we re-derive these results in an independent way and extend the result to arbitrary interaction.

Our strategy is as follows. We first project the interaction onto the lowest Landau level (LLL) in each valley and introduce the Landau-gauge orbital basis on the cylinder and its periodization to the torus, in which the interaction matrix elements take a separable form in terms of Haldane pseudopotentials. We then perform a particle-hole transformation on the filled $\uparrow$ valley, which recasts the magnon excitations as single particle--hole pairs on top of the filled-Landau-level reference state. In the pseudopotential basis, this one-hole--one-particle problem becomes exactly solvable on the torus, yielding a closed-form magnon spectrum that we compare to the numerically obtained magnon dispersion of MoTe$_2$ in the main text.

We consider the single-particle Hamiltonian for electrons of charge $-e$ in a 2D plane and minimally coupled to a valley (or spin)-dependent vector potential $\mbf{A}^\eta(\mbf{r})$
\begin{equation}
 \hat{H}_0^\eta=\frac{(\mbf{p}+e\mbf{A}^\eta(\mbf{r}))^2}{2m},
\end{equation}
where $\eta=\uparrow,\downarrow$ labels the two valleys. Additionally, $\nabla\times\mbf{A}^\eta=(0,0,B^\eta)^T$, $B^\eta$ is the magnetic field felt in valley $\eta$. In the following, we will be interested in the two specific configurations, $B^\eta=B$ and $B^\eta=\eta B$, which we label as the $\mathrm{SU}(2)$ and Ising setup. $\mbf{p}$ and $\mbf{r}$ are the 2D momentum and position operators, respectively. We define the magnetic length $\ell_\mathrm{B}=\sqrt{\frac{\hbar}{eB}}$ which will often be set to unity. For both configurations, the Hamiltonian has a $\mathrm{U}(1)$ valley conservation symmetry whose generator is $2\hat{S}_z=\hat{N}_\uparrow-\hat{N}_\downarrow$, where $\hat{N}_\eta$ is the number operator in valley $\eta$. For $B^\eta=B$, this symmetry is enhanced to a full $\mathrm{SU}(2)$ symmetry. Furthermore, in the Ising case, the system preserves time-reversal symmetry $\mathcal{T}$ since the magnetic fields in the two valley sectors are equal and opposite, and the corresponding Landau levels have opposite Chern numbers $C^\eta=\eta$. We construct the single-particle Hilbert space by projecting to the lowest Landau levels (LLLs) in the two valley sectors.

The $\mathrm{SU}(2)$ setup corresponds to the well-studied fractional quantum Hall (FQH) bilayer, which is realized in the FQHE with negligible Zeeman field$\,$\cite{Halperin_1983,Halperin_1984}, whilst the time-reversal symmetric Ising setup is the toy model for the highest valence band of tMoTe$_2$ around $3.7^\circ$ discussed in the main text.

\subsection{Interaction projected to the lowest Landau levels}\label{app:subsec:LLL_setup}
We consider $\mathcal{T}$ and $\hat{S}_z$-preserving density-density interactions
\begin{equation}\label{app:eq:LLL:dens_dens}
 \hat{H}_\mathrm{int}=\sum_{\eta \eta'}\int \frac{\dd^2\mbf{q}}{(2\pi)^2}V(\mbf{q}):\rho_\eta (\mbf{q})\rho_{\eta'}(-\mbf{q}):\,,
\end{equation}
where $:\dots:$ denotes normal ordering with respect to $\nu=0$. This differs from the normal-ordering of $H_I$ in Eq.$\,$\ref{eq:H_V_general}. The reason is that in order to solve the magnon problem analytically by transforming it into a $2$-body problem, we want to normal-order with respect to $\nu_\downarrow=0$ and $\nu_\uparrow=1$, which is achieved by particle-hole transforming the $\uparrow$-component of the above Hamiltonian. Then, the magnon problem is a two-body problem consisting of a single hole in the $\uparrow$-LLL and a single electron in the $\downarrow$-LLL.

We now project the Hamiltonian onto the LLL for valley $\eta=\uparrow$ and $\downarrow$. Let $\ket{n_\eta}$ label the basis states of the LLL for each valley $\eta$ and $\gamma_{n,\eta}^{(\dag)}$ the corresponding annihilation(creation) operator. The full valley-dependent Hamiltonian then reads
\begin{equation}\label{app:eq:Ham_valley}
 \hat{H}=\sum_{\eta,\eta'}\sum_{m_1,m_2,n_1,n_2}V^{\eta \eta'}_{m_1,m_2;n_1,n_2}\Cdag{\eta}{m_2}\Cdag{\eta'}{m_1}\C{\eta'}{n_1}\C{\eta}{n_2}
\end{equation}
with interaction matrix elements
\begin{equation}\label{app:eq:matrix_elements}
 V^{\eta \eta'}_{m_1,m_2;n_1,n_2}=\bra{m_{1,\eta'},m_{2,\eta}}{\hat{V}}\ket{n_{1,\eta'},n_{2,\eta}}\,,
\end{equation}
where $\hat{V}$ is the $2$-body interaction corresponding to $V(\mbf{q})$; the valley dependence of the matrix elements stems solely from the valley dependence of the LLL wave functions. The matrix elements obey the symmetries
\begin{equation}\label{app:eq:LLL:symmetries}
 V^{\eta \eta'}_{m_1,m_2;n_1,n_2}=V^{\eta' \eta}_{m_2,m_1;n_2,n_1}=V^{\eta \eta'}_{n_1,n_2;m_1,m_2}\,,
\end{equation}
where the first equality follows from relabeling the two particles in Eq.$\,$\ref{app:eq:matrix_elements}, and the second from hermiticity of $\hat{H}$ together with the reality of the matrix elements in the Landau-gauge basis introduced below. These relations will be used repeatedly to combine terms after the particle-hole transformation.

For the FTI model with opposite Chern numbers, both valleys are related by time-reversal symmetry, so their wave functions $\psi_{n,\eta}(\mbf{r})$ are related by complex conjugation:
\begin{equation}
 \psi_{n,-\eta}(\mbf{r})=\psi^{*}_{n,\eta}(\mbf{r})\,,
\end{equation}
where $-\eta$ denotes the opposite valley. For the matrix elements of the intervalley interaction, this implies
\begin{equation}
 V_{m_1,m_2;n_1,n_2}^{\eta,-\eta}=V_{n_1,m_2;m_1,n_2}^{\eta\eta}\,.
\end{equation}
By defining $V_{m_1,m_2;n_1,n_2}\coloneqq V_{m_1,m_2;n_1,n_2}^{\uparrow\uparrow}$, we can conclude
\begin{equation}
\begin{split}
 V_{m_1,m_2;n_1,n_2}^{\eta\eta}&=V_{m_1,m_2;n_1,n_2}\,,\\
 V_{m_1,m_2;n_1,n_2}^{\eta,-\eta}&=V_{n_1,m_2;m_1,n_2}\,,
\end{split}
\end{equation}
i.e.$\,$the inter-valley matrix elements are identical to the intra-valley matrix elements up to the permutation of indices $m_1\leftrightarrow n_1$.

\subsection{Landau-gauge basis on the cylinder and torus}\label{app:subsec:LLL_geometry}
To obtain explicit matrix elements, we place the system on a cylinder of circumference $L_y$ pierced by $N_\phi$ magnetic flux quanta. In the Landau gauge $\mbf{A}=\ell_\mathrm{B}^{-2}(0,-x)$, the LLL eigenstates in valley $\eta=\uparrow$ are given by
\begin{equation}\label{app:eq:LLL:cylinder_orbitals}
 \phi_m(x,y)\propto e^{iky}e^{-\frac{1}{2\ell_\mathrm{B}^2}\left(x-\ellB k\right)^2},\quad k=\frac{2\pi m \ellB}{L_y}\,,
\end{equation}
with the orbital index $m=0,\dots,N_\phi-1$; the orbitals of valley $\downarrow$ are their complex conjugates. These orbitals constitute the basis states $\ket{m_\eta}$ used in Eq.$\,$\ref{app:eq:matrix_elements}.

In the LLL, it is convenient to parametrize the interaction in terms of Haldane pseudopotentials$\,$\cite{Haldane_1983}
\begin{equation}\label{app:eq:Haldane_PP}
 V_\alpha\equiv \int \frac{\dd^2\mbf{q}}{(2\pi)^2}V(\mbf{q})\mathcal{L}_\alpha(\mbf{q}^2\ell_\mathrm{B}^2)e^{-\mbf{q}^2\ell_\mathrm{B}^2}\,,
\end{equation}
where $\mathcal{L}_\alpha$ are the Laguerre polynomials. On the cylinder, the matrix elements of Eq.$\,$\ref{app:eq:matrix_elements} take a closed, separable form$\,$\cite{Ortiz_2013,Lee_2013,Lee_2015}
\begin{equation}\label{app:eq:LLL:cylinder_matrix_elements}
 V_{m_1,m_2;n_1,n_2}=\delta_{m_1+m_2,n_1+n_2}\sum_{\alpha=0}^{\infty} V_{\alpha}\,\eta^\alpha_{m_2-m_1}\eta^\alpha_{n_2-n_1}\,,
\end{equation}
with the form factors
\begin{equation}\label{app:eq:LLL:form_factors}
 \eta^\alpha_{k}=\frac{2^{-1/4-\alpha/2}\sqrt{\ell_{\mathrm{B}}}}{\sqrt{R\cdot \alpha! \sqrt{\pi}}}\exp\left(-\frac{k^2\ellB^2}{4R^2}\right)H_\alpha\left(\frac{k\ellB}{\sqrt{2}R}\right)\,,
\end{equation}
where $H_{\alpha}$ are the physicist's Hermite polynomials and $R=L_y/(2\pi)$ is the radius of the cylinder. Since $H_\alpha$ is an even (odd) function for $\alpha$ even (odd), the form factors satisfy the parity relation
\begin{equation}\label{app:eq:LLL:parity}
 \eta^\alpha_{-k}=(-1)^\alpha\,\eta^\alpha_{k}\,.
\end{equation}
Moreover, they are normalized as
\begin{equation}
\begin{split}
 \sum_{k} (\eta^\alpha_k)^2&=\sum_k \frac{\ell_\mathrm{B}}{2^\alpha\sqrt{2}R\cdot \alpha! \sqrt{\pi}}e^{-\frac{k^2\ell_\mathrm{B}^2}{2R^2}}H_\alpha\left(\frac{k \ell_\mathrm{B}}{\sqrt{2}R}\right)^2\\
 &\approx \int \dd q\, H_\alpha(q)^2\frac{e^{-q^2}}{\sqrt{\pi}2^\alpha\alpha!}=1\,,
\end{split}
\label{eq:eta_k_norm}
\end{equation}
where we introduced the continuum variable $q=k\ellB/(\sqrt{2}R)$, and the approximation becomes exact in the large-$R$ limit. With this normalization, the finite-energy $2$-body eigenstate of the pseudopotential $V_\alpha$ has energy $V_\alpha$ in the thermodynamic limit, as we will see explicitly below.

As we are ultimately interested in comparing the nature of the magnon modes in the LLL with realistic models, which are studied on a torus, and to reduce finite-size effects due to boundaries, we periodize the interaction over mirror copies of the finite cylinder$\,$\cite{Ortiz_2013,Lee_2013}. On the torus, the matrix elements retain the separable form of Eq.$\,$\ref{app:eq:LLL:cylinder_matrix_elements} with two modifications. First, momentum is only conserved modulo $N_\phi$,
\begin{equation}\label{app:eq:LLL:mod_conservation}
 V_{m_1,m_2;n_1,n_2}=0 \quad \text{if } m_1+m_2 \not\equiv n_1+n_2 \;\mathrm{mod}\; N_\phi\,.
\end{equation}
Second, the form factors are periodized,
\begin{equation}\label{app:eq:LLL:periodized_ff}
 \eta^\alpha_{k}\rightarrow \bar{\eta}^\alpha_{k}=\sum_{s\in\mathbb{Z}}\eta^\alpha_{k+2sN_\phi}\,.
\end{equation}
The periodized form factors satisfy the parity relation Eq.$\,$\ref{app:eq:LLL:parity} exactly, as seen by relabeling $s\rightarrow -s$ in Eq.$\,$\ref{app:eq:LLL:periodized_ff}. Furthermore, as the form factors $\eta_{k}^\alpha$ exhibit a gaussian decay beyond $|k|>2\sqrt{\alpha}$, the effect of the mirror copies ($s\neq 0$) can be neglected in the thermodynamic limit, $\bar{\eta}^\alpha_k\approx \eta^\alpha_k$, and the normalization Eq.$\,$\ref{eq:eta_k_norm} continues to hold. In the following, we work on the torus and keep the notation $\eta^\alpha_k$ for the periodized form factors.

\subsection{Particle-hole transformation}

As we are interested in studying the magnon excitation out of the fully filled $\eta=\uparrow$ band, it is convenient to perform a particle-hole (PH) transformation
\begin{equation}
 \h{\uparrow}{n}=\Cdag{\uparrow}{n}, \quad \Hdag{\uparrow}{n}=\C{\uparrow}{n}\,,
\end{equation}
on the $\uparrow$ valley, whilst keeping the unoccupied valley in the electronic basis.

We begin with normal-ordering the intravalley term for $\eta=\uparrow$ in the hole-basis, where we drop $\eta$ for readability:
\begin{align}
 &\Cdag{m_2}\Cdag{m_1}\C{n_1}\C{n_2}= \h{m_2}\h{m_1}\Hdag{n_1}\Hdag{n_2}=\Hdag{n_2}\Hdag{n_1}\h{m_1}\h{m_2}\nonumber\\
\begin{split}
 &\quad-\Hdag{n_2}\h{m_2}\delta_{m_1,n_1}-\Hdag{n_1}\h{m_1}\delta_{m_2,n_2}\\
 &\quad+\Hdag{n_2}\h{m_1}\delta_{m_2,n_1}+\Hdag{n_1}\h{m_2}\delta_{m_1,n_2}
\end{split}\\
 &\quad-\delta_{m_1,n_2}\delta_{m_2,n_1}+\delta_{m_2,n_2}\delta_{m_1,n_1}\,.\nonumber
\end{align}
The intervalley terms transform as
\begin{equation}
\begin{split}
 &\Cdag{\uparrow}{m_2}\Cdag{\downarrow}{m_1}\C{\downarrow}{n_1}\C{\uparrow}{n_2}=\h{\uparrow}{m_2}\Cdag{\downarrow}{m_1}\C{\downarrow}{n_1}\Hdag{\uparrow}{n_2}\\
 &\quad=-\Hdag{\uparrow}{n_2}\Cdag{\downarrow}{m_1}\C{\downarrow}{n_1}\h{\uparrow}{m_2}+\Cdag{\downarrow}{m_1}\C{\downarrow}{n_1}\delta_{n_2,m_2}\,
\end{split}
\end{equation}
and
\begin{equation}
\begin{split}
 &\Cdag{\downarrow}{m_2}\Cdag{\uparrow}{m_1}\C{\uparrow}{n_1}\C{\downarrow}{n_2}= \Cdag{\downarrow}{m_2}\h{\uparrow}{m_1}\Hdag{\uparrow}{n_1}\C{\downarrow}{n_2}\\
 &\quad=-\Cdag{\downarrow}{m_2}\Hdag{\uparrow}{n_1}\h{\uparrow}{m_1}\C{\downarrow}{n_2}+\Cdag{\downarrow}{m_2}\C{\downarrow}{n_2}\delta_{m_1,n_1}\,.
\end{split}
\end{equation}

Overall, the Hamiltonian normal-ordered with respect to $\C{\downarrow}$ and $\h{\uparrow}$ is given by:
\begin{equation}
\begin{split}\label{app:eq:LLL:ph_transformed}
 \hat{H}&=\sum_{m_1,m_2,n_1,n_2}V_{m_1,m_2;n_1,n_2}\Hdag{\uparrow}{m_2}\Hdag{\uparrow}{m_1} \h{\uparrow}{n_1} \h{\uparrow}{n_2}\\
 &\quad +\sum_{m_1,m_2,n_1,n_2}V_{m_1,m_2;n_1,n_2}\Cdag{\downarrow}{m_2}\Cdag{\downarrow}{m_1} \C{\downarrow}{n_1} \C{\downarrow}{n_2}\\
 &\quad -2\sum_{m_1,m_2,n_1,n_2}V_{m_1,m_2;n_1,n_2}\Hdag{\uparrow}{m_2}\Cdag{\downarrow}{m_1} \C{\downarrow}{n_1} \h{\uparrow}{n_2}\\
 &\quad+\hat{H}_{1\text{-}\mathrm{body}},
\end{split}
\end{equation}
where lines $1$ and $3$ were brought to this form by relabeling summation indices and using the matrix-element symmetries of Eq.$\,$\ref{app:eq:LLL:symmetries}. In particular, the two intervalley terms yield identical contributions, producing the factor of $2$ in line $3$. The $1$-body term reads
\begin{widetext}
\begin{equation}
\begin{split}\label{app:eq:LLL:1_body}
 \hat{H}_{1\text{-}\mathrm{body}}&=\sum_{n_1,n_2}\left(-2(V_{n_1,n_2;n_1,n_2}-V_{n_2,n_1;n_1,n_2})\Hdag{\uparrow}{n_2}\h{\uparrow}{n_2}+2V_{n_1,n_2;n_1,n_2}\Cdag{\downarrow}{n_2}\C{\downarrow}{n_2} \right)\\
 &\qquad+\sum_{n_1,n_2}(V_{n_1,n_2;n_1,n_2}-V_{n_2,n_1;n_1,n_2})\,.
\end{split}
\end{equation}
\end{widetext}
We utilized the modulo-$N_\phi$ momentum conservation of Eq.$\,$\ref{app:eq:LLL:mod_conservation} to simplify the $1$-body term: each contraction fixes one pair of orbital indices to be equal, so that Eq.$\,$\ref{app:eq:LLL:mod_conservation} forces the remaining pair to be equal modulo $N_\phi$, and since all orbital indices lie in $\lbrace 0,\dots,N_\phi-1\rbrace$, they are equal. The $1$-body term is thus diagonal in the orbital basis.

\subsection{Magnon Hamiltonian}
To calculate the magnon spectrum with respect to the ground state, we begin by calculating the energy of the fully filled LLL for $\eta=\uparrow$,
\begin{equation}
 \ket{\uparrow}\coloneqq\prod_{n=0}^{N_\phi-1}\Cdag{\uparrow}{n}\ket{0}\,,
\end{equation}
where $\ket{0}$ is the vacuum.
By construction, $\ket{\uparrow}$ is the vacuum of the hole and $\downarrow$-electron operators, i.e. both $\h{\uparrow}{n}$ and $\C{\downarrow}{n}$ annihilate this state for all orbitals $n$, so only the last term of $\hat{H}_{1\text{-body}}$ in Eq.\ref{app:eq:LLL:1_body} contributes to the energy
\begin{equation}\label{app:eq:LLL:E0_def}
 E_0 \coloneqq \langle \uparrow | \hat{H}|\uparrow\rangle= \sum_{n_1,n_2}(V_{n_1,n_2;n_1,n_2}-V_{n_2,n_1;n_1,n_2})\,.
\end{equation}

The magnon Hamiltonian $\hat{H}_{\text{magnon}}$ is then defined as $\hat{H}-E_0$ projected onto the space of magnon states

\bea
\mathcal{H}_\text{magnon}=\operatorname{span}\left\{ \Cdag{\downarrow}{n} \Hdag{\uparrow}{m}\ket{\uparrow}: n,m\in \lbrace 0,\dots,N_\phi-1\rbrace\right\}
\label{eq:Hilmagnon_lll}
\eea
that is,
\begin{equation}
 \hat{H}_{\text{magnon}}=P(\hat{H}-E_0)P\,,
\end{equation}
where $P$ is the projector onto $\mathcal{H}_\text{magnon}$.

We now evaluate $\hat{H}_{\text{magnon}}$ term by term in Eq.\ref{app:eq:LLL:ph_transformed}. The first two lines contain two annihilation operators of the same species and therefore annihilate every state in $\mathcal{H}_\text{magnon}$, which hosts only a single hole in the $\uparrow$-LLL and a single electron in the $\downarrow$-LLL. The intervalley term (line $3$) restricted to $\mathcal{H}_\text{magnon}$ defines a genuine $2$-body problem of one hole and one electron. Since the $\uparrow$-LLL wave functions are the complex conjugates of the $\downarrow$-LLL wave functions, the hole in the $\uparrow$-LLL couples to the interaction like a particle in the $\downarrow$-LLL, so this electron--hole problem maps onto the $2$-body problem of two (distinguishable) particles in a single LLL, which is diagonalized by the pseudopotential decomposition of Eqs.\ref{app:eq:Haldane_PP} and \ref{app:eq:LLL:cylinder_matrix_elements}.

As shown in Ref.$\,$\cite{PhysRevB.90.115132}, on the torus each pseudopotential $V_\alpha$ has rank $2$: in each momentum sector, the corresponding two-body problem has two finite-energy states, with energies $V_\alpha$ in the thermodynamic limit, while all remaining states have zero energy. Moreover, the finite-energy eigenstates associated with different pseudopotentials are mutually orthogonal. The doubled rank compared to e.g. the disk, sphere or cylinder geometry appears because the finite-energy state of a pseudopotential $V_\alpha$ on e.g. the cylinder and the same state translated by $N_\phi/2$ orbitals lie in the same momentum sector on the torus after momentum backfolding. In finite size, they scatter into each other, leading to a splitting of the degeneracy, which is recovered in the thermodynamic limit \cite{PhysRevB.90.115132}. Consequently, line $3$ of Eq.$\,$\ref{app:eq:LLL:ph_transformed} projected on $\mathcal{H}_\text{magnon}$ is diagonalized by the two finite-energy eigenstates of each pseudopotential $V_\alpha$ with eigenvalue $-V_\alpha$ and the orthogonal complement of these finite-energy manifolds with eigenvalue $0$.

We are thus left with evaluating $\hat{H}_{1\text{-body}}$ in Eq.$\,$\ref{app:eq:LLL:1_body} on the subspace $\mathcal{H}_\text{magnon}$ in Eq.$\,$\ref{eq:Hilmagnon_lll}. From the separable form Eq.$\,$\ref{app:eq:LLL:cylinder_matrix_elements} and the normalization Eq.$\,$\ref{eq:eta_k_norm}, the direct matrix elements satisfy
\begin{equation}\label{app:eq:LLL:direct_norm}
 \sum_{n_2}V_{n_1,n_2;n_1,n_2}=\sum_{\beta}V_\beta\sum_{k}(\eta^\beta_k)^2=\sum_{\beta}V_\beta\,, \quad \forall n_1\,,
\end{equation}
in the thermodynamic limit, whilst by the parity relation Eq.$\,$\ref{app:eq:LLL:parity}, the exchange matrix elements carry an alternating sign,
\begin{equation}\label{app:eq:LLL:exchange_norm}
 \sum_{n_2}V_{n_2,n_1;n_1,n_2}=\sum_{\beta}(-1)^\beta V_\beta\,, \quad \forall n_1\,.
\end{equation}
Consequently, only the odd pseudopotentials act on a pair of identical fermions, consistent with the antisymmetry of their relative wave function, and the ground state energy Eq.$\,$\ref{app:eq:LLL:E0_def} evaluates to $E_0=2N_\phi\sum_{\beta\text{ odd}}V_\beta$.
Since $\hat{H}_{1\text{-body}}$ is diagonal in the orbital basis, it reduces to a constant on $\mathcal{H}_\text{magnon}$: by Eqs.$\,$\ref{app:eq:LLL:direct_norm} and \ref{app:eq:LLL:exchange_norm}, the single hole in the $\uparrow$-LLL contributes $-2\sum_{\beta}\left(1-(-1)^\beta\right)V_\beta=-4\sum_{\beta\text{ odd}}V_\beta$, the single electron in the $\downarrow$-LLL contributes $2\sum_\beta V_\beta$, and the last line of Eq.$\,$\ref{app:eq:LLL:1_body} equals $E_0$, i.e.
\begin{equation}
 P\hat{H}_{1\text{-body}}P=\left(-4\sum_{\beta \text{ odd}}V_\beta +2\sum_{\beta}V_\beta + E_0\right)P\,.
\end{equation}
After subtracting $E_0$ and combining the even and odd sums, the magnon spectrum in each momentum sector consists of the two-fold degenerate discrete levels
\begin{equation}
 -V_\alpha + 2\sum_{\beta \text{ even}} (V_\beta - V_{\beta+1})\,,
\end{equation}
$\alpha\geq 0$, corresponding to the two finite-energy eigenstates of $V_\alpha$, and a continuum formed by all magnon states orthogonal to these finite-energy manifolds, residing at energy $2\sum_{\beta \text{ even}} (V_\beta - V_{\beta+1})$.

The pseudopotentials for the Coulomb interaction satisfy $V_0/2=V_1$ and $V_m<V_n$ for $m>n$. Assuming that gating decreases the ratio of consecutive pseudopotentials $V_{m+1}/V_m$, this inequality is preserved and we have $V_0/2\geq V_1$, as e.g. for the dual-gate screened Coulomb interaction in Eq.$\,$\ref{eq:dual_gate}$\,$\cite{FTI_Yves}. Therefore, for a screened Coulomb interaction, the lowest two magnon modes are given by the $V_0$ finite-energy eigenstates and the magnon spectrum is gapped with a gap
\begin{equation}
 V_0-2V_1+2\sum_{\beta \geq 2 \text{, even}}(V_\beta-V_{\beta+1})>V_0-2V_1\,.
\end{equation}
For a pure Hubbard interaction $V_0=U/4$, we have two magnon modes at $E=U/4$ and a continuum of modes at $E=U/2$, as shown in Fig.$\,$\ref{fig:4}(a).

\section{Flat Band Hubbard models}

\label{App:FlatBandHubbard_details}

In this Appendix, we discuss in more detail the flat band Hubbard models discussed in the main text, in Sec.$\,$\ref{subsec:Flatband_Hubbard}.

Let us start from a general tight-binding Hamiltonian with $S_z$ conservation and time-reversal symmetry $\mathcal{T}$. We will consider the full model
\bea
H &= \sum_{\mbf{k} \al \be \eta} c^\dag_{\mbf{k},\al \eta} h_{\al \be}^\eta(\mbf{k}) c_{\mbf{k},\be \eta} + \sum_{\mbf{R} \al} V_\al n_{\mbf{R},\al, \uparrow} n_{\mbf{R},\al, \downarrow} \,,
\eea
which describes onsite repulsion at each orbital site $\al=1,\cdots,N_{orb}$ ($V_{\alpha}\geq0$). We assume $h(\mbf{k})$ has $N_f$ gapped degenerate flat bands at the bottom of the spectrum. We choose the zero of energy such that these $N_f$ flat bands have
energy zero. In this case, there are two degenerate ground-states
\bea
\ket{\eta} &= \prod_{\mbf{k},n \in \text{flat}} \gamma^\dag_{\mbf{k},n,\eta} \ket{0}
\eea
which are annihilated by the Hamiltonian:
\bea
H \ket{\eta} = 0 \\
\eea
because $n_{\mbf{R},\al, -\eta} \ket{\eta} = 0$. Since $H$ is positive semi-definite, $\ket{\eta}$, with $\eta=\up,\down$ should be ground-states. With mild conditions that guarantee the connectedness of the lattice, these are the unique ground-states \cite{1999JPhA...32.8411M,1993CMaPh.158..341M,2003CMaPh.242..445T,tasaki2020manybody,2016PhRvB..94x5149T,kang2024quantum}.

To compute exact excitations on top of these ground-states, we must project to the flat band. This yields
\bea
\bar{H} &= \frac{1}{N} \sum_{\mbf{q} \al} V_\al \bar{\rho}_{\mbf{q},\al,\uparrow} \bar{\rho}_{-\mbf{q},\al,\downarrow} \,, \\
\bar{\rho}_{\mbf{q},\al,\eta} &= \sum_{\mbf{k} mn} M^{\al,\eta}_{mn}(\mbf{k},\mbf{q}) \gamma^\dag_{\mbf{k}+\mbf{q},m,\eta} \gamma_{\mbf{k},n,\eta} \ .
\eea
where we defined
\bea
 M^{\al,\eta}_{mn}(\mbf{k},\mbf{q}) &= U_{\al m}^{\eta *}(\mbf{k}+\mbf{q})U_{\al n}^{\eta}(\mbf{k}), \\
 M^{\eta}_{mn}(\mbf{k},\mbf{q}) &= \sum_\al M^{\al,\eta}_{mn}(\mbf{k},\mbf{q}) = [U^{\eta \dag}(\mbf{k}+\mbf{q}) U^\eta(\mbf{k})]_{mn} \ ,
 \label{eq:fb_hubbard_models_Hproj}
\eea
and we restrict the band indices $m$ and $n$ to the flat bands - a notation that we will use for all the remaining calculations in this Appendix.

Let us assume the system polarizes into the up sector. We will frequently use the identity
\bea
\frac{1}{N} \bar{\rho}_{\mbf{q},\al,\eta} \ket{\eta} &= \frac{1}{N} \sum_{\mbf{k} mn} M_{mn}^{\al,\eta}(\mbf{k},\mbf{q}) \gamma^\dag_{\mbf{k}+\mbf{q},m,\eta} \gamma_{\mbf{k},n,\eta} \ket{\eta} \\
&= \frac{1}{N} \sum_{\mbf{k} mn} M_{mn}^{\al,\eta}(\mbf{k},\mbf{q}) \delta_{\mbf{q},0} \delta_{mn} \ket{\eta} \\
&= \frac{\delta_{\mbf{q},0}}{N} \sum_{\mbf{k} m} M_{mm}^{\al,\eta}(\mbf{k},\mbf{0}) \ket{\eta} \\
&=\frac{\delta_{\mbf{q},0}}{N} \sum_{\mbf{k}} P^\eta_{\al \al}(\mbf{k}) \ket{\eta} \\
&\equiv \delta_{\mbf{q},0} \epsilon_\al \ket{\eta} \\
\eea
where we defined the groundstate density on the $\al$ orbital:
\bea
\epsilon_\al = \frac{1}{N} \sum_\mbf{k} U_\al(\mbf{k}) U^*_\al(\mbf{k}) = \frac{1}{N} \sum_\mbf{k} P_{\al \al}(\mbf{k}) \in (0,1) \ .
\label{eq:eps_alpha_P_aa}
\eea
Note that $\sum_\al \epsilon_\al = N_f$. Thus if all orbitals are related by symmetry, $\epsilon_\al = \epsilon \equiv N_f/N_{orb}$ and $V_\al = V$ since all interactions are symmetry-related. We call such a lattice a uniform lattice. We shall obtain a series of fully analytical results when the condition $\epsilon_\al V_\al$ is constant, i.e. $\epsilon_\al V_\al = U \epsilon $ for all $\al$.

We list a number of useful identities below:
\bea
[ &\bar{\rho}_{\mbf{q},\al,\eta'}, \gamma^\dag_{\mbf{k},n,\eta}] \\ &=\sum_{\mbf{k}' m'n'} M_{m'n'}^{\al,\eta'}(\mbf{k}',\mbf{q}) [\gamma^\dag_{\mbf{k}'+\mbf{q},m',\eta'} \gamma_{\mbf{k},n',\eta'}, \gamma^\dag_{\mbf{k},n,\eta}] \\
&= \delta_{\eta,\eta'} \sum_{m'} \gamma^\dag_{\mbf{k}+\mbf{q},m',\eta} M_{m'n}^{\al,\eta}(\mbf{k},\mbf{q}) \\
&[ \bar{\rho}_{-\mbf{q},\al,\eta'}, \gamma_{\mbf{k},n,\eta}] \\ &= \sum_{\mbf{k}' m'n'} M_{m'n'}^{\al,\eta'}(\mbf{k}',-\mbf{q}) [\gamma^\dag_{\mbf{k}'-\mbf{q},m',\eta'} \gamma_{\mbf{k}',n',\eta'}, \gamma_{\mbf{k},n,\eta}] \\
&= - \delta_{\eta \eta'} \sum_{n'} M_{nn'}^{\al,\eta'}(\mbf{k}+\mbf{q},-\mbf{q}) \gamma_{\mbf{k}+\mbf{q},n',\eta'} \\
\eea

Finally, let us comment on the relation to the interaction Hamiltonian $H_I$ used in the Main Text, which is manifestly particle-hole symmetric. Note that both Hamiltonians $\bar{H}$ and $H_I$ have zero-energy ferromagnetic ground states at $\nu=1$. We will show that they can only differ by a chemical potential and constant term, in analogy to the usual identity $\frac{1}{2}\sum_\mbf{R} (n_{\mbf{R} \uparrow}+n_{\mbf{R} \downarrow}-1)^2 = n_{\mbf{R} \uparrow} n_{\mbf{R} \downarrow} - \frac{n_{\mbf{R} \uparrow} + n_{\mbf{R} \downarrow}}{2} + \frac{1}{2}$. This is a result of the uniform condition $n_\al V_\al = U$. In the projected case, the analogous identity is
\bea
\bar{H}' &= \frac{1}{2N} \sum_{\mbf{q} \al} V_\al (\bar{\rho}^\dag_{\mbf{q},\al} - \frac{N}{2} \epsilon_\al \delta_{\mbf{q},0})(\bar{\rho}_{\mbf{q},\al} - \frac{N}{2} \epsilon_\al \delta_{\mbf{q},0}) \\
&= \frac{1}{2N} \sum_{\mbf{q} \al} V_\al \bar{\rho}^\dag_{\mbf{q},\al} \bar{\rho}_{\mbf{q},\al} - \frac{1}{2} \sum_\al V_\al \epsilon_\al \bar{\rho}_{\mbf{0},\al} + \text{const} \\
&= \frac{1}{2N} \sum_{\mbf{q} \al,\eta \eta'} V_\al \bar{\rho}^\dag_{\mbf{q},\al,\eta} \bar{\rho}_{\mbf{q},\al,\eta'} - \frac{1}{2} \sum_\al \epsilon_\al V_\al \bar{\rho}_{\mbf{0},\al} + \text{const} \\
&= \frac{1}{2N} \sum_{\mbf{q} \al,\eta \eta'} V_\al \bar{\rho}^\dag_{\mbf{q},\al,\eta} \bar{\rho}_{\mbf{q},\al,\eta'} - \frac{U}{2} \bar{N}+ \text{const} \\
&= \sum_{\mbf{q} \al} V_\al \bar{\rho}^\dag_{\mbf{q},\al,\uparrow} \bar{\rho}_{\mbf{q},\al,\downarrow} - \frac{U}{2} \bar{N} + \text{const} \\
&=\bar{H} - \frac{U}{2} \bar{N} + \text{const} \\
\eea
We emphasize that both $\bar{H}$ and $\bar{H}'$ annihilate the $\nu=1$ ground states.

\subsection{Charge Excitations}
\label{app:hubbardcharge}

We now compute the particle and hole excitations and impose a condition for analytical solvability of the charge gap. We first examine the charge $-1$ excitations. The candidate wave function is $\gamma_{\mbf{k},m,\uparrow} \ket{\uparrow}$ since $\gamma_{\mbf{k},m,\downarrow} \ket{\uparrow} = 0$. In fact, $\bar{H} \gamma_{\mbf{k},m,\uparrow} \ket{\uparrow} = 0$ since the state is fully polarized. Hence the charge gap (which is independent of the chemical potential) is determined entirely by the charge $+1$ excitations spanned by the basis $\gamma^\dag_{\mbf{k},m,\downarrow} \ket{\uparrow}$. Acting with the projected Hamiltonian in Eq.$\,$\ref{eq:fb_hubbard_models_Hproj} yields
\bea
\bar{H} \gamma^\dag_{\mbf{k},m,\downarrow} \ket{\uparrow} &= \frac{1}{N} \sum_{\mbf{q} \al} V_\al \bar{\rho}_{\mbf{q},\al,\downarrow} \bar{\rho}_{-\mbf{q},\al,\uparrow} \gamma^\dag_{\mbf{k},m,\downarrow} \ket{\uparrow} \\
&= \frac{1}{N} \sum_{\mbf{q} \al} V_\al \bar{\rho}_{\mbf{q},\al,\uparrow} [\bar{\rho}_{-\mbf{q},\al,\downarrow}, \gamma^\dag_{\mbf{k},m,\downarrow}] \ket{\uparrow} \\
&= \frac{1}{N} \sum_{\mbf{q} \al} V_\al \Big( [\bar{\rho}_{\mbf{q},\al,\uparrow}, [\bar{\rho}_{-\mbf{q},\al,\downarrow}, \gamma^\dag_{\mbf{k},m,\downarrow}] ] \\
&\quad + [\bar{\rho}_{-\mbf{q},\al,\downarrow}, \gamma^\dag_{\mbf{k},m,\downarrow}] \bar{\rho}_{\mbf{q},\al,\uparrow} \Big) \ket{\uparrow} \\
&= \frac{1}{N} \sum_{\mbf{q} \al} V_\al \Big( [\bar{\rho}_{\mbf{q},\al,\uparrow}, [\bar{\rho}_{-\mbf{q},\al,\downarrow}, \gamma^\dag_{\mbf{k},m,\downarrow}] ] \ket{\uparrow} \\
&\quad + [\bar{\rho}_{\mbf{0},\al,\downarrow}, \gamma^\dag_{\mbf{k},m,\downarrow}] \epsilon_\al \Big) \ket{\uparrow} \ , \\
\eea
which together with $ [\bar{\rho}_{\mbf{q},\al,\uparrow}, [\bar{\rho}_{-\mbf{q},\al,\downarrow}, \gamma^\dag_{\mbf{k},m,\downarrow}] ] = 0$, leads to
\bea
\bar{H} \gamma^\dag_{\mbf{k},m,\downarrow} \ket{\uparrow} &= \sum_{m'} \gamma^\dag_{\mbf{k},m',\downarrow} \ket{\uparrow} R^\downarrow_{m' m}(\mbf{k}), \\
R^\downarrow_{m' m}(\mbf{k}) &= \sum_\al M_{m' m}^{\al \downarrow}(\mbf{k},0) V_\al \epsilon_\al \ . \\
\eea
The spectrum of $R^\downarrow (\mbf{k})$ is gauge-invariant and can be easily calculated given the wave functions. However, if we enforce the condition $V_\al \epsilon_\al = \epsilon U$, then we have
\bea
R^\downarrow_{m' m}(\mbf{k}) &= \sum_\al M_{m' m}^{\al \downarrow}(\mbf{k},0) V_\al \epsilon_\al \\ &= \epsilon U M_{m' m}^{\downarrow}(\mbf{k},0) = \epsilon U \delta_{m'm}
\eea
and the spectrum is perfectly flat. Note that this result is independent of the wave functions of the flat band. This shows that in the Hubbard limit and when $\epsilon_\al V_\al$ is constant, there is no direct relationship between the charge excitation dispersion and the topology/quantum geometry of the band.

\subsection{Magnons}
\label{app:magnonandSU2}

Next we consider the Ising magnons. Acting the projected Hamiltonian in the magnon basis, we get
\bea
& \bar{H} \gamma^\dag_{\mbf{k}+\mbf{p},m,\downarrow} \gamma_{\mbf{k},n,\uparrow} \ket{\uparrow} \\
&= [\bar{H},\gamma^\dag_{\mbf{k}+\mbf{p},m,\downarrow} \gamma_{\mbf{k},n,\uparrow}] \ket{\uparrow} \\
&= [\bar{H}, \gamma^\dag_{\mbf{k}+\mbf{p},m,\downarrow}] \gamma_{\mbf{k},n,\uparrow} \ket{\uparrow} + \gamma^\dag_{\mbf{k}+\mbf{p},m,\downarrow} [\bar{H}, \gamma_{\mbf{k},n,\uparrow}] \ket{\uparrow} \\
&= \frac{1}{N} \sum_{\mbf{q} \al} V_\al [\bar{\rho}_{\mbf{q},\al,\downarrow} \bar{\rho}_{-\mbf{q},\al,\uparrow}, \gamma^\dag_{\mbf{k}+\mbf{p},m,\downarrow}] \gamma_{\mbf{k},n,\uparrow} \ket{\uparrow} \\
&= \frac{1}{N} \sum_{\mbf{q} \al} V_\al [\bar{\rho}_{\mbf{q},\al,\downarrow}, \gamma^\dag_{\mbf{k}+\mbf{p},m,\downarrow}] \bar{\rho}_{-\mbf{q},\al,\uparrow} \gamma_{\mbf{k},n,\uparrow} \ket{\uparrow} \\
&= \frac{1}{N} \sum_{\mbf{q} \al} V_\al [\bar{\rho}_{\mbf{q},\al,\downarrow}, \gamma^\dag_{\mbf{k}+\mbf{p},m,\downarrow}] [\bar{\rho}_{-\mbf{q},\al,\uparrow}, \gamma_{\mbf{k},n,\uparrow}] \ket{\uparrow} \\
&\quad +\frac{1}{N} \sum_{\mbf{q} \al} V_\al [\bar{\rho}_{\mbf{q},\al,\downarrow}, \gamma^\dag_{\mbf{k}+\mbf{p},m,\downarrow}] \gamma_{\mbf{k},n,\uparrow} \bar{\rho}_{-\mbf{q},\al,\uparrow} \ket{\uparrow} \\
&= \frac{1}{N} \sum_{\mbf{q} \al} V_\al [\bar{\rho}_{\mbf{q},\al,\downarrow}, \gamma^\dag_{\mbf{k}+\mbf{p},m,\downarrow}] [\bar{\rho}_{-\mbf{q},\al,\uparrow}, \gamma_{\mbf{k},n,\uparrow}] \ket{\uparrow} \\
&\quad + \sum_{\al} V_\al [\bar{\rho}_{\mbf{0},\al,\downarrow}, \gamma^\dag_{\mbf{k}+\mbf{p},m,\downarrow}] \gamma_{\mbf{k},n,\uparrow}\epsilon_\al \ket{\uparrow} \\
&= \sum_{\mbf{k}'m'n'} \gamma^\dag_{\mbf{k}'+\mbf{p},m',\downarrow} \gamma_{\mbf{k},n',\uparrow} \ket{\uparrow} \\
&\qquad \times (-\mathcal{S}_{\mbf{k}'m'n'}^{\mbf{k}mn} + \delta_{\mbf{k},\mbf{k}'} \delta_{n'n} R^\downarrow_{m' m}(\mbf{k}+\mbf{p}))
\eea
where the scattering matrix is
\bea
\mathcal{S}_{\mbf{k}'m'n'}^{\mbf{k}mn} &= \sum_\al\frac{ V_\al}{N} M^{\al \downarrow}_{m' m}(\mbf{k}+\mbf{p},\mbf{k}'-\mbf{k}) M_{nn'}^{\al \uparrow}(\mbf{k}',\mbf{k}-\mbf{k}') \ .
\eea
The whole excitation matrix can be diagonalized numerically, but it can also be solved analytically when the charge 1 bands are flat and the second term above is proportional to the identity. This is because the scattering matrix is exactly solvable, as we now show. We first observe that
\bea
\mathcal{S}_{\mbf{k}'m'n'}^{\mbf{k}mn} &= \frac{1}{N} \sum_\al V_\al M^{\al \downarrow}_{m' m}(\mbf{k}+\mbf{p},\mbf{k}'-\mbf{k}) M_{nn'}^{\al \uparrow}(\mbf{k}',\mbf{k}-\mbf{k}') \\
 =\sum_\al& \frac{V_\al}{N} U^{\downarrow *}_{\al m'}(\mbf{k}'+\mbf{p}) U^\downarrow_{\al m}(\mbf{k}+\mbf{p}) U^{\uparrow*}_{\al n}(\mbf{k}) U^{\uparrow}_{\al n'}(\mbf{k}') \\
&= \frac{1}{N} \sum_\al \mathcal{U}_{\mbf{k}' m'n', \al}(\mbf{p}) \mathcal{U}^*_{\mbf{k} mn,\al}(\mbf{p}),\\
\eea
where we defined $\mathcal{U}_{\mbf{k} mn,\al}(\mbf{p}) = \sqrt{V_\al} U^{\downarrow *}_{\al m}(\mbf{k}+\mbf{p}) U^{\uparrow}_{\al n}(\mbf{k})$.
The spectrum of an outer product $\mathcal{U}\mathcal{U}^\dag$ has the same nonzero eigenvalues as $\mathcal{U}^\dag\mathcal{U}$, with all the rest zero. This is because $\mathcal{U} \mathcal{U}^\dag = 0$ has $N_f^2 L^2 - N_{orb}$ solutions, where $L^2$ is the number of unit cells. The nonzero eigenvalues come from the eigenvectors $\mathcal{U}u$ where
\bea
\mathcal{S} \mathcal{U}u = \frac{1}{N} \mathcal{U}\mathcal{U}^\dag\mathcal{U}u = \mathcal{U} h u = E \mathcal{U} u
\eea
and we defined the binding energy Hamiltonian $\mathcal{W} = \frac{1}{N} \mathcal{U}^\dag \mathcal{U}$. Explicitly, we have
\bea
\, &\mathcal{W}_{\al \be}(\mbf{p}) = \frac{1}{N} \sum_{\mbf{k} mn} \mathcal{U}^*_{\mbf{k} mn,\al}(\mbf{p}) \mathcal{U}_{\mbf{k} mn,\be}(\mbf{p}) \\
&= \sqrt{V_\al V_\be} \frac{1}{N} \sum_{\mbf{k} mn} U^\downarrow_{\al m}(\mbf{k}+\mbf{p}) U^{\uparrow *}_{\al n}(\mbf{k}) U^{\downarrow *}_{\be m}(\mbf{k}+\mbf{p}) U^{\uparrow}_{\be n}(\mbf{k}) \\
&= \sqrt{V_\al V_\be} \frac{1}{N} \sum_{\mbf{k}} P^\downarrow_{\al \be}(\mbf{k}+\mbf{p}) P^{\uparrow}_{\be \al}(\mbf{k}) \ . \\
\eea
We denote the energies/eigenvectors of $\mathcal{W}(\mbf{p})$ by $\mathcal{W}(\mbf{p}) u_\mu(\mbf{p}) = \epsilon_\mu(\mbf{p}) u_\mu(\mbf{p})$.
In the SU(2) case when $P^\up(\mbf{k}) = P^\down(\mbf{k})$, we can show explicitly that there is a Goldstone mode at $\mbf{p}=0$. In this case, we can use the exact eigenvector $u_\be(\mbf{0}) = \sqrt{\epsilon_\be}$ and Eq.$\,$\ref{eq:eps_alpha_P_aa} to find
\bea
\sum_\be \mathcal{W}_{\al \be}(\mbf{0})u_\be(\mbf{0})&= \frac{1}{N} \sum_{\be \mbf{k}} \sqrt{V_\al V_\be \epsilon_\be} P_{\al \be}(\mbf{k}) P_{\be \al}(\mbf{k}) \\
&= \frac{1}{N} \sum_{\be \mbf{k}} \sqrt{V_\al \epsilon U} P_{\al \be}(\mbf{k}) P_{\be \al}(\mbf{k}) \\
&= \frac{1}{N} \sum_{\mbf{k}} \sqrt{V_\al \epsilon U} [P(\mbf{k}) P(\mbf{k})]_{\al \al} \\
&= \sqrt{V_\al \epsilon U} \epsilon_\al \\
&= \sqrt{V_\al \epsilon_\al \epsilon U} \sqrt{\epsilon_\al} \\
&= \epsilon U \sqrt{\epsilon_\al} \\
&= \epsilon U u_\al
\eea
and thus the binding energy is $\epsilon U$. Since the particle-hole continuum is at $\epsilon U$, this is a many-body zero mode.

\subsection{Eigenstates}
\label{sec:magnoneigenstates}

We now obtain the many-body magnon eigenstates:

\bea
\, &\ket{S_{\mbf{p},\mu}} = \sum_{\mbf{k} mn \al}\gamma^\dag_{\mbf{k}+\mbf{p},m,\downarrow} \gamma_{\mbf{k},n,\uparrow} \ket{\uparrow} \mathcal{U}_{\mbf{k} mn,\al}(\mbf{p}) u^\mu_\al(\mbf{p}) \\
&= \sum_{\mbf{k} mn \al}\gamma^\dag_{\mbf{k}+\mbf{p},m,\downarrow} \gamma_{\mbf{k},n,\uparrow} \ket{\uparrow}\sqrt{V_\al} U^{\downarrow *}_{\al m}(\mbf{k}+\mbf{p}) U^{\uparrow}_{\al n}(\mbf{k}) u^\mu_\al(\mbf{p}) \\
&= \sum_{\mbf{k} \al} \sqrt{V_\al} u^\mu_\al(\mbf{p}) \bar{c}^\dag_{\mbf{k}+\mbf{p},\al,\downarrow} \bar{c}_{\mbf{k},\al,\uparrow} \ket{\uparrow} \\
\eea
so we see that $\sqrt{V_\al} u_\al^\mu(\mbf{p})$ is the magnon wave function, where $\mu=1,\cdots,N_{orb}$. To normalize this state, we compute
\bea
\braket{S_{\mbf{p},\mu}}{S_{\mbf{p},\nu}} &= \sum_{\mbf{k} mn \al \be}
U^{\downarrow}_{\be m}(\mbf{k}+\mbf{p}) U^{\uparrow *}_{\be n}(\mbf{k}) u^{\mu *}_\be(\mbf{p}) \\
&\quad \times
\sqrt{V_\be V_\al}
U^{\downarrow *}_{\al m}(\mbf{k}+\mbf{p}) U^{\uparrow}_{\al n}(\mbf{k}) u^\nu_\al(\mbf{p}) \\
&= \sum_{\mbf{k} \al \be} \sqrt{V_\be V_\al} P^{\downarrow}_{\be \al}(\mbf{k}+\mbf{p}) P^{\uparrow}_{ \al \be}(\mbf{k}) u^{\mu *}_\be(\mbf{p}) u^\nu_\al(\mbf{p}) \\
&= N \sum_{\al \be} u^{\mu *}_\be(\mbf{p}) \mathcal{W}_{\be \al}(\mbf{p}) u^\nu_\al(\mbf{p}) \\
&= N \delta_{\mu,\nu} \epsilon_\mu(\mbf{p}) \ .
\eea
We will now study the effective magnon Hamiltonian in more detail.

\subsection{Symmetries}

In this section, we assume all orbitals form an orbit of the space group so that $V_\al =U$. We call such lattices uniform
lattices, as before$\,$\cite{2022arXiv220900007H}. We follow the notation in Ref.$\,$\cite{2022PhRvL.128h7002H}. Then we obtain
\bea
\mathcal{W}_{\al \be}(\mbf{p}) &= \frac{U}{N} \sum_{\mbf{k}} P^\downarrow_{\al \be}(\mbf{k}+\mbf{p}) P^{\uparrow}_{\be \al}(\mbf{k}) \\
\mathcal{W}_{\al \be}(g \mbf{p}) &= \frac{U}{N} \sum_{\mbf{k}} P^\downarrow_{\al \be}(\mbf{k}+g\mbf{p}) P^{\uparrow}_{\be \al}(\mbf{k}) \\
&= \frac{U}{N} \sum_{\mbf{k}} P^\downarrow_{\al \be}(g\mbf{k}+g\mbf{p}) P^{\uparrow}_{\be \al}(g\mbf{k}) \\
&= \frac{U}{N} \sum_{\mbf{k},\al' \be',\al'' \be''} D^{\downarrow}_{\al \al'}[g] P^\downarrow_{\al' \be'}(\mbf{k}+\mbf{p}) D^{\downarrow \dag}_{\be' \be}[g] \\
&\qquad \times D^\uparrow_{\be \be''}[g] P^{\uparrow}_{\be'' \al''}(\mbf{k}) D^{\uparrow \dag}_{\al'' \al}[g] \ . \\
\eea
where $g$ denotes a non-magnetic space group symmetry in a fixed $S_z$ sector (ie a spin-less representation) and $D^{\eta}[g]$ the corresponding representation matrices. We now notice that
\bea
D^{\downarrow \dag}_{\be' \be}[g] D^\uparrow_{\be \be''}[g] &= D^{\downarrow}_{\be \be' }[g]^* D^\uparrow_{\be \be''}[g] \\
&= D_{\be \be' }[g] D_{\be \be''}[g] \equiv \Gamma^\dag_{\be' \be}[g] \delta_{\be \be''} \ .
\eea
The fact that the matrices $\Gamma[g]$ defined above form a unitary representation of the (spin-less) space group was shown in uniform lattices in Ref. \cite{2022arXiv220900007H}. Briefly, this is because the representation matrices on uniform lattices take the form of complex permutation matrices. We get
\bea
\mathcal{W}_{\al \be}(g \mbf{p}) &= \frac{U}{N} \sum_{\mbf{k},\al' \be'} \Gamma_{\al \al'}[g] P^\downarrow_{\al' \be'}(\mbf{k}+\mbf{p}) P^{\uparrow}_{\be' \al'}(\mbf{k}) \Gamma^\dag_{\be' \be }[g] \\
&= \sum_{\al' \be'} \Gamma_{\al \al'}[g] \mathcal{W}_{\al' \be'}(\mbf{p}) \Gamma^\dag_{\be' \be }[g] \\
\eea
showing that the magnons are also symmetric under the non-magnetic (e.g. $S_z$-preserving) space group symmetries.

We now focus on the particular case of two bands with a $C_2$ symmetry represented by $\sigma_x$, relevant to the Main Text. We will show that in this case there is an additional magnon symmetry
\bea
\label{eq:TRSmagnon}
\mathcal{W}_{\al \be}(\mbf{p}) = \mathcal{W}^*_{\al \be}(-\mbf{p})
\eea
which is a bosonic spinless time-reversal symmetry of the magnon. We emphasize that this symmetry does not require a spin-less time-reversal for the electron bands.

To show Eq. \ref{eq:TRSmagnon}, we will use the identity $\sigma_3 P^*(-\mbf{k}) \sigma_3 = 1 - P(\mbf{k})$ which follows from $d_x(-\mbf{k}) = d_x(\mbf{k})$ and $d_i(-\mbf{k}) = - d_i(\mbf{k})$ for $i= y,z$ as enforced by the $C_2 = \sigma_x$ symmetry of the electron Hamiltonian $h(\mbf{k}) = \mbf{d}(\mbf{k}) \cdot \sigma$.
In indices, this is
\bea
P(\mbf{k}) = \sigma_3(\mathbb{1} - P^*(-\mbf{k})) \sigma_3 \\
P_{\al \be}(\mbf{k}) = (-1)^\al (\delta_{\al \be} - P_{\be \al}(-\mbf{k})) (-1)^\be \ . \\
\eea
We then have
\bea
\mathcal{W}_{\al \be}(\mbf{p}) &= \frac{U}{N} \sum_{\mbf{k}} P_{\be \al}(-\mbf{k}-\mbf{p}) P_{\be \al}(\mbf{k}) \\
&= \frac{U}{N} \sum_{\mbf{k}} (\delta_{\al \be} - P_{\al \be }(\mbf{k}+\mbf{p}))(\delta_{\al \be} - P_{\al \be}(-\mbf{k})) \\
&= \frac{U}{N} \sum_{\mbf{k}} \delta_{\al \be}(1 - P_{\al \al }(\mbf{k}+\mbf{p}) - P_{\al \al}(-\mbf{k})) \\
&\qquad + P_{\al \be}(\mbf{k}+\mbf{p}) P_{\al \be}(-\mbf{k}) \\
&= \frac{U}{N} \sum_{\mbf{k}} \delta_{\al \be}(1 -2\epsilon) + P_{\al \be}(\mbf{k}+\mbf{p}) P_{\al \be}(-\mbf{k}) \\
&= \frac{U}{N} \sum_{\mbf{k}} P_{\al \be}(\mbf{k}+\mbf{p}) P_{\al \be}(-\mbf{k}) \\
&= \frac{U}{N} \sum_{\mbf{k}} P_{\al \be}(-\mbf{k}+\mbf{p}) P_{\al \be}(\mbf{k}) \\
&= \mathcal{W}_{\be \al}(-\mbf{p}) \ .
\eea
With hermiticity, this completes our proof of Eq. \ref{eq:TRSmagnon}. This effective time-reversal symmetry, along with the $C_2$ and $C_3$ symmetries, explains the exact degeneracy of the magnon bands at the $K$ point in the Haldane model example discussed in Secs.$\,$\ref{subsec:Flatband_Hubbard_models} and \ref{sec:boundApp} of the Main Text.
In realistic models with more than 2 bands, this symmetry will only be approximate and the splitting will open a gap, as is the case in the continuum model of MoTe$_2$, see Fig.$\,$\ref{fig:App_MoTe2_SMA_energies}. If inversion is preserved, then the magnon bands will acquire nonzero Chern number $\pm 1$ since this model corresponds to a decomposable elementary band representation \cite{2017Natur.547..298B,PhysRevE.96.023310}. The topology of the magnon \cite{JianXinTopoMagnonsMoTe2,FengchengTopoMagnonsMoTe2} and more general exciton \cite{PhysRevLett.133.176601,jtgq-vc7n,shrm-4swg,2026arXiv260421643H,2026arXiv260612525D}bands has been previously explored and is not our focus here.

\section{Additional details on ideal-band BHZ model}

\label{sec:Additional_details_BHZ}

In this section we provide some additional calculations and details on the ideal-band BHZ model explored in the main text in Sec.$\,$\ref{subsec:Ideal_band_BHZ_model}.

\subsection{Berry curvature}

\label{subsec:BHZ_berry_curvature}

The Berry curvature for the flat band eigenfunction defined in Eq.$\,$\ref{eq:psi_n_eta_BHZ}
is given by

\begin{equation}
\Omega_{n,\eta}(k,\theta)=-\eta\frac{2n^{2}(kV)^{2n}}{k^{2}(k^{2n}+V^{2n})^{2}}\,.\label{eq:berry_curvature_idealBandModel}
\end{equation}

We note that because the flat band eigenfunction is ideal, the absolute
value of the Berry curvature is exactly equal to the trace of the
quantum geometry, $G({\bf k})$.
For $n>1$, the Berry curvature is maximized at
$k(V)=\big[(n-1)/(n+1)\big]^{1/(2n)}V$, while for $n=1$ the maximum
occurs at $k=0$. The corresponding maximum is
\begin{equation}
\max|\Omega_n({\bf k})|
=
\begin{cases}
2/V^2, & n=1,\\[4pt]
\displaystyle
\frac{n^2-1}{2V^2}
\left(\frac{n+1}{n-1}\right)^{1/n}, & n>1.
\end{cases}
\end{equation}

\subsection{Derivation of charge-1 gap}

\label{subsec:BHZ_charge_gap}

We will now derive Eq.$\,$\ref{eq:Rmat_BHZ_general} in detail. We start by rewriting the flat band Hamiltonian in Eq.$\,$\ref{eq:H_nI_BHZ} here for convenience,

\begin{equation}
H_{I}=\frac{1}{2N}\sum_{{\bf q}}\overline{\delta\rho}_{{\bf q}}\overline{\delta\rho}_{-{\bf q}}\,,\label{eq:H_nI_BHZ_SM}
\end{equation}
where $N=L^2$ and we will discretize our momentum mesh in units of $2\pi/L$ on a $L\times L$ system. The choice of interaction strength in Eq.$\,$\ref{eq:H_nI_BHZ_SM} is the equivalent of taking $V({\bf q})=1$ for all ${\bf q}$ (i.e. Hubbard interaction) in Eq.$\,$\ref{eq:H_V_general} used for the continuum models of MoTe$_2$. We also rewrite the projected density operators, that are given by
\bea
\overline{\delta\rho}_{{\bf q}}&=\sum_{{\bf k},\eta}M_n^{\eta}({\bf k},{\bf q})\big(\gamma_{{\bf k}+{\bf q},\eta}^{\dagger}\gamma_{{\bf k},\eta}-\frac{1}{2}\delta_{{\bf q},{\bf 0}}\big)\,\,\,,\\
& M_n^{\eta}({\bf k},{\bf q})=\psi_{n,\eta}({\bf k}+{\bf q})^{\dagger}\psi_{n,\eta}(\bf k) \ .
\eea
where
\begin{equation}
 \psi_{n,\eta}(k_v,\theta)=
 \frac{1}{\sqrt{1+k_v^{2n}}} \left(\begin{array}{c}
 (\eta e^{i\eta\theta}k_v)^{n}\\
1
\end{array}\right) \,.
\end{equation}
and where we defined $k_v=k/V$.
The charge-1
gap can be rewritten in terms of the flat band projectors $P_{n}^{\eta}({\bf k})=\psi_{n,\eta}(k_v,\theta)\psi_{n,\eta}^{\dagger}(k_v,\theta)$ as

\begin{equation}
R^{\eta}_n({\bf k})=\frac{1}{2 N}\sum_{{\bf q}}\Tr[P_{n}^{\eta}({\bf k}+{\bf q})P_{n}^{\eta}({\bf k})].
\end{equation}

We can write $P_{n}^{\eta}({\bf k})$ in terms of $P_{1}^{\eta}({\bf k})$
by noting that

\begin{equation}
\begin{array}{c}
\psi_{1,\eta}\big(k_v^{n},n\theta\big)=\frac{1}{\sqrt{1+k_v^{2n}}}\left(\begin{array}{c}
\eta e^{i\eta n\theta} k_v^{n}\\
1
\end{array}\right)\\
=\left(\begin{array}{cc}
\eta^{1-n} & 0\\
0 & 1
\end{array}\right)\psi_{n,\eta}\big(k_v,\theta\big)\\
= (-\sigma_{z})^{(\frac{1-\eta}{2})(n+1)} \psi_{n,\eta}\big(k_v,\theta\big)\,,
\end{array}\label{eq:eigenstate_relation_n_1}
\end{equation}
which implies that

\begin{equation}
\begin{array}{c}
P_{n}^{\eta}\big(k_v,\theta\big)=(-\sigma_{z})^{(\frac{1-\eta}{2})(n+1)}P_{1}^{\eta}\big(k_v^{n},n\theta\big)(-\sigma_{z})^{(\frac{1-\eta}{2})(n+1)}\,.\end{array}\label{eq:Pn_P1_nmodel}
\end{equation}

Since the charge gap is the product of two equal-spin projectors,
using $(-\sigma_{z})^{2(\frac{1-\eta}{2})(n+1)}=1$ and the cutoff prescription $|{\bf k}+{\bf q}|\equiv|{\bf k}'|,|{\bf k}|<\Lambda$,

\begin{equation}
\begin{array}{c}
R_{n}^{\eta}({\bf k})=\frac{1}{8\pi^2}\int_{|{\bf k}'|<\Lambda}d^{2}{\bf k}'\,\,\Tr[P_{n}^{\eta}({\bf k}')P_{n}^{\eta}({\bf k})]\\
=\frac{V^{\frac{2n-2}{n}}}{8\pi^2 n}\Tr\Big[P_{1}^{\eta}\big(k_v^n,n\theta\big)\\
\int_{0}^{V(\Lambda/V)^{n}}\tilde{k}^{\frac{2-n}{n}}d\tilde{k}\int_{0}^{2\pi}d\tilde{\theta}\,\,P_{1}^{\eta}\big(\tilde{k}_v,\tilde{\theta}\big)\Big]\,,
\end{array}\label{eq:R_k_nmodel_first}
\end{equation}

\noindent where we used the change of variables $\tilde{k}_v=\tilde{k}/V=(k'/V)^{n}$ and $\tilde{\theta}=n\theta'$, and also used

\begin{equation}
 \sum_{\bf{q}} \rightarrow \frac{N}{(2\pi)^2} \int d^2\bf{q}\,.
\end{equation}

Using the explicit expression for the projectors,

\begin{equation}
P_{1}^{\eta}\big(k_v,\theta\big)=\frac{1}{1+k_v^{2}}\left(\begin{array}{cc}
k_v^2 & \eta e^{i\eta\theta}k_v\\
\eta e^{-i\eta\theta}k_v & 1
\end{array}\right)\,,
\end{equation}
we arrive at Eq.$\,$\ref{eq:Rmat_BHZ_general}, that we reproduce here for convenience,
\bea
\label{eq:Rmat_BHZ_general_sup}
&R_n^\eta({\bf k})\equiv R_n({\bf k})\equiv R_n(k_{v};V,x) \\ &=
 \frac{(-1)^{-1/n}V^2}{8 \pi n(1+k_{v}^{2})} \Big(\mathcal{B}_{-x^{-n}}(\frac{1}{n},0)-k_{v}^{2}\mathcal{B}_{-x^{-n}}(\frac{n+1}{n},0)\Big).
\eea
where the dependence on $\eta$ and $\theta$ is lost.

The particular cases
of $n=1,2$ are given by:

\bea
R_1(k_{v};V,x)= V^2\frac{k_{v}^{2}/x+(1-k_{v}^{2}) \log(1+\frac{1}{x})}{8\pi(1+k_{v}^{2})} \,,\\
\eea

\bea
R_2(k_{v};V,x)=V^2\frac{k_{v}^{2}/x+(1-k_{v}^{2})\textrm{arccot}(x)}{8\pi(1+k_{v}^{2})}\,.
 \,.
\eea

We will now obtain Eqs.$\,$\ref{eq:BHZ_R_smallx} by expanding for
small $x$. Using $\mathcal{B}_{-1/x}(a,0)=(-1)^{a}\pi\csc(\pi a)+x^{-a}\Big(\frac{(-1)^{a}x}{a-1}+\mathcal{O}(x^{2})\Big)$,
we arrive at

\begin{equation}
R_1(k_{v};V,x)\approx\frac{k_{v}^{2}\Lambda^2+(k_{v}^{2}-1)V^2\log(x)+\mathcal{O}(x^{2})}{8\pi(1+k_{v}^{2})},\label{eq:R_n1_smallx}
\end{equation}

\begin{equation}
R_{n>1}(k_{v};V,x)\approx\frac{\pi[1-k_{v}^{2}] V^2 \csc(\pi/n)+nk_{v}^{2}\Lambda^2+\mathcal{O}(x^{2})}{8\pi n(1+k_{v}^{2})},\label{eq:R_anu_n_smallx}
\end{equation}

Eqs.$\,$\ref{eq:R_n1_smallx}
and \ref{eq:R_anu_n_smallx} also imply that for finite but small
$x$, the minimum charge gap occurs at $k=0$, and is given by Eq.$\,$\ref{eq:BHZ_R_smallx}.

\begin{figure*}
\begin{centering}
\includegraphics[width=1\textwidth]{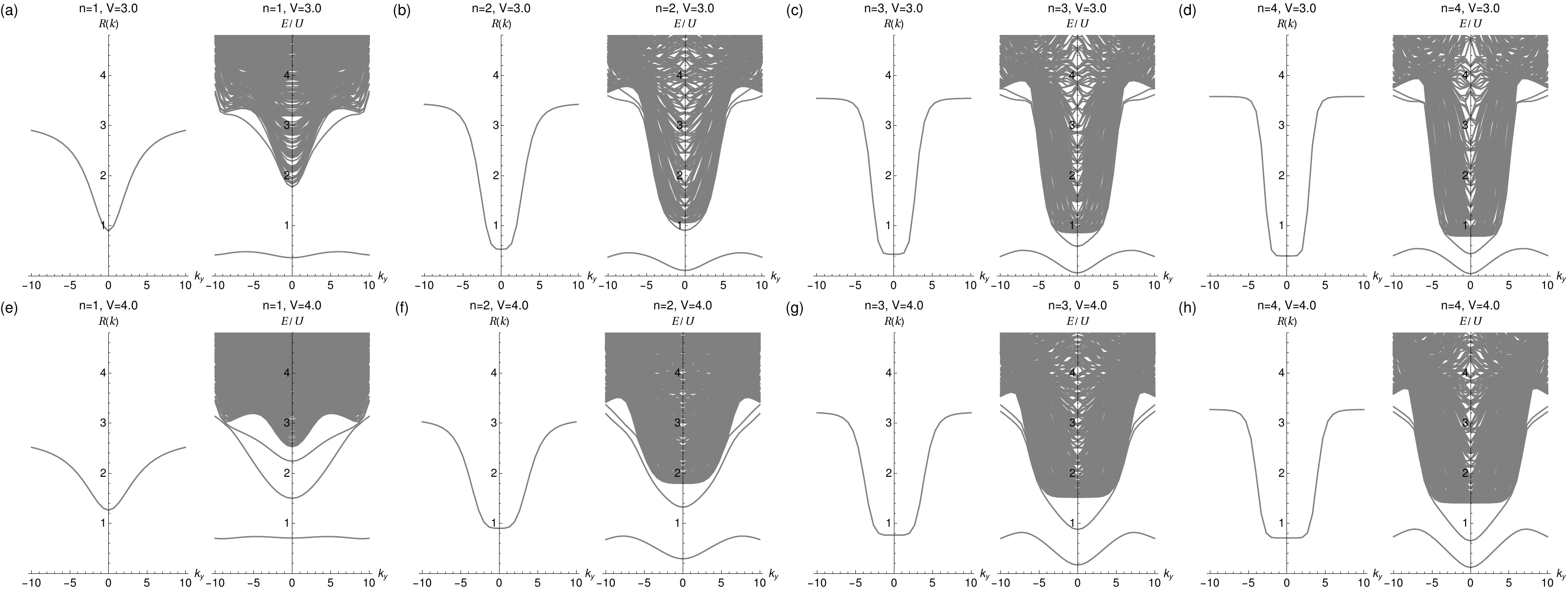}
\par\end{centering}
\caption{Numerical results for the magnon spectrum of the ideal band BHZ model,
for $\Lambda=10,L=15$ and different $n$ and $V$ indicated above
each figure. The charge gap ($R(k)$ in the figures) is shown alongside
the magnon spectrum ($E_{\textrm{magnon}}$).\label{fig:n_model_Smatrix}}
\end{figure*}

To illustrate the momentum dependence of the charged and magnon
excitations, Fig.$\,$\ref{fig:n_model_Smatrix} shows the charge-excitation
dispersion $R_n(k_v;V,x)$ together with the exact magnon spectrum obtained
numerically by diagonalizing the S-matrix in Eq.$\,$\ref{eq:Smatrix_BHZ}, for
 $n$ between 1 and 4 and $V=3,4$.

\subsection{Derivation of exact expressions for the bound magnon energies}

\label{subsec:BHZ_exact_magnon_energies}

Our starting point is Eq.$\,$\ref{eq:Tp_kernel}, which we rewrite here
for convenience:
\bea
T_{n,\lambda}({\bf p})
&=
\frac{1}{L^2}
\left[
\mathcal{U}_n^{\dagger}({\bf p})
\frac{1}{\mathcal{R}_n({\bf p})-\lambda}
\mathcal{U}_n({\bf p})
\right]
\\
&=
\int \frac{d^2 k}{(2\pi)^2}\,
\frac{
P_n^\down({\bf k}+{\bf p})
\otimes
P_n^{\up,T}({\bf k})
}{
R_n^\down({\bf k}+{\bf p})
+
R_n^\up({\bf k})
-\lambda
}.
\label{eq:Tp_kernel_sup}
\eea

We first introduce the coefficients
\bea
a_n(V,x)
&=
\frac{(-1)^{-1/n}V^2}{4\pi n}
\mathcal{B}_{-x^{-n}}
\left(\frac{1}{n},0\right),
\\
b_n(V,x)
&=
\frac{(-1)^{-1/n}V^2}{4\pi n}
\mathcal{B}_{-x^{-n}}
\left(\frac{n+1}{n},0\right),
\eea
in terms of which Eq.$\,$\ref{eq:Rmat_BHZ_general_sup} can be written
as
\begin{equation}
R_n(k_v;V,x)
=
\frac{a_n(V,x)-b_n(V,x)k_v^2}
{2(1+k_v^2)}.
\label{eq:Rmat_BHZ_general_rewrite}
\end{equation}
In what follows, the dependence of $a_n$ and $b_n$ on $V$ and $x$
will be left implicit for clarity.

At ${\bf p}=0$, Eq.$\,$\ref{eq:Tp_kernel_sup} becomes
\begin{equation}
T_{n,\lambda}({\bf 0})
=
\int\frac{d^2{\bf k}}{(2\pi)^2}
\frac{
P_n^\down({\bf k})
\otimes
P_n^{\up,T}({\bf k})
}{
2R_n({\bf k})-\lambda
}.
\label{eq:Tnlambda_p0_initial}
\end{equation}
We now use Eq.$\,$\ref{eq:Pn_P1_nmodel} and perform the same change of
variables employed in the charge-gap calculation in
Eq.$\,$\ref{eq:R_k_nmodel_first}, $\widetilde{k}/V=(k/V)^n$ and $\widetilde{\theta}=n\theta$.

The radial measure transforms according to
\begin{equation}
k\,dk
=
\frac{V^{\frac{2(n-1)}{n}}}{n}
\widetilde{k}^{\frac{2-n}{n}}d\widetilde{k}.
\end{equation}

Equation~\ref{eq:Tnlambda_p0_initial} therefore becomes
\begin{equation}
\begin{aligned}
T_{n,\lambda}({\bf 0})
={}&
\frac{V^{\frac{2(n-1)}{n}}}{n}
\int_0^{V(\Lambda/V)^n}
\frac{d\widetilde{k}\,
\widetilde{k}^{\frac{2-n}{n}}}{(2\pi)^2}
\int_0^{2\pi}d\widetilde{\theta}
\\
&\times
\frac{
\left[
\sigma_z^{n+1}
P_1^\down(\widetilde{k}/V,\widetilde{\theta})
\sigma_z^{n+1}
\right]
\otimes
P_1^{\up,T}(\widetilde{k}/V,\widetilde{\theta})
}{
2R_n(\widetilde{k}/V;V,x)-\lambda
}.
\end{aligned}
\label{eq:Tnlambda_transformed}
\end{equation}

Relabeling $\widetilde{k}\rightarrow k$, recalling the definition
$k_v=k/V$, and defining $y=k_v^2$, the two projectors entering
Eq.$\,$\ref{eq:Tnlambda_transformed} are
\begin{equation}
\begin{aligned}
&\sigma_z^{n+1}
P_1^\down(\sqrt{y},\widetilde{\theta})
\sigma_z^{n+1}
\\
&\qquad=
\frac{1}{1+y}
\begin{pmatrix}
y &
(-1)^n\sqrt{y}\,e^{-i\widetilde{\theta}}
\\
(-1)^n\sqrt{y}\,e^{i\widetilde{\theta}} &
1
\end{pmatrix},
\end{aligned}
\end{equation}
and
\begin{equation}
P_1^{\up,T}(\sqrt{y},\widetilde{\theta})
=
\frac{1}{1+y}
\begin{pmatrix}
y &
\sqrt{y}\,e^{-i\widetilde{\theta}}
\\
\sqrt{y}\,e^{i\widetilde{\theta}} &
1
\end{pmatrix}.
\end{equation}
It follows that
\begin{equation}
\begin{aligned}
&\int_0^{2\pi}d\widetilde{\theta}\,
\left[
\sigma_z^{n+1}
P_1^\down(\sqrt{y},\widetilde{\theta})
\sigma_z^{n+1}
\right]
\otimes
P_1^{\up,T}(\sqrt{y},\widetilde{\theta})
\\
&\qquad=
\frac{2\pi}{(1+y)^2}\widetilde{T}_n(y),
\end{aligned}
\label{eq:Ttilde_angular_integral}
\end{equation}
where
\begin{equation}
\widetilde{T}_n(y)
=
\begin{pmatrix}
y^2 & 0 & 0 & 0\\
0 & y & -(-1)^{n+1}y & 0\\
0 & -(-1)^{n+1}y & y & 0\\
0 & 0 & 0 & 1
\end{pmatrix}.
\label{eq:Ttilde_n}
\end{equation}

The radial measure becomes
\begin{equation}
\frac{V^{\frac{2(n-1)}{n}}}{n}
k^{\frac{2-n}{n}}dk
=
\frac{V^2}{2n}
y^{\frac{1-n}{n}}dy,
\label{eq:T_y_jacobian}
\end{equation}
while the upper integration limit becomes $y_{\mathrm{max}}
=(\Lambda/V)^{2n}=x^{-n}$.

Finally, using Eq.$\,$\ref{eq:Rmat_BHZ_general_rewrite}, we can write
\begin{equation}
\begin{aligned}
2R_n(k_v;V,x)-\lambda
&=
\frac{a_n-b_ny}{1+y}-\lambda
\\
&=
\frac{
a_n-\lambda-y(\lambda+b_n)
}{
1+y
}.
\end{aligned}
\label{eq:T_denominator}
\end{equation}
Combining Eqs.$\,$\ref{eq:Ttilde_angular_integral},
\ref{eq:T_y_jacobian}, and \ref{eq:T_denominator}, we obtain
\begin{equation}
T_{n,\lambda}({\bf 0})
=
\frac{V^2}{4\pi n}
\int_0^{x^{-n}}dy\,
\frac{
y^{\frac{1-n}{n}}\widetilde{T}_n(y)
}{
(1+y)
\left[
a_n-\lambda-y(\lambda+b_n)
\right]
}.
\label{eq:BHZ_binding_matrix}
\end{equation}

For $y>0$, the matrix $\widetilde{T}_n(y)$ has rank three.
The nonzero eigenvectors of $\widetilde{T}_n(y)$ may be
chosen as
\begin{equation}
\begin{aligned}
v_{2n}&=(1,0,0,0)^T,
\\
v_n&=(0,(-1)^n,1,0)^T,
\\
v_0&=(0,0,0,1)^T,
\end{aligned}
\end{equation}
with corresponding eigenvalues $y^2$, $2y$, and $1$, while
\begin{equation}
v_{\mathrm{null}}=(0,-(-1)^n,1,0)^T
\end{equation}
has eigenvalue zero.

Since these eigenvectors are independent of $y$, they also diagonalize
the integrated kernel $T_{n,\lambda}({\bf 0})$. Defining
\begin{equation}
F_{n,m}(\lambda)
=
\frac{V^2}{4\pi n}
\int_0^{x^{-n}}dy\,
\frac{
y^{\frac{1-n}{n}+m}
}{
(1+y)
\left[
a_n-\lambda-y(\lambda+b_n)
\right]
},
\label{eq:Fnm_definition}
\end{equation}
we obtain
\begin{equation}
\begin{aligned}
T_{n,\lambda}({\bf 0})v_{2n}
&=
F_{n,2}(\lambda)v_{2n},
\\
T_{n,\lambda}({\bf 0})v_n
&=
2F_{n,1}(\lambda)v_n,
\\
T_{n,\lambda}({\bf 0})v_0
&=
F_{n,0}(\lambda)v_0,
\\
T_{n,\lambda}({\bf 0})v_{\mathrm{null}}
&=
0.
\end{aligned}
\label{eq:Tnlambda_eigenvalues}
\end{equation}
Therefore,
\bea
\label{eq:Tnlambda_determinant_factorization}
\det\left[
 \mathbb{1}-T_{n,\lambda}({\bf 0})
\right] =
&\left[1-F_{n,2}(\lambda)\right] \\
&\left[1-2F_{n,1}(\lambda)\right]
\left[1-F_{n,0}(\lambda)\right].
\eea
The bound-state energies are the values of $\lambda$ satisfying
\begin{equation}
\det\left[
\mathbb{1}-T_{n,\lambda}({\bf 0})
\right]
=0,
\end{equation}
or, equivalently,
\begin{equation}
F_{n,2}(\lambda)=1,
\qquad
F_{n,1}(\lambda)=\frac{1}{2},
\qquad
F_{n,0}(\lambda)=1.
\label{eq:BHZ_bound_state_equations}
\end{equation}

The roots in Eq.$\,$\ref{eq:BHZ_bound_state_equations} are determined by performing the integral in Eq.$\,$\ref{eq:Fnm_definition} using Wolfram Mathematica \cite{Mathematica} to be
\bea
F_{n,m}(\lambda)&=\frac{V^2}{4\pi n}\int_{0}^{x^{-n}}dy\,\,\frac{y^{\frac{(1-n)}{n}+m}}{(1+y)[a_{n}-\lambda-y(\lambda+b_{n})]}\\
&=\Lambda^2 [4 \pi x^{nm}(a_{n}+b_{n})(a_{n}-\lambda)(1+mn)]^{-1}\\
& \Big[(a_{n}-\lambda)\,_{2}F_{1}\Big(1,m+\frac{1}{n};m+1+\frac{1}{n};-\frac{1}{x^{n}}\Big)\\
+(b_{n}+\lambda)& \,_{2}F_{1}\Big(1,m+\frac{1}{n};m+1+\frac{1}{n};\frac{(b_{n}+\lambda)}{x^{n}(a_{n}-\lambda)}\Big)\Big]\,,
\eea
where $\,_{2}F_{1}(a,b;c;z)$ is the hypergeometric
function$\,$\cite{WolframHypergeometric2F1}.
The three bound modes can be obtained by solving $F_{n,m}(\lambda)=1,$
with $m=0,2$ and $F_{n,1}(\lambda)=1/2$. The special point where the continuum becomes exactly flat corresponds to setting $a_{n}=-b_{n}$, so that $R_n(k_v;V,x)=a_{n}(V,x)/2$.

\subsection{Derivation of SyMA expressions for the magnon gaps}

\label{subsec:BHZ_magnon_gaps_SyMA}

We start by deriving the angular-momentum eigenvalues of the SyMA modes in the continuum model. To do so, it will be convenient to pick a new gauge for $\psi_{n,\eta}$ by
making $\psi_{n,\eta}\rightarrow e^{-i\frac{\eta n\theta}{2}}\psi_{n,\eta}$
with respect to the definition in the main text,
\begin{equation}
\psi_{n,\eta}(k_v,\theta)
=
\frac{1}{\sqrt{1+k_v^{2n}}}
\left(
\eta^n k_v^n e^{i\eta n\theta/2},
e^{-i\eta n\theta/2}
\right)^T.
\label{eq:BHZ_rot_gauge}
\end{equation}
In this gauge,
\begin{equation}
\psi_{n,\eta}(k_v,\theta+\phi)
=
e^{i\eta n\phi\sigma_z/2}
\psi_{n,\eta}(k_v,\theta),
\label{eq:BHZ_spinor_rotation}
\end{equation}
and therefore
\begin{equation}
\psi^\dagger_{n,\eta}(k_v,\theta+\phi)
=
\psi^\dagger_{n,\eta}(k_v,\theta)
e^{-i\eta n\phi\sigma_z/2}.
\label{eq:BHZ_spinor_rotation_dagger}
\end{equation}

Recalling that the orbital annihilation operator is defined as
\begin{equation}
c_{\eta,\mathbf k}
=
\begin{pmatrix}
c_{0,\eta,\mathbf k}\\
c_{1,\eta,\mathbf k} \,
\end{pmatrix},
\end{equation}
where $\alpha=0,1$ is the orbital label, the projected flat band operator is
\begin{equation}
\gamma_{\eta,\mathbf k}
=
\psi^\dagger_{n,\eta}(k_v,\theta)c_{\eta,\mathbf k}
\label{eq:BHZ_gamma_def_rot_gauge}
\end{equation}

We define the SO(2) rotation as
\begin{equation}
W(\phi)=e^{-i\phi J_z},
\end{equation}
where
\begin{equation}
J_z
=
\sum_{\eta,\mathbf k}
c^\dagger_{\eta,\mathbf k}
\left(
-i\partial_\theta
-\frac{\eta n}{2}\sigma_z
\right)
c_{\eta,\mathbf k}.
\label{eq:BHZ_Jz_def}
\end{equation}
Here \(\partial_\theta\) acts on the momentum label of the orbital
operator \(c_{\eta,\mathbf k}\), with
\(\mathbf k=(k_v,\theta)\). The second term in Eq.~\eqref{eq:BHZ_Jz_def}
is the internal orbital angular momentum.

To derive the action of \(W(\phi)\), we define the one-body operator
\begin{equation}
h_\eta
=
-i\partial_\theta
-\frac{\eta n}{2}\sigma_z.
\end{equation}
For a second-quantized one-body operator
\(\sum c^\dagger h c\), we have
\begin{equation}
\left[
\sum_{\mathbf q} c^\dagger_{\eta,\mathbf q}
h_\eta
c_{\eta,\mathbf q},
c_{\eta,\mathbf k}
\right]
=
-h_\eta c_{\eta,\mathbf k}.
\end{equation}
Therefore,
\begin{equation}
[J_z,c_{\eta,\mathbf k}]
=
-\left(
-i\partial_\theta
-\frac{\eta n}{2}\sigma_z
\right)c_{\eta,\mathbf k}.
\label{eq:Jz_c_commutator}
\end{equation}
Expanding \(W(\phi)c_{\eta,\mathbf k}W^\dagger(\phi)\) to first order,
we obtain
\begin{align}
W(\phi)c_{\eta,\mathbf k}W^\dagger(\phi)
&=
e^{-i\phi J_z}
c_{\eta,\mathbf k}
e^{i\phi J_z}
\nonumber\\
&=
c_{\eta,\mathbf k}
-i\phi[J_z,c_{\eta,\mathbf k}]
+O(\phi^2)
\nonumber\\
&=
c_{\eta,\mathbf k}
+i\phi
\left(
-i\partial_\theta
-\frac{\eta n}{2}\sigma_z
\right)c_{\eta,\mathbf k}
+O(\phi^2)
\nonumber\\
&=
c_{\eta,\mathbf k}
+
\phi
\left(
\partial_\theta
-\frac{i\eta n}{2}\sigma_z
\right)c_{\eta,\mathbf k}
+O(\phi^2).
\label{eq:BHZ_c_rotation_inf}
\end{align}
This is precisely the infinitesimal expansion of the finite transformation
\begin{equation}
W(\phi)c_{\eta,\mathbf k}W^\dagger(\phi)
=
e^{-i\eta n\phi\sigma_z/2}
c_{\eta,R_\phi\mathbf k},
\label{eq:BHZ_c_rotation}
\end{equation}
where \(R_\phi\mathbf k\) denotes the momentum with polar angle
\(\theta+\phi\), which we can readily check,
\begin{align}
e^{-i\eta n\phi\sigma_z/2}
c_{\eta,R_\phi\mathbf k}
&=
\left(
1-\frac{i\eta n\phi}{2}\sigma_z
\right)
\left(
c_{\eta,\mathbf k}
+
\phi\partial_\theta c_{\eta,\mathbf k}
\right)
+O(\phi^2)
\nonumber\\
&=
c_{\eta,\mathbf k}
+
\phi
\left(
\partial_\theta
-\frac{i\eta n}{2}\sigma_z
\right)c_{\eta,\mathbf k}
+O(\phi^2).
\end{align}

We can now derive the rotation of the projected operator,
\begin{align}
W(\phi)\gamma_{\eta,\mathbf k}W^\dagger(\phi)
&=
W(\phi)
\left[
\psi^\dagger_{n,\eta}(k_v,\theta)c_{\eta,\mathbf k}
\right]
W^\dagger(\phi)
\nonumber\\
&=
\psi^\dagger_{n,\eta}(k_v,\theta)
W(\phi)c_{\eta,\mathbf k}W^\dagger(\phi)
\nonumber \\
&=
\psi^\dagger_{n,\eta}(k_v,\theta)
e^{-i\eta n\phi\sigma_z/2}
c_{\eta,R_\phi\mathbf k}
\nonumber\\
&=
\psi^\dagger_{n,\eta}(k_v,\theta+\phi)
c_{\eta,R_\phi\mathbf k}
\nonumber\\
&=
\psi^\dagger_{n,\eta}(R_\phi\mathbf k)
c_{\eta,R_\phi\mathbf k}
\nonumber\\
&=
\gamma_{\eta,R_\phi\mathbf k}.
\label{eq:BHZ_gamma_rotation}
\end{align}
where in the second line we used that the operator $W(\phi)$ only acts on the Fock-space operators $c_{\eta,{\bf k}}$ (\(\psi^*_{n,\eta,\alpha}(k_v,\theta)\) are coefficients/c-numbers), in the third line we used Eq.$\,$\ref{eq:BHZ_c_rotation} and in the fourth line we used Eq.~\eqref{eq:BHZ_spinor_rotation_dagger}.

Acting on the magnon modes we have, using Eq.$\,$\ref{eq:gap_exact_gs_projectors} and writing $M_{n,D}^{(\down,\up)}({\bf k},{\bf 0})=M_{n,D}^{(\down,\up)}(k,\theta)$
in polar coordinates,

\begin{equation}
\begin{array}{c}
W(\phi)S_{D;n,{\bf 0}}^{\dagger}W^{\dagger}(\phi)\\
=\int d^{2}{\bf k}\,\,M_{n,D}^{(\down,\up)}(k,\theta-\phi)\gamma_{{\bf k},\down}^{\dagger}\gamma_{{\bf k},\up}\\
=e^{-iw(D)\phi}S_{D;n,{\bf 0}}^{\dagger}
\end{array}
\end{equation}
where $w(D)$ is the $J_{z}$ eigenvalue corresponding to the winding
of each mode, shown in Table \ref{tab:n_model_winding_table}. We
can then label our bound magnon modes as $S_{n,{\bf 0}}^{(w)\dagger}$
through their $J_{z}$ eigenvalues $w$, i.e., $W(\phi)S_{n,{\bf 0}}^{(w)\dagger}W^{\dagger}(\phi)=e^{-iw\phi}S_{n,{\bf 0}}^{(w)\dagger}$.

The magnon gap expression in Eq.$\,$\ref{eq:Ising_gap_1BPV} can also be written in terms of the projector $P_n^{\eta}({\bf k})=\psi_{n,\eta}(k_v,\theta)\psi_{n,\eta}^{\dagger}(k_v,\theta)$ defined above (or the projectors
$P^{\eta}({\bf k})=U_{\eta,{\bf k}}U_{\eta,{\bf k}}^{\dagger}$ in the case of the continuum model of MoTe$_2$ discussed in Sec.$\,$\ref{sec:moire_magnons}) as

\begin{widetext}
\bea
\Delta^{\down,\up}({\bf p})=\Big(\sum_{{\bf k}}\Tr[P^{\down}({\bf k}+{\bf p})DP^{\up}({\bf k})D^{\dagger}]\Big)^{-1}\times \qquad\qquad\qquad\qquad\qquad\qquad\null \\
\frac{1}{2\Omega_{\textrm{tot}}}\sum_{{\bf k},{\bf q} \in \mathbb{R}^2}V({\bf q})\Tr\Big[P^{\up}({\bf k})\Big(D^{\dagger}P^{\down}({\bf k}+{\bf p})-P^{\up}({\bf k}+{\bf q})D^{\dagger}\Big)P^{\down}({\bf k}+{\bf q}+{\bf p})\Big(P^{\down}({\bf k}+{\bf p})D-DP^{\up}({\bf k}+{\bf q})\Big)\Big]\,.
\label{eq:gap_exact_gs_projectors}
\eea

\end{widetext}
where we dropped the index $n$ from the projectors for readability. Using $\Omega_{\textrm{tot}}=N=L^2$ and $V({\bf q})=1$ for the ideal-band BHZ model with Hubbard interaction, we have at ${\bf p}=0$,

\begin{widetext}

\begin{equation}
\Delta_{n}^{\down\up}({\bf 0})=\frac{\frac{1}{8\pi^2}\int_{|{\bf k}|,|{\bf k}'|<\Lambda}d^{2}{\bf k}\,d^{2}{\bf k}'\,\,\Tr\Big[P_{n}^{\up}({\bf k})\Big(D^{\dagger}P_{n}^{\down}({\bf k})-P_{n}^{\up}({\bf k}')D^{\dagger}\Big)P_{n}^{\down}({\bf k}')\Big(P_{n}^{\down}({\bf k})D-DP_{n}^{\up}({\bf k}')\Big)\Big]}{\int_{|{\bf k}|<\Lambda}d^{2}{\bf k}\,\Tr[P_{n}^{\down}({\bf k})DP_{n}^{\up}({\bf k})D^{\dagger}]}\,.\label{eq:n_model_SMAgap_p=00003D0}
\end{equation}

\end{widetext}

Using that $D^{T}P_n^{\eta}({\bf k})D^{*}=DP_n^{\eta}({\bf k})D^{\dagger}$,
which is satisfied for all the ansatz $D$ matrices since $D=\zeta D^{T}$,
with $\zeta$ the $\mathcal{C}$ eigenvalue, Eq.$\,$\ref{eq:n_model_SMAgap_p=00003D0}
can be simplified to

\begin{widetext}

\begin{equation}
\Delta_{n}^{\down\up}({\bf 0})=\frac{\frac{1}{4\pi^2}\int_{|{\bf k}|,|{\bf k}'|<\Lambda}d^{2}{\bf k}\,d^{2}{\bf k}'\,\,\Tr\Big[P_{n}^{\up}({\bf k})D^{\dagger}P_{n}^{\down}({\bf k})P_{n}^{\down}({\bf k}')P_{n}^{\down}({\bf k})D-P_{n}^{\up}({\bf k})D^{\dagger}P_{n}^{\down}({\bf k})P_{n}^{\down}({\bf k}')DP_{n}^{\up}({\bf k}')\Big]}{\int_{|{\bf k}|<\Lambda}d^{2}{\bf k}\,\Tr[P_{n}^{\down}({\bf k})DP_{n}^{\up}({\bf k})D^{\dagger}]}\,.\label{eq:Delta_SMA_p0_2idealband}
\end{equation}

\end{widetext}

\subsubsection{Denominators}

We start by evaluating the denominator in Eq.$\,$\ref{eq:Delta_SMA_p0_2idealband}.
We have

\begin{equation}
\begin{array}{c}
\mathcal{Z}_n=\int_{|{\bf k}|<\Lambda}d^{2}{\bf k}\,\Tr[P_{n}^{\down}({\bf k})DP_{n}^{\up}({\bf k})D^{\dagger}]\\
=\int_{0}^{\Lambda}dk \, k \, \,\int_{0}^{2\pi}d\theta\,\,\Tr\Big[D^{\dagger}(-\sigma_{z})^{(n+1)}P_{1,\down}\big(k_v^{n},n\theta\big)\\
(-\sigma_{z})^{(n+1)}DP_{1,\up}\big(k_v^{n},n\theta\big)\Big]\,.
\end{array}\label{eq:denominator_2idealband_first}
\end{equation}

For the $D$ matrices in Eq.$\,$\ref{eq:Mk_different_modes_nmodel}
we have $D^{\dagger}(-\sigma_{z})^{(n+1)}=\pm D$ for $D=(1\pm\sigma_{z})/2$
The mode with $D=\sigma_{y}$ only exists for odd $n$ and therefore
$(-\sigma_{z})^{(n+1)}=1\,(n\,\,\textrm{odd})$. On the other hand,
the mode with $D=\sigma_{x}$ only exists for even $n$ and we have
$D^{\dagger}(-\sigma_{z})^{(n+1)}=i\sigma_{y}\,(n\,\,\textrm{even})$.
We can therefore simplify Eq.$\,$\ref{eq:denominator_2idealband_first}
to

\begin{equation}
\begin{aligned}
\mathcal{Z}_n
={}&
\frac{V^{\frac{2(n-1)}{n}}}{n}
\int_{0}^{V(\Lambda/V)^n}
dk'\,
(k')^{\frac{2-n}{n}}
\\
&\times
\int_{0}^{2\pi}d\theta'\,
\Tr\left[
B_D^\dagger
P_1^\down(k'_v,\theta')
B_D
P_1^\up(k'_v,\theta')
\right].
\end{aligned}
\end{equation}
where we defined

\begin{equation}
B_{D}=\begin{cases}
\frac{1\pm\sigma_{z}}{2} &,D=\frac{1\pm\sigma_{z}}{2}\\
\sigma_{y} &,D=\sigma_{x},\sigma_{y}
\end{cases}\,.
\end{equation}

Defining $s=(V/\Lambda)^{2n}$, the denominator for the different modes yields

\begin{widetext}

\begin{equation}
\mathcal{Z}_n\big(\Lambda,s\big)=\begin{cases}
\frac{\pi\Lambda^{2}}{n}\Big(\frac{1}{s(s+1)}-\frac{(n+1)\,\,\,_{2}F_{1}(1,2+1/n;3+1/n;-1/s)}{s^{2}(2n+1)}\Big) &,w=2n\\
\frac{4\pi\Lambda^{2}}{n}\Big(\frac{1}{s+1}-\frac{\,_{2}F_{1}(1,1+1/n;2+1/n;-1/s)}{s(n+1)}\Big) &,w=n\\
\frac{\pi\Lambda^{2}}{n}\Big(\frac{s}{s+1}+(n-1)\,\,\,_{2}F_{1}(1,1/n;1+1/n;-1/s)\Big) &,w=0
\end{cases}\label{eq:nmodel_SMA_denom}
\end{equation}
\end{widetext}

The special cases of $n=1,2$
 are given by

\begin{equation}
\mathcal{Z}_1\big(\Lambda,x\big)=\begin{cases}
\pi\Lambda^{2}\Big(\frac{1+2x}{1+x}-2x\log(1+1/x)\Big) &,w=2n\\
4\pi\Lambda^{2}x\Big(-\frac{1}{1+x}+\log(1+1/x)\Big) &,w=n\\
\pi\Lambda^{2}\frac{x}{1+x} &,w=0
\end{cases}
\end{equation}

\begin{equation}
\mathcal{Z}_2\big(\Lambda,x\big)=\begin{cases}
\pi\Lambda^{2}\Big(1-\frac{3x}{2}\textrm{arccot}(x)+\frac{x^{2}}{2(1+x^{2})}\Big) &,w=2n\\
2\pi\Lambda^{2}x\Big(\textrm{arccot}(x)-\frac{x}{1+x^{2}}\Big) &,w=n\\
\pi\Lambda^{2}x\Big(\frac{1}{2}\textrm{arccot}(x)+\frac{x}{2(1+x^{2})}\Big) &,w=0
\end{cases}
\end{equation}

\subsubsection{Numerators}

For the numerator in Eq.$\,$\ref{eq:Delta_SMA_p0_2idealband}, we
have

\begin{equation}
\begin{array}{cc}
\mathcal{N}_{n}=\frac{1}{4\pi^2}\int_{|{\bf k}|<\Lambda}d^{2}{\bf k}\,\,\Tr\Big[P_{n}^{\up}({\bf k})D^{\dagger}P_{n}^{\down}({\bf k})\bar{P}_{n}^{\down}P_{n}^{\down}({\bf k})D\\
-P_{n}^{\up}({\bf k})D^{\dagger}P_{n}^{\down}({\bf k})\bar{P_{n}}^{\down\up}\Big]\,
\end{array}
\end{equation}
where $\bar{P}_{n}^{\down}=\int_{|{\bf k}'|<\Lambda}\,d^{2}{\bf k}'\,\,P_{n}^{\down}({\bf k}')$
and $\bar{P}_{n}^{\down\up}=\int_{|{\bf k}'|<\Lambda}\,d^{2}{\bf k}'\,\,P_{n}^{\down}({\bf k}')DP_{n}^{\up}({\bf k}')$.
As for the denominator, all these integrals can be computed by writing
$P_{n}^{\eta}({\bf k})$ in terms of $P_{1}^{\eta}({\bf k})$. We
can get closed-form expressions for all the ansatze. Again defining $s=(V/\Lambda)^{2n}$ and $\mathcal{F}(a,b)=\,_{2}F_{1}(1,a;b;-1/s)$ we have,

\begin{widetext}

\begin{equation}
\begin{array}{cc}
 & \underline{w=2n\,\,\textrm{and}\,\,w=0} \\ &\mathcal{N}_{n}\big(s\big)=\frac{\Lambda^{4}}{8 n^{3}(1+s)^{2}}\Big[-2ns^{2}+(-1)^{-1/n}s^{1/n}\mathcal{B}_{-1/s}(1+1/n,0)\Big(1+s-(2s+1)n+\frac{(n-1)(1+s)^{2}\mathcal{F}(1+1/n,2+1/n)}{(n+1)s}\Big)\\
 & +n \mathcal{F}(1/n,1+1/n)\Big(3s(s+1)-ns(s+2)+(n-1)(s+1)^{2}\,\mathcal{F}(1/n,1+1/n)\Big)\Big]\,,
\end{array}\label{eq:nmodel_SMA_num_w_2n}
\end{equation}

\begin{equation}
\begin{array}{cc}
 & \underline{w=n}\\ &\mathcal{N}_{n}\big(\Lambda,s\big)=\frac{\Lambda^{4}}{2n^{3}s^{2}(1+s)^{2}}\Big[-\frac{4n\big(s(1+n)-(s+1)\,\mathcal{F}(1+1/n,2+1/n)\big)^{2}}{(n+1)^{2}}\\
 & -(-1)^{-1/n}s^{1/n}\mathcal{B}_{-1/s}(1+1/n,0)\Big(-s+(n-1)s^{2}+\frac{(n+1)(1+s)^{2}\,\mathcal{F}(2+1/n,3+1/n)]}{2n+1}\Big)\\
 & -n s\,_{2}F_{1}(1,1/n;1+1/n;-1/s)]\Big(s-ns-(2n-1)s^{2}+\frac{(n-1)(1+s)^{2}\,\mathcal{F}(1+1/n,2+1/n)]}{n+1}\Big)\Big]\,.
\end{array}\label{eq:nmodel_SMA_num_w_n}
\end{equation}

\end{widetext}

Interestingly, the numerator for $w=0$ and $w=2n$ is exactly the
same (which is not the case for the denominator). Particular cases
for $n=1,2$ follow:

\begin{widetext}

\begin{equation}
\mathcal{N}_{1}\big(\Lambda,x=(V/\Lambda)^{2}\big)=\begin{cases}
\Lambda^{4}\frac{x[1-2x+2x^{2}\log(1+1/x)]}{8(1+x)^{2}} &,w=0,2n\\
\Lambda^{4}x\frac{\Big(-6x-3+2\log(1+1/x)\big(1+2x(3x+4)-3x(x+1)^{2}\log(1+1/x)\big)\Big)}{2(1+x)^{2}} &,w=n
\end{cases}
\end{equation}

\begin{equation}
\mathcal{N}_{2}\big(\Lambda,x=(V/\Lambda)^{2}\big)=\begin{cases}
\frac{\Lambda^{4}x}{8}\frac{[x-3x^{3}+(1+3x^{4})\textrm{arccot}(x)]}{4(1+x^{2})^{2}} &,w=0,2n\\
\frac{\Lambda^{4}x}{8}\Big(\frac{\big(-x(5+7x^{2})+\textrm{arccot}(x)[3+20x^{2}+13x^{4}]\big)}{(1+x^{2})^{2}}-6x\,\textrm{arccot}^{2}(x)\Big) &,w=n
\end{cases}
\end{equation}

\end{widetext}

It is possible to greatly simplify the SyMA gap expression of the different
modes by carrying out a small $x=(V/\Lambda)^{2}$ expansion. The
leading-order terms for the different modes are given by:

\begin{equation}
\Delta_{n,w}^{\textrm{SyMA}}({\bf 0})
=
\begin{cases}
\displaystyle
\frac{V^2}{8\pi},
& w=2n,\quad n=1,
\\[6pt]
\displaystyle
\frac{(n-1)\csc(\pi/n)}{8n^3}V^2,
& w=2n,\quad n>1,
\\[6pt]
\displaystyle
\Lambda^2\frac{n+1}{8\pi n},
& w=n,
\\[6pt]
\displaystyle
\frac{\Lambda^2}{8\pi n},
& w=0.
\end{cases}
\label{eq:nmodel_Isinggap_smallV}
\end{equation}

This implies that, for sufficiently large $\Lambda$, the $w=2n$ mode
necessarily has the lowest energy. Moreover, it is the only mode that remains isolated below the
continuum in the small-$x$ limit for any fixed finite $\Lambda$. To see this, we
compare the SyMA energies with the lowest energy of the particle-hole
continuum in the small-$x$ limit.

As directly follows from Eqs.$\,$\ref{eq:R_n1_smallx} and
\ref{eq:R_anu_n_smallx}, for sufficiently small $x$ the minimum charge
gap occurs at $k_v=0$. The energy minimum of the continuum is therefore
given by

\begin{equation}
E_{\textrm{min}}^{\textrm{c}}(n,x)
=
2R_n(k_v=0;V,x).
\end{equation}

For $n=1$, Eq.$\,$\ref{eq:BHZ_R_smallx} implies that
$E_{\textrm{min}}^{\textrm{c}}(1,x)$ scales as
$V^2\log(1/x)$ for small $x$. It therefore vanishes more slowly than
the SyMA energy of the $w=2n$ mode, which scales as $V^2$ and provides
an upper bound on the exact Ising gap\footnote{As mentioned in the
main text, the exact magnon gap vanishes even faster,
$\Delta_{n=1}^{\textrm{exact}}({\bf 0})
\approx V^{2}x\log^{2}(x)/(4\pi)$. By contrast, Fig.$\,$\ref{fig:BHZ} shows that
the remaining exact modes enter the continuum for sufficiently small
$V$.}. Consequently, the $w=2n$ mode remains below the continuum,
whereas the minimum of the continuum eventually becomes smaller than the
energies of the remaining modes, see
Eq.$\,$\ref{eq:nmodel_Isinggap_smallV}.

For $n>1$, this distinction follows directly by comparing the minimum of the
continuum with the three SyMA energies:

\begin{equation}
\frac{E_{\textrm{min}}^{\textrm{c}}(n,x)}
{\Delta_{n,w}^{\textrm{SyMA}}({\bf 0})}
\longrightarrow
\begin{cases}
\displaystyle
\frac{\frac{\csc(\pi/n)}{4n}}
{\frac{(n-1)\csc(\pi/n)}{8n^3}}
=
\frac{2n^2}{n-1}>1,
& w=2n,
\\[8pt]
\displaystyle
\frac{2\pi x\csc(\pi/n)}{n+1}
\longrightarrow 0,
& w=n,
\\[8pt]
\displaystyle
2\pi x\csc(\pi/n)
\longrightarrow 0,
& w=0.
\end{cases}
\end{equation}
Thus, as $x\rightarrow0$, the minimum of the continuum remains above the ${\bf p}=0$ energy of the
$w=2n$ mode. For
the $w=n$ and $w=0$ modes, instead, the ratios vanish, showing that
they merge in the continuum at sufficiently small $x$. The $w=2n$ mode is
therefore the only mode that remains isolated below the continuum in
the small-$x$ regime and consequently determines the magnon gap, as
stated in the main text.

\section{Additional results at fractional filling}

\label{sec:Additional_results_Fractional}

In this Appendix, we present complementary results for the magnon spectrum at fractional filling, for the FH model of twisted MoTe$_2$, at $\theta=3.7^{\circ}$ and $V_z=0$.

The momentum meshes used for the ED calculations in Fig.$\,$\ref{fig:Fractional_Fillings} of the main text and in this Appendix are explicitly shown in Fig.$\,$\ref{fig:App_momentum_meshes} (see Appendix.$\,$\ref{sec:sec_details_magnon_exact} for precise definition of the momentum meshes).

\begin{figure}[h!]
 \centering
 \includegraphics[width=\columnwidth]{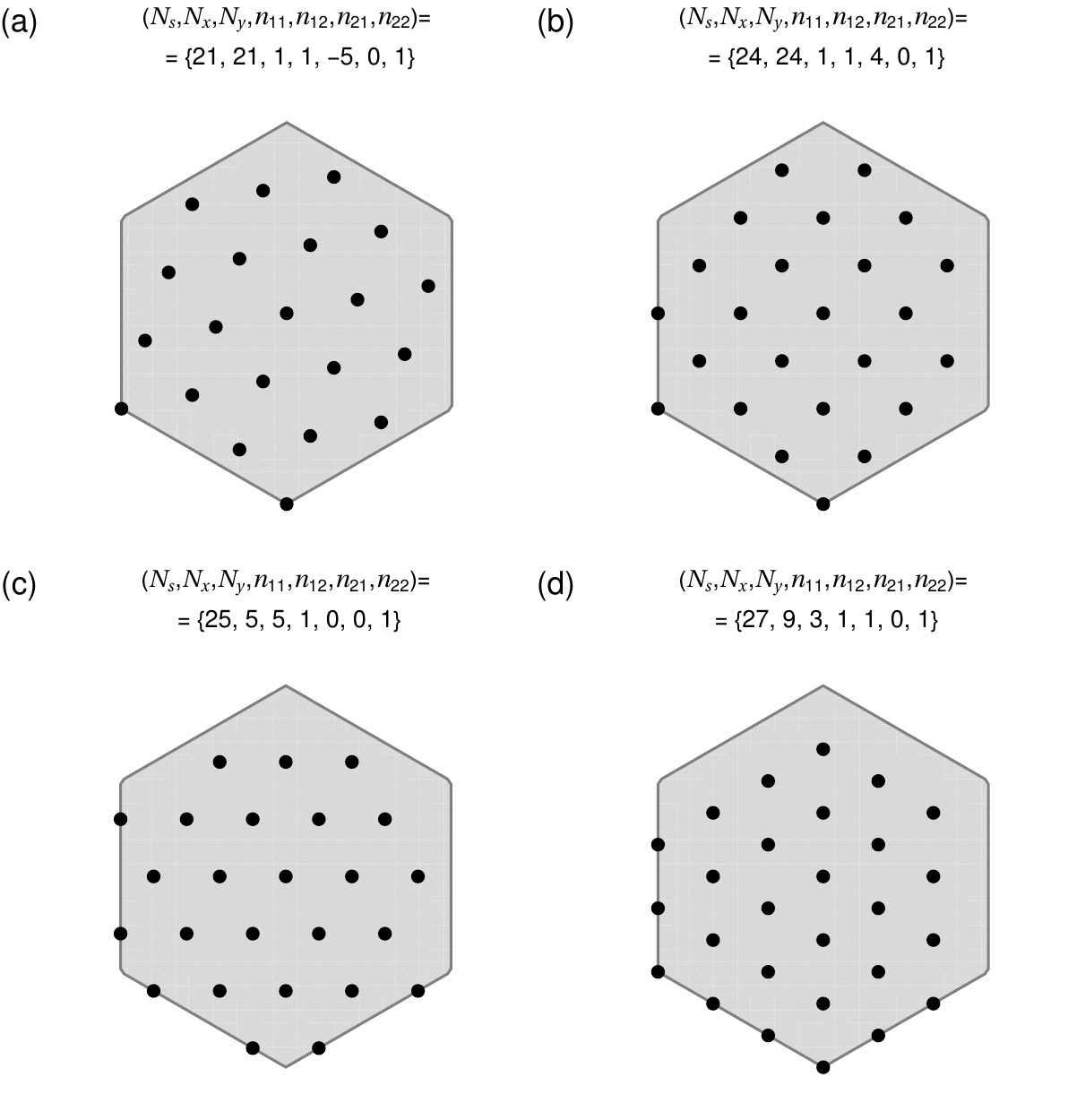}
 \caption{Momentum meshes used in the ED calculations at fractional fillings (see Appendix.$\,$\ref{sec:sec_details_magnon_exact} for precise definition of the momentum meshes).}
 \label{fig:App_momentum_meshes}
\end{figure}

\begin{figure}[h!]
\centering
\includegraphics[width=\columnwidth]{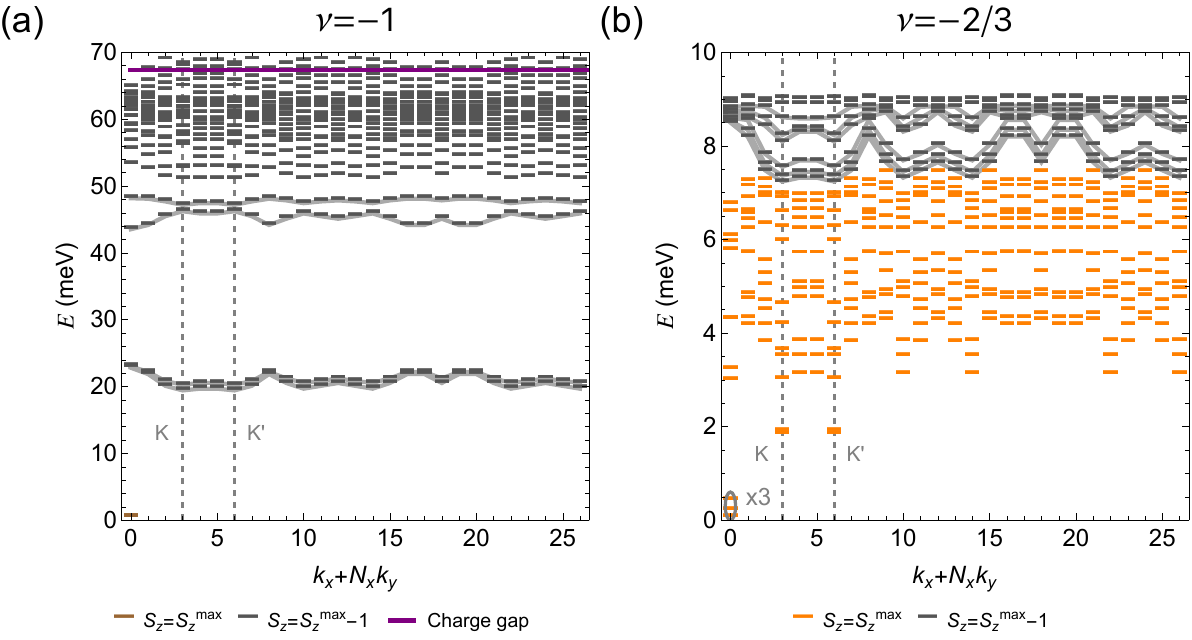}
\caption{
Spin-polarized (orange) and magnon (gray) spectra for the FH model of twisted MoTe$_2$ obtained from the full Hamiltonian in Eq.~\ref{eq:H_full_band_basis}, for $u=0.9$ and $\theta=3.7^\circ$.
(a) Filling $\nu=-1$. The spinful charge gap computed from Eq.~\ref{eq:DeltaS_c} is shown in purple. The four lowest-energy magnon modes are highlighted by solid gray lines.
(b) Filling $\nu=-2/3$. The six lowest-energy magnon modes are highlighted by solid gray lines. They form two branches of three quasi-degenerate states each. The spinful charge gap computed from Eq.~\ref{eq:DeltaS_c} is $23\,\mathrm{meV}$ and is therefore outside the plotted energy range. All results are for the system size defined by $(N_x,N_y,n_{11},n_{12},n_{21},n_{22})=(9,3,1,1,0,1)$.
\label{fig:nu1_vs_nu2o3}
}
\end{figure}

We start by presenting exact diagonalization (ED) results for the spin-polarized and magnon neutral excitation spectra in Fig.~\ref{fig:nu1_vs_nu2o3}, for $\theta=3.7^\circ$. To obtain the spin-polarized spectrum, we perform ED on systems with $N_h$ holes in the spin sector $S_z=S_z^{\textrm{max}}\equiv N_h/2$. The magnon spectrum is obtained by performing ED on a system with the same number of holes but in the spin sector $S_z=S_z^{\textrm{max}}-1$. We also compute the spinful charge gap, given by
\bea
\Delta^{S}_c(N_h) &= E\!\left(N_h + 1,\frac{N_h-1}{2}\right) + E\!\left(N_h - 1,\frac{N_h-1}{2}\right) \nonumber\\
&\quad - 2 E\!\left(N_h,\frac{N_h}{2}\right),
\label{eq:DeltaS_c}
\eea
where $E(N_P,S_z)$ is the ground-state energy in the sector with $N_P$ holes and spin $S_z$. This gap corresponds to the energy cost of creating an unbound particle-hole excitation, consisting of a quasihole with spin $\down$ and a quasiparticle with spin $\up$, with respect to the fully polarized state $\ket{\uparrow}$.

For $\nu=-1$, Fig.~\ref{fig:nu1_vs_nu2o3}(a) shows that the four lowest-energy magnon modes captured in Fig.~\ref{fig:1}(b) can already be clearly identified in ED for the smaller 27-unit-cell system size. The spinful charge gap (Eq.$\,$\ref{eq:DeltaS_c}) is inside the continuum, as expected.

The results for $\nu=-2/3$ using the same system size are shown in Fig.~\ref{fig:nu1_vs_nu2o3}(b). The spin-polarized spectrum reveals three quasi-degenerate ground states separated from the excited states by a gap of approximately $2\,\mathrm{meV}$, consistent with an FCI and reproducing the results of Ref.~\cite{goncalves2025spinlessspinfulchargeexcitations} for the same model and parameters. The spectrum of magnon excitations shows that the six lowest-energy states form two low-energy branches, each consisting of three quasi-degenerate states at fixed momentum. These six lowest-energy magnon modes are highlighted in Fig.$\,$\ref{fig:nu1_vs_nu2o3}. These six lowest-energy modes have an overall dispersion very similar to that of the lowest magnon modes at $\nu=-1$.
Note that these results differ from those shown in Fig.~\ref{fig:Fractional_Fillings}(a), because in the latter the single-particle dispersion was neglected, i.e. the single-particle part of the Hamiltonian in Eq.~\ref{eq:H_full_band_basis} was dropped. Including the single-particle dispersion leads to stronger finite-size effects and significantly increases the spread of the three quasi-degenerate FCI ground states. As a consequence, the spread of the three quasi-degenerate magnon states within each of the two lowest-energy magnon branches is also larger than in the flat band case shown in Fig.~\ref{fig:Fractional_Fillings}(a).

\begin{figure}[h!]
\centering{}\includegraphics[width=1\columnwidth]{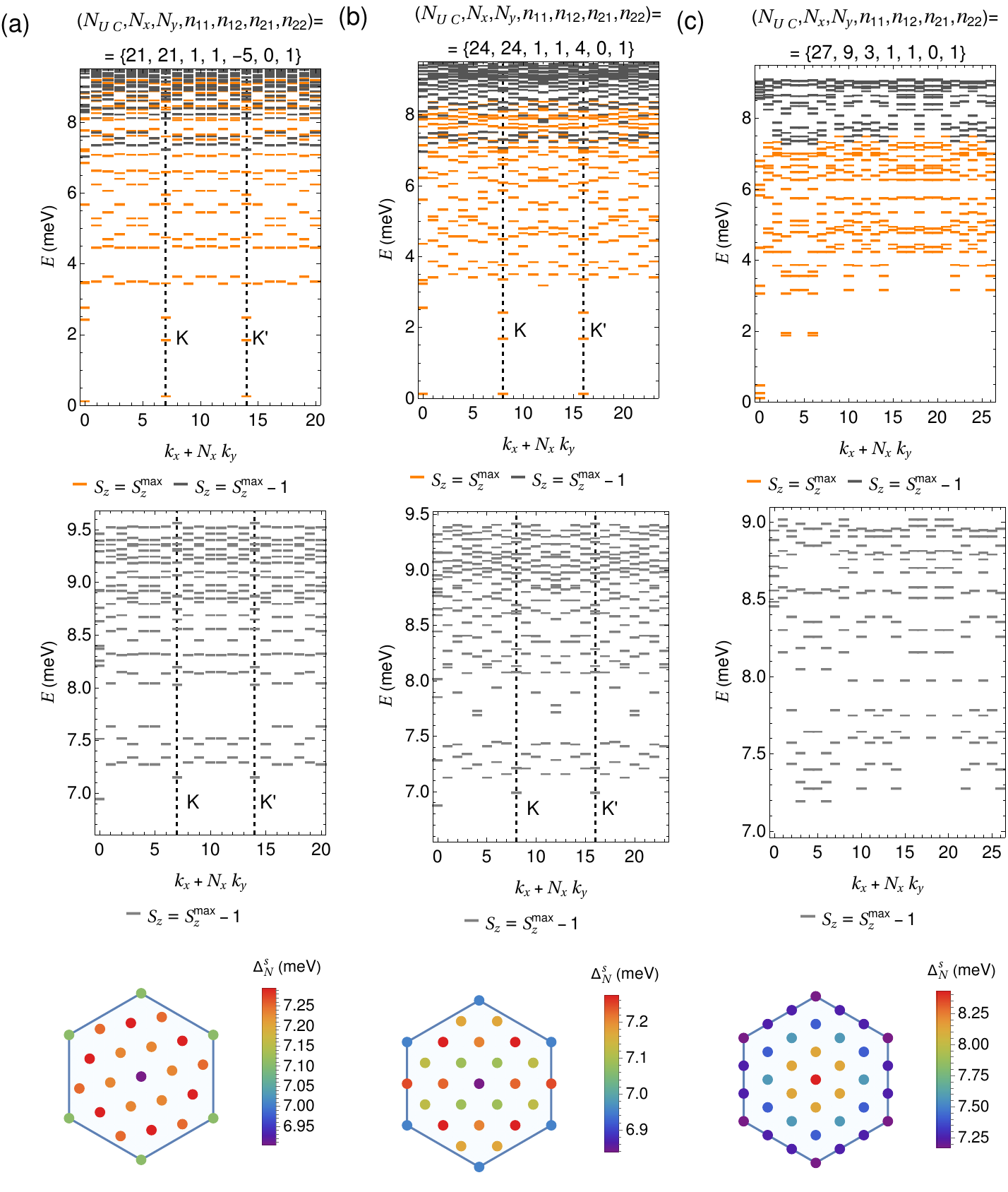}\caption{Spin-polarized (orange) vs. magnon (gray) spectrum for the FH model of MoTe$_2$, for $u=0.9$, $\theta=3.7^{\circ}$ and different system sizes indicated on top of the figures (a-c). The
momentum-resolved magnon gap in the moiré Brillouin zone is also shown in the bottom panels. Note that for
the $N_{UC}=27$ system, the FCI ground-states all occur at $\bf{k}=\Gamma$,
in contrast with the remaining system sizes. This leads to significant
differences in the magnon spectrum and gap. 16 eigenvalues per momentum sector are always shown, except for the magnon spectra in (c), that contains 8 eigenvalues per momentum. \label{fig:Neutral-vs.-spin-flip_ED_var_Sizes}}
\end{figure}

In order to provide a better understanding of the magnon spectrum at $\nu=-2/3$, we compare it for different system sizes in Fig.$\,$\ref{fig:Neutral-vs.-spin-flip_ED_var_Sizes}.
Although low-lying magnon modes can be identified in all cases, they show a quite different dispersion for different system sizes. The bottom row of Fig.$\,$\ref{fig:Neutral-vs.-spin-flip_ED_var_Sizes} shows the magnon gap dispersion in the moiré Brillouin zone, also revealing significant differences for the different system sizes - such as a minimum or maximum at $\bsl{k}=\bsl{\Gamma}$ depending on the size.
In fact, the results for $N_{UC}=27$ - for which the three FCI ground-states all occur in the same momentum sector - are very different from those for the remaining system sizes, where the FCI ground-states occur in different momentum sectors. This difference - FCI ground-states occurring in the same momentum sector or in different momentum sectors - is crucial to understand the results for the dispersion of the magnon modes, as we now explain.

When the FCI ground-states lie in the same momentum sector, as in the $N_{UC}=27$ lattice, the magnon excitation spectrum can be obtained by creating spin-flip excitations on top of each of the three quasi-degenerate ground-states. Since all three ground-states have momentum $\bsl{\Gamma}$, applying a spin-flip operator with momentum ${\bf p}$ to any one of them produces an excitation with total momentum ${\bf p}$. Therefore, at each momentum ${\bf p}$, three magnon states with almost the same energy can be clearly observed, one originating from each FCI ground-state. These modes are not exactly degenerate because of finite-size effects. In the no-dispersion limit shown in Fig.$\,$\ref{fig:Fractional_Fillings}(a), the energy spread of the FCI ground-states is significantly smaller, as is the spread of the three magnon modes within each of the two lowest-energy branches.

When the FCI ground-states occur in different momentum sectors, as for the $N_{UC}=21$ and $N_{UC}=24$ lattices, the construction of the magnon excitations is unchanged: spin-flip excitations are again created on top of each of the three FCI ground-states. More precisely, if an FCI ground-state has momentum ${\bf P}$, acting with a spin-flip bilinear carrying momentum ${\bf p}$ produces a state with total momentum ${\bf P}+{\bf p}$. For $N_{UC}=27$, all three ground-states have ${\bf P}=\bsl{\Gamma}$. Therefore, the lowest-energy excitations, that have momentum transfer $\bsl{K}$ or $\bsl{K}'$, appear at total momentum $\bsl{K}$ or $\bsl{K}'$, respectively, for all three ground-states. By contrast, for $N_{UC}=21$ and $N_{UC}=24$, the three ground-states occur at $\bsl{\Gamma}$, $\bsl{K}$, and $\bsl{K}'$. The lowest-energy excitations created on top of these ground-states therefore occur at different total momenta: those originating from the $\bsl{\Gamma}$ ground-state appear at $\bsl{K}$ and $\bsl{K}'$, those originating from the $\bsl{K}$ ground-state appear at $\bsl{K}'$ and $\bsl{\Gamma}$, and those originating from the $\bsl{K}'$ ground-state appear at $\bsl{\Gamma}$ and $\bsl{K}$. As a consequence, for $N_{UC}=27$, the magnon excitations constructed on top of the three FCI ground-states occur at the same total momenta and form nearly degenerate triplets.
But, when the three parent FCI ground states occur at different momenta, otherwise similar spin-flip excitations generated from them are shifted relative to one another in total momentum. This produces
the more complex momentum-resolved spectrum observed for $N_{UC}=21$ and $N_{UC}=24$.

We will now provide evidence for the interpretation that the low-energy magnon states are obtained by applying spin-flip bilinears to each of the FCI ground-states, with their total momenta determined by the momentum of the parent ground-state, by diagonalizing the Hamiltonian in the subspace generated by all the bilinear operators $\gamma^\dag_{{\bf k}+{\bf p},\down}\gamma_{{\bf k},\up}$ acting on the polarized FCI ground-state(s).

Let $\ket{\Psi_i}$, with $i=1,2,3$, denote the three quasi-degenerate FCI ground-states, and let ${\bf P}_i$ denote their respective momenta. First, for each FCI ground-state separately, we diagonalize the Hamiltonian in the subspace
\begin{equation}
\mathcal{H}_{i}({\bf p})
=
\operatorname{span}_{\bf k}
\left\{
\gamma_{{\bf k}+{\bf p}-{\bf P}_{i},\down}^{\dagger}
\gamma_{{\bf k},\up}
\ket{\Psi_i}
\right\},
\qquad i=1,2,3.
\label{eq:FCI_bilinear_subspace_i}
\end{equation}
The spin-flip bilinear carries momentum ${\bf p}-{\bf P}_i$, so the resulting state has total momentum ${\bf p}$. In this calculation, the excitations constructed on top of the three FCI ground-states are treated independently.

Second, we diagonalize the Hamiltonian in the combined subspace
\begin{equation}
\mathcal{H}_{\mathrm{bil}}({\bf p})
=
\operatorname{span}_{\substack{i=1,2,3\\{\bf k}}}
\left\{
\gamma_{{\bf k}+{\bf p}-{\bf P}_{i},\down}^{\dagger}
\gamma_{{\bf k},\up}
\ket{\Psi_i}
\right\}.
\label{eq:FCI_bilinear_subspace_full}
\end{equation}

In the first procedure, the subspaces $\mathcal{H}_{i}({\bf p})$ are diagonalized separately. Therefore, overlaps and Hamiltonian matrix elements between states generated from different FCI ground-states are not included. In the second procedure, all such states are included in the same combined subspace $\mathcal{H}_{\mathrm{bil}}({\bf p})$. Since states generated from different FCI ground-states are not necessarily orthogonal, we orthonormalize the combined basis before diagonalizing the Hamiltonian.

\begin{figure}[h!]
\begin{centering}
\includegraphics[width=\columnwidth]{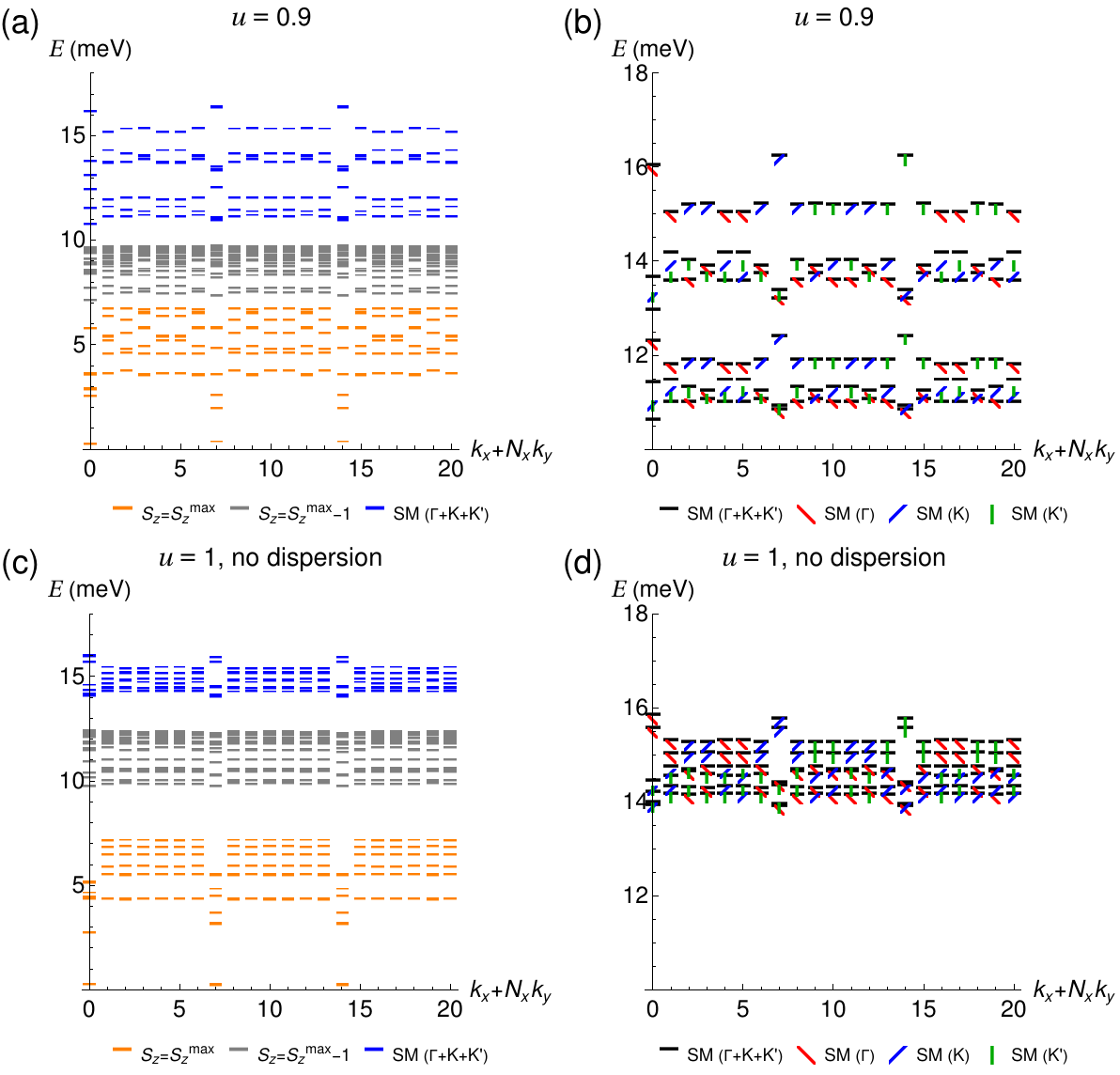}\caption{
Example application of S-matrix (SM) technique at filling $\nu=-2/3$ for the FH model of MoTe$_2$, for $\theta=3.7^{\circ}$, in the $N_{UC}=21$ lattice defined by $(N_x,N_y,n_{11},n_{12},n_{21},n_{22})=(21,1,1,-5,0,1)$. (a) Comparison between exact results and S-matrix spectrum obtained by diagonalizing the Hamiltonian in the combined bilinear subspace
$\mathcal H_{\rm bil}({\bf p})$ defined in
Eq.$\,$\ref{eq:FCI_bilinear_subspace_full}, on the ground-states at $\bsl{\Gamma},\bsl{K}$ and $\bsl{K}'$, for $u=0.9$. (b) Comparison between the S-matrix results in (a) with the results obtained by diagonalizing the Hamiltonian in the individual subspaces obtained by acting the bilinears on the individual ground-states at $\bsl{\Gamma}$,$\bsl{K}'$ and $\bsl{K}$. (c,d) Same as (a,b), but in the no-dispersion limit, where we only consider the two-body term in Eq.$\,$\ref{eq:H_full_band_basis}, and take $u=1$.
 \label{fig:ED_vs_Smat_2o3_21site}}
\par\end{centering}
\end{figure}

Example results for a $N_{UC}=21$ lattice are shown in Fig.$\,$\ref{fig:ED_vs_Smat_2o3_21site}. This geometry supports three nearly degenerate FCI ground-states at $\bsl{k} = \bsl{\Gamma}, \bsl{K}, \bsl{K'}$, as illustrated in Fig.$\,$\ref{fig:Neutral-vs.-spin-flip_ED_var_Sizes}(b).
In Fig.$\,$\ref{fig:ED_vs_Smat_2o3_21site}(a), we compare the exact results with those from the S-matrix method using the subspace of bilinears applied to all the ground-states, for $u = 0.9$. The S-matrix method overestimates the magnon gap by 53\%. The origin of this sizable discrepancy is not yet fully understood and will be better explored in a future work. However, the dispersion of the lowest modes is approximately captured.
We also study the no-dispersion limit that was considered in the main text, where we neglect the energy band dispersion in Eq.$\,$\ref{eq:H_full_band_basis}, and take $u=1$.
In this limit, shown in Fig.$\,$\ref{fig:ED_vs_Smat_2o3_21site}(c), the overestimation of the magnon gap is reduced to 45\%, and the exact dispersion is also captured reasonably well. For completeness, in Table \ref{tab:overlaps_gaps_2o3} we provide the comparison of the S-matrix and exact results for the magnon gap and for the overlaps between the lowest-energy exact magnon eigenstates and the full S-matrix subspace obtained by acting with all the bilinears in the quasi-degenerate FCI ground-states. The lowest-energy exact magnon eigenstates therefore have a squared
projection of approximately $80\%$ onto the bilinear subspace, a reasonably large fraction, with
similar values for all three system sizes.

\begin{table}[h!]
\begin{centering}
\begin{tabular}{|c|c|c|c|c|}
\hline
\multirow{2}{*}{} & \multicolumn{3}{c|}{$\bra{\psi_{\textrm{exact}}({\bf k})}
P_{\textrm{bil}}({\bf k})
\ket{\psi_{\textrm{exact}}({\bf k})}$} & \multirow{2}{*}{$\Delta(\textrm{SM})/\Delta(\textrm{Exact})$}\tabularnewline
\cline{2-4}
 & ${\bf k}=\Gamma$ & ${\bf k}={\bf K}$ & ${\bf k}={\bf K}'$ & \tabularnewline
\hline
\hline
$N_{UC}=21$ & 0.81 & 0.78 & 0.78 & 1.45\tabularnewline
\hline
$N_{UC}=24$ & 0.79 & 0.76 & 0.76 & 1.46\tabularnewline
\hline
$N_{UC}=27$ & - & 0.77 & 0.77 & 1.45\tabularnewline
\hline
\end{tabular}
\par\end{centering}
\caption{Eigenstate overlaps and magnon gap ratios between the S-matrix calculations
and the exact results for the FH model of MoTe$_2$, at $\theta=3.7^{\circ}$,
and in the no-dispersion limit, where we only consider the two-body term in
Eq.$\,$\ref{eq:H_full_band_basis} and take $u=1$. The overlaps $\bra{\psi_{\textrm{exact}}({\bf k})}
P_{\textrm{bil}}({\bf k})
\ket{\psi_{\textrm{exact}}({\bf k})}$ are
computed for the lowest-energy exact magnon states
$\ket{\psi_{\textrm{exact}}({\bf k})}$, obtained with ED in the spin sector
$S_z=S_z^{\textrm{max}}-1$, that lie in the
${\bf k}=\Gamma,{\bf K},{\bf K}'$ sectors for $N_{UC}=21,24$, and in the
${\bf k}={\bf K},{\bf K}'$ sectors for $N_{UC}=27$ (see
Fig.$\,$\ref{fig:Neutral-vs.-spin-flip_ED_var_Sizes}).
Here, $P_{\textrm{bil}}({\bf k})$ denotes the projector onto the combined
bilinear subspace $\mathcal{H}_{\mathrm{bil}}({\bf k})$ defined in
Eq.$\,$\ref{eq:FCI_bilinear_subspace_full}.}
\label{tab:overlaps_gaps_2o3}
\end{table}

Figures$\,$\ref{fig:ED_vs_Smat_2o3_21site}(b,d) compare the full S-matrix spectrum with the spectra obtained by diagonalizing the Hamiltonian separately in the three subspaces $\mathcal{H}_i({\bf p})$. The low-energy part of the full S-matrix spectrum is approximately reproduced by collecting the states obtained from these three separate diagonalizations. For $N_{UC}=21$, the excitations constructed on top of the three FCI ground-states have their minima at different total momenta. This confirms the interpretation given above: the different momentum sectors of the parent FCI ground-states determine how the corresponding low-energy magnon states are distributed in momentum space. It also explains why the magnon dispersion differs substantially from the $N_{UC}=27$ case, where all three FCI ground-states occur at $\bsl{\Gamma}$.

\begin{figure*}[h!]
\centering{}\includegraphics[width=1\textwidth]{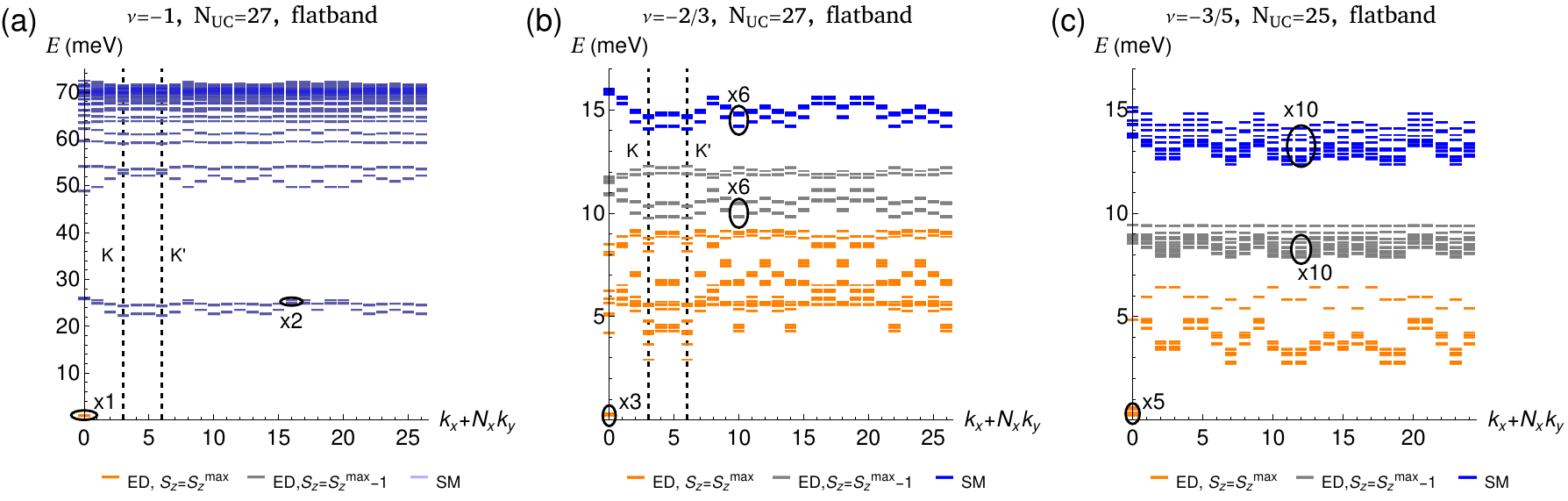}\caption{ Exact polarized (orange, $S_z=S_z^{\textrm{max}}$) and magnon (gray, $S_z=S_z^{\textrm{max}}-1$) spectra, together with S-matrix (SM) results (blue) for fillings $\nu=-1$ (a), $\nu=-2/3$ (b) and $\nu=-3/5$ (c), for the FH model of MoTe$_2$ with $\theta=3.7^{\circ}$. All the results were obtained in the no-dispersion limit - where we only consider the two-body term in Eq.$\,$\ref{eq:H_full_band_basis}, and take $u=1$ - and for the system sizes defined by
$(N_x,N_y,n_{11},n_{12},n_{21},n_{22})=(9,3,1,1,0,1)$ (a,b) and $(N_x,N_y,n_{11},n_{12},n_{21},n_{22})=(5,5,1,0,0,1)$ (c).
For $\nu=-2/3$ ($\nu=-3/5$), 12 (6) eigenstates were computed per momentum sector for $S_z=S_z^{\textrm{max}}$ and 8 (12) eigenstates were computed for $S_z=S_z^{\textrm{max}}-1$.
The exact and S-matrix results are exactly superimposed in (a), because S-matrix is exact at $\nu=-1$.
\label{fig:filling_1_vs_2o3_vs_3o5_flatband}}
\end{figure*}

We next investigate whether the same picture applies at another fractional filling. Figure$\,$\ref{fig:filling_1_vs_2o3_vs_3o5_flatband} compares the exact and S-matrix spectra at $\nu=-1$, $\nu=-2/3$, and $\nu=-3/5$ in the no-dispersion limit. At $\nu=-3/5$, the fully polarized sector contains five quasi-degenerate FCI ground-states at ${\bf k}=\bsl{\Gamma}$, as shown in Fig.$\,$\ref{fig:filling_1_vs_2o3_vs_3o5_flatband}(c). This FCI state was first reported for the FH model of MoTe$_2$ at $\theta=3.7^{\circ}$ in Ref.$\,$\cite{goncalves2025spinlessspinfulchargeexcitations}.

\begin{figure*}[h!]
\centering{}\includegraphics[width=1\textwidth]{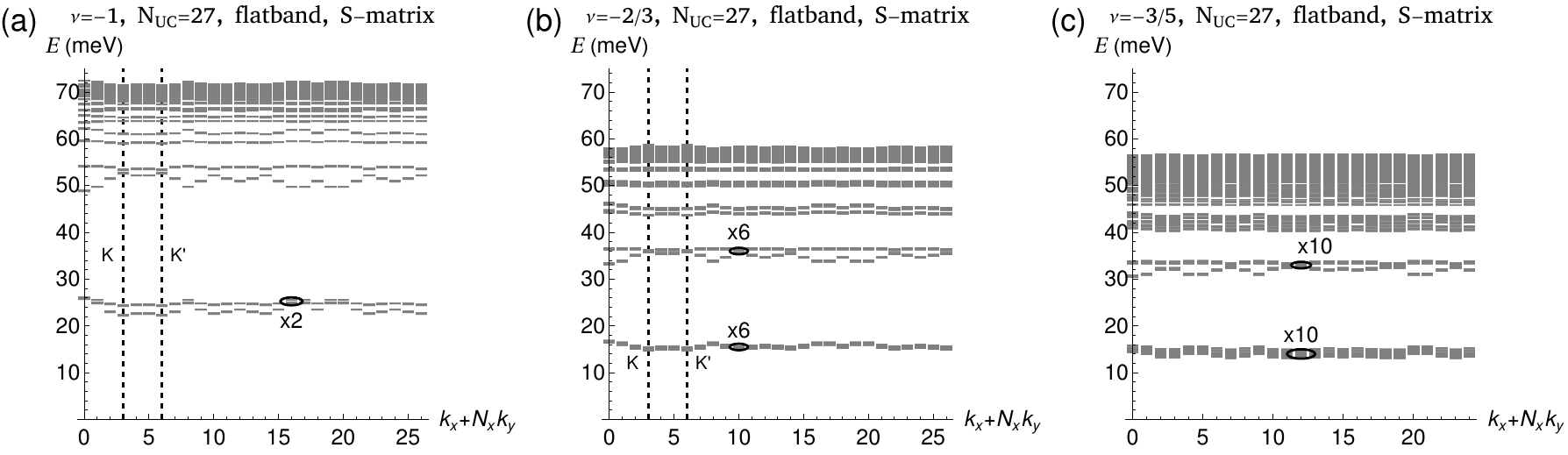}\caption{
Full S-matrix results for fillings $\nu=-1$ (a), $\nu=-2/3$ (b) and $\nu=-3/5$ (c), for the FH model of MoTe$_2$ with $\theta=3.7^{\circ}$, in the no-dispersion limit (where we only consider the two-body term in Eq.$\,$\ref{eq:H_full_band_basis}, and take $u=1$). The results are for the system sizes defined by
$(N_x,N_y,n_{11},n_{12},n_{21},n_{22})=(9,3,1,1,0,1)$ (a,b) and $(N_x,N_y,n_{11},n_{12},n_{21},n_{22})=(5,5,1,0,0,1)$ (c).
\label{fig:filling_1_vs_2o3_vs_3o5_flatband_Smatrixonly}}
\end{figure*}

Because the five FCI ground-states at $\nu=-3/5$ all have momentum $\bsl{\Gamma}$, spin-flip excitations constructed on top of them occur at the same total momenta. Consequently, each of the two lowest-energy magnon branches contains five nearly degenerate states, giving ten low-energy states in total. The S-matrix results for these states are shown together with the exact ED spectrum in Fig.$\,$\ref{fig:filling_1_vs_2o3_vs_3o5_flatband}. The complete S-matrix spectrum is shown separately in Fig.$\,$\ref{fig:filling_1_vs_2o3_vs_3o5_flatband_Smatrixonly}.

The momentum dependence of the ten lowest-energy states at $\nu=-3/5$ is reasonably reproduced by the S-matrix calculation. However, as at $\nu=-2/3$, the magnon gap is substantially overestimated - at $\nu=-3/5$, the error is $59\%$. In addition, the exact ED spectrum contains further excitations immediately above these ten low-energy states, with no clear energy gap separating them. These additional low-energy excitations are absent from the bilinear subspace. As a result, the S-matrix calculation incorrectly predicts that the ten low-energy states are well isolated from higher-energy states.

\begin{figure*}[h!]
\centering{}\includegraphics[width=1\textwidth]{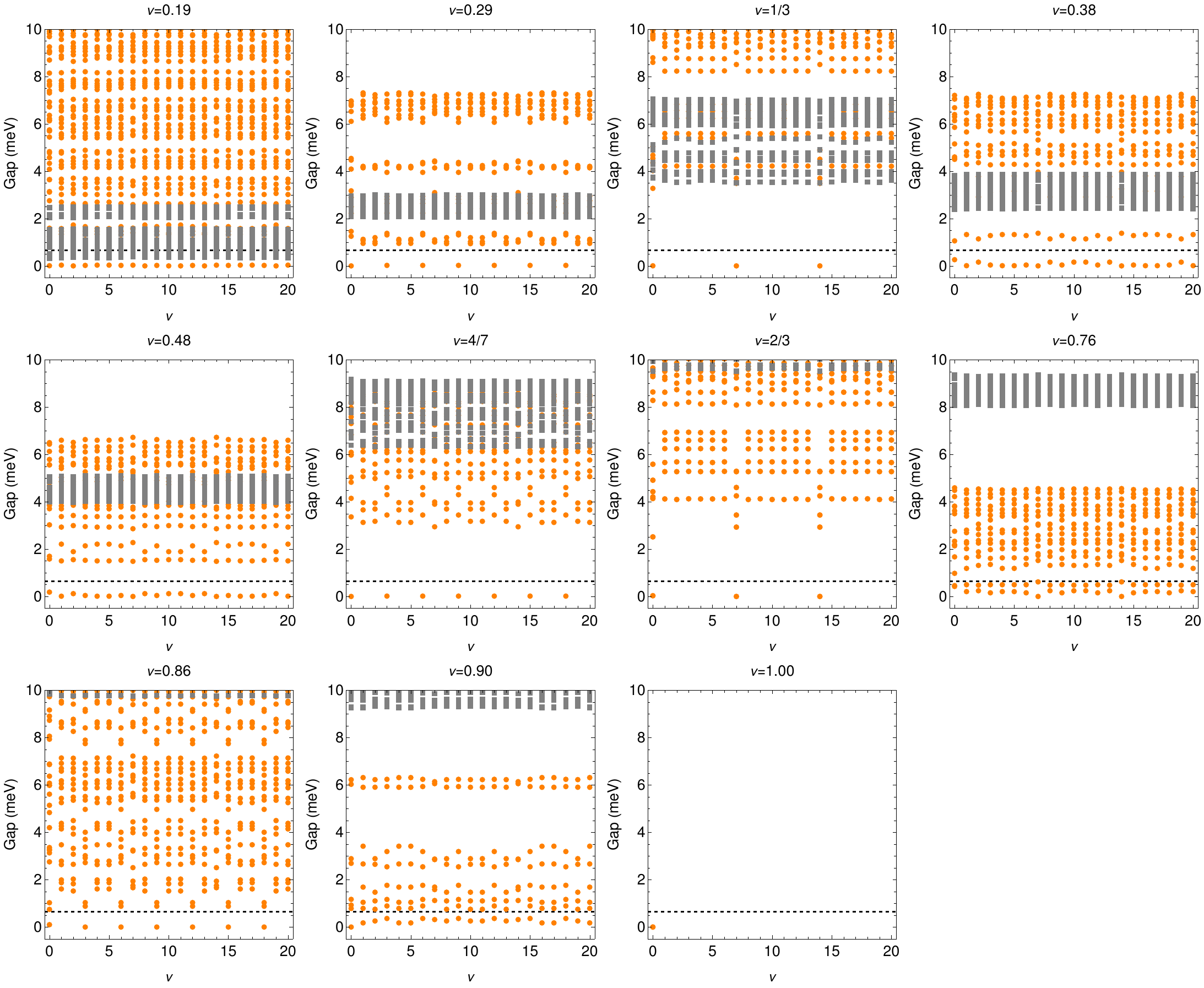}\caption{
Exact results for variable filling, for the FH model of MoTe$_2$ with $\theta=3.7^{\circ}$ in the no-dispersion limit (where we only consider the two-body term in Eq.$\,$\ref{eq:H_full_band_basis}, and take $u=1$) and for the system size defined by $(N_x,N_y,n_{11},n_{12},n_{21},n_{22})=(21,1,1,-5,0,1)$. The dashed line bounds the low-energy states selected to apply the bilinears $\gamma^\dag_{\bf{k}+\bf{p},\down} \gamma_{\bf{k},\up}$ in order to obtain the S-matrix subspace in which the Hamiltonian was diagonalized to obtain the results in Fig.$\,$\ref{fig:Fractional_Fillings}(b).
\label{fig:exact_vs_sm_variable_filling_21site_selection}}
\end{figure*}

Finally, we explain in detail how the S-matrix results at variable filling in Fig.$\,$\ref{fig:Fractional_Fillings}(b) were obtained. At generic non-commensurate fillings, the fully polarized spectrum does not exhibit an isolated quasi-degenerate ground-state multiplet on which to base the bilinear construction. We therefore select a set of low-energy states in the fully polarized sector and apply all the bilinears $\gamma_{{\bf k}+{\bf p},\down}^{\dagger}\gamma_{{\bf k},\up}$ to each selected state. The selected states are shown explicitly in Fig.$\,$\ref{fig:exact_vs_sm_variable_filling_21site_selection}, being those enclosed by the dashed line. We then diagonalize the Hamiltonian in the combined subspace generated by these states and obtain the S-matrix data shown in Fig.$\,$\ref{fig:Fractional_Fillings}(b).

\renewcommand{\addcontentsline}[3]{}
\bibliographystyle{apsrev4-2}
\bibliography{refs}
\let\addcontentsline\oldaddcontentsline

\renewcommand{\thetable}{S\arabic{table}}
\renewcommand{\thefigure}{S\arabic{figure}}
\renewcommand{\theequation}{S\arabic{section}.\arabic{equation}}
\onecolumngrid
\pagebreak
\thispagestyle{empty}

\cleardoublepage



\end{document}